\documentclass[a4paper,11pt]{article}
\pdfoutput=1 

\usepackage{jheppub} 

\usepackage[T1]{fontenc} 
\usepackage{rotating}
\usepackage{amsmath,amssymb,graphicx,slashed,xcolor,cancel}
\usepackage{array,booktabs,colortbl,multirow,tablefootnote}
\usepackage{colortbl,colordvi,xcolor,color}

\newcommand{\mhiggs}{\ensuremath{m_{\mathrm{H}}}\,}

\newcommand{\sqrts}{\sqrt{s}}

\def\lsim{\mathrel{\rlap{\lower3pt\hbox{\hskip0pt$\sim$}}
   \raise1pt\hbox{$<$}}}         
\def\gsim{\mathrel{\rlap{\lower4pt\hbox{\hskip1pt$\sim$}}
   \raise1pt\hbox{$>$}}}         
\newcommand{\luv}{\Lambda_{UV}}

\newcommand{\crowcolorA}{\rowcolor[rgb]{0.9,0.9,0.9}}

\newcommand{\lrD}{~\!\overset{\leftrightarrow}{\hspace{-0.1cm}D}\!}
\newcommand{\lrDa}{~\!\overset{\leftrightarrow}{\hspace{-0.1cm}D}\!^{\!~a}}
\newcommand{\hc}{\mathrm{h.c.}}

\newcommand{\SM}{\mathrm{SM}}
\newcommand{\vtext}[1]{\begin{sideways}\small{#1}\end{sideways}}
\newcommand{\mt}[1]{\mathrm{#1}}

\title{\boldmath Higgs Boson studies at future particle colliders}

\author[1,2]{J.~de~Blas}
\author[3]{M.~Cepeda}
\author[4]{J.~D'Hondt}
\author[5]{R.~K.~Ellis}
\author[6,7]{C.~Grojean}
\author[6,8]{B.~Heinemann}
\author[9,10]{F.~Maltoni}
\author[11]{A.~Nisati, ,\note{Corresponding author.}}
\author[12]{E.~Petit}
\author[13]{R.~Rattazzi}
\author[14]{W.~Verkerke}
\affiliation[1]{Dipartimento di Fisica e Astronomia Galileo Galilei, Universita di Padova, Via Marzolo 8, I-35131 Padova, Italy}
\affiliation[2]{Istituto Nazionale di Fisica Nucleare (INFN), Sezione di Padova,Via Marzolo 8, I-35131 Padova, Italy }
\affiliation[3]{Centro de Investigaciones Energ\'eticas, Medioambientales y Tecnol\'ogicas (CIEMAT), Avda. Complutense 40, 28040, Madrid, Spain}
\affiliation[4]{Inter-University Institute for High Energies (IIHE), Vrije Universiteit Brussel, Brussels, 1050, Belgium}
\affiliation[5]{IPPP, University of Durham, Durham DH1 3LE, UK}
\affiliation[6]{Deutsches Elektronen-Synchrotron (DESY), Hamburg, 22607, Germany}
\affiliation[7]{Institut f\"ur Physik, Humboldt-Universit\"at,  Berlin, 12489, Germany}
\affiliation[8]{Albert-Ludwigs-Universit\"at Freiburg, Freiburg, 79104, Germany}
\affiliation[9]{Centre for Cosmology, Particle Physics and Phenomenology, Université catholique de Louvain, Louvain-la-Neuve,1348, Belgium}
\affiliation[10]{Dipartimento di Fisica e Astronomia, Universit\`a di Bologna and INFN, Sezione di Bologna, via Irnerio 46, 40126 Bologna, Italy}
\affiliation[11]{Istituto Nazionale di Fisica Nucleare (INFN), Sezione di Roma, P.le A. Moro 2, I-00185 Roma, Italy}
\affiliation[12]{Aix Marseille Univ, CNRS/IN2P3, CPPM, Marseille, France}
\affiliation[13]{Theoretical Particle Physics Laboratory (LPTP),EPFL, Lausanne, Switzerland}
\affiliation[14]{Nikhef and University of Amsterdam, Science Park 105, 1098XG Amsterdam, the Netherlands}

\emailAdd{nisati@cern.ch}

\abstract{
This document aims to provide an assessment of the potential of future colliding beam facilities to perform
Higgs boson studies. The analysis builds on the submissions made by the proponents of future colliders
to the European Strategy Update process, and takes as its point of departure the results expected at the completion of the HL-LHC program. This report presents quantitative results on many aspects of Higgs physics for future collider projects of sufficient maturity using uniform methodologies. 
}

\begin{document} 
\maketitle
\flushbottom

\section{Introduction} \label{introduction}
This article presents the results of the Standard Model (SM) Higgs boson studies performed by the \textit{Higgs@FutureColliders} group based on the input submitted to the Update of the European Strategy by the various proponents of new high-luminosity energy-frontier particle accelerator projects beyond the \textit{High Luminosity LHC} (HL-LHC). This report fulfils part of the mandate given to this group by the restricted ECFA (REFCA) committee, see Appendix~\ref{app:mandate}.  
The exploration of the Higgs boson through direct searches and precision measurements at future colliders is among the most important aspects of their scientific programmes.

The colliders considered for this document are High-Energy LHC (HE-LHC), Future Circular Colliders (FCC-{ee,eh,hh})~\cite{Abada:2019lih}, the Circular Electron-Positron Collider (CEPC)~\cite{CEPCStudyGroup:2018ghi}, the International Linear Collider (ILC)~\cite{Bambade:2019fyw,Fujii:2019zll}, the Compact Linear Collider (CLIC)\cite{Charles:2018vfv}, and the Large Hadron electron Collider~\cite{AbelleiraFernandez:2012cc} (LHeC or HE-LHeC~\footnote{For HE-LHeC no analysis was performed here, but it is expected that the relative improvements w.r.t. LHeC are expected to be similar as from HL-LHC to HE-LHC} ).  
The physics results that are expected by the completion of HL-LHC
are assumed to represent the scenario from where these future colliders would start. Furthermore, a muon collider is also briefly illustrated, but given the less advanced stage, it is not part of the default analyses performed. The potential of a $\gamma\gamma$ collider (based on an $e^+e^-$ collider and laser beams) for Higgs boson physics has been studied a while ago~\cite{Asner:2001ia,Bogacz:2012fs} and more recently again in context of plasma-wakefield driven accelerators~\cite{ALEGRO:2019alc}. Plasma-wakefield driven accelerators also offer promise to provide multi-TeV $e^+e^-$ colliders (e.g.~\cite{ALEGRO:2019alc}) and are addressed briefly later in this report. 

\begin{figure}[tbh]
\centering
\includegraphics[width=\linewidth]{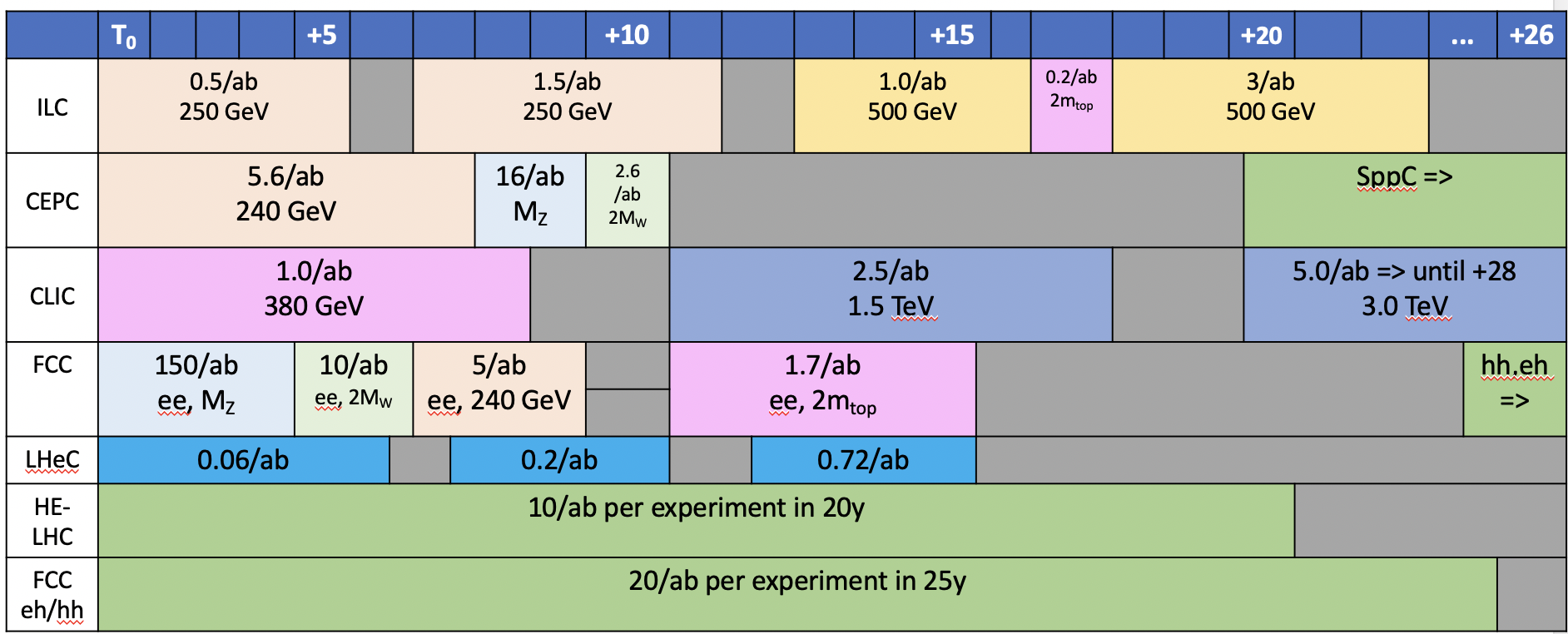}
\caption{\label{fig:timeline}
Time line of various collider projects starting at time $T_0$ as submitted to the European Strategy Update process. Some possible extensions beyond these baseline run plans have been discussed and are presented in more detail in Appendix~\ref{app:addons}.
For the clarification of the meaning of a year of running, see the caption of Table~\ref{tab:colliders}.
Figure~\ref{fig:timelineabs} in Appendix~\ref{app:inputs} shows an alternative version of this figure using the earliest possible start date (i.e. the calendar date of $T_0$) given by the proponents.
}
\end{figure}

A table of the colliders and their parameters (type, $\sqrt{s}$, polarisation $\cal{P}$, integrated luminosity $\cal{L}$, the run time) is given in Table~\ref{tab:colliders}. A graphical display of the time line and luminosity values is shown in Fig.~\ref{fig:timeline}. The parameters used are taken from the references also given in that table. For the purpose of this study, only inputs as provided by the various collaborations are used, and there is no attempt to make any judgement on the validity of the assumptions made in estimating the projected measurement uncertainties (see also mandate in Appendix~\ref{app:mandate}). 
In addition to the collider runs shown in Table~\ref{tab:colliders}, a few other scnearios are considered such as FCC-hh with $\sqrt{s}=37.5$~TeV~\cite{lowefcc} and ${\cal L}=15$~fb$^{-1}$, FCC-ee with 4 instead of 2 IPs (doubling the total integrated luminosity), and CLIC and ILC with a dedicated running period of 1-3 years to collect ${\cal L}=100$~fb$^{-1}$ at $\sqrt{s}\approx M_Z$~\cite{Fujii:2019zll,gigazclic}. These are discussed in Appendix~\ref{app:addons}. 

\begin{sidewaystable}
\caption{Summary of the future colliders considered in this report. The number of detectors given is the number of detectors running concurrently, and only counting those relevant to the entire Higgs physics programme. The instantaneous and integrated luminosities provided are those used in the individual reports, and for $e^+e^-$ colliders the integrated luminosity corresponds to the sum of those recorded by the detectors. For HL-LHC this is also the case while for HE-LHC and FCC$_{hh}$ it corresponds to 75\% of that. The values for $\sqrt{s}$ are approximate, e.g. when a scan is proposed as part of the programme  this is included in the closest value (most relevant for the $Z$, $W$ and $t$ programme). For the polarisation, the values given correspond to the electron and positron beam, respectively. For HL-LHC, HE-LHC, FCC, CLIC and LHeC the instantaneous and integrated luminosity values are taken from Ref.~\cite{Bordry:2018gri}. For these colliders the number of seconds per year is $1.2\times 10^{7}$ based on CERN experience~\cite{Bordry:2018gri}. CEPC (ILC) assumes $1.3\times 10^{7}$ ($1.6\times 10^{7}$) seconds for the annual integrated luminosity calculation. When two values for the instantaneous luminosity are given these are before and after a luminosity upgrade planned.
The last column gives the abbreviation used in this report in the following sections. When the entire programme is discussed, the highest energy value label is used, e.g. ILC$_{1000}$ or CLIC$_{3000}$. It is always inclusive, i.e. includes the results of the lower-energy versions of that collider. Also given are the shutdowns (SDs) needed between energy stages of the machine. SDs planned during a run at a given energy are included in the respective energy line.(*) For FCC-hh a value of $\sqrt{s}=37.5$~TeV is also considered, see App.~\ref{app:addons}. Additional scenarios where ILC/CLIC accumulate $100$~fb$^{-1}$ on the $Z$-pole, and where FCC-ee has 4 IPs are also discussed in Appendix~\ref{app:addons}.
\label{tab:colliders}}
\begin{center}
\small
\begin{tabular}{|l|ccccc|cc|cc|} 
\toprule
\rule{0pt}{1.0em}%
Collider & Type & $\sqrt{s}$ & $\cal{P}$ [\%] & N(Det.)& $\cal{L}_{\rm inst}$ [$10^{34}$]& $\cal{L}$ & Time  & Refs. & Abbreviation\\
& &  & [$e^-$/$e^+$] & & cm$^{-2}$s$^{-1}$ & [ab$^{-1}$] & [years] & & \\\hline
\rule{0pt}{1.0em}%
HL-LHC & $pp$ & 14\,TeV & - & 2 & 5 & 6.0 & 12 & \cite{Cepeda:2019klc} & HL-LHC\\\hline
HE-LHC & $pp$ & 27\,TeV & - & 2 & 16 & 15.0 & 20 & \cite{Cepeda:2019klc} & HE-LHC\\\hline
FCC-hh$^{(*)}$ & $pp$ & 100\,TeV & - & 2 & 30 & 30.0 & 25 & \cite{Abada:2019lih} & FCC-hh\\\hline
FCC-ee & $ee$ & $M_Z$ & 0/0 & 2 & 100/200 & 150 & 4 & \cite{Abada:2019lih} & \\
  & & $2M_W$ & 0/0 & 2 & 25 & 10 & 1-2 &  & \\
  & & $240$\,GeV & 0/0 & 2 & 7 & 5 & 3 &  & FCC-ee$_{240}$ \\
  & & $2m_{top}$ & 0/0 & 2 & 0.8/1.4 & 1.5 & 5 &  & FCC-ee$_{365}$ \\
    & & & & & & & (+1) & \multicolumn{2}{c|}{(1y SD before $2m_\textrm{top}$ run)}\\\hline
  ILC & $ee$ & $250$~GeV & $\pm 80$/$\pm 30$ & 1 & 1.35/2.7 & 2.0 & 11.5 & \cite{Fujii:2017vwa,Bambade:2019fyw} & ILC$_{250}$\\
  & & $350$~GeV & $\pm 80$/$\pm 30$ & 1 & 1.6 & 0.2 & 1 & & ILC$_{350}$\\
  & & $500$~GeV & $\pm 80$/$\pm 30$ & 1 & 1.8/3.6 & 4.0 & 8.5 & & ILC$_{500}$\\
  & & & & & & & (+1) & \multicolumn{2}{c|}{(1y SD after 250 GeV run)} \\
  & & $1000$~GeV & $\pm 80$/$\pm 20$ & 1 & 3.6/7.2 & 8.0 & 8.5 & \cite{Fujii:2019zll} & ILC$_{1000}$\\
  & & & & & & & (+1-2) & \multicolumn{2}{c|}{(1-2y SD after 500 GeV run)} \\\hline
   CEPC & $ee$ & $M_Z$ & 0/0 & 2 & 17/32 & 16 & 2 & \cite{CEPCStudyGroup:2018ghi} & CEPC\\
  & & $2M_W$ & 0/0 & 2 & 10 & 2.6 & 1 &  & \\
  & & $240$~GeV & 0/0 & 2 & 3 & 5.6 & 7 &  & \\\hline
    CLIC & $ee$ & $380$~GeV & $\pm 80$/$0$ & 1 & 1.5 & 1.0 & 8 & \cite{Roloff:2018dqu} & CLIC$_{380}$\\
  & & $1.5$~TeV & $\pm 80$/$0$ & 1 & 3.7 & 2.5 & 7 & & CLIC$_{1500}$\\
  & & $3.0$~TeV & $\pm 80$/$0$ & 1 & 6.0 & 5.0 & 8 & & CLIC$_{3000}$ \\
  & & & & & & & (+4) & \multicolumn{2}{c|}{(2y SDs between energy stages)} \\\hline
   LHeC & $ep$ & 1.3\,TeV & - & 1 & 0.8 & 1.0 & 15 & \cite{Bordry:2018gri} & LHeC \\\hline
   HE-LHeC & $ep$ & 1.8 \,TeV & - & 1 & 1.5 & 2.0 & 20 &  \cite{Abada:2019lih} & HE-LHeC \\\hline
   FCC-eh & $ep$ & 3.5\,TeV & - & 1 & 1.5 &  2.0 & 25 & \cite{Abada:2019lih} & FCC-eh\\
\bottomrule
\end{tabular}
\end{center}
\end{sidewaystable}

For the following sections the tables and plots are labelled using the acronyms given in Table~\ref{tab:colliders}. The energy subscript indicates the highest energy stage of the given collider, and the results always assume that it is combined with results from the lower energy stages.

At the heart of the Higgs physics programme is the question of how the Higgs boson couples to Standard Model elementary particles. 
Within the SM itself, all these couplings are uniquely determined. But new physics beyond the SM (BSM) can modify these couplings in many different ways. The structure of these deformations is in general model-dependent. One important goal of the Higgs programme at the future colliders is to identify, or at least constrain, these deformations primarily from the measurements the Higgs production cross section, $\sigma$, times decay branching ratio, BR\footnote{The Higgs couplings could be constrained less directly from processes with no Higgs in the final state or without even a non-resonant Higgs. But the main focus of the study presented in this report will be on the information obtained from the measured $\sigma \times {\rm BR}$. Still, note that, at lepton colliders, the $ZH$ associated production can be measured without the decay of the Higgs.}. 
Ultimately, these studies will be used to assess the fundamental parameters of the new physics models. For the time being, in the absence of knowledge of new physics, 
we need to rely on a parametrisation of our ignorance in terms of continuous deformations of the Higgs boson couplings. Different assumptions allow to capture different classes of new physics dynamics. First, in the so-called $\kappa$-framework~\cite{LHCHiggsCrossSectionWorkingGroup:2012nn,LHCHXSWG3}, often used to interpret the LHC measurements, the Higgs couplings to the SM particles are assumed to keep the same helicity structures as in the SM. While it offers a convenient exploration tool that does not require other computations than the SM ones and still captures the dominant effects of well motivated new physics scenarios on a set of on-shell Higgs observables, the $\kappa$-framework suffers from some limitations that will be discussed later and it includes some biases that will prevent to put the Higgs programme in perspective with other measurements, see e.g. the discussion in Ref.~\cite{Barklow:2017suo} and at the beginning of Section~\ref{couplings}. An alternative approach, based on \textit{Effective Field Theory} (EFT), considers new Higgs couplings with different helicity structures, with different energy dependence or with different number of particles. They are not present in the SM but they can potentially be generated by new heavy degrees of freedom. 

Furthermore, the sensitivity of the data to the Higgs self-coupling is analysed based on single-Higgs and di-Higgs production measurements by future colliders. Due to lack of access to the simulated data of the collaborations, in particular differential kinematical distributions, it is not possible in this case to perform a study for the Higgs self-coupling with similar rigour as the analysis of the single-Higgs-coupling described in the previous paragraph.

The Higgs width determination is also discussed as is the possible decay of the Higgs bosons into new particles that are either "invisible" (i.e. observed through missing energy - or missing transverse energy) or "untagged", to which none of the Higgs analyses considered in the study are sensitive. Rare decays and CP aspects are also discussed.

All colliders have provided extensive documentation on their Higgs physics programme. However, sometimes different choices are made e.g. on which parameters to fit for and which to fix, what theoretical uncertainties to assume, which operators to consider in e.g. the EFT approach. This would lead to an unfair comparison of prospects from different future colliders, with consequent confusing scientific information. In this report, we aim to have a single clear and reasonable approach to the assumptions made when comparing the projections for the future. 

In general, one should not over-interpret 20\% differences between projected sensitivities for partial widths of different future projects. In many cases, these are likely not significant. For instance, CEPC and FCC-ee at $\sqrt{s}=240$~GeV expect to acquire a very similar luminosity and should obtain very similar results if both use two optimized detectors and analyses. Differences between the projected sensitivities, when considering only results from the $\sqrt{s}=240$~GeV run, originate likely from different choices made in the analyses at this stage or, in some cases, because an analysis has not yet been performed. For the EFT analysis the measurements at different $\sqrt{s}$ values play an important role, and this results in significant differences as CEPC and FCC-ee have proposed different integrated luminosities at the different $\sqrt{s}$ values and CEPC lacks a run at $\sqrt{s}=365$~GeV. It is also useful to keep in mind that the target luminosity values have some uncertainties, and historically colliders have sometimes exceeded them (e.g. LHC by a factor of about two in instantaneous luminosity) and sometimes fallen short.

In this document only inclusive cross section times branching ratio measurements, and in some cases ratios of inclusive measurements, are used. It is well known that probing the Higgs boson at high $p_T$ enhances the sensitivity to new physics and the analysis presented here does not capture this. As a result of this, the true power of high-energy colliders (where $\sqrt{\hat{s}}\gg m_H$) for probing Higgs physics is underestimated. 

This document is organised as follows. Section~\ref{sec:method} discusses the methodology, including the systematic uncertainties on the theoretical calculations which are common to all colliders. Section~\ref{couplings} presents the study made on the Higgs boson couplings to SM elementary particles.
The results found in the context of the \textit{$\kappa$}-framework (briefly summarised in Section~\ref{kappa}) are presented in Section~\ref{kappa-results}. 
Likewise, the results from the EFT fits (summarised in Section~\ref{eft}) are collected in Section \ref{eft-results}. The impact of theory uncertainties on the Higgs projections is discussed in Section~\ref{sec:SMThUnc}.
Particular attention is dedicated to the Higgs self-coupling in Section~\ref{selfcoupling} and the Higgs boson rare decays, in Section~\ref{raredecays}.
The prospects for measurements of Higgs boson CP properties are given in Section~\ref{higgscp}, and  
the prospects for precision measurements of the mass and width are summarized in Section~\ref{mass}.

Section~\ref{futureprospects} presents future studies that would be important to deepen to get a more complete view of the Higgs physics potential at future colliders. The Muon Collider (Section~\ref{muoncollider}) and plasma-wakefield accelerators (Section~\ref{sec:MultiTeVee}) are discussed first, and then phenomenological studies that relate the precision measurements to new physics models are discussed (Section~\ref{futurestudies}). 

In the appendix, all theoretical and experimental input parameters related to the Higgs observables are provided, and some results that seemed too detailed for the main body, are also shown.

\section{Methodology}\label{sec:method}

The various colliders measure values for the cross sections  times branching ratios, $\sigma\times$ BR. At hadron colliders the main processes are gluon-gluon-fusion (ggF), vector boson fusion (VBF), Higgs-strahlung (VH, where $V=Z,W$) and $t\bar{t}H$ production. At lepton colliders, Higgs-strahlung ($ZH$) dominates at low values of $\sqrt{s}$, while at high $\sqrt{s}$ values the VBF process becomes dominant. At lepton-hadron colliders, the Higgs boson is dominantly produced via $WW$ or $ZZ$ fusion in the $t$-channel.

The extraction of the couplings of the Higgs boson relies on a simultaneous fit of all the projected measurements of $\sigma\times$BR, and their comparison to the SM values. As such, it is sensitive to both the experimental uncertainties as well as theoretical uncertainties on the production cross sections and branching ratios.

At the HL-LHC, these theoretical uncertainties are taken from the S2 scenario of the HL-LHC~\cite{Cepeda:2019klc}, which assumes that the current uncertainties can be reduced by a factor of two by the end of the HL-LHC running in twenty years.~\footnote{Apart from improved theoretical calculations, part of this reduction would require a more precise knowledge of PDFs and $\alpha_S$, which could be possible with an $ep$ machine such as the LHeC.}
For the studies at future lepton colliders we use the future projections for the theory uncertainties on the partial width values given in Table~\ref{tab:widths}. 
At the FCC-hh a 1\% total uncertainty is assumed, combined for the luminosity and cross section normalisation~\cite{Abada:2019lih}. It is expected that this 1\% is dominated by the luminosity uncertainty and that theory uncertainties will be negligible in comparison, also thanks to the PDF uncertainty reduction from FCC-eh and/or LHeC. For HL-LHC and HE-LHC a luminosity uncertainty of 1\% is assumed.
For LHeC the theoretical uncertainties on the charged-current and neutral-current production processes are taken to be 0.5\%~\cite{lhecdoc}.
For the decays the uncertainties as given in Table~\ref{tab:widths} are used. 

Some caution must be taken when studying the HE-LHC results provided here. They are derived from the same inputs as the HL-LHC ones evolved with integrated luminosity and increased cross section.
This is a  simplified approach, and all the HE-LHC results are thus approximations. 
As in Ref.\cite{Cepeda:2019klc}, we consider 2 scenarios: one 
where we use the same S2 assumptions as for the HL-LHC; and a second
scenario, denoted S2$^\prime$, which assumes
a further reduction in the signal theoretical systematic uncertainties by another factor of two with respect to the S2 scenario at the HL-LHC, i.e. roughly four times smaller than current studies from Run 2.
It must be noted that such reduction of the uncertainty is not motivated on solid theoretical grounds and it is simply a hypothesis, based on the reasoning that the time available to make progress is significantly longer for HE-LHC than HL-LHC. 
When combined with the HL-LHC, the theory systematics are assumed to be fully correlated between HL-LHC and HE-LHC, using the same uncertainties, S2 or S2$^\prime$, for both colliders.

For the purpose of the analyses presented here, it is assumed that all observables of relevance have the SM value and there are no new physics effects present in the simulated data. If new physics effects are observed e.g. in the data from the 2nd or 3rd LHC run, either in the Higgs sector or otherwise, the analysis method and assumptions made will likely change significantly.

The combination of any future data with HL-LHC results is done assuming no correlations between the colliders, except for those between HL-LHC and HE-LHC which are treated as discussed above. 

In the input HL-LHC predictions it is assumed that the intrinsic theory systematic uncertainties for the various production modes are uncorrelated. A consequence of this assumption is that, when Higgs model parameters are constructed that represent (directly or indirectly) an average over independent measurements with independent theory uncertainties that are all interpreted to measure the same physics quantity (e.g. the global signal strength $\mu$, or the EFT parameter $c_{\phi}$ in Eq.~(\ref{eq:LSilh}) below), such averages can have smaller (theory) uncertainties than the component measurements. This reduced average uncertainty occurs by virtue of the choices:
\begin{itemize}
    \item to consider the input systematic uncertainties to be independent;
    \item to interpret different classes of measurement to measure the same physics. 
\end{itemize}
The impact of the choice of correlation between theory systematic uncertainties should be carefully investigated in the future, but was not possible with the set of inputs provided for the preparation of this document. Where the effect of theory error averaging was observed to be prominent in the presented results, e.g. in Section \ref{eft-results}, it is mentioned. 

Electroweak precision observables also contribute significantly in the EFT-based analysis. At present, LEP still provides the best constraints in many cases, and these are used here, except when new higher precision measurements are expected to be made by the given collider. For instance, for the HL-LHC and HE-LHC projections, LEP values are used for the constraints on electroweak precision observables, whilst all FCC machines use the values expected from FCC-ee. 

The fits presented in this report have been produced using the fitting framework of the {\tt HEPfit} package~\cite{hepfit}, a general tool to combine information from direct and indirect searches and test the Standard Model and its extensions~\cite{deBlas:2016ojx,deBlas:2016nqo,deBlas:2017wmn,deBlas:2018tjm}. We use the Markov-Chain Monte-Carlo implementation provided by the Bayesian Analysis Toolkit~\cite{Beaujean:2015bwl}, to perform a Bayesian statistical analysis of the sensitivity to deformations from the SM at the different future collider projects. The experimental projections for the different observables included in the fits are implemented in the likelihood assuming Gaussian distributions, with SM central values and standard deviations given by the corresponding projected uncertainties. Model parameters are assumed to have flat priors. Finally, theory uncertainties, when included, are introduced via nuisance parameters with Gaussian priors. 

The projected uncertainties of all measurements of observables, relevant to the various analyses presented in this article, are listed in Appendix~\ref{app:inputs}. 


\section{The Higgs boson couplings to fermions and vector bosons} \label{couplings}

Within the SM, all the Higgs couplings are uniquely fixed in terms of the Fermi constant and the masses of the different particles.  Measuring the Higgs couplings thus requires a parametrization of the deviations from the SM induced by new physics. The $\kappa$-framework is the simplest parametrization directly related to experimental measurements of the Higgs boson production and decay modes. For this reason, it has been widely used by the community. It only compares the experimental measurements to their best SM predictions and does not require any new BSM computations {\it per se}. From a more theoretical perspective, its relevance arises from the fact that it actually fully captures the leading effects in single Higgs processes of well motivated scenarios. For instance, in the minimal supersymmetric standard model with R-parity, all dominant corrections to the Higgs couplings induced by the new super-partners are of order $m_H^2/m_{\rm SUSY}^2$ relative to the SM predictions ($m_{\rm SUSY}$ is the mass of the new particles) and they appear as shifts of the Higgs couplings with the \textit{same} SM helicity structures while new helicity structures are only generated as subleading effects further suppressed by a loop factor. In scenarios where the Higgs boson arises from a strongly-interacting sector as a composite (pseudo-Goldstone) boson, the  leading deformations to the SM scale like $\xi= (g_\star^2/g_{\rm SM}^2)\, m_W^2/m_\star^2$ ($m_\star$ and $g_\star$ are the overall mass scale of the strong sector resonances and  their mutual coupling respectively) and they all preserve the helicity structure of the interactions already present in the SM.  The constraints derived in the $\kappa$ analysis can thus be readily exploited to derive constraints on the new physics parameters. This kappa-framework has, however, its own limitations when Higgs measurements need to be put in perspective and compared to processes with different particle multiplicities or combined with other measurements done in different sectors or at different energies. An effective field theory (EFT) approach naturally extends the kappa-framework. First, it allows to exploit polarisation- and angular-dependent observables that a $\kappa$-analysis will remain blind to. Second, an EFT analysis constitutes a useful tool to probe the Higgs boson in the extreme kinematical regions relevant for colliders operating far above the weak scale, exploring the tails of kinematical distributions, even though these observables have not been fully exploited yet in the studies presented by the different future collider collaborations.  Third, the EFT offers a consistent setup where predictions can be systematically improved via the inclusion of both higher loop corrections in the SM couplings and further new physics corrections encoded in operators of even higher dimensions. 

Both approaches will be studied in this document and we will report the fits to the experimental projected measurements obtained in these two frameworks. As an illustration, a concrete interpretation of the results obtained will be done in the context of composite Higgs models.


\subsection{The kappa framework} \label{kappa}

\subsubsection{Choice of parametrization} The kappa framework, described in detail in Ref.~ \cite{LHCHiggsCrossSectionWorkingGroup:2012nn, LHCHXSWG3}, facilitates the characterisation of Higgs coupling properties in terms of a series of Higgs coupling strength modifier parameters $\kappa_{i}$, which are defined as the ratios of the couplings of the Higgs bosons to particles $i$ to their corresponding Standard Model values.
The kappa framework assumes a single narrow resonance so that the zero-width approximation can be used to decompose the cross section as follows
\begin{equation}
( \sigma \cdot \text{BR} ) ( i \to \text{H }\to f ) =  \frac{ \sigma_{i} \cdot \Gamma_{f}}{\Gamma_H},
\label{eq:kappa_zw}
\end{equation}
\noindent where $\sigma_{i}$ is the production cross section through the initial state $i$, $\Gamma_{f}$ the partial decay width into the final state $f$ and $\Gamma_H$ the total width of the Higgs boson. The $\kappa$ parameters are introduced by expressing each of the components of Eq. (\ref{eq:kappa_zw}) as their SM expectation multiplied by the square of a coupling strength modifier for the corresponding process at leading order: 

\begin{equation}
( \sigma \cdot \text{BR} ) ( i \to \text{H }\to f ) =  \frac{ \sigma_{i}^{SM} \kappa_i^2 \cdot \Gamma_{f}^{SM}\kappa_f^2 }{\Gamma_H^{SM} \kappa_H^2} ~~~\to~~~ 
\mu_i^f \equiv \frac{\sigma \cdot {\rm BR}}{\sigma_{\rm SM}\cdot{\rm BR_{\rm SM}}}
    = \frac{\kappa_i^2\cdot\kappa_f^2}{\kappa_H^2} \quad ,
\end{equation}
\noindent where $\mu_i^f$ is the rate relative to the SM expectation  (as given in Tables~\ref{tab:xsecs} and~\ref{tab:widths}) and $\kappa_H^2$ is an expression that adjusts the SM Higgs width to take into account of modifications $\kappa_i$ of the SM Higgs coupling strengths:

\begin{equation}
\kappa_{H}^2 \equiv \sum\limits_{j} \frac{\kappa_j^2 \Gamma_{j}^\text{SM}}
{\Gamma_{H}^\text{SM}}\,.
\label{eq:CH2_def}
\end{equation}

 \noindent When all $\kappa_i$ are set to 1, the SM is reproduced. For loop-induced processes, e.g.  $H \to \gamma \gamma$, there is a choice of either resolving the coupling strength modification in its SM expectation, i.e. $\kappa_{\gamma}(\kappa_t,\kappa_W)$ or keeping $\kappa_{\gamma}$ as an effective coupling strength parameter. 
 
\noindent For the results presented in the document, we choose to describe loop-induced couplings with effective couplings, resulting in a total of 10 $\kappa$ parameters: $\kappa_W$, $\kappa_Z$, $\kappa_c$, $\kappa_b$, $\kappa_t$, $\kappa_\tau$, $\kappa_\mu$, and the effective coupling modifiers  $\kappa_\gamma$, $\kappa_g$ and $\kappa_{Z\gamma}$. The couplings $\kappa_s, \kappa_d, \kappa_u$ and $\kappa_e$ that are only weakly constrained from very rare decays are not included in the combined $\kappa$-framework fits presented in this section, their estimated limits are discussed separately in Section \ref{raredecays}.
We note the parameter $\kappa_t$ is only accessible above the $tH$ threshold as the processes involving virtual top quarks are all described with effective coupling modifiers ($\kappa_g, \kappa_{Z\gamma}, \kappa_{\gamma}$),
hence standalone fits to low-energy (lepton) colliders have no sensitivity to $\kappa_t$ in the $\kappa$-framework fits considered here.\footnote{ At high Higgs/jet $p_T$, $gg\to H$ becomes directly sensitive to $\kappa_t$. However, high-$p_T$ regions
are not separately considered in the $\kappa$-framework fits reported here. Furthermore, there is no sensitivity to the sign of the $\kappa$ parameters as the loop-induced processes with sensitivity to the sign have all been described with effective modifiers. Single top production is sensitive to the sign but not used in the $\kappa$ fits presented here (but used in the CP studies). Finally, note that, for vector-boson-fusion, the small interference effect between W- and Z boson fusion is neglected.
 }

\subsubsection{Modeling of invisible and untagged Higgs decays} 

The $\kappa$-framework can be extended to allow for the possibility of Higgs boson decays to invisible or untagged BSM particles. The existence of such decays increases the total width $\Gamma_H$ by a factor $1/(1-BR_{BSM})$, where $BR_{BSM}$ is the Higgs branching fraction to such BSM particles. Higgs boson decays to BSM particles can be separated in two classes: decays into invisible particles, which are experimentally directly constrained at all future colliders (e.g $ZH, H \to \text{invisible}$), and decays into all other 'untagged' particles. 

Reflecting this distinction we introduce two branching fraction parameters $BR_{inv}$ and $BR_{unt}$ so that:
\begin{equation}
\Gamma_H = \frac{ \Gamma_{H}^\text{SM} \cdot \kappa_H^2 }{ 1-(BR_{inv}+BR_{unt}) }\,,
\label{eq:width}
\end{equation}
where $\kappa_H^2$ is defined in Eq.~(\ref{eq:CH2_def}).

\noindent For colliders that can directly measure the Higgs width, $BR_{unt}$ can be constrained together with $\kappa_i$ and $BR_{inv}$ from a joint fit to the data. For standalone fits to colliders that cannot, such as the HL-LHC, either an indirect measurement can be included, such as from off-shell Higgs production, or additional theoretical assumptions must be introduced.  A possible assumption is $|\kappa_V|\leq$1 $(V=W,Z)$, which is theoretically motivated as it holds in a wide class of BSM models albeit with some exceptions~\cite{Falkowski:2012vh} (for more details see \cite{LHCHXSWG3}, Section~10).

\subsubsection{Fitting scenarios}

To characterise the performance of future colliders in the $\kappa$-framework, we defined four benchmark scenarios, which are listed in Table \ref{tab:kappa_scenarios}. The goal of the kappa-0 benchmark is to present the constraining power of the $\kappa$-framework under the assumption that there exist no light BSM particles to which the Higgs boson can decay. The goal of benchmarks kappa-1,2 is to expose the impact of allowing BSM Higgs decays, in combination with a measured or assumed constraint on the width of the Higgs, on the standalone $\kappa$ results. Finally, the goal of the kappa-3 benchmark is to show the impact of combining the HL-LHC data with each of the future accelerators. In all scenarios with BSM branching fractions, these branching fractions are constrained to be positive definite. 

Experimental uncertainties -- defined as statistical uncertainties and, when provided, experimental systematic uncertainties, background theory uncertainties and signal-acceptance related theory uncertainties -- are included in all scenarios. Theory uncertainties on the Higgs branching fractions predictions for all future colliders and  uncertainties on production cross section predictions for hadron colliders, as described in Section \ref{sec:method}, are partially included; intrinsic theory uncertainties, arising from missing higher-order corrections, are {\em not} included in any of the benchmarks, while parametric theory uncertainties arising from the propagation of experimental errors on SM parameters {\em are} included in all scenarios. A detailed discussion and assessment of the impact of theory uncertainties is given in Section \ref{sec:SMThUnc}.

\begin{table}[hbt]
\caption{\label{tab:kappa_scenarios} Definition of the benchmark scenarios used to characterize future colliders in the $\kappa$-framework.}
\begin{center}
\begin{tabular}{cccccc}
\toprule
Scenario & $BR_{inv}$ & $BR_{unt}$ & include HL-LHC \\
\midrule
kappa-0  & fixed at 0 & fixed at 0 & no \\
\midrule
kappa-1  &  measured & fixed at 0 & no \\
kappa-2  &  measured & measured & no \\
\midrule 
kappa-3  &  measured & measured & yes \\
\bottomrule 
\end{tabular}
\end{center}
\end{table}


\subsection{Results from the kappa-framework studies and comparison} \label{kappa-results}

The $\kappa$-framework discussed in the previous section was validated comparing the  results obtained 
with the scenarios described as kappa-0 and kappa-1 to the original results presented by the Collaborations to the European Strategy. In general, good agreement is found. 

The results of the kappa-0 scenario described in the previous section are reported in Table ~\ref{tab:resultsKappa0}. 
In this scenario, no additional invisible or untagged branching ratio is allowed in the fits, and colliders are considered independently. This is the simplest scenario considered in this report, and illustrates the power of the kappa framework to constrain new physics in general, and in particular the potential to constrain new physics at the proposed new colliders discussed in this report. In general the precision is at the per cent level, In the final stage of the future colliders a precision of the order of a few per-mille would be reachable for several couplings, for instance $\kappa_W$ and $\kappa_Z$. Cases in which a particular parameter has been fixed to the SM value due to lack of sensitivity are shown with a dash (-). Examples of this are $\kappa_c$, not accessible at HL-LHC and HE-LHC, and $\kappa_t$, only accessible above the $ttH$/$tH$ threshold. Not  all colliders reported results for all possible decay modes in the original reference documentation listed in~Table\ref{tab:colliders}, the most evident example of this being the $Z\gamma$ channel. In this standalone collider scenario, the corresponding parameters were left to float in the fits. They are  indicated with $\ast$ in the tables.   

This kappa-0 scenario can be expanded to account for invisible decays (kappa-1) and invisible and untagged decays (kappa-2), still considering individual colliders in a standalone way. The overall effect of this additional width is a slight worsening of the precision of the kappa parameters from the kappa-0 scenario to the kappa-1, and further on to the  kappa-2. It is most noticeable for $\kappa_W$, $\kappa_Z$ and $\kappa_b$. For comparison of the total impact, the kappa-2 scenario results can be found in Tables~\ref{tab:resultsLHCKappa2} and ~\ref{tab:resultsKappa2}  in Appendix~\ref{kappabenchmarks}.

\begin{table}[ht]
\centering
\caption{ \label{tab:resultsKappa0}
Expected relative precision (\%) of the $\kappa$ parameters in the kappa-0
scenario  described in
Section~\ref{tab:kappa_scenarios} for future accelerators. Colliders are considered independently, not in combination with the HL-LHC. No BSM width is allowed in the fit: both $\mathrm{BR}_{\mathrm{unt}}$ and $\mathrm{BR}_{\mathrm{inv}}$ are set to 0, and therefore $\kappa_V$ is not constrained.
Cases in which a particular parameter has been fixed to the SM value due to lack of sensitivity are shown with a dash (-). A star ($\star$) indicates the cases in which a parameter
has been left free in the fit due to lack of input in the reference documentation. 
The integrated luminosity and running conditions considered for each collider in this comparison are described in Table~\ref{tab:colliders}. FCC-ee/eh/hh corresponds to the combined performance of FCC-ee$_{240}$+FCC-ee$_{365}$, FCC-eh and FCC-hh. In the case of HE-LHC, two theoretical uncertainty scenarios (S2 and S2$^\prime$)~\cite{Cepeda:2019klc} are given for comparison.
}
\small 
\setlength\tabcolsep{2pt}
\begin{tabular}{ c | c | c | c  c | c cc | c c c | c |  c c |  c}
\toprule
kappa-0   &HL-LHC   &LHeC   &\multicolumn{2}{|c|}{HE-LHC}  & \multicolumn{3}{|c|}{ILC}   & \multicolumn{3}{|c|}{CLIC}     &CEPC  & \multicolumn{2}{|c|}{FCC-ee}  &FCC-ee/eh/hh  \\
&   &   &S2 & S2$^\prime$   &250 & 500 & 1000 & 380 & 15000 & 3000 & & 240 & 365 & 
\\
\bottomrule
$\kappa_{W}$ [\%] & $1.7$   & 0.75 & $1.4$   & 0.98 & $1.8$   & $0.29$ & 0.24 & $0.86$   & $0.16$   & $0.11$   & $1.3$   & $1.3$   & $0.43$   & $0.14$  \\
\crowcolorA$\kappa_{Z}$ [\%] & $1.5$   & $1.2$ &  $1.3$ &  0.9 & $0.29$   & $0.23$ & 0.22  & $0.5$   & $0.26$   & $0.23$   & $0.14$   & $0.20$   & $0.17$   & $0.12$  \\
$\kappa_{g}$ [\%]  & $2.3$   & $3.6$   & $1.9$ & 1.2 & $2.3$   & $0.97$ & 0.66  & $2.5$   & $1.3$   & $0.9$   & $1.5$   & $1.7$   & $1.0$   & $0.49$  \\
\crowcolorA$\kappa_{\gamma}$ [\%]  & $1.9$   & $7.6$   & $1.6$ & 1.2  & $6.7$   & $3.4$ & 1.9  & $98\star$  & $5.0$   & $2.2$   & $3.7$   & $4.7$   & $3.9$   & $0.29$  \\
$\kappa_{Z\gamma}$  [\%]  & $10.$   & $-$   & 5.7 & 3.8 & $99\star$   & $86\star$ & 85$\star$  & $120\star$   & $15$   & $6.9$   & $8.2$   & $81\star$   & $75\star$   & $0.69$  \\
\crowcolorA$\kappa_{c}$  [\%] & $-$   & $4.1$   & $-$ & $-$   & $2.5$   & $1.3$ & 0.9   & $4.3$   & $1.8$   & $1.4$   & $2.2$   & $1.8$   & $1.3$   & $0.95$  \\
$\kappa_{t}$ [\%]  & $3.3$   & $-$   & 2.8 & 1.7 & $-$   & $6.9$ & 1.6  & $-$   & $-$   & $2.7$   & $-$   & $-$   & $-$   & $1.0$  \\
\crowcolorA$\kappa_{b}$  [\%]  & $3.6$   & $2.1$   & 3.2 & 2.3 & $1.8$   & $0.58$ & 0.48  & $1.9$   & $0.46$   & $0.37$   & $1.2$   & $1.3$   & $0.67$   & $0.43$  \\
$\kappa_{\mu}$  [\%]  & $4.6$   & $-$   & 2.5 & 1.7 & $15$   & $9.4$ & 6.2  & $320\star$   & $13$   & $5.8$   & $8.9$   & $10$   & $8.9$   & $0.41$  \\
\crowcolorA$\kappa_{\tau}$ [\%]   & $1.9$   & $3.3$   & $1.5$ & 1.1 & $1.9$   & $0.70$  & 0.57  & $3.0$   & $1.3$   & $0.88$   & $1.3$   & $1.4$   & $0.73$   & $0.44$  \\
\bottomrule
\end{tabular}
\end{table}

Table~\ref{tab:resultsLHCKappa3} shows the expected precision of the $\kappa$ parameters in the final benchmark scenario discussed in this paper in which 95\% CL limits
on $\mathrm{BR}_{\mathrm{unt}}$ and $\mathrm{BR}_{\mathrm{inv}}$ are set,  for the three possibilities using the LHC tunnel: HL-LHC, LHeC, and HE-LHC.  The results correspond to the kappa-3 scenario.
  
As discussed before, for these hadron colliders a constraint on $|\kappa_V|\leq$1 is applied in this case, as no direct access to the Higgs width is possible.

Table~\ref{tab:resultsKappa3} shows the corresponding kappa-3 scenario for the different lepton colliders and a final FCC-ee/eh/hh combination, all combined with the HL-LHC results. 
The integrated luminosity and running conditions considered for each collider in this comparison are taken for Table~\ref{tab:colliders}.  
The constraints on $\Gamma_H$ derived from the fit parameters using 
Eq.~\ref{eq:width} are discussed in detail in Section~\ref{mass}. 
In this case when HL-LHC is combined with a lepton collider the assumption $|\kappa_V|\leq$1 is no longer necessary, and therefore it is not used as a constrain in these kappa-3  fits. 
For those particular analyses  not reported in the original reference documentation listed in Table~\ref{tab:colliders} (e.g. $\kappa_{Z\gamma}$) the HL-LHC prospects drive the combination. They are  indicated with $\ast$ in the tables. 

We have examined the correlations of the lepton collider kappa-3 fits. 
In the initial stage of ILC (ILC$_{250}$), $\kappa_W$, $\kappa_g$, $\kappa_b$, $\kappa_t$ and $\kappa_\tau$ show sizeable correlations ($>70\%$), with the largest corresponding to $\kappa_b$ and $\kappa_\tau$ (93$\%$). There is practically no correlation between $\kappa_W$ and $\kappa_Z$ (8$\%$). The untagged branching fraction is not particularly correlated with the couplings, with the largest correlation corresponding to $\kappa_Z$ (50\%), and an anti correlation (-20\%) seen for $\kappa_{Z\gamma}$ where the only information comes from the HL-LHC data. In the case of FCC-ee$_{365}$, we see a slight correlation between $\kappa_Z$ and $\kappa_W$ (30\%), and a similar correlation between these and the untagged  branching fraction (30-50$\%$). The correlations between $\kappa_b$,  $\kappa_\tau$,  $\kappa_g$ and  $\kappa_W$ are mild, with the largest value corresponding once again to  $\kappa_b$ and $\kappa_\tau$ (74\%). In this case there is also no strong correlation between the untagged branching fraction and the couplings, with the largest correlation corresponding to $\kappa_Z$  (50\%), followed by $\kappa_b$ (30$\%$). Again an anti correlation (-20$\%$) is seen for $\kappa_{Z\gamma}$. For CLIC$_{3000}$ the situation is markedly different, with large correlations between $\kappa_Z$ and $\kappa_W$ (80\%), and between the untagged branching fraction and $\kappa_Z$, $\kappa_W$ and  $\kappa_b$ (90\%, 80\%, 70\% respectively). The correlations between  $\kappa_b$,  $\kappa_Z$, $\kappa_\tau$,  $\kappa_g$ and  $\kappa_W$ are not negligible, with the highest corresponding to $\kappa_b$ and $\kappa_W$  (70\%). In this case,  $\kappa_b$ and $\kappa_\tau$ are correlated to 45\%. These correlations can be seen graphically in Figure~\ref{fig:Kappa3Correlation} in the Appendix. 

\begin{figure}[t]
\centering
\includegraphics[width=1\linewidth]{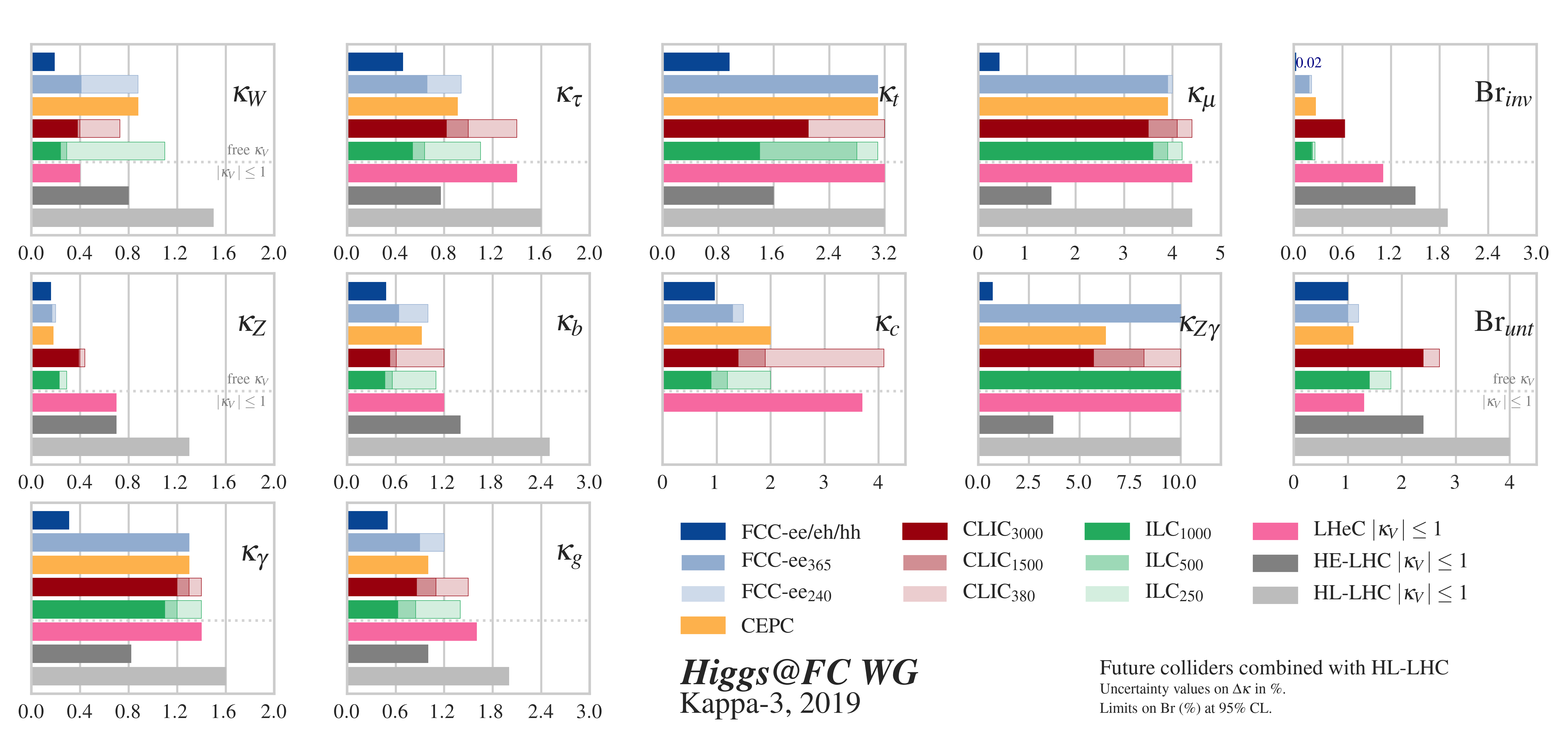}
\caption{\label{fig:Kappa3Summary}
Expected relative precision (\%) of the $\kappa$ parameters in the kappa-3  scenario  described in
Section~\ref{tab:kappa_scenarios}. For details, see  Tables~\ref{tab:resultsLHCKappa3} and ~\ref{tab:resultsKappa3}. For HE-LHC, the S2' scenario is displayed. For LHeC, HL-LHC and HE-LHC a constrained $\kappa_V\leq 1$ is applied.
}
\end{figure}

The results of the kappa-3 benchmark scenario are also presented graphically in Figure~\ref{fig:Kappa3Summary}. Note that while hadron colliders and lepton colliders are shown together, the caveat that a bound on $|\kappa_V|\leq 1$ is required for HL-LHC, HE-LHC and LHeC still applies. Parameters fixed to the Standard Model value are not displayed.  

Intrinsic theoretical uncertainties for future lepton colliders are omitted in Tables~\ref{tab:resultsKappa0}, \ref{tab:resultsLHCKappa3} and \ref{tab:resultsKappa3}. Their effect is discussed in detail in Section~\ref{sec:SMThUnc}.

\begin{table}[hbt]
\centering
\caption{ \label{tab:resultsLHCKappa3}
Expected relative precision (\%) of the $\kappa$ parameters in the kappa-3  scenario  described in
Section~\ref{tab:kappa_scenarios} for the HL-LHC, LHeC, and HE-LHC. A bound on $|\kappa_V|\leq 1$ is applied since no direct access to the Higgs width is possible, thus the uncertainty on $\kappa_W$ and $\kappa_{Z}$ is one-sided. For the remaining kappa parameters one standard deviation is provided in $\pm$. 
The corresponding 95\%CL upper limit on $\mathrm{BR}_{\mathrm{inv}}$
is also given. In this kappa-3 scenario $\mathrm{BR}_{\mathrm{unt}}$ is a floating parameter in the fit, to propagate the effect of an assumed uncertain total width on the measurement of the other $\kappa_{i}$. Based on this constraint the reported values on $\mathrm{BR}_{\mathrm{unt}}$ are inferred.   Cases in which a particular parameter has been fixed to the SM value due to lack of sensitivity are shown with a dash ($-$). An asterisk (*) indicates the cases in which there is no analysis input in the reference documentation, and HL-LHC dominates the combination. In the case of $\kappa_t$ sensitivity at the LHeC, note that the framework relies as input on $\mu_{ttH}$, and does not take into consideration $\mu_{tH}$.  The integrated luminosity and running conditions considered for each collider in this comparison are described in Table~\ref{tab:colliders}. In the case of HL-LHC and HE-LHC, both the S2 and the S2' uncertainty models~\cite{Cepeda:2019klc} are given for comparison. }
\begin{tabular}{| c | c  | c  | c c |}
\toprule
kappa-3    &HL-LHC  & \multicolumn{3}{c|} {HL-LHC  \&}\\
& & LHeC &  HE-LHC (S2) & HE-LHC (S2')\\
\bottomrule
$1\geq\kappa_{W}>$ (68\%)  & 0.985 & 0.996 & 0.988 & 0.992\\  
\crowcolorA$1\geq\kappa_{Z}>$ (68\%)  & 0.987 & 0.993 & 0.989 & 0.993 \\  
$\kappa_{g}$ ($\%$)   & $\pm 2.$    & $\pm1.6$    & $\pm1.6$   & $\pm1.$  \\
\crowcolorA$\kappa_{\gamma}$ ($\%$)  & $\pm1.6$   & $\pm1.4$   & $\pm1.2$   & $\pm0.82$  \\
$\kappa_{Z\gamma}$ ($\%$)  & $\pm10.$   & $\pm10.$ $\ast$    & $\pm5.5$ & $\pm3.7$  \\
\crowcolorA$\kappa_{c}$ ($\%$)  & $-$    & $\pm 3.7$    & $-$   & $-$ \\
$\kappa_{t}$ ($\%$)  & $\pm3.2$    & $\pm3.2$ $\ast$  & $\pm2.6$ & $\pm1.6$   \\
\crowcolorA$\kappa_{b}$ ($\%$)  & $\pm2.5$     & $\pm1.2$   & $\pm2.$  & $\pm1.4$ \\
$\kappa_{\mu}$ ($\%$)  & $\pm4.4$    & $\pm4.4$ $\ast$  & $\pm2.2$ & $\pm1.5$  \\
\crowcolorA$\kappa_{\tau}$ ($\%$)  & $\pm1.6$   & $\pm1.4$   & $\pm1.2$ & $\pm0.77$    \\
\arrayrulecolor{black}\bottomrule
$\mathrm{BR}_{\mathrm{inv}}$ ($<$\%, 95\% CL)   & $1.9$   & $1.1$   & $1.8$ $\ast$ &$1.5$ $\ast$    \\
\crowcolorA $\mathrm{BR}_{\mathrm{unt}}$ ($<$\%, 95\% CL)      & \multicolumn{4}{c|}{inferred using constraint $|\kappa_V|\leq 1$} \\
\crowcolorA &  $4.$    & $1.3$    & $3.3$ & $2.4$ \\
\arrayrulecolor{black}\bottomrule
\end{tabular}
\end{table}

\begin{sidewaystable}[ht]
\centering
\caption{ \label{tab:resultsKappa3}
Expected relative precision (\%) of the $\kappa$ parameters
in the kappa-3 (combined with HL-LHC) scenario  described in
Section~\ref{tab:kappa_scenarios} for future accelerators beyond the LHC era.
The corresponding 95\%CL upper limits on $\mathrm{BR}_{\mathrm{unt}}$ and $\mathrm{BR}_{\mathrm{inv}}$
and the derived constraint on the Higgs width (in \%) are also given. No requirement on
$\kappa_V$ is applied in the combination with HL-LHC, since the lepton colliders provide
the necessary access to the Higgs width.  Cases in which a particular parameter has been
fixed to the SM value due to lack of sensitivity are shown with a dash ($-$). An asterisk ($\ast$) indicates the cases in which there is no analysis input in the reference documentation, and HL-LHC dominates the combination.
The integrated luminosity and running conditions considered for each collider in this
comparison are described in Table~\ref{tab:colliders}. 
FCC-ee/eh/hh corresponds to the combined performance of FCC-ee$_{240}$+FCC-ee$_{365}$, FCC-eh and FCC-hh. 
}
\setlength\tabcolsep{4.5pt}
\setlength\tabcolsep{2.5pt}
\begin{tabular}{ c |  c c c | c c c | c | c c c }
\toprule
\multirow{2}{*}{kappa-3 } & \multicolumn{9}{c}{HL-LHC \&} \\
                                   & ILC$_{250}$   & ILC$_{500}$ & ILC$_{1000}$ & CLIC$_{380}$ &  CLIC$_{1500}$ & CLIC$_{3000}$  & CEPC & FCC-ee$_{240}$ & FCC-ee$_{365}$ & FCC-ee/eh/hh      \\
  \bottomrule
$\kappa_{W}$ [\%]          & 1.0   & 0.29 & 0.24 & 0.73  & 0.40   & 0.38  & 0.88  & 0.88  & 0.41  & 0.19 \\
\crowcolorA$\kappa_{Z}$[\%]           & 0.29  & 0.22 & 0.23 & 0.44  & 0.40  & 0.39  & 0.18  & 0.20  & 0.17  & 0.16  \\
$\kappa_{g}$[\%]           & 1.4   & 0.85 & 0.63 & 1.5    & 1.1   & 0.86  & 1.   & 1.2   & 0.9  & 0.5  \\
\crowcolorA$\kappa_{\gamma}$ [\%]     & 1.4   & 1.2  & 1.1  & 1.4$\ast$   & 1.3   & 1.2   & 1.3   & 1.3   & 1.3   & 0.31  \\
$\kappa_{Z\gamma}$ [\%]    & 10.$\ast$   & 10.$\ast$  & 10.$\ast$ & 10.$\ast$   & 8.2   & 5.7   & 6.3   & 10.$\ast$   & 10.$\ast$   & 0.7  \\
\crowcolorA$\kappa_{c}$ [\%]          & 2.    & 1.2 & 0.9  & 4.1   & 1.9   & 1.4   & 2.    & 1.5   & 1.3   & 0.96  \\
$\kappa_{t}$ [\%]          & 3.1   & 2.8 & 1.4 & 3.2   & 2.1   & 2.1   & 3.1   & 3.1   & 3.1  & 0.96  \\
\crowcolorA$\kappa_{b}$ [\%]          & 1.1   & 0.56 & 0.47 & 1.2   & 0.61  & 0.53  & 0.92  & 1.    & 0.64  & 0.48  \\
$\kappa_{\mu}$ [\%]        & 4.2   & 3.9 & 3.6  & 4.4$\ast$   & 4.1   & 3.5   & 3.9   & 4.    & 3.9   & 0.43  \\
\crowcolorA$\kappa_{\tau}$ [\%]       & 1.1   & 0.64 & 0.54 & 1.4   & 1.0  & 0.82  & 0.91  & 0.94  & 0.66  & 0.46  \\
\arrayrulecolor{black}\midrule
$\mathrm{BR}_{\mathrm{inv}}$ ($<$\%, 95\% CL)                    & 0.26  & 0.23 & 0.22 & 0.63  & 0.62  & 0.62  & 0.27  & 0.22  & 0.19  & 0.024  \\
\crowcolorA$\mathrm{BR}_{\mathrm{unt}}$ ($<$\%, 95\% CL)                     & 1.8   & 1.4 & 1.4  & 2.7   & 2.4   & 2.4   & 1.1   & 1.2   & 1.    & 1.  \\
\arrayrulecolor{black}\bottomrule
\end{tabular}
\end{sidewaystable}

\clearpage


\subsection{Effective field theory description of Higgs boson couplings} \label{eft}

As already discussed, the $\kappa$-framework provides a convenient first parametrization of new physics in single Higgs processes.
By construction, it is perfectly suitable to spot a deviation from the SM, but  it does not provide a systematic description of new physics. As such it does not permit to correlate different processes  nor to describe their energy dependence, which is certainly a drawback when trying to develop a strategic perspective. When aiming at a more systematic approach one must distinguish the two cases of light and heavy new physics.
In the first case, there is no simple and systematic description. One must proceed  case by case, perhaps with the aid of simplified models.
While we are not aware of any attempt at a general analysis, it should be noted that light degrees of freedom carrying  electroweak quantum numbers seem disfavored, and that the less constrained options involve portal type interactions of the Higgs to SM singlets. Examples in the latter class can involve mixing with a new light CP-even scalar $S$, or the trilinear couplings to scalar ( $h S^2$) or fermion ($h\bar\psi\psi$) bilinears. In these cases, besides the presence of new processes, e.g. the production of a sequential Higgs-like scalar, the effect on single Higgs production and decay  are  well described by the $\kappa$-framework, including the option for an invisible width into new physics states.
Heavy new physics can instead be  systematically described in the effective Lagrangian approach. This fact, and the richer set of consistent and motivated heavy new physics  options, gives particular prominence to the effective Lagrangian approach.  One can distinguish two broad classes of heavy new physics depending on the origin of the corresponding mass scale, which we henceforth indicate by $\Lambda$. In the first class, $\Lambda$  is controlled by the Higgs VEV ($v$) and is expected to be bounded to be less than $4\pi v\sim 3$\,TeV .
The effective Lagrangian corresponds here to the so-called Higgs-EFT, which cannot be written as a polynomial 
expansion in gauge invariant operators \cite{Falkowski:2019tft,Chang:2019vez}. In this scenario, which is in this respect analogous to Technicolor, deviations in Higgs couplings and EWPT are expected to exceed their present bounds, unless the new physics effect can be tuned to be small for each and every coupling, which makes it rather implausible. 
  In the second class,  basically including all the more plausible scenarios, $\Lambda$ is not controlled by the Higgs VEV, and can virtually be arbitrarily large. In that case the effective Lagrangian corresponds to the so-called SMEFT. It 
 is polynomial in gauge invariant operators and organised as an expansion in inverse powers of $\Lambda$:

\begin{equation}
\label{eq:EFT_Lops}
{\cal L}_{\mathrm{Eff}}={\cal L}_{\rm SM}
+  \frac{1}{\Lambda}{\cal L}_{5} \
+  \frac{1}{\Lambda^2}{\cal L}_{6} \
+  \frac{1}{\Lambda^3}{\cal L}_{7} \
+  \frac{1}{\Lambda^4}{\cal L}_{8}  + \cdots\,,~~~~~~~
{\cal L}_{d}= \sum_{i}  c_{i}^{(d)} {\cal O}_{i}^{(d)}\, .
\end{equation}
In the previous equation, each ${\cal O}_i^{(d)}$ is a local  $SU(3)_c \times SU(2)_L \times U(1)_Y$-invariant operator of canonical mass dimension $d$, built using only fields from the light particle spectrum. Moreover,
${\cal L}_{\rm SM}$ represents the renormalizable SM Lagrangian that nicely complies with basically all the
measurements made so far in particle physics, with the exception of the tiny neutrino masses, which are however  nicely described  by the next term, ${\cal L}_{5}$.
The contribution of the higher order terms ${\cal L}_{d\geq 5}$ to physical amplitudes is suppressed by $(E/\Lambda)^{d-4}$, where $E$ is the relevant energy scale of  the process. The {\em Wilson  coefficients} $c_i^{(d)}$ encode the virtual effects of the heavy new physics in low-energy observables. Their precise form in terms of masses and couplings of the new particles can be obtained via {\it matching} with an ultraviolet (UV) completion of the SM~\cite{deBlas:2017xtg}, or inferred using {\it power-counting} rules~\cite{Giudice:2007fh,Liu:2016idz}.

The success of ${\cal L}_{\rm SM}$ in explaining the data  indicates that either
the scale of new physics $\Lambda$ is large, or that the structure of the terms ${\cal L}_{d\geq 5}$ is particularly elaborate, or perhaps a combination of both. Moreover it is important to stress that in general we  expect new physics at multiple and even widely separated scales, and that the parametrization in terms of a single scale $\Lambda$ is a simplification. It is however clear that given the good but limited precision of future high energy experiments  only the lowest scale is expected to matter. In particular, given the observed suppression of lepton and baryon number violation, the operators mediating such violation, which appear already in ${\cal L}_5$ and ${\cal L}_6$,
must be further suppressed if $\Lambda$ is as low as to be interesting in collider physics. That suppression could be due to approximate symmetries or simply because the dynamics generating these processes is $\gg \Lambda$. The same remarks apply to flavour and CP violation.

Assuming lepton and baryon numbers are conserved independently, all relevant operators in the previous expansion are of even dimension. Therefore  new physics effects start at dimension $d=6$.  In this report we work under the assumption that $\Lambda$ is large enough for $d=6$ to dominate over $d\geq 8$ (but see comment below) and restrict our studies to  the effective Lagrangian truncated part $d=6$. The resulting Lagrangian is that of the so-called dimension-6 Standard Model Effective Field Theory (SMEFT). In a bottom-up approach, one can write a complete basis for the dimension-6 SMEFT Lagrangian using a total of 59 types of operators~\cite{Grzadkowski:2010es}, for a total of 2499 taking into account flavour indices~\cite{Alonso:2013hga}. For most  of the calculations presented in this report we use the dimension-six basis first presented in~\cite{Grzadkowski:2010es}, the so-called {\it Warsaw basis}, with minor modifications.\footnote{By using a perturbative field redefinition we trade the operators ${\cal O}_{\phi WB}$ and ${\cal O}_{\phi D}$ in Ref.~\cite{Grzadkowski:2010es} for the operators $i D^\mu \phi^\dagger \sigma_a D^\nu \phi  W_{\mu\nu}^{a}$ and $i D^\mu \phi^\dagger D^\nu \phi  B_{\mu\nu}$.} In the discussion presented in this section, however, we will use a different parameterization, which is usually deemed to be more transparent from the point of view of Higgs physics.

One must notice that in any realistic situation there will be  structure in 
the coefficients of the 2499 operators of dimension $6$. For instance, if they were a set of structureless $O(1)$ numbers, then  the experimental constraints from flavour and/or CP violation on $\Lambda$ would already be much stronger than from any foreseeable study of Higgs and EW processes. Moreover in realistic situations we should also expect structure in the coefficients of flavour preserving operators. In other words some coefficients may be significantly smaller than others.
 This remark, together with a sensitivity limited to $\Lambda$'s  that are not very much above the energy of the processes, implies that it may in principle happen that operators of dimension 8 are equally or more important than the dimension 6 operators.  We shall later mention a natural example of this phenomenon. In  structured scenarios like the SILH~\cite{Giudice:2007fh}, it is easy to address this caveat, also thanks to the fact that the operators that matter in ${\cal L}_6$ are much fewer than in the general case (even after imposing flavour violation). The message here is that the reduction to pure dimension $6$, with full neglect of dimension $8$, while reasonable and useful, contains nonetheless an assumption which may not be universally  true for all observables even in simple motivated models. 

When considering Higgs data, one can reasonably focus on
a relatively small subset of the 2499 operators  in ${\cal L}_6$. In particular the vast subset of 4-fermion operators, whether flavour and CP preserving or not, can be more strongly constrained by other processes.
Thus, it makes  sense to neglect this whole class, with the exception of one particular four-fermion interaction that contributes to the muon decay and thus directly affects the Fermi constant, see caption in Table~\ref{tab:d6BasisBF}. The dipole operators, instead do directly affect Higgs production, however under very general and plausible assumptions on the flavour structure of new physics, the coefficients of these operators display the same structure and the same chiral suppression of Yukawa couplings. The consequence is that, with the possible exception of processes involving the top quark, their effect in Higgs production is expected to be negligible given that the leading SM contribution (for instance in $e^+e^-\to ZH$) as well as the other new physics effects are not chirally  suppressed. Furthermore, as far as Higgs decays are concerned, the dipole operators only contribute to three (or more)-body final states (for instance $H\to \bar b b \gamma$) and as such they are easily seen to be negligible. 
In what follows we shall thus neglect  this whole class, and leave the consideration of their effect in top sector to future studies.
Eliminating these two classes, there remain  three other classes:
1) purely bosonic operators, 2) generalized Yukawas, 3) Higgs-fermion current operators.
Neglecting CP violating operators in class 1, the corresponding structures  are shown in Table~\ref{tab:d6BasisBF}.
Operators in class 2 and 3, per se, can still contain CP- or flavour-violating terms, on which experimental constraints are rather strong. In order to proceed we shall consider two alternative scenarios to minimize the remaining flavour and/or CP violations:
\begin{enumerate}
\item  {\it Flavour Universality},  corresponding  to 
\begin{equation}
 Y_u^{(6)}\propto Y_u,\quad Y_d^{(6)}\propto Y_d,\quad Y_e^{(6)}\propto Y_e,\quad\mbox{and}\quad \Delta_{ij}^{q,u,d,l, e, ud}, {\Delta'}_{ij}^{q,l},
\propto \delta_{ij},
\label{eq:FU}
 \end{equation}

where $Y_{f}^{(6)}$ are the coefficients of dimension-6 operators of class 2, which control the flavour structure of the modifications to the SM Yukawa matrices $Y_f$.
Similarly, $\Delta^{f}$ and $\Delta^{\prime~\!f}$ represent the combinations of dimension-6 operators of class 3, which induce flavour-dependent modifications of the neutral and charged current couplings of the fermions to the EW vector bosons. 
In terms of the Wilson coefficients of the operators in Table \ref{tab:d6BasisBF} one has $Y_{f}^{(6)}=c_{f\phi}$ $(f=u,d,e)$; 
$\Delta^{f}=c_{\phi f}$ for the operators involving the right-handed fermion multiplets $(f=u,d,e,ud)$; and $\Delta^{f}=c_{\phi f}^{(1)}$, $\Delta^{\prime~f}=c_{\phi f}^{(3)}$ for the left-handed ones ($f=q,l$).
The choice in (\ref{eq:FU})  corresponds to  {\it Minimal Flavour Violation} (MFV)~\cite{DAmbrosio:2002vsn}  in the limit where terms only up to linear in the Yukawa matrices are considered. Notice that {\it Minimal Flavour Violation} corresponds to the assumption that the underlying dynamics respects the maximal flavour symmetry group $SU(3)^5$. A more appropriate name would then perhaps be {\it Maximal Flavour Conservation}.
  
\item {\it Neutral Diagonality}, corresponding to a scenario where $Y^{(6)}_{u,d,e}$ while  not proportional to the corresponding Yukawa matrices are nonetheless diagonal in the same basis. That eliminates all flavour-changing couplings to the Higgs boson. Similarly the 
$\Delta_{ij}^{q,u,d,\ell, e, ud}, {\Delta'}_{ij}^{q,\ell}$, while not universal, are such that no flavour-changing couplings to the Z-boson are generated.
In fact we shall work under the specific assumption where flavour universality is respected by the first two quark families, and  violated by the third quark family and by leptons. This choice, per se, does not correspond to any motivated
or even plausible scenario (it is rather cumbersome to produce sizeable flavour non-universality without any flavour violation). 
We consider it principally to test the essential constraining power of future machines and because it is widely studied by the community. Moreover non-universality limited to the third quark family is an often recurring feature of  scenarios motivated by the hierarchy problem. 
That is simply because the large top Yukawa makes it intricately  involved in the EW symmetry breaking dynamics and calls for the existence of various top partners.
\end{enumerate}

\begin{table}[ht]
\caption{Dimension six operators considered in the SMEFT analysis. The hermitian derivatives $\lrD$ and $\lrDa$ are defined as: 
$\lrD_\mu \equiv \overset{\rightarrow}{D}_\mu - \overset{\leftarrow}{D}_\mu$ and $\lrDa_\mu \equiv \sigma_a \overset{\rightarrow}{D}_\mu - \overset{\leftarrow}{D}_\mu \sigma_a$, while $B_{\mu\nu}$, $W_{\mu\nu}^a$ and $G_{\mu\nu}^A$ denote the SM gauge boson field-strengths. See text for details. 
Apart from these, the effects of the four-lepton operator $({\cal O}_{ll})_{1221}=\left(\bar{l}_1 \gamma_\mu l_2\right)\left(\bar{l}_2 \gamma^\mu l_1\right)$, which modifies the prediction for the muon decay amplitude, must also be included in the fit since we use the Fermi constant as one of the SM input parameters.
\label{tab:d6BasisBF}}
  \begin{center}
    \begin{tabular}{c cclccl}
      \toprule
      & & Operator & Notation & & Operator & Notation \\
      \cmidrule{3-4} \cmidrule{6-7}
      \multirow{8}{*}{\vtext{Class 1}}&\multirow{1}{*}{$X^3$}
      &
      $\varepsilon_{abc} W^{a\,\nu}_\mu W^{b\,\rho}_\nu W^{c\,\mu}_\rho$ &
      $\mathcal{O}_{W}$ &
      &
      &
      \\
       \vspace{-3mm} & & & & & \\
      \cmidrule{2-7}
      &$\phi^6$ &
      $\left(\phi^{\dagger} \phi\right)^3$ &
      $\mathcal{O}_{\phi}$ &
      & & \\[1mm]
       \vspace{-3mm} & & & & & \\
      \cmidrule{2-7}
      &$\phi^4 D^2$ &
      $\left(\phi^{\dagger} \phi\right)\square
      \left(\phi^{\dagger} \phi\right)$ &
      $\mathcal{O}_{\phi \square}$ &
      &
      $\left(\phi^{\dagger}D_\mu \phi\right)
      (\left(D^\mu \phi\right)^{\dagger}\phi)$ &
      $\mathcal{O}_{\phi D}$ \\[1mm]
       \vspace{-3mm} & & & & & \\
      \cmidrule{2-7}
      &\multirow{2}{*}{$X^2 \phi^2$} &
      $\phi^\dagger \phi B_{\mu\nu} B^{\mu\nu} $ &
      $\mathcal{O}_{\phi B}$ &
      &
      $\phi^\dagger \phi W_{\mu\nu}^a W^{a\,\mu\nu} $ &
      $\mathcal{O}_{\phi W}$ \\
      &
      &$\phi^\dagger \sigma_a \phi W^a_{\mu\nu} B^{\mu\nu}$ &
      $\mathcal{O}_{\phi WB}$ &
      &
      $\phi^\dagger \phi G_{\mu\nu}^A G^{A\,\mu\nu} $ &
      $\mathcal{O}_{\phi G}$ \\[1mm]
       \vspace{-3mm} & & & & & \\
      \midrule
      \multirow{2}{*}{\vtext{Class 2}}&\multirow{2}{*}{$\psi^2 \phi^2$} &
      $\left(\phi^{\dagger} \phi\right)
      (\bar{l}_L^i \phi e_R^j)$ &
      $\left(\mathcal{O}_{e \phi}\right)_{ij}$ & & \\
      &
      &$\left(\phi^{\dagger} \phi\right)
      (\bar{q}_L^i \phi d_R^j)$ &
      $\left(\mathcal{O}_{d \phi }\right)_{ij}$ &
      & 
      $\left(\phi^{\dagger} \phi\right)
      (\bar{q}_L^i \tilde{\phi} u_R^j)$ &
      $\left(\mathcal{O}_{u \phi}\right)_{ij}$ \\
       \vspace{-3mm} & & & & & \\
      \midrule
      \multirow{6}{*}{\vtext{Class 3}}&\multirow{5}{*}{$\psi^2 \phi^2 D$} &
      $(\phi^{\dagger} i\overset{\leftrightarrow}{D}_\mu \phi)
      (\bar{l}_L^i \gamma^\mu l_L^j)$ &
      $\left(\mathcal{O}_{\phi l}^{(1)}\right)_{ij}$ &
      &
      $(\phi^{\dagger} i\lrDa_\mu \phi)
      (\bar{l}_L^i \gamma^\mu \sigma_a l_L^j)$ &
      $(\mathcal{O}_{\phi l}^{(3)})_{ij}$ \\
      &
      &$(\phi^{\dagger} i\overset{\leftrightarrow}{D}_\mu \phi)
      (\bar{e}_R^i \gamma^\mu e_R^j)$ &
      $\left(\mathcal{O}_{\phi e}\right)_{ij}$ &
      & & \\
      &
      &$(\phi^{\dagger} i\overset{\leftrightarrow}{D}_\mu \phi)
      (\bar{q}_L^i \gamma^\mu q_L^j)$ &
      $(\mathcal{O}_{\phi q}^{(1)})_{ij}$ &
      &
      $(\phi^{\dagger} i \lrDa_\mu \phi)
      (\bar{q}_L^i \gamma^\mu\sigma_a q_L^j)$ &
      $(\mathcal{O}_{\phi q}^{(3)})_{ij}$ \\
      &
      &$(\phi^{\dagger} i\overset{\leftrightarrow}{D}_\mu \phi)
      (\bar{u}_R^i \gamma^\mu u_R^j)$ &
      $\left(\mathcal{O}_{\phi u}\right)_{ij}$ &
      &
      $(\phi^{\dagger} i\overset{\leftrightarrow}{D}_\mu \phi)
      (\bar{d}_R^i \gamma^\mu d_R^j)$ &
      $\left(\mathcal{O}_{\phi d}\right)_{ij}$ \\
      &
      &$(\tilde{\phi}^{\dagger} iD_\mu \phi)
      (\bar{u}_R^i \gamma^\mu d_R^j)$ &
      $\left(\mathcal{O}_{\phi ud}\right)_{ij}$ &
      & & \\[1mm]
      \bottomrule
    \end{tabular}
\end{center}
\end{table}

Working in the unitary gauge and performing suitable redefinition of fields and input parameters the effective Lagrangian
 can be conveniently expressed in the parameterization of~\cite{Falkowski:2001958,Falkowski:2015fla}, the so-called {\it Higgs basis}. Considering only the terms that are relevant for our analysis, we can identify five classes of terms\footnote{In this paper we shall refer to the doublet Higgs field as $\phi$. After symmetry breaking the field for the Higgs boson will be referred to as $h$. The Higgs particle will be referred to as $H$.}
 \vskip0.2truecm
 \noindent{\bf{-- Higgs trilinear:}}
\begin{equation}
\Delta {\cal L}^{\rm h,self}_{6}= -\delta \lambda_3\, v h^3.
\label{eq:Lh3}
 \end{equation}  
The impact of this coupling in single Higgs processes and its extraction from Higgs pair production will be discussed in Section~\ref{selfcoupling}.
 \vskip0.2truecm
  \noindent{\bf{-- Higgs couplings to vector bosons:}}
\begin{eqnarray}
\label{eq:LhVV}
\Delta {\cal L}^{\rm hVV}_{6}   &= & {\frac{h}{v}} \left [ 
\vphantom{\frac{1}{2}}
2 \delta c_w\,   m_W^2 W_\mu^+ W_\mu^- +   \delta c_z\,   m_Z^2 Z_\mu Z_\mu
\right . \nonumber\\ & & \left . 
+ c_{ww}\,  {\frac{g^2}{2}} W_{\mu \nu}^+  W_{\mu\nu}^-  
+ c_{w \Box}\, g^2 \left (W_\mu^- \partial_\nu W_{\mu \nu}^+ + {\rm h.c.} \right )  
\right . \nonumber\\ & & \left . 
+  c_{gg}\, {\frac{g_s^2}{4}} G_{\mu \nu}^a G_{\mu \nu}^a   + c_{\gamma \gamma}\, {\frac{e^2}{4}} A_{\mu \nu} A_{\mu \nu} 
+ c_{z \gamma}\, {\frac{e \sqrt{g^2+g^{\prime~\!2}}}{2}} Z_{\mu \nu} A_{\mu\nu} 
\right . \nonumber\\ & & \left .
+ c_{zz}\, {\frac{g^2+g^{\prime~\!2}}{4}} Z_{\mu \nu} Z_{\mu\nu}+ c_{z \Box}\, g^2 Z_\mu \partial_\nu Z_{\mu \nu} + c_{\gamma \Box}\, g g^\prime Z_\mu \partial_\nu A_{\mu \nu}
\vphantom{\frac{1}{2}}
\right ],
\end{eqnarray}
where only $c_{gg}, \  \delta c_z,  \ c_{\gamma \gamma}, \ c_{z \gamma},  \ c_{zz},  \ c_{z \Box}$ are independent parameters:
\begin{eqnarray}
\label{eq:dep_pars}
\delta  c_{w} &=&  \delta c_z + 4 \delta m , 
\nonumber\\
c_{ww} &=&  c_{zz} + 2 \sin^2{\theta_w} c_{z \gamma} + \sin^4{\theta_w} c_{\gamma \gamma}, 
\nonumber\\
c_{w \Box}  &= & {\frac{1}{g^2 - g^{\prime~\!2}}} \left [ 
g^2 c_{z \Box} + g^{\prime~\!2} c_{zz}  - e^2 \sin^2{\theta_w}   c_{\gamma \gamma}  -(g^2 - g^{\prime~\!2}) \sin^2{\theta_w}  c_{z \gamma} 
\right ],  
\nonumber\\
  c_{\gamma \Box}  &= &  
  {\frac{1}{g^2 - g^{\prime~\!2}}} \left [ 
2 g^2 c_{z \Box} + (g^2+ g^{\prime~\!2}) c_{zz}  - e^2  c_{\gamma \gamma}  -(g^2 - g^{\prime~\!2})   c_{z \gamma} 
\right ], 
\end{eqnarray}
where $\theta_w$ denotes the weak mixing angle while $\delta m$ is an independent parameter from ${\cal L}_6$ controlling  the deviation of $m_W^2$ with respect to its tree level SM value.
\vskip0.2truecm
  \noindent{\bf{-- Trilinear Gauge Couplings:}}
  \begin{eqnarray}
   \label{eq:LaTGC}
\Delta {\cal L}^{\mathrm{aTGC}} &= & i e \delta \kappa_\gamma\, A^{\mu\nu} W_\mu^+ W_\nu^- \nonumber\\
+
&&i g  \cos{\theta_w} \left[ \delta g_{1Z}\, (W_{\mu\nu}^+ W^{-\mu} - W_{\mu\nu}^- W^{+\mu})Z^\nu + (\delta g_{1Z}-\frac{g^{\prime~\!2}}{g^2}\delta \kappa_\gamma)\, Z^{\mu\nu} W_\mu^+ W_\nu^- \right] 
\nonumber\\
&&+\frac{ig \lambda_z}{m_W^2}\left( \sin{\theta_w} W_{\mu}^{+\nu} W_{\nu}^{-\rho} A_{\rho}^{\mu} + \cos{\theta_w} W_{\mu}^{+\nu} W_{\nu}^{-\rho} Z_{\rho}^{\mu} \right),
  \end{eqnarray}
  where of the three coefficients $g_{1,z}$ and $ \delta \kappa_\gamma$ depend on $c_{gg}, \  \delta c_z,  \ c_{\gamma \gamma}, \ c_{z \gamma},  \ c_{zz},  \ c_{z \Box}$:
\begin{eqnarray}
 \delta  g_{1,z} &=& 
{\frac{1}{2} (g^2 - g^{\prime~\!2})} \left [   c_{\gamma\gamma} e^2 g^{\prime~\!2} + c_{z \gamma} (g^2 - g^{\prime~\!2}) g^{\prime~\!2}  - c_{zz} (g^2 + g^{\prime~\!2}) g^{\prime~\!2}  - c_{z\Box} (g^2 + g^{\prime~\!2}) g^2 \right ], 
 \nonumber\\
 \delta \kappa_\gamma  &=& - {\frac{g^2}{2}} \left ( c_{\gamma\gamma} {\frac{e^2 }{g^2 + g^{\prime~\!2}}}   + c_{z\gamma} \frac{g^2  - g^{\prime~\!2}}{ g^2 + g^{\prime~\!2}} - c_{zz} \right ), 
 \end{eqnarray}  
while $\lambda_z$ is  an independent parameter.

\vskip0.2truecm
  \noindent{\bf{-- Yukawa couplings:}}
\begin{equation}
\label{eq:hff}
\Delta {\cal L}^{\rm hff}_{6}  =  - {\frac{h}{v}} \sum_{f \in u,d,e}    \hat \delta y_f \, m_f  \bar{f} f  + {\rm h.c.} , 
\end{equation}
where $\hat \delta y_f \, m_f$ should be thought as $3\times 3$ matrices in flavour space. FCNC are avoided when $\hat \delta y_f$ is diagonal in the same basis as $m_f$.
Under the assumption of {\it Flavour Universality}
 $(\hat \delta y_f)_{ij} \equiv \delta y_f \times \delta_{ij}$, corresponding to a total of three parameters $\delta y_u,\,\delta y_d,\, \delta y_e$. The assumption of {\it Neutral Diagonality} corresponds instead to $(\hat \delta y_f)_{ij} \equiv \delta (y_f)_i \times \delta_{ij}$ (no summation) corresponding to 9 parameters $\delta_u,\,\delta_c,\,\delta_t$ for the ups and similarly for downs and charged leptons.
 In practice only $\delta_{t,c}$, $\delta_b$ and $\delta_{\tau,\mu}$ are expected to matter in plausible models and in the experimental situations presented by all future colliders. This adds two parameters with respect to  {\it Flavour Universality}.
 \vskip0.2truecm
  \noindent{\bf{-- Vector couplings to fermions:}}
  \begin{eqnarray} 
\label{eq:vff}
\Delta {\cal L}^{vff,hvff}_{6}&=& \frac{g}{\sqrt{2}} \left(1+2 \frac{h}{v}\right)   W_\mu^+   \left (
 \hat \delta g^{W \ell }_L \bar{\nu}_L \gamma^\mu  e_L
+   \hat \delta g^{Wq}_L \bar{u}_L \gamma^\mu  d_L
+ \hat \delta g^{Wq}_R  \bar{u}_R  \gamma^\mu   d_R
 + \hc  \right )
\nonumber\\ 
&+ &\sqrt{g^2 + g^{\prime~\!2}}  \left(1+2 \frac{h}{v} \right)  Z_\mu 
\left [ \sum_{f = u,d,e,\nu}  \hat  \delta g^{Zf}_L  \bar{f}_L \gamma^\mu f_L  +  
\sum_{f = u,d,e}  \hat  \delta g^{Zf}_R  \bar{f}_R \gamma^\mu f_R  \right ]
\nonumber\\ 
&&
\end{eqnarray}
where, again, not all terms are independent\footnote{Here we choose a slightly different convention for the dependent couplings with respect to~\cite{Falkowski:2001958,Falkowski:2015fla}, and we express everything in terms of the modifications of the neutral currents.}:
\begin{equation} 
\hat \delta g^{W \ell }_L =\hat  \delta g^{Z\nu}_L - \hat \delta g^{Ze}_L,~~~~~~~\hat \delta g^{Wq}_L = \hat \delta g^{Z u}_L V_{CKM} - V_{CKM} \hat \delta g^{Z d}_L.
\end{equation} 
In the case of {\it Flavour Universality}, all the $\hat \delta g$ are proportional to the identity corresponding to a total of 8 parameters:
$(\hat \delta g^{Z u}_L)_{ij}\equiv \delta g^{Z u}_L \times \delta_{ij}$, etc.
However the right handed charged current, associated with $\hat \delta g^{Wq}_R$ does not interfere with the SM amplitudes in the limit $m_q\to 0$ and can be neglected, reducing the number of parameters to 7.

In the case of {\it Neutral Diagonality}, the  
assumption  ${\hat \delta g}_{ij}\propto \delta_{ij}$ is relaxed, allowing for the four coefficients associated with the third quark family $(\hat \delta g^{Z u}_L)_{33},\, (\hat \delta g^{Z d}_L)_{33},\,(\hat \delta g^{Z u}_R)_{33},\,(\hat \delta g^{Z d}_R)_{33}$ as well as all diagonal coefficients associated with leptons to be different.
This adds 10 further  parameters with respect to the flavour Universal case.

In conclusion considering single Higgs and EW processes (i.e. neglecting the Higgs trilinear)
in the scenarios of {\it Flavour Universality} and {\it Neutral Diagonality} we end up with respectively 18 and 30 independent parameters~\footnote{The impact at NLO of the relatively poorly constrained Higgs self-coupling on the determination of the single-Higgs couplings will be discussed in Section~\ref{selfcoupling}.}: 
\begin{eqnarray}
{\rm SMEFT}_{\rm FU}
&\equiv &\left\{\delta m, \ c_{gg}, \  \delta c_z,  \ c_{\gamma \gamma}, \ c_{z \gamma},  \ c_{zz},  \ c_{z \Box},  \  \delta y_u, \   \delta y_d,  \ \delta y_e, \ \lambda_z\right\} \nonumber\\
& & +\left\{\delta g^{Zu}_L, \delta g^{Zd}_L, \delta g^{Z\nu}_L, \delta g^{Ze}_L, \delta g^{Zu}_R, \delta g^{Zd}_R, \delta g^{Ze}_R \right\},
\label{eq:SMEFT_FU}
\\
{\rm SMEFT}_{\rm ND}&\equiv& \left\{\delta m, \ c_{gg}, \  \delta c_z,  \ c_{\gamma \gamma}, \ c_{z \gamma},  \ c_{zz},  \ c_{z \Box},  \  \delta y_{t}, \  \delta y_{c}, \   \delta y_b,  \ \delta y_{\tau}, \ \delta y_{\mu}, \ \lambda_z\right\} \nonumber\\
&&+\left\{(\delta g^{Zu}_L)_{q_i}, (\delta g^{Zd}_L)_{q_i}, (\delta g^{Z\nu}_L)_{\ell}, (\delta g^{Ze}_L)_{\ell}, (\delta g^{Zu}_R)_{q_i}, (\delta g^{Zd}_R)_{q_i}, (\delta g^{Ze}_R)_{\ell} \right\}_{q_1=q_2\not=q_3,~\ell = e, \mu, \tau}.
\label{eq:SMEFT_ND}
\end{eqnarray}
While we have chosen to present the degrees of freedom used in the different fitting scenarios described above using the parameterization of the Higgs basis, one can of course do the same in any other basis. In particular, the mapping between the Higgs basis parameters in the previous Lagrangians and the Wilson coefficients in other popular dimension-6 bases in the literature can be found in Section 3 and appendices A and B in \cite{Falkowski:2001958}. 

The previous two scenarios will be used to study the sensitivity at future colliders to general departures from the SM in the global fit to EW precision observabkles (EWPO), Higgs boson rates and diboson production. 
We will, however, also consider another more simplified scenario, designed exclusively to study (1) the interplay between the EW and Higgs constraints, and (2) the impact of the SM theory uncertainties in Higgs boson processes. 
The impact of the EW precision constraints on Higgs boson measurements will be illustrated comparing the results of the fit in the SMEFT$_{\rm ND}$ scenario, with the analogous ones assuming the electroweak precision observables are known with infinite accuracy, both from experiment and theory. We will refer to this idealized case as a scenario with {\it perfect EW} constraints. In practice, this means that any new physics contributions to the EWPO are bounded to be exactly zero. This includes all possible corrections to the $Vff$ vertices as well as any possible modification to the $W$ mass, i.e.
\begin{equation}
\left\{\delta m, (\delta g^{Zu}_L)_{q_i}, (\delta g^{Zd}_L)_{q_i}, (\delta g^{Z\nu}_L)_{\ell}, (\delta g^{Ze}_L)_{\ell}, (\delta g^{Zu}_R)_{q_i}, (\delta g^{Zd}_R)_{q_i}, (\delta g^{Ze}_R)_{\ell} \right\}\equiv 0.
\end{equation}
As also mentioned above, in this scenario it is also implicit that the SM theory uncertainties on EWPO are negligible, which makes it suitable to isolate the effect of the SM theory uncertainties in Higgs processes in the fit. Imposing the previous constraints in Eq.~(\ref{eq:SMEFT_ND}) we are thus left with a total of 12 parameters for this scenario assuming {\it perfect EW} constraints:
\begin{eqnarray}
{\rm SMEFT}_{\rm PEW}&\equiv& \left\{\ c_{gg}, \  \delta c_z,  \ c_{\gamma \gamma}, \ c_{z \gamma},  \ c_{zz},  \ c_{z \Box},  \  \delta y_{t}, \ \delta y_c, \   \delta y_b,  \ \delta y_{\tau}, \ \delta y_\mu, \ \lambda_z\right\}.
\label{eq:SMEFT_PEW}
\end{eqnarray}

Finally, while the setup described above aims at some generality, it makes sense to add some perspective on  the nature of the UV theory and  to frame the EFT results in terms of particularly well-motivated scenarios. Understandably, heavy new physics is more visible in low energy observables the more strongly it is coupled. In this respect models with a Composite Higgs (CH) are the natural arena in which to perform indirect studies of new physics. The basic idea of CH models is that  all the degrees of freedom of the SM apart from the Higgs are elementary.
The Higgs instead arises as a bound state from a strong dynamics. In the simplest possible situation such dynamics is roughly described by two parameters, the overall  mass scale  and its overall coupling strength,  respectively   $m_*$ and  $g_*$. The prototypical template for such a two-parameter description is offered by large $N$ gauge theories, which are characterized by the overall mass of their resonances ($m_*$) and by their mutual coupling $g_* \sim 4\pi /\sqrt N$. Concrete and largely calculable realizations of the scenario have been constructed 
in the context of warped compactifications and of their holographic interpretation, for reviews see e.g.~\cite{Contino:2010rs,Panico:2015jxa} (there are also attempts to build explicit composite models in 4D, see e.g.~\cite{Ferretti:2013kya,Cacciapaglia:2014uja}).
Of course, as in all matters, it is easy to imagine more elaborate situations, but at the very least the minimal case can provide a first perspective on future machines. Indeed a more interesting variation concerns the top quark, which in motivated scenarios can become partially and even fully composite. 
Under the assumptions described in \cite{Giudice:2007fh,Liu:2016idz}, the low energy signatures of these kind of models can be parameterized in terms of the following effective Lagrangian:
\begin{equation}
\begin{split}
{\cal L}_{\mathrm{SILH}}=&
\frac{c_\phi }{\Lambda^2} \frac 12 \partial_\mu (\phi^\dagger \phi)\partial^\mu (\phi^\dagger \phi)+ 
\frac{c_T}{\Lambda^2} \frac 12(\phi^\dagger \lrD_\mu \phi)(\phi^\dagger \lrD^\mu \phi) - 
\frac{c_6}{\Lambda^2} \lambda (\phi^\dagger \phi)^3\\
& +\left(\frac{c_{y_f} }{\Lambda^2} y^f_{ij} \phi^\dagger \phi  \bar{\psi}_{Li} \phi \psi_{Rj} + \hc \right)\\
&
+\frac{c_W}{\Lambda^2} \frac{ig}{2}\left(\phi^\dagger \lrDa_\mu \phi\right) D_\nu W^{a~\!\mu\nu} +\frac{c_B}{\Lambda^2} \frac{ig^\prime}{2}\left(\phi^\dagger \lrD_\mu \phi\right) \partial_\nu B^{\mu\nu}\\
& +\frac{c_{\phi W}}{\Lambda^2} i g D_\mu\phi^\dagger \sigma_a D_\nu \phi W^{a~\!\mu\nu}+\frac{c_{\phi B} }{\Lambda^2} i g^\prime D_\mu\phi^\dagger \sigma_a D_\nu \phi B^{\mu\nu}\\
&+\frac{c_{\gamma} }{\Lambda^2} g^{\prime~\!2}  \phi^\dagger \phi B^{\mu\nu}B_{\mu\nu}+\frac{c_{g}}{\Lambda^2} g^{2}_s  \phi^\dagger \phi G^{A~\!\mu\nu}G^A_{\mu\nu}\\
&-\frac{c_{2W}}{\Lambda^2} \frac{g^2}{2} (D^\mu W_{\mu\nu}^a)(D_\rho W^{a~\!\rho\nu})-\frac{c_{2B}}{\Lambda^2} \frac{g^{\prime~\!2}}{2}(\partial^\mu B_{\mu\nu})(\partial_\rho B^{\rho\nu})-\frac{c_{2G}}{\Lambda^2}\frac{g_S^2}{2} (D^\mu G_{\mu\nu}^A)(D_\rho G^{A~\!\rho\nu})\\
&+\frac{c_{3W}}{ \Lambda^2}g^3\varepsilon_{abc}W_{\mu}^{a~\nu}W_\nu^{b~\rho}W_\rho^{c~\mu}+\frac{c_{3G}}{ \Lambda^2} g_S^3 f_{ABC}G_{\mu}^{A~\nu}G_\nu^{B~\rho}G_\rho^{C~\mu},
\end{split}
\label{eq:LSilh}
\end{equation}
where the different Wilson coefficients can be written in terms of the couplings and masses of the resonances, denoted in short by $g_\star \lesssim 4\pi$ and $m_\star$, as
\begin{equation}
\begin{split}
\frac{c_{\phi,6,y_f}}{\Lambda^2}&\sim \frac{g_\star^2}{ m_\star^2}\equiv \frac{1}{f^2},~~~~~~~~~~~~~~~~~~~~~~~
\frac{c_{T}}{\Lambda^2}\sim \frac{y_t^4}{16\pi^2}\frac{1}{m_\star^2},~~~~~~~~~~~~~~~~
\\
\frac{c_{W,B,\phi W,\phi B,\gamma,g}}{\Lambda^2}&\sim \frac{1}{m_\star^2},~~~~~~~~~~~~~~~
\frac{c_{2W,2B,2G,3W,3G}}{\Lambda^2}\sim \frac{1}{g_\star^2}\frac{1}{m_\star^2},~~~~~~~~~
\end{split}
\label{eq:SILHpc}
\end{equation}
up to $O(1)$ factors.
The expression for $c_T$ has been derived under the most favorable hypothesis where the new physics preserves custodial symmetry. Note also that, for the  relevant case of a pseudo-Nambu-Goldstone-boson (pNGb) Higgs, $c_{g,\gamma}$ benefit from a further suppression $\sim y_t^2/16\pi^2$. Moreover, in explicit constructions  based on warped compactifications $c_{\phi W,\phi B,3W,3G}$ arise at "loop level" and have a further suppression $\sim g_\star^2/16\pi^2$, which of course matters only when $g_\star$ is not maximally strong.

A few remarks concerning the above effective Lagrangian are in order. First, notice that the only effects enhanced by the strong coupling $g_*$ are those on the first line and  involving non linearities in the Higgs field. 
That is not surprising given that in CH, the Higgs itself is strongly interacting while the other SM degrees of freedom are not. In view of that, see discussion in Section~\ref{futurestudies}, in CH the measurements of Higgs couplings compete very well with much more precise measurements, like EWPT, which are  not directly zooming in on the strongly coupled nature of the Higgs boson. 
Second, notice that in CH the whole set  $\psi^2\phi^2 D$ is subdominant and neglected in lowest approximation. However, the operator basis used above, which is the one naturally dictated by the structure of the model, is not precisely the one we used for our global analysis. In particular, the operators associated with $c_{2W,2B,2G}$ can be turned, by a field redefinition, into a particular combination of 4-fermion operators and one particular and flavour universal  combination of the $\psi^2\phi^2 D$. 
Third, the CH models, when considering $gg \to HH $ at high energy,  offer a nice example of dim-8 operators potentially winning over dim-6 ones. Indeed, as mentioned above, when the Higgs is a composite pNGb, the coefficient of the dim-6 operator is further suppressed by a top loop factor $y_t^2/16\pi^2$ \cite{Azatov:2015oxa}. However that is not the case for the dim-8 operator $D_\rho \phi^\dagger D^\rho \phi G_{\mu\nu}^A G^{A~\!\mu\nu}$ which simply comes with coefficient $\sim g_s^2/{m_*^4}$. 
One can then easily see that when the experimental accuracy in the measurement of $gg\to HH$  is worse than $O(y_t^2/16\pi^2)$, the sensitivity on $m_*$ is dominated by the dim-8 operator.

Although the particular structure of the previous Lagrangian is not fully general, it provides a theoretically sound
benchmark to interpret the results of our studies from a more BSM-oriented perspective. The contributions from the different SILH Wilson coefficients in the Lagrangian~(\ref{eq:LSilh}) to the parameters of the Higgs basis can be found in~\cite{Falkowski:2001958}.



\subsection{Results from the EFT framework studies} \label{eft-results}

In the previous section we have detailed the counting of the degrees of freedom that enter in the different SMEFT fit scenarios using the so-called Higgs basis. While physical results do not depend on the
choice of basis, in some cases a particular basis may be convenient for computational, presentational or interpretational purposes (note that the physical interpretation of each dimension-six operator does depend on the basis). 
From the point of view of the results presented in this section, however, we are mostly interested in comparing the sensitivity to deformations with respect to the SM in the Higgs couplings at the different future collider projects. 
To assess these deformations with respect to the SM
in a basis-independent way one can {\it project} the results of the SMEFT fit onto a set of on-shell properties of the Higgs boson, via the following {\it Higgs effective couplings}:
\begin{equation}
g_{HX}^{\mathrm{eff}~2}\equiv \frac{\Gamma_{H\to X}}{\Gamma_{H\to X}^{\SM}}.
\label{eq:gHeff}
\end{equation}
By definition, these quantities, constructed from physical observables, are basis independent. These definitions are also convenient to compare in a straightforward manner the SMEFT results with those of the $\kappa$ framework for the single Higgs couplings. 
Such definition is, however, not phenomenologically possible for the top-Higgs coupling and the Higgs self-interaction. 
For the present report we will sidestep these issues by: 
(1) defining the effective top coupling in a similar way to all other fermions; 
(2) to connect and compare with all current studies of the Higgs self-interaction, we will define $g_{HHH}\equiv \lambda_3 /\lambda_3^{\SM}$.

Note that, at the dimension-six level and truncating the physical effects at order $1/\Lambda^2$ one can always express the previous effective couplings in terms of the dimension-six operators via a linear transformation. Provided one has a large enough set of such effective couplings, one can then map the effective coupling result into Wilson coefficients, and viceversa (of course, the former are not a basis per se and the connection is only well-defined at a fixed order in perturbation theory and in the EFT expansion).
The single Higgs couplings plus $g_{HHH}$ are however not enough to match the number of free parameters in the SMEFT fits, even in the simplified scenario SMEFT$_{\rm PEW}$ in eq.~(\ref{eq:SMEFT_PEW}).
In particular, the on-shell couplings $g_{HZZ,HWW}^{\mathrm{eff}}$ in eq.~(\ref{eq:gHeff}) do not capture all possible linear combinations of the different types of EFT interactions contributing to the $HZZ$ and $HWW$ vertices.\footnote{We note, however, that, from the point of view of the interpretation in terms of motivated scenarios like those described below Eq.~(\ref{eq:SILHpc}), the contributions to such interactions are dominated only by $c_\phi$, unless $g_\star~\sim g$.
}
For that reason we will also present our results by adding the predictions for the anomalous Triple Gauge Coupling (aTGC), a (pseudo)-observable obtained from the di-boson analysis. These extra parameters offer a measure of the Higgs couplings to gauge bosons with a non-SM Lorentz structure. As long as we restrict the analysis to observables around the Higgs mass scale, this approach with on-shell effective couplings and aTGC is perfectly appropriate. When high-energy observables are considered, like in Section~\ref{sec:SILHfit}, it would have to be revisited. (In that section, however, we will present the results directly in terms of the Wilson coefficients, for easier interpretation in terms of BSM scenarios.) 
Even after adding the aTGC, in the SMEFT$_{\rm PEW}$ scenario where $\delta m\equiv 0$ the $g_{HZZ,HWW}^{\mathrm{eff}}$ couplings are not independent, and therefore we will present the results reporting only the coupling to $Z$ bosons. 

In the global fit scenarios SMEFT$_{\rm FU}$ and SMEFT$_{\rm ND}$, where we also add those combinations of operators that can contribute to EWPO, extra information needs to be added to illustrate the constraints on the different degrees of freedom included in the fit. Since $\delta m$ is now a free parameter, we report separately the $g_{HZZ,HWW}^{\mathrm{eff}}$ couplings.
Following a similar approach as for the Higgs couplings, one can report the sensitivity to modifications in the effective couplings of the $Z$ to fermions, which can be defined from the $Z$-pole measurements of the $Z$ decays and asymmetries, e.g.
\begin{equation}
\Gamma_{Z\to e^+ e^-}=\frac{\alpha ~\! M_Z}{6 \sin^2{\theta_w} \cos^2{\theta_w}}(|g_L^e|^2 + |g_R^e|^2) ,\quad\quad A_e=\frac{|g_L^e|^2 - |g_R^e|^2}{|g_L^e|^2 + |g_R^e|^2}.
\label{eq:gZeff}
\end{equation}

In what follows, we discuss the results of the SMEFT fit from the point of view of the expected sensitivity to modifications of the Higgs couplings in the scenarios SMEFT$_{\rm FU}$ and SMEFT$_{\rm ND}$. As was done in the fits in the $\kappa$ framework, we will present the results assuming that at future colliders only the SM theory uncertainties associated with the knowledge of the SM input parameters are non-negligible.
(As also discussed there, for the HL-LHC and HE-LHC scenarios we always consider the uncertainties adopted by the studies in \cite{Cepeda:2019klc}.)
The impact of these and other SM theory uncertainties in Higgs processes will be discussed afterwards in Section~\ref{sec:SMThUnc}, using for that purpose the results in the benchmark SMEFT$_{\rm PEW}$.

\subsubsection{SMEFT fit results \label{sec:SMEFTfit}} 

\begin{figure}[!ht]
\centering
\includegraphics[width=0.9\linewidth]{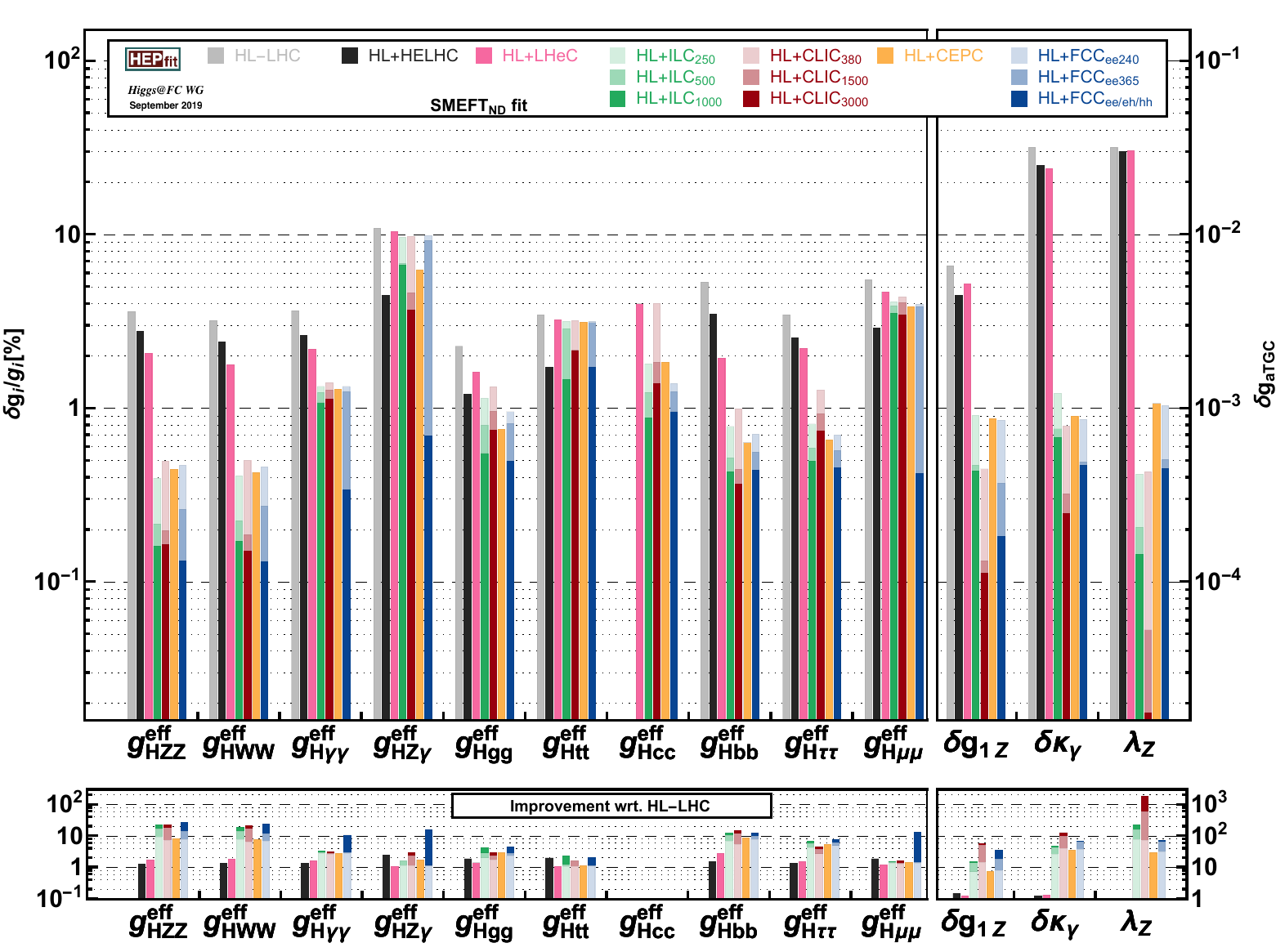}
\caption{\label{fig:SMEFT2prime}
Sensitivity at 68\% probability to deviations in the different effective Higgs couplings and aTGC from a global fit to the projections available at each future collider project. Results obtained within the SMEFT framework in the benchmark SMEFT$_{\rm ND}$. The HE-LHC results correspond to the $S_2^\prime$ assumptions for the theory systematic uncertainties in Higgs processes~\cite{Cepeda:2019klc}.
}
\end{figure}

The main results of this section are summarised in Table~\ref{tab:eft-global}, where we compare the 68$\%$ probability sensitivity to deviations in the Higgs couplings from the global SMEFT fit to Higgs, di-boson and EWPO at future colliders. We show the projections for the fits with and without flavour universality assumptions, given by the scenarios SMEFT$_{\rm FU}$ and SMEFT$_{\rm ND}$, respectively. 
Note that the SMEFT$_{\rm ND}$ scenario not only has $g_{Htt}^{\rm eff}\not = g_{Hcc}^{\rm eff}$, $g_{H\tau\tau}^{\rm eff}\not = g_{H\mu\mu}^{\rm eff}$, but also treats in a family-dependent way the corrections to $Z f\bar{f}$ couplings, which typically leads to less stringent constraints from EWPO. The impact of the EWPO in the fit will be discussed below.
The results for the more general scenario SMEFT$_{\rm ND}$ are also shown in Figure~\ref{fig:SMEFT2prime} 
where we compare the results across colliders. In the lower panel of Figure~\ref{fig:SMEFT2prime}
we also show the relative improvement compared to the HL-LHC results. In both
table and figure we illustrate the impact of the data taking at different energy stages at each collider. As in the previous sections, we distinguish between the initial energy stage when each collider can start operating as a Higgs factory, and subsequent upgrades to higher energies. In the case of FCC, we also consider the results in combination with the other collider options foreseen as part of the FCC integrated program.

\begin{figure}[!ht]
\centering
\includegraphics[width=0.9\linewidth]{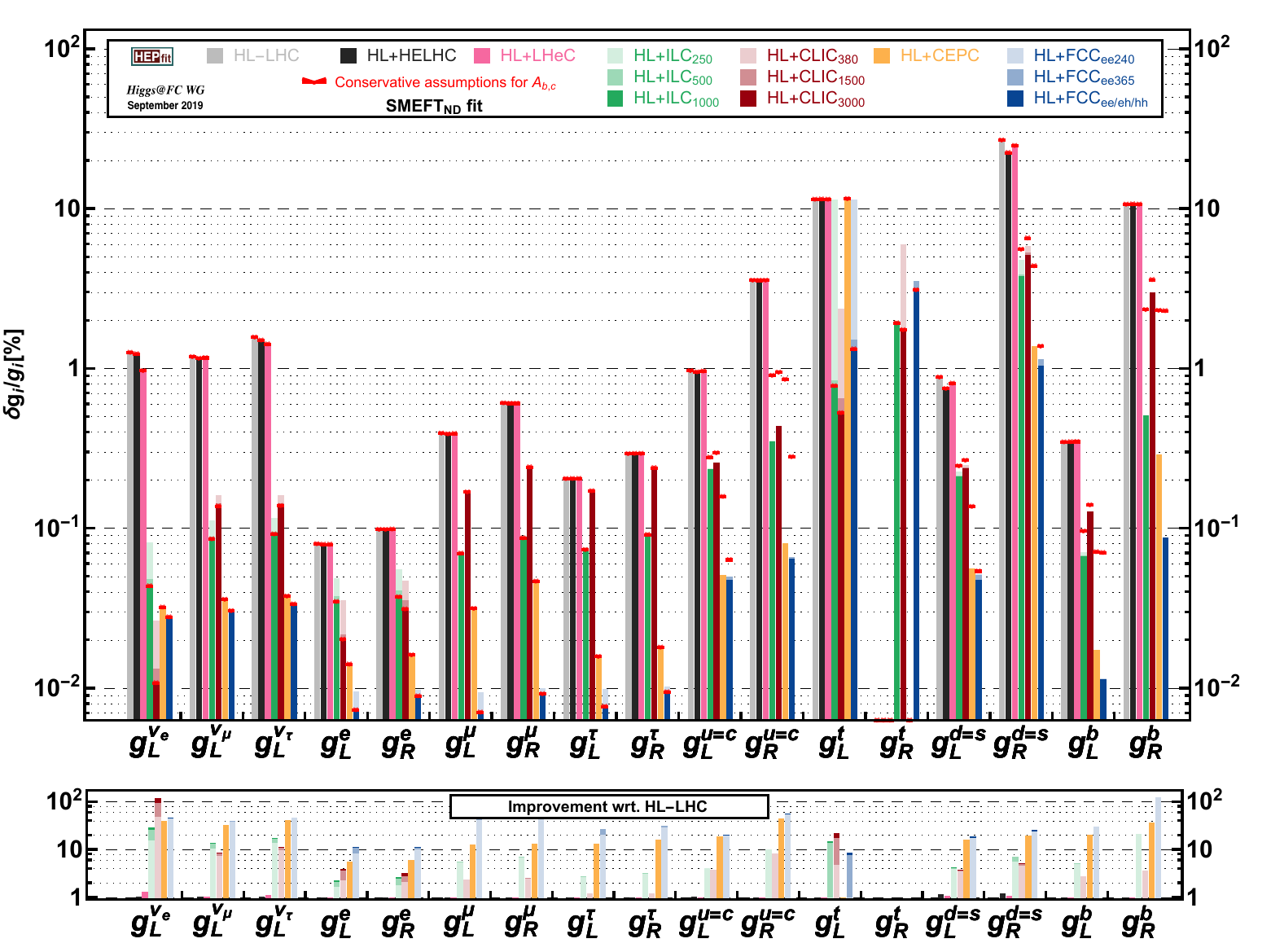}
\caption{\label{fig:SMEFT2primeEW}
Sensitivity at 68\% probability to deviations in the different EW couplings from a global fit to the projections available at each future collider project. Results obtained within the SMEFT framework in the benchmark SMEFT$_{\rm ND}$. Note that $Z$-radiative return measurements at ILC and CLIC are included in the fit. Two different assumptions are considered for the systematic errors. The HE-LHC results correspond to the $S_2^\prime$ assumptions for the theory systematic uncertainties in Higgs processes~\cite{Cepeda:2019klc}.
See text for details. 
}
\end{figure}

Although in this section we will be mainly interested in the comparison of the sensitivities to modifications of Higgs couplings, for completeness we show in Figure~\ref{fig:SMEFT2primeEW} and Table~\ref{tab:eft-global-EW-Zff} the results of the remaining degrees of freedom included in the SMEFT$_{\rm ND}$ fit, i.e. the precisions for the corresponding $Z\bar{f}f$ couplings. 
These are constrained mainly by the future projections for EWPO. In this regard, it must be noted that, unlike most of the Higgs results, where the uncertainties are expected to be controlled by the statistical component, the future projections for EWPO are expected to be dominated, in most cases, by systematic errors. 
Because of that, the results for the $Zff$ couplings have a significant dependence on what assumptions are made by the different collider projects in terms of these systematics. 
Whenever large differences between these assumptions were identified, we tried to unify them in order to provide a more coherent comparison. This is the case of the results for heavy flavour measurements of the $Z$ properties ($A_{b,c}$ and $R_{b,c}$), where clearly different assumptions were made in terms of the expected size of future theory uncertainties associated with QCD corrections. These are expected to be collider independent (i.e. apply equally to linear or circular collider) and greatly affect the projections for the heavy flavor asymmetries $A_f$. Because of this, we chose 2 different scenarios for the systematics applied to these observables. We take as a base scenario one where the systematic uncertainties on the asymmetries are given by the main ``collider-dependent'' uncertainty quoted by each project. For linear colliders, where $A_f$ are determined from a left-right forward backward asymmetry, this is the uncertainty on the knowledge of the beam polarization. In absence of polarization, at circular colliders the $A_f$ parameters are derived from an unpolarized forward-backward asymmetry, $A_{FB}^f =\frac{3}{4} A_e A_f$, and therefore are subject to the uncertainty associated to the knowledge of $A_e$. To illustrate the impact of the QCD uncertainties, in Figure~\ref{fig:SMEFT2primeEW} we compare the result of this first scenario 
with a different one, obtained assuming the QCD uncertainties  at future lepton colliders will be reduced
by a factor of 2 compared to LEP. (The results for this latter scenario are indicated with the red marks in the figure.)
In any case, the difference in the results between similar machines must therefore be interpreted with caution. For instance, the final CEPC capabilities from the point of view of the EWPO should not be significantly different than those for FCC-ee, at least regarding those measurement possible below 240\,GeV.\footnote{The absence of a run around the $t\bar{t}$ threshold would, however, prevent measuring the top quark mass with increased precision, which is also a key observable in the EWPO analysis.}
Finally, the scenarios considered here for linear colliders correspond to the baseline presented by the corresponding projects, which do not foresee a $Z$-pole run. Some results including that possibility, i.e. the Giga-Z factory, are presented in Appendix~\ref{app:addons}. 

Focusing our attention on the results for the Higgs couplings, from the results we observe that the LHeC and HE-LHC would help in pushing the knowledge of some of the Higgs couplings close to the 2\% level.
This may be surprising compared to the results obtained in the $\kappa$ framework (kappa-0), especially for the LHeC case, where the couplings to $W$ and $Z$ bosons were obtained with slightly below 1\% accuracy. This deterioration in the precision of the EFT results is due to the absence of projections for improved measurements of the aTGC. This limits the constraining power on the non-SM tensor structures that are present in the EFT formalism but not in the $\kappa$ framework.
One must also note that the improvement at the HE-LHC $S_2^\prime$ on the Higgs couplings is mostly dominated by the assumptions on the reduction of theory and systematics with respect to HL-LHC which, as explained in Section~\ref{sec:method}, are reduced by fiat, 
rather than by a detailed workplan for the reduction of uncertainties. If such hypothesised improvement is not realised, the HE-LHC reach would be, with a few exceptions, not far from the HL-LHC one, as illustrated by the HE-LHC $S_2$ results in Table~\ref{tab:eft-global}.
A future lepton collider could achieve below 1-percent accuracy for several of the $g_{HX}^{\rm eff}$ parameters.

Even at a low energy run, all future lepton colliders can bring the precision of the Higgs coupling to vector bosons to the $0.5\%$ level or below (note also that lepton colliders are the only type of Higgs factory able to provide an absolute normalization for the Higgs couplings, via the measurement of the $e^+ e^- \to ZH$ cross section using the recoil mass method). With similar luminosities collected at 240 GeV, the overall performances of CEPC and the 240 GeV run of FCC-ee are expected to be comparable.\footnote{The differences between the CEPC and FCC-ee results at 240 GeV are simply due to the details of the available projections from each collider project. In particular, the better sensitivity to the $HZ\gamma$ coupling at CEPC is simply due to the absence of a projections for the $H\to Z\gamma$ channel at the FCC-ee. 
} 
In particular, both machines would be able to measure the effective $HZZ$ coupling with a precision of $\sim 0.5\%$.
After running at 365\,GeV and completing the 14 year physics program of the FCC-ee collider\footnote{Note that this also includes the runs at the $Z$ pole and $WW$ threshold, which are crucial for the EW precision program. The total run time as Higgs factory is 8 years.} the precision of the $HZZ$ coupling would be further reduced to $\sim 0.3\%$, nearly a factor of 2 improvement. 
This of course could also be achieved at the CEPC, if a similar run at such energies were included in their physics program.
For the ILC, running at 250 GeV would bring a precision of $\sim 0.4\%$ for $g_{HZZ}^{\rm eff}$. This would be pushed down to $0.2\%$ with an increase of the centre-of-mass energy to 500 GeV and after collecting 4 ab$^{-1}$ of data,
with a total combined run time of 22 years. A further ILC upgrade to energies of 1 TeV would bring an extra $\sim 30\%$ gain in precision. 
Finally, the determination of the different Higgs couplings to $W$ and $Z$ bosons obtained from the 380\,GeV run of CLIC would be comparable to that of the circular colliders at 240 GeV. As in the ILC case, the CLIC data taken at high-energies would help to reach/surpass the two per-mille accuracy on the Higgs coupling to vector bosons after the 1.5 TeV/3 TeV run, concluding a 23-year program.

Turning our attention to the Higgs couplings to fermions, a similar pattern of improvements can be observed for the couplings to bottom quark and $\tau$ lepton.
The top quark Yukawa is not directly accessible for lepton colliders running below the $ttH$ threshold. Indeed, below threshold the top quark coupling can only be accessed via its contribution to the SM loop-induced processes, e.g. $H\to gg$. In the EFT framework, however, these can also receive corrections from new local operators, preventing the extraction of $g_{Htt}^{\rm eff}$. In these cases, only a minor improvement can be achieved in the SMEFT$_{\rm ND}$ scenario~\footnote{We remind that in the SMEFT$_{\rm FU}$ scenario, the corrections to the Yukawa interactions of the different fermion families are universal. Therefore, in that scenario, the apparent improvement on the top coupling is in most instances directly linked to the percent level precision of the measurement of the coupling to charm quarks.}, due to the more precise determinations of the other couplings involved in the extraction of $g_{Htt}^{\rm eff}$ from the $ttH$ channels at the HL-LHC.
The high-energy runs of the lepton machines would give access to the $ttH$ threshold.
ILC studies at 500 GeV --included in this study-- project a determination of $g_{Htt}^{\rm eff}$ with a precision $\sim 6$-$7\%$. This could be significantly improved by running slightly above threshold, at 550 GeV, where due to the increased statistics it would be possible to access the same coupling at the $3\%$ level~\cite{Bambade:2019fyw}. 
Similar precision is projected for the CLIC run at 1500 GeV.
Note that in order to take full advantage of these studies 
it is necessary to also have an adequate determination of
the $Zt\bar{t}$ couplings. These also contribute to the $ttH$ process and are not precisely constrained by current data. Here we use the results from \cite{Amjad:2015mma} for ILC at 500 GeV and from \cite{Bambade:2019fyw,Abramowicz:2018rjq} for CLIC.
In any case, these projected uncertainties for $g_{Htt}^{\rm eff}$ 
would still be similar to the one from the HL-LHC determination of the top Yukawa coupling.
Only the FCC project would be able to surpass that precision on its own, after including in the picture the measurements possible at the 100\,TeV $pp$ collider. The improvement in this case comes from the measurement of the $ttH/ttZ$ cross sections, which then also relies on a precise measurement of the $Z\bar{t}t$ coupling. For the FCC this would come from the FCC-ee run at 365 GeV~\cite{Janot:2015yza}.
It should be recalled that in all these studies of the $ttH$ or $ttZ$ processes, both at hadron and lepton colliders, we are making explicit use of the assumption that other interactions such as four-fermion or dipole operators can be neglected. A fully global analysis of these processes has to include those operators as well, including the corresponding constraints.

Finally, even after the full physics program of any of the future leptonic machines, there are several couplings whose precision are still above the one percent threshold, mainly those associated to rare decays and that are statistically limited. Only a future lepton collider combined with a high-luminosity hadron machine like the FCC-hh would be able to bring down all the main Higgs couplings below 1\%, as can be seen in the last column in Table~\ref{tab:eft-global}. 
In this regard, we also note the role of the FCC-eh measurements, which would help to further increase the precision in the determination of the couplings to vector bosons and $b$ quarks, after the completion of the FCC-ee program.

\begin{figure}[t]
\centering
\includegraphics[width=0.9\linewidth]{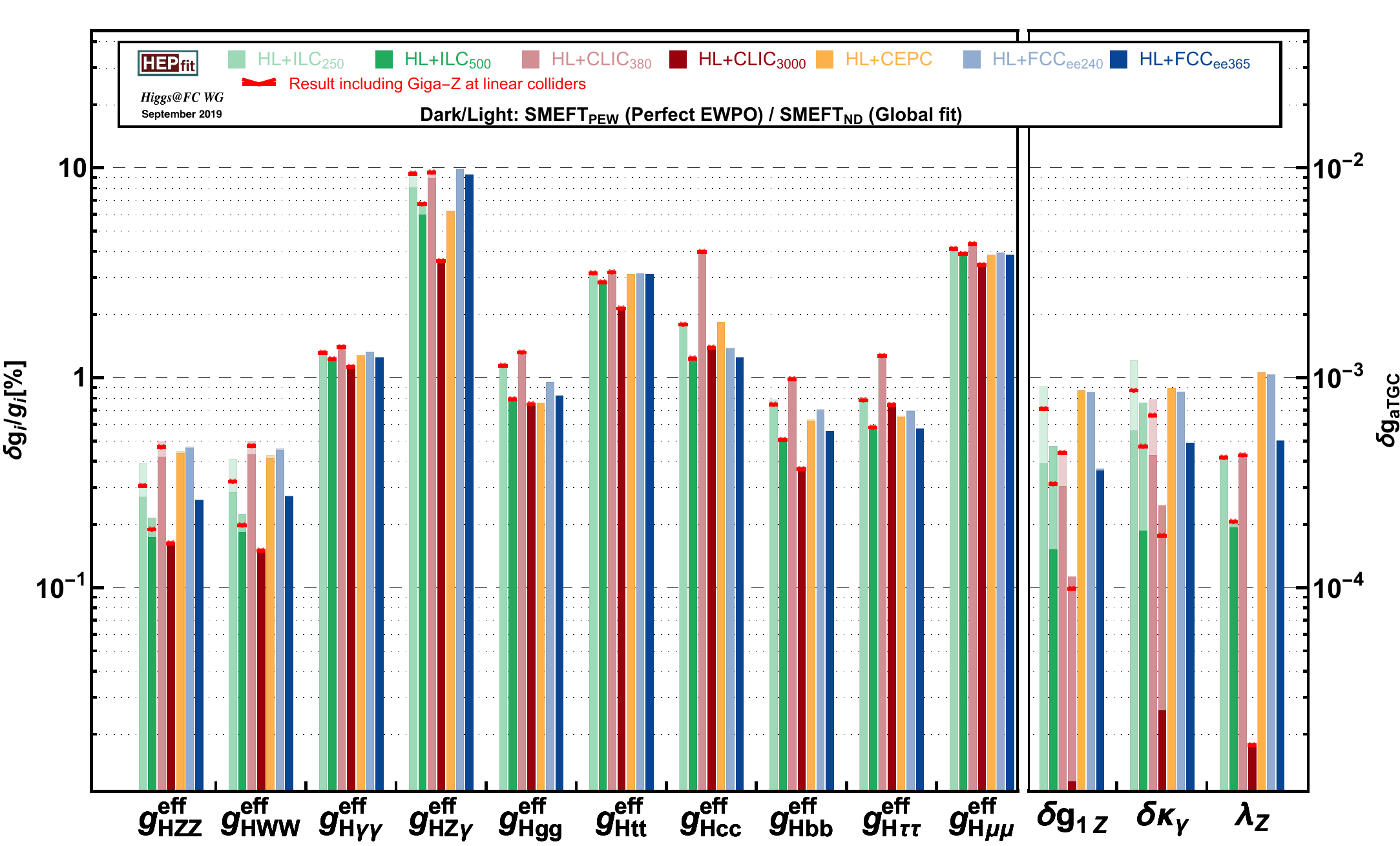}
\caption{\label{fig:SMEFTGlobalPerfect}
68$\%$ probability reach on Higgs couplings and aTGC values for the different lepton colliders from the Global fit SMEFT$_{\rm ND}$, compared with the results obtained assuming infinite precision for the EWPO (scenario SMEFT$_{\rm PEW}$). The difference (partially) illustrates the impact of the EW constraints on the Higgs results. See text for discussion and caveats which apply to this study. The measurements based on $Z$ bosons from radiative return at ILC and CLIC are included in the default fit, and the horizontal red marks indicate the coupling reach when additionally a dedicated $Z$-pole run is taken.
}
\end{figure}

A comparison between the results of the global fit with those obtained assuming perfect EW measurements --scenario SMEFT$_{\rm PEW}$-- illustrates the relative importance of the EWPO in the extraction of the different Higgs couplings from the global fit~\cite{deBlas:2019wgy}. 
Figure~\ref{fig:SMEFTGlobalPerfect} compares the two results for the future Higgs factories at lepton colliders. For what concerns the Higgs couplings, in most cases the impact is quite mild and, in the case of FCC-ee and CEPC, almost nonexistent due to the rich program for measuring the EWPO at the $Z$ pole and above. The default analysis presented in this report includes the preliminary studies of the radiative return process $e^+ e^- \to Z\gamma$ at 250 GeV (380 GeV) with polarized beams at ILC (CLIC). The results are also shown for the case when a Giga-Z run is also included, with on 100~fb$^{-1}$ of data at $\sqrt{s}\sim m_Z$. It is seen that for ILC and CLIC$_{380}$ there is a clear degredation of the uncertainty on the $g_{HVV}$ without the Giga-Z run , which is largely reduced by a dedicated Giga-$Z$ run ~\cite{Bambade:2019fyw,Fujii:2019zll,gigazclic} since the uncertainties on the fermion asymmetries and partial width ratios are reduced by a factor of $\sim 10$ (see Table~\ref{tab:ewkpar}). For these $W$ and $Z$ couplings, such loss of precision can also be minimized by including the information from a high-energy run, as can be seen for CLIC$_{3000}$, where there is little impact on the precision of the same $HVV$ effective couplings. However, for the aTGC parameters $\delta g_{1Z}$ and $\delta \kappa_\gamma$, there is still a substantial degradation compared to perfect knowledge of the EWPO values. A significant improvement in the measurements of the electron EW interactions is therefore still needed, if one wants to extract the maximum precision across all the different couplings at $e^+ e^-$ colliders~\cite{deBlas:2019wgy}.

One must take into account that, with the set of projections available from each future collider project, the global fit results presented here are, in some cases, not entirely consistent, due to some approximations present in the projections for $e^+ e^- \to W^+ W^-$. Indeed, these are typically reported in terms of the precision on the aTGC but, except for the CLIC studies presented in \cite{deBlas:2018mhx}, they are obtained assuming that new physics can only modify $\delta g_{1z}$, $\delta \kappa_\gamma$ and $\lambda_Z$, but not the other couplings involved in the production or decays of the $WW$ pairs \footnote{For the ILC studies~\cite{Barklow:2017awn,Barklow:2017suo,Bambade:2019fyw} part of this dependence is taken into account, adding those contributions from dimension-6 operator coefficients that are enhanced by a factor $s/2 m_W^2$. This approximation, justified in the high-energy limit, may not be a good assumption for the ILC run at 250\,GeV, but should work well for the aTGC projections at 500 GeV. 
(These were not available in \cite{Bambade:2019fyw} and we take them from \cite{Barklow:2017suo}.)}.
 This explains the large difference for those parameters in the CLIC results
 between the global fit and the ones computed under the assumption of perfect EW measurements, see Figure~\ref{fig:SMEFTGlobalPerfect}. The {\it aTGC dominance} assumption was a good approximation at LEP2, due to the comparatively more precise constraints from the Z-pole measurements at LEP/SLD, but is something to be tested at future colliders, especially for those projects where a run at the Z-pole will not happen. In those cases, the results presented here must therefore be interpreted with caution~\cite{deBlas:2019wgy}.  

\begin{table}[ht]
\caption{\label{tab:eft-global} 
Sensitivity at 68\% probability to deviations in the different effective Higgs couplings and aTGC from a global fit to the projections available at each future collider project. Results obtained for the Global SMEFT fit benchmarks denoted as SMEFT$_{\rm FU}$ and SMEFT$_{\rm ND}$ in the text. The numbers for all future colliders are shown {\it in combination} with the HL-LHC results. For ILC and CLIC results from the $Z$ boson radiative return events are included.
}
\centering
\rotatebox{90}{
{\scriptsize
\begin{tabular}{ |c | c | c | c | c c | c c c | c c c | c | c c | c|}
\toprule
        &     &       &\multicolumn{13}{c}{HL-LHC +}  \\
   &Benchmark   &HL-LHC   &LHeC   &\multicolumn{2}{c|}{HE-LHC}   &\multicolumn{3}{c|}{ILC}   &\multicolumn{3}{c|}{CLIC} &CEPC   &\multicolumn{2}{c|}{FCC-ee}   &FCC-ee/eh/hh  \\
   &    &    &    & $S_{2}$   & $S_{2}^\prime$   &  250  & 500   & 1000   & 380   &  1500  & 3000  &    & 240  & 365 &   \\
\midrule
\hspace{-0.2cm}$g_{HZZ}^{\mathrm{eff}}[\%]$& SMEFT$_{\rm FU}$   & $3.2$   & $1.9$   & $2.8$   & $2.4$   & $0.37$   & $0.21$   & $0.16$   & $0.48$   & $0.2$   & $0.16$   & $0.44$   & $0.47$   & $0.26$   & $0.13$  \\
 & SMEFT$_{\rm ND}$   & $3.6$   & $2.1$   & $3.2$   & $2.8$   & $0.39$   & $0.22$   & $0.16$   & $0.5$   & $0.2$   & $0.16$   & $0.45$   & $0.47$   & $0.26$   & $0.13$  \\
\crowcolorA\hspace{-0.2cm}$g_{HWW}^{\mathrm{eff}}[\%]$& SMEFT$_{\rm FU}$   & $2.9$   & $1.6$   & $2.5$   & $2.1$   & $0.38$   & $0.22$   & $0.17$   & $0.49$   & $0.19$   & $0.15$   & $0.42$   & $0.45$   & $0.27$   & $0.13$  \\
\crowcolorA & SMEFT$_{\rm ND}$   & $3.2$   & $1.8$   & $2.8$   & $2.4$   & $0.41$   & $0.22$   & $0.17$   & $0.5$   & $0.19$   & $0.15$   & $0.43$   & $0.46$   & $0.27$   & $0.13$  \\
\hspace{-0.2cm}$g_{H \gamma \gamma}^{\mathrm{eff}}[\%]$& SMEFT$_{\rm FU}$   & $3.3$   & $2.$   & $2.8$   & $2.3$   & $1.3$   & $1.2$   & $1.1$   & $1.4$   & $1.3$   & $1.1$   & $1.3$   & $1.3$   & $1.2$   & $0.33$  \\
 & SMEFT$_{\rm ND}$   & $3.6$   & $2.2$   & $3.1$   & $2.6$   & $1.3$   & $1.2$   & $1.1$   & $1.4$   & $1.3$   & $1.1$   & $1.3$   & $1.3$   & $1.2$   & $0.34$  \\
\crowcolorA\hspace{-0.2cm}$g_{HZ \gamma}^{\mathrm{eff}}[\%]$& SMEFT$_{\rm FU}$   & $11.$   & $10.$   & $6.1$   & $4.2$   & $8.8$   & $6.8$   & $6.6$   & $9.6$   & $4.6$   & $3.6$   & $6.2$   & $9.9$   & $9.3$   & $0.66$  \\
\crowcolorA & SMEFT$_{\rm ND}$   & $11.$   & $10.$   & $6.3$   & $4.5$   & $9.6$   & $6.8$   & $6.7$   & $9.7$   & $4.6$   & $3.7$   & $6.3$   & $9.8$   & $9.3$   & $0.7$  \\
\hspace{-0.2cm}$g_{Hgg}^{\mathrm{eff}}[\%]$& SMEFT$_{\rm FU}$   & $2.3$   & $1.6$   & $1.8$   & $1.2$   & $1.1$   & $0.79$   & $0.55$   & $1.3$   & $0.95$   & $0.74$   & $0.75$   & $0.95$   & $0.81$   & $0.42$  \\
 & SMEFT$_{\rm ND}$   & $2.3$   & $1.6$   & $1.8$   & $1.2$   & $1.1$   & $0.79$   & $0.55$   & $1.3$   & $0.96$   & $0.75$   & $0.76$   & $0.95$   & $0.82$   & $0.49$  \\
\crowcolorA\hspace{-0.2cm}$g_{Htt}^{\mathrm{eff}}[\%]$& SMEFT$_{\rm FU}$   & $3.5$   & $2.5$   & $2.9$   & $1.8$   & $1.6$   & $1.1$   & $0.75$   & $2.5$   & $1.4$   & $1.2$   & $1.6$   & $1.3$   & $1.2$   & $0.66$  \\
\crowcolorA & SMEFT$_{\rm ND}$   & $3.5$   & $3.2$   & $2.8$   & $1.7$   & $3.2$   & $2.9$   & $1.5$   & $3.2$   & $2.2$   & $2.1$   & $3.1$   & $3.1$   & $3.1$   & $1.7$  \\
\hspace{-0.2cm}$g_{Hcc}^{\mathrm{eff}}[\%]$& SMEFT$_{\rm FU}$   & &\multicolumn{7}{c}{Same as $g_{Htt}^{\mathrm{eff}}$} & \multicolumn{6}{c}{Same as $g_{Htt}^{\mathrm{eff}}$}\\ 
 & SMEFT$_{\rm ND}$   & $-$   & $4.$   & $-$   & $-$   & $1.8$   & $1.2$   & $0.88$   & $4.$   & $1.8$   & $1.4$   & $1.8$   & $1.4$   & $1.2$   & $0.95$  \\
\crowcolorA\hspace{-0.2cm}$g_{Hbb}^{\mathrm{eff}}[\%]$& SMEFT$_{\rm FU}$   & $4.9$   & $1.7$   & $4.1$   & $3.1$   & $0.77$   & $0.51$   & $0.42$   & $0.97$   & $0.44$   & $0.37$   & $0.62$   & $0.7$   & $0.56$   & $0.39$  \\
\crowcolorA & SMEFT$_{\rm ND}$   & $5.3$   & $1.9$   & $4.4$   & $3.5$   & $0.78$   & $0.52$   & $0.43$   & $0.99$   & $0.44$   & $0.37$   & $0.63$   & $0.71$   & $0.56$   & $0.44$  \\
\hspace{-0.2cm}$g_{H\tau\tau}^{\mathrm{eff}}[\%]$& SMEFT$_{\rm FU}$   & $3.1$   & $2.$   & $2.6$   & $2.2$   & $0.79$   & $0.58$   & $0.49$   & $1.2$   & $0.91$   & $0.73$   & $0.64$   & $0.69$   & $0.57$   & $0.3$  \\
 & SMEFT$_{\rm ND}$   & $3.4$   & $2.2$   & $2.9$   & $2.5$   & $0.81$   & $0.59$   & $0.5$   & $1.3$   & $0.93$   & $0.74$   & $0.66$   & $0.7$   & $0.57$   & $0.46$  \\
\crowcolorA\hspace{-0.2cm}$g_{H\mu\mu}^{\mathrm{eff}}[\%]$& SMEFT$_{\rm FU}$   & &\multicolumn{7}{c}{Same as  $g_{H\tau\tau}^{\mathrm{eff}}$} & \multicolumn{6}{c}{Same as  $g_{H\tau\tau}^{\mathrm{eff}}$} \\
\crowcolorA & SMEFT$_{\rm ND}$   & $5.5$   & $4.7$   & $3.6$   & $2.9$   & $4.1$   & $3.9$   & $3.5$   & $4.4$   & $4.1$   & $3.5$   & $3.8$   & $4.$   & $3.8$   & $0.42$  \\
\midrule
\hspace{-0.2cm}$\delta g_{1Z}[\times 10^{2}]$& SMEFT$_{\rm FU}$   & $0.63$   & $0.48$   & $0.45$   & $0.39$   & $0.068$   & $0.043$   & $0.04$   & $0.044$   & $0.013$   & $0.011$   & $0.087$   & $0.085$   & $0.036$   & $0.017$  \\
 & SMEFT$_{\rm ND}$   & $0.66$   & $0.52$   & $0.49$   & $0.45$   & $0.091$   & $0.047$   & $0.044$   & $0.045$   & $0.013$   & $0.011$   & $0.087$   & $0.085$   & $0.037$   & $0.018$  \\
\crowcolorA\hspace{-0.2cm}$\delta \kappa_{ \gamma}[\times 10^{2}]$& SMEFT$_{\rm FU}$   & $2.9$   & $2.2$   & $2.4$   & $2.2$   & $0.098$   & $0.069$   & $0.062$   & $0.078$   & $0.032$   & $0.025$   & $0.089$   & $0.086$   & $0.049$   & $0.047$  \\
\crowcolorA & SMEFT$_{\rm ND}$   & $3.2$   & $2.4$   & $2.7$   & $2.5$   & $0.12$   & $0.076$   & $0.068$   & $0.079$   & $0.032$   & $0.025$   & $0.089$   & $0.086$   & $0.049$   & $0.047$  \\
\hspace{-0.2cm}$\lambda_{Z}[\times 10^{2}]$& SMEFT$_{\rm FU}$   & $3.2$   & $3.$   & $3.$   & $3.$   & $0.041$   & $0.02$   & $0.014$   & $0.043$   & $0.0053$   & $0.0018$   & $0.11$   & $0.1$   & $0.05$   & $0.045$  \\
 & SMEFT$_{\rm ND}$   & $3.2$   & $3.$   & $3.$   & $3.$   & $0.042$   & $0.021$   & $0.014$   & $0.043$   & $0.0053$   & $0.0018$   & $0.11$   & $0.1$   & $0.051$   & $0.045$  \\
\bottomrule
\end{tabular}
}
}
\end{table}

\begin{table}[ht]
\caption{\label{tab:eft-global-EW-Zff} 
Sensitivity at 68\% probability to deviations in the different effective EW couplings from a global fit to the projections available at each future collider project. Results obtained for the Global SMEFT fit benchmark denoted as SMEFT$_{\rm ND}$ in the text. The numbers for all future colliders are shown {\it in combination} with the HL-LHC results.
The results for ILC and CLIC are shown without a dedicated Giga-Z run. Appendix~\ref{app:addons} includes the results with a Giga-Z run in  Table~\ref{tab:eft-global-gigazew}.  
}
\centering
\rotatebox{90}{
{\scriptsize
\begin{tabular}{ |c | c | c | c c | c c c | c c c | c | c c | c|}
\toprule
        &       &\multicolumn{13}{c}{HL-LHC +}  \\
   &HL-LHC   &LHeC   &\multicolumn{2}{c|}{HE-LHC}   &\multicolumn{3}{c|}{ILC}   &\multicolumn{3}{c|}{CLIC} &CEPC   &\multicolumn{2}{c|}{FCC-ee}   &FCC  \\
   &    &    & $S_{2}$   & $S_{2}^\prime$   &  250  & 500   & 1000   & 380   &  1500  & 3000  &    & 240  & 365 &  ee/eh/hh \\
\midrule
$g_{L}^{\nu_{e}}[\%]$   & $1.3$   & $0.97$   & $1.3$   & $1.2$   & $0.082$   & $0.048$   & $0.043$   & $0.027$   & $0.013$   & $0.011$   & $0.032$   & $0.028$   & $0.028$   & $0.028$  \\
\crowcolorA$g_{L}^{\nu_{\mu}}[\%]$   & $1.2$   & $1.2$   & $1.2$   & $1.2$   & $0.11$   & $0.088$   & $0.085$   & $0.16$   & $0.14$   & $0.14$   & $0.036$   & $0.03$   & $0.03$   & $0.03$  \\
$g_{L}^{\nu_{\tau}}[\%]$   & $1.6$   & $1.4$   & $1.6$   & $1.5$   & $0.12$   & $0.095$   & $0.092$   & $0.16$   & $0.14$   & $0.14$   & $0.038$   & $0.034$   & $0.034$   & $0.033$  \\
\midrule
\crowcolorA$g_{L}^{e}[\%]$   & $0.08$   & $0.079$   & $0.079$   & $0.079$   & $0.048$   & $0.037$   & $0.035$   & $0.035$   & $0.022$   & $0.02$   & $0.014$   & $0.0095$   & $0.0073$   & $0.0072$  \\
$g_{R}^{e}[\%]$   & $0.098$   & $0.098$   & $0.098$   & $0.098$   & $0.055$   & $0.041$   & $0.037$   & $0.047$   & $0.035$   & $0.031$   & $0.016$   & $0.0097$   & $0.0089$   & $0.0088$  \\
\crowcolorA$g_{L}^{\mu}[\%]$   & $0.39$   & $0.39$   & $0.39$   & $0.39$   & $0.072$   & $0.07$   & $0.069$   & $0.17$   & $0.17$   & $0.17$   & $0.031$   & $0.0094$   & $0.007$   & $0.007$  \\
$g_{R}^{\mu}[\%]$   & $0.61$   & $0.6$   & $0.6$   & $0.6$   & $0.09$   & $0.087$   & $0.087$   & $0.24$   & $0.24$   & $0.24$   & $0.047$   & $0.0099$   & $0.0092$   & $0.0091$  \\
\crowcolorA$g_{L}^{\tau}[\%]$   & $0.2$   & $0.2$   & $0.2$   & $0.2$   & $0.076$   & $0.073$   & $0.073$   & $0.17$   & $0.17$   & $0.17$   & $0.016$   & $0.0099$   & $0.0076$   & $0.0076$  \\
$g_{R}^{\tau}[\%]$   & $0.29$   & $0.29$   & $0.29$   & $0.29$   & $0.094$   & $0.091$   & $0.091$   & $0.24$   & $0.24$   & $0.24$   & $0.018$   & $0.01$   & $0.0094$   & $0.0094$  \\
\midrule
\crowcolorA$g_{L}^{u=c}[\%]$   & $0.97$   & $0.96$   & $0.95$   & $0.95$   & $0.24$   & $0.23$   & $0.23$   & $0.26$   & $0.26$   & $0.26$   & $0.051$   & $0.05$   & $0.05$   & $0.047$  \\
$g_{R}^{u=c}[\%]$   & $3.6$   & $3.6$   & $3.5$   & $3.5$   & $0.35$   & $0.35$   & $0.35$   & $0.43$   & $0.44$   & $0.43$   & $0.08$   & $0.066$   & $0.066$   & $0.064$  \\
\crowcolorA$g_{L}^{t}[\%]$   & $11.$   & $11.$   & $11.$   & $11.$   & $11.$   & $0.84$   & $0.78$   & $2.4$   & $0.65$   & $0.52$   & $11.$   & $11.$   & $1.5$   & $1.3$  \\
$g_{R}^{t}[\%]$   & $-$   & $-$   & $-$   & $-$   & $-$   & $2.$   & $1.9$   & $6.$   & $1.6$   & $1.7$   & $-$   & $-$   & $3.5$   & $3.1$  \\
\midrule
\crowcolorA$g_{L}^{d=s}[\%]$   & $0.88$   & $0.81$   & $0.77$   & $0.75$   & $0.23$   & $0.21$   & $0.21$   & $0.25$   & $0.24$   & $0.24$   & $0.056$   & $0.051$   & $0.051$   & $0.047$  \\
$g_{R}^{d=s}[\%]$   & $27.$   & $25.$   & $23.$   & $22.$   & $4.8$   & $3.9$   & $3.8$   & $5.9$   & $5.3$   & $5.1$   & $1.4$   & $1.1$   & $1.1$   & $1.$  \\
\crowcolorA$g_{L}^{b}[\%]$   & $0.35$   & $0.35$   & $0.35$   & $0.35$   & $0.071$   & $0.068$   & $0.067$   & $0.13$   & $0.13$   & $0.13$   & $0.017$   & $0.011$   & $0.011$   & $0.011$  \\
$g_{R}^{b}[\%]$   & $11.$   & $11.$   & $11.$   & $11.$   & $0.51$   & $0.51$   & $0.51$   & $3.$   & $3.$   & $3.$   & $0.29$   & $0.088$   & $0.088$   & $0.087$  \\
\bottomrule
\end{tabular}
}
}
\end{table}

\subsubsection{Results for BSM-motivated effective Lagrangians} 
\label{sec:SILHfit}

In this subsection, we adopt a more BSM-oriented perspective and present the global fit results in a way that can be easily matched to theory-motivated scenarios, such as composite Higgs models. For that purpose, we will restrict the results to the set of dimension-6 interactions in the effective Lagrangian in eq.~(\ref{eq:LSilh}) and adopt the usual presentation of results in terms of the bounds on the dimension-6 operator coefficients.
We will also extend the global fits presented in previous sections, adding further studies available in the literature about high-energy probes of the EFT. These are designed to benefit from the growth with energy of the contributions of certain dimension-6 operators in physical processes, leading to competitive constraints on new physics, without necessarily relying on extreme experimental precision. In this regard, we note that these studies are usually not performed in a fully global way within the EFT framework, but rather focus on the most important effects at high energies. Therefore, the results when such processes dominate in the bounds on new physics should be considered with a certain amount of caution, although they should offer a reasonable approximation under the assumptions in (\ref{eq:LSilh}) and (\ref{eq:SILHpc}). 
In particular, we will add the following high-energy probes using di-boson and di-fermion processes: 
\begin{itemize}
\item The constraints on the $W$ and $Y$ oblique parameters\cite{Barbieri:2004qk} (which can be mapped into $c_{2W,2B}$) from fermion pair production at the HL-LHC, HE-LHC~\cite{Cepeda:2019klc}, FCC-hh~\cite{Farina:2016rws}, ILC at 250, 500 and 1000\,GeV~\cite{Fujii:2019zll} and CLIC~\cite{deBlas:2018mhx}\footnote{
The studies in~\cite{deBlas:2018mhx} and \cite{Fujii:2019zll} make use of significantly different assumptions for the systematic uncertainties and efficiencies for each $e^+ e^- \to f \bar{f}$ channel. The apparent small difference in terms of reach at the highest energy stages for CLIC/ILC is, however, due to the high luminosity assumed at ILC, as well as the use of positron polarization, which allow to partially compensate the lower energy achievable compared to CLIC.}. 

It must be noted that, for the HE-LHC, only the sensitivity to $W$ and $Y$ from $pp\to \ell^+ \ell^-$ is available in~\cite{Cepeda:2019klc}. There is no sensitivity reported from charged-current process, which can constrain $W$ independently. 
No studies on the reach for the $W$ and $Y$ parameters were available for CEPC or the FCC-ee. For this section for these two lepton colliders it has been estimated following the studies in Ref.~\cite{deBlas:2018mhx,Fujii:2019zll}~\footnote{We obtain alues of $\delta W_{\rm CEPC}\sim 5.3\times 10^{-5}$, $\delta Y_{\rm CEPC}\sim 4.7\times 10^{-5}$, with a correlation of -0.5; $\delta W_{\rm FCC-{ee} (240)}\sim 5.4\times 10^{-5}$, $\delta Y_{\rm \rm FCC-{ee} (240)}\sim 4.9\times 10^{-5}$, with the same -0.5 correlation; and $\delta W_{\rm FCC-{ee}}\sim 3.2\times 10^{-5}$, $\delta Y_{\rm \rm FCC-{ee}}\sim 2.9\times 10^{-5}$, with a correlation of -0.53.}.

\item The study in Ref.~\cite{Banerjee:2018bio} of the $M_{ZH}$ distribution in $pp\to ZH, H\to b\bar{b}$ in the boosted regime for the HL-LHC~\cite{Cepeda:2019klc} and FCC-hh~\cite{Abada:2019lih}. (This was not available for the HE-LHC.) Note that both CLIC (and to a lesser extent ILC) have access to similar physics in the leptonic case, from the $ZH$ measurements at 1.5/3\,TeV (500/1000\,GeV). Current ILC projections for Higgs production at 1 TeV~\cite{Fujii:2019zll} are only available for the $W$ boson fusion channel.
For the fits presented in this section, for $\sigma_{ZH}\times BR(H\to bb)$ at ILC at 1 TeV an uncertainty of 1.3$\%$ is assumed for each polarization~\cite{listprivate}.

\item The $p_{TV}$ distribution in $pp\to WZ$ from Ref.~\cite{Franceschini:2017xkh} for the HL-LHC, HE-LHC and FCC-hh.
\end{itemize}
These are of course only a sample of the high-energy precision probes that could be tested at future colliders (and at HL-LHC) so the results presented are not an exhaustive study of the potential of the different machines in this regard (see e.g. \cite{Henning:2018kys,Maltoni:2019aot}.)

\begin{figure}[ht]
\centering
\includegraphics[width=\linewidth]{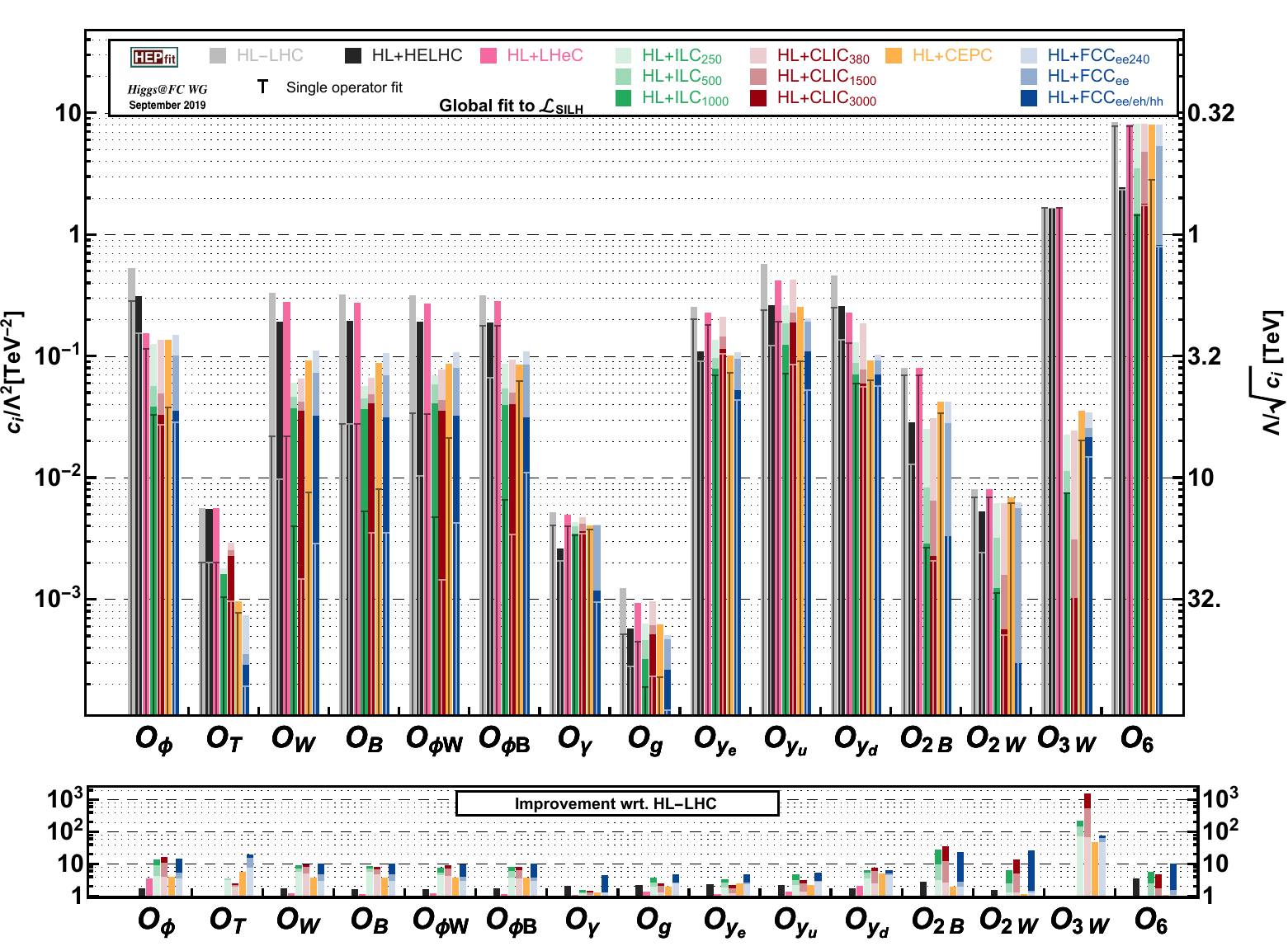}
\caption{\label{fig:SILH}
Global fit to the EFT operators in the Lagrangian (\ref{eq:LSilh}). We show the marginalized 68\% probability reach for each Wilson coefficient $c_i/\Lambda^2$ in Eq.~(\ref{eq:LSilh}) from the global fit (solid bars). The reach of the vertical ``T'' lines indicate the results assuming only the corresponding operator is generated by the new physics.
The HE-LHC results correspond to the $S_2^\prime$ assumptions for the theory systematic uncertainties in Higgs processes~\cite{Cepeda:2019klc}.}
\end{figure}

The results of this fit are shown in Figure~\ref{fig:SILH} after the full run of each future collider project, and in Table \ref{tab:eft-silh}. Apart from the 68\% probability bounds for each operator from the global fit, we also present the results assuming only one operator is generated by the UV dynamics. The difference between both results is indicative of the correlations between the different operators in the fit. These can, in some cases, be rather large. A full study of such correlations goes beyond the scope of this report, but it is worth mentioning that some of the largest correlations typically occur between ${\cal O}_{\gamma}$, ${\cal O}_{\phi W}$, ${\cal O}_{\phi B}$, ${\cal O}_{W}$, ${\cal O}_{B}$ where all contribute to the Higgs interactions with neutral vector bosons. Large correlations also connect ${\cal O}_{g}$ and ${\cal O}_{y_u}$.
These are typically constrained along the $H\to gg$ direction with better precision than the one obtained for ${\cal O}_{y_u}$ from the corresponding $ttH$ process at the different colliders.

For those operators whose effects are mainly constrained by Higgs observables, e.g. ${\cal O}_\phi$ and ${\cal O}_{y_f}$, the evolution
of the results in the table follows essentially the same pattern as in the discussion of the Higgs coupling results of the SMEFT fit. Likewise, similar 
considerations must be taken into account when comparing the results across colliders, in particular regarding the dependence of the HE-LHC results on the assumptions of the reduction of the theory/systematic uncertainties, which control most of the improvement with respect to HL-LHC. (See comment on the S2$^\prime$ assumptions in Section \ref{sec:method}.)
Also regarding the results at high luminosity/energy upgrades of the LHC, some of the numbers in Table \ref{tab:eft-silh}, namely
those involving a single operator fit to $c_\phi$, may look surprising, given that the projections for most Higgs observables at such machines are expected to be dominated by the theory/systematic uncertainties. These results are marked with a $^\dagger$ in the table. For instance, the HL-LHC result corresponds to a precision in an overall Higgs coupling modification at the level of $0.8\%$. This is below the dominant signal theory uncertainties assumed in the HL-LHC S2 hypothesis. As explained in Section \ref{sec:method}, this is a consequence of the assumptions in the treatment of theory/systematic uncertainties in the simplified set of inputs used in this report for the HL-LHC fits. A rough estimate of the bound that would result from assuming 100\% correlated signal theory uncertainties would return, for the same case, $c_\phi/\Lambda^2 \sim 0.42$ TeV$^{-2}$, illustrating the impact of the choice of assumption in the treatment of these theory systematics. Given the implications of these bounds in terms of constraining BSM scenarios (as will be illustrated below, $c_\phi$ sets some of the most important constraints in composite Higgs models), this is an issue that should be carefully studied at hadron colliders, as it will become (even more) relevant at the end of the HL-LHC era. 
There is another caveat affecting the results presented in the Table \ref{tab:eft-silh} that concerns the HE-LHC limits for $c_{2B}$ and $c_{2W}$, also marked with a $^\dagger$. In this case, the reaches for $c_{2B}$ and $c_{2W}$, which can be mapped into the $W$ and $Y$ oblique parameters, are limited by the lack of constraints from the charged current channel at HE-LHC since no projections were provided at this time. The charged current channel is sensitive to the $W$ parameter and, via its correlation with $Y$, can also affect the results for the latter in the global fit.

\begin{table}[ht]
\caption{\label{tab:eft-silh} 
68\% probability reach on the different Wilson coefficients in the Lagrangian Eq.~(\ref{eq:LSilh}) from the global fit.
In parenthesis we give the corresponding results from a fit assuming only one operator is generated by the UV physics.
See text for details, in particular regarding the results marked with a $^\dagger$. 
}
\centering
\rotatebox{90}{
{\scriptsize
\begin{tabular}{ |c | c | c | c c | c c c | c c c | c | c c | c|}
\toprule
     &       &\multicolumn{13}{c}{HL-LHC +}  \\
 $[\mbox{TeV}^{-2}]$     &HL-LHC   &LHeC   &\multicolumn{2}{c|}{HE-LHC}   &\multicolumn{3}{c|}{ILC}   &\multicolumn{3}{c|}{CLIC} &CEPC   &\multicolumn{2}{c|}{FCC-ee}   &FCC \\
   &    &    & $S_{2}$   & $S_{2}^\prime$   &  250  & 500   & 1000   & 380   &  1500  & 3000  &    & 240  & 365 &   ee/eh/hh \\
   
\midrule
$\frac{c_{\phi}}{\Lambda^2}$   & $0.53$   & $0.15$   & $0.43$   & $0.31$   & $0.13$   & $0.057$   & $0.038$   & $0.14$   & $0.049$   & $0.033$   & $0.14$   & $0.15$   & $0.1$   & $0.036$  \\
   & $(0.28)^\dagger$   & $(0.11)$   & $(0.21)^\dagger$   & $(0.16)^\dagger$   & $(0.061)$   & $(0.041)$   & $(0.033)$   & $(0.076)$   & $(0.04)$   & $(0.027)$   & $(0.038)$   & $(0.044)$   & $(0.038)$   & $(0.029)$  \\
\crowcolorA$\frac{c_{T}}{\Lambda^2}$   & $0.0056$   & $0.0056$   & $0.0056$   & $0.0055$   & $0.0018$   & $0.0016$   & $0.0016$   & $0.0029$   & $0.0025$   & $0.0023$   & $0.00097$   & $0.0007$   & $0.0004$   & $0.0003$  \\
\crowcolorA   & $(0.002)$   & $(0.002)$   & $(0.002)$   & $(0.002)$   & $(0.0013)$   & $(0.0011)$   & $(0.001)$   & $(0.001)$   & $(0.001)$   & $(0.001)$   & $(0.0008)$   & $(0.0007)$   & $(0.0002)$   & $(0.0002)$  \\
$\frac{c_{W}}{\Lambda^2}$   & $0.33$   & $0.28$   & $0.24$   & $0.19$   & $0.06$   & $0.046$   & $0.037$   & $0.065$   & $0.042$   & $0.035$   & $0.092$   & $0.11$   & $0.072$   & $0.032$  \\
   & $(0.022)$   & $(0.022)$   & $(0.0098)$   & $(0.0098)$   & $(0.011)$   & $(0.0073)$   & $(0.004)$   & $(0.011)$   & $(0.0037)$   & $(0.0015)$   & $(0.0076)$   & $(0.0051)$   & $(0.0036)$   & $(0.0029)$  \\
\crowcolorA$\frac{c_{B}}{\Lambda^2}$   & $0.32$   & $0.27$   & $0.24$   & $0.19$   & $0.057$   & $0.045$   & $0.037$   & $0.066$   & $0.048$   & $0.041$   & $0.088$   & $0.11$   & $0.069$   & $0.031$  \\
\crowcolorA   & $(0.028)$   & $(0.028)$   & $(0.028)$   & $(0.028)$   & $(0.011)$   & $(0.0084)$   & $(0.0053)$   & $(0.013)$   & $(0.0079)$   & $(0.0035)$   & $(0.0081)$   & $(0.005)$   & $(0.0035)$   & $(0.0035)$  \\
$\frac{c_{\phi W}}{\Lambda^2}$   & $0.32$   & $0.27$   & $0.24$   & $0.19$   & $0.07$   & $0.058$   & $0.041$   & $0.078$   & $0.044$   & $0.036$   & $0.086$   & $0.11$   & $0.08$   & $0.032$  \\
   & $(0.034)$   & $(0.033)$   & $(0.01)$   & $(0.01)$   & $(0.026)$   & $(0.012)$   & $(0.0047)$   & $(0.02)$   & $(0.0039)$   & $(0.0014)$   & $(0.021)$   & $(0.021)$   & $(0.015)$   & $(0.0043)$  \\
\crowcolorA$\frac{c_{\phi B}}{\Lambda^2}$   & $0.32$   & $0.28$   & $0.24$   & $0.19$   & $0.086$   & $0.054$   & $0.039$   & $0.093$   & $0.05$   & $0.04$   & $0.086$   & $0.11$   & $0.086$   & $0.031$  \\
\crowcolorA   & $(0.18)$   & $(0.18)$   & $(0.099)$   & $(0.067)$   & $(0.048)$   & $(0.016)$   & $(0.0066)$   & $(0.035)$   & $(0.0092)$   & $(0.0034)$   & $(0.062)$   & $(0.066)$   & $(0.042)$   & $(0.011)$  \\
$\frac{c_{\gamma}}{\Lambda^2}$   & $0.0052$   & $0.0049$   & $0.0042$   & $0.0026$   & $0.0043$   & $0.004$   & $0.0035$   & $0.0048$   & $0.0042$   & $0.0036$   & $0.004$   & $0.0041$   & $0.004$   & $0.0012$  \\
   & $(0.004)$   & $(0.004)$   & $(0.0031)$   & $(0.0021)$   & $(0.0039)$   & $(0.0038)$   & $(0.0033)$   & $(0.004)$   & $(0.0039)$   & $(0.0035)$   & $(0.0038)$   & $(0.0039)$   & $(0.0038)$   & $(0.0010)$  \\
\crowcolorA$\frac{c_{g}}{\Lambda^2}$   & $0.0012$   & $0.0009$   & $0.001$   & $0.0006$   & $0.0006$   & $0.0005$   & $0.0003$   & $0.001$   & $0.0006$   & $0.0005$   & $0.0006$   & $0.0005$   & $0.0005$   & $0.0003$  \\
\crowcolorA   & $(0.0005)$   & $(0.0005)$   & $(0.0004)$   & $(0.0003)$   & $(0.0004)$   & $(0.0003)$   & $(0.0002)$   & $(0.0004)$   & $(0.0003)$   & $(0.0002)$   & $(0.0002)$   & $(0.0003)$   & $(0.0003)$   & $(0.0001)$  \\
$\frac{c_{y_{e}}}{\Lambda^2}$   & $0.25$   & $0.23$   & $0.18$   & $0.11$   & $0.14$   & $0.097$   & $0.079$   & $0.21$   & $0.15$   & $0.11$   & $0.1$   & $0.11$   & $0.094$   & $0.052$  \\
   & $(0.2)$   & $(0.18)$   & $(0.13)$   & $(0.091)$   & $(0.096)$   & $(0.079)$   & $(0.07)$   & $(0.17)$   & $(0.13)$   & $(0.1)$   & $(0.072)$   & $(0.078)$   & $(0.071)$   & $(0.044)$  \\
\crowcolorA$\frac{c_{y_{u}}}{\Lambda^2}$   & $0.57$   & $0.42$   & $0.44$   & $0.26$   & $0.26$   & $0.19$   & $0.12$   & $0.42$   & $0.23$   & $0.19$   & $0.25$   & $0.2$   & $0.19$   & $0.11$  \\
\crowcolorA   & $(0.24)$   & $(0.19)$   & $(0.19)$   & $(0.12)$   & $(0.14)$   & $(0.099)$   & $(0.072)$   & $(0.16)$   & $(0.11)$   & $(0.085)$   & $(0.091)$   & $(0.11)$   & $(0.099)$   & $(0.052)$  \\
$\frac{c_{y_{d}}}{\Lambda^2}$   & $0.46$   & $0.23$   & $0.37$   & $0.26$   & $0.13$   & $0.088$   & $0.071$   & $0.18$   & $0.077$   & $0.059$   & $0.091$   & $0.1$   & $0.092$   & $0.071$  \\
   & $(0.25)$   & $(0.13)$   & $(0.19)$   & $(0.14)$   & $(0.084)$   & $(0.066)$   & $(0.059)$   & $(0.098)$   & $(0.063)$   & $(0.055)$   & $(0.064)$   & $(0.068)$   & $(0.064)$   & $(0.057)$  \\
\crowcolorA$\frac{c_{2B}}{\Lambda^2}$   & $0.08$   & $0.08$   & $0.028^\dagger$   & $0.028^\dagger$   & $0.025$   & $0.0083$   & $0.0029$   & $0.031$   & $0.0064$   & $0.0023$   & $0.042$   & $0.042$   & $0.028$   & $0.0034$  \\
\crowcolorA   & $(0.069)$   & $(0.069)$   & $(0.013)$   & $(0.013)$   & $(0.023)$   & $(0.0078)$   & $(0.0027)$   & $(0.028)$   & $(0.0059)$   & $(0.0021)$   & $(0.034)$   & $(0.029)$   & $(0.021)$   & $(0.0034)$  \\
$\frac{c_{2W}}{\Lambda^2}$   & $0.008$   & $0.008$   & $0.0053^\dagger$   & $0.0053^\dagger$   & $0.0062$   & $0.0032$   & $0.0012$   & $0.0062$   & $0.0016$   & $0.0006$   & $0.0069$   & $0.0062$   & $0.0056$   & $0.0003$  \\
   & $(0.0069)$   & $(0.0069)$   & $(0.0024)^\dagger$   & $(0.0024)^\dagger$   & $(0.0058)$   & $(0.003)$   & $(0.0011)$   & $(0.0058)$   & $(0.0014)$   & $(0.0005)$   & $(0.0062)$   & $(0.0057)$   & $(0.0049)$   & $(0.0003)$  \\
\crowcolorA$\frac{c_{3W}}{\Lambda^2}$   & $1.7$   & $1.6$   & $1.6$   & $1.6$   & $0.023$   & $0.011$   & $0.0076$   & $0.024$   & $0.0031$   & $0.001$   & $0.036$   & $0.034$   & $0.026$   & $0.021$  \\
\crowcolorA   & $(1.6)$   & $(1.6)$   & $(1.6)$   & $(1.6)$   & $(0.022)$   & $(0.011)$   & $(0.0075)$   & $(0.024)$   & $(0.0031)$   & $(0.001)$   & $(0.02)$   & $(0.019)$   & $(0.015)$   & $(0.015)$  \\
$\frac{c_{6}}{\Lambda^2}$   & $8.4$   & $8.1$   & $2.5$   & $2.4$   & $8.1$   & $3.5$   & $1.5$   & $8.1$   & $4.8$   & $1.8$   & $8.$   & $8.$   & $5.3$   & $0.81$  \\
   & $(7.8)$   & $(7.7)$   & $(2.4)$   & $(2.3)$   & $(4.7)$   & $(3.1)$   & $(1.4)$   & $(7.7)$   & $(4.5)$   & $(1.7)$   & $(2.8)$   & $(3.2)$   & $(3.1)$   & $(0.79)$  \\
\bottomrule
\end{tabular}
}
}
\end{table}

\begin{figure}[ht]
\centering
\begin{tabular}{c c}
\includegraphics[width=0.47\linewidth]{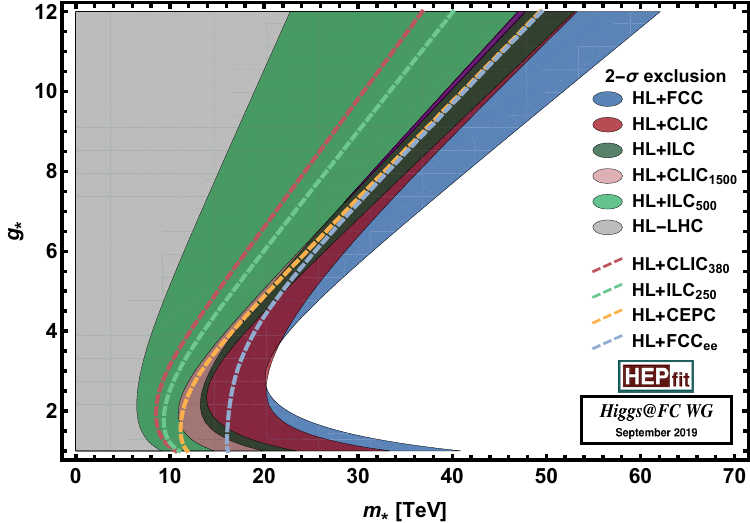}
&
\includegraphics[width=0.47\linewidth]{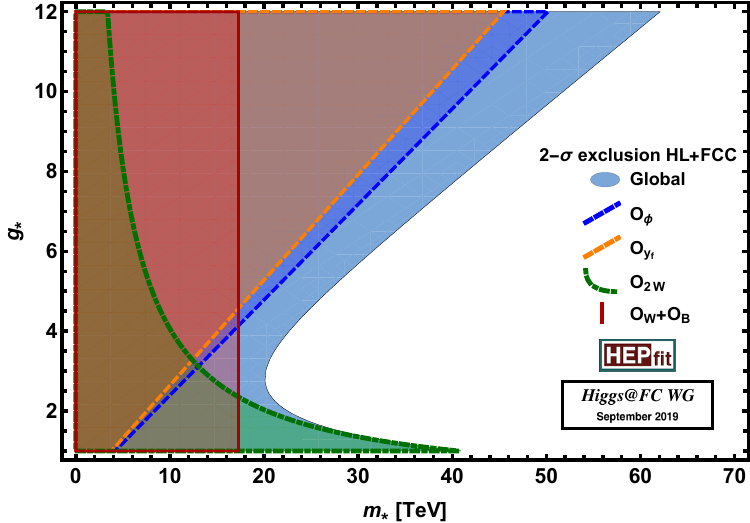}
\end{tabular}
\caption{\label{fig:SILH_gstar}
(Left) 2-$\sigma$ exclusion regions in the $(g_\star, m_\star)$ plane from the fit presented in Figure~\ref{fig:SILH}, using the SILH power-counting described in Eq.~(\ref{eq:SILHpc}) and below (solid regions). Dashed lines indicate the regions constrained by the corresponding low-energy runs (or FCC-ee only for the case of the FCC project). (Right) The same comparing the results from the global fit with the constraints set by some of the operators individually, for the illustrative case of the HL-LHC+FCC-{ee/eh/hh}. In this case, the constraints from the on-shell Higgs measurements mainly affect ${\cal O}_\phi$ and ${\cal O}_{y_f}$.
}
\end{figure}

A meaningful interpretation of these results in terms of a broad class of composite Higgs models can be obtained under the assumptions leading to the dependence of the Wilson coefficients on new physics coupling, $g_\star$, and mass, $m_\star$, described in Eq.~(\ref{eq:SILHpc}) and below 
(i.e. we assume $c_{g,\gamma}$ and $c_{\phi V, 3V}$ are loop suppressed in $y_t$ and $g_\star$, respectively).
In Figure~\ref{fig:SILH_gstar} we translate the results of the fit in Figure~\ref{fig:SILH} in terms of the 95\% probability constraints in the $(g_\star, m_\star)$ plane under such assumptions, and setting all $O(1)$ coefficients exactly to 1, i.e. 

\begin{equation}
\begin{split}
\frac{c_{\phi,6,y_f}}{\Lambda^2}&= \frac{g_\star^2}{ m_\star^2},~~~~~~~~
\frac{c_{W,B}}{\Lambda^2}= \frac{1}{m_\star^2}, ~~~~~~~~~
\frac{c_{2W,2B,2G}}{\Lambda^2}= \frac{1}{g_\star^2}\frac{1}{m_\star^2},~~~~~
\frac{c_{T}}{\Lambda^2}= \frac{y_t^4}{16\pi^2}\frac{1}{m_\star^2}, 
\\
\frac{c_{\phi W,\phi B}}{\Lambda^2}&= \frac{g_\star^2}{16\pi^2}\frac{1}{m_\star^2},~~~
\frac{c_{\gamma,g}}{\Lambda^2}= \frac{y_t^2}{16\pi^2}\frac{1}{m_\star^2},~~~~~~
\frac{c_{3W,3G}}{\Lambda^2}= \frac{1}{16\pi^2}\frac{1}{m_\star^2}.
\end{split}
\label{eq:SILHpc2}
\end{equation}
We focus the comparison, again, on the full physics program at each future collider project (solid regions), but also show the region delimited by the low energy runs, or the FCC-ee for the case of the FCC project (the boundaries are indicated by the dashed lines). 
In the right panel of that figure we also show, for illustration purposes, the individual constraints set by several of the operators in (\ref{eq:LSilh}) for the FCC fit. The modifications of the on-shell Higgs properties discussed in this report are mainly controlled, within the SILH assumptions, by the contributions to the operators ${\cal O}_{\phi}$ and ${\cal O}_{y_f}$, both of which set similar constraints in the global fit for this collider. These give the leading constraints in strongly coupled scenarios. Electroweak precision measurements, on the other hand, are more affected by a combination of ${\cal O}_{W,B}$ and set bounds independently of the new physics coupling. Finally, some of the high-energy probes included in the analysis provide the most efficient way of testing weakly coupled scenarios.

\subsection{Impact of Standard Model theory uncertainties in Higgs calculations\label{sec:SMThUnc}} 

As important as it is to have very precise experimental measurements of the different Higgs processes, it is also fundamental from the point of view of their physical interpretation to have theoretical calculations for the predictions of such processes with comparable or better precision. In this sense, to quantify to what extent an experimental measurement with uncertainty $\delta_\mt{exp}$ can be translated into a constraint on new physics,\footnote{Or, equivalently, to what extent a measurement agrees with the SM.} one needs to know the corresponding uncertainty $\delta_\mt{SM}$ for the SM prediction. In order to extract the maximum experimental information, ideally, $\delta_\mt{SM} \ll \delta_\mt{exp}$. The sources of the SM uncertainty are typically separated in two types of contributions:
\begin{itemize}
{\item {\it Parametric theory uncertainties}~(Th$_\mt{Par}$). For a given observable $O$, this is the error associated to the propagation of the experimental error of the SM input parameters to the SM prediction $O_\SM$.}
{\item The second source of uncertainty is due to the fact that, in practice, $O_\mt{SM}$ is only known to a finite order in perturbation theory. The estimate of the net size associated with the contribution to $O_\mt{SM}$ from missing higher-order corrections is usually referred to as {\it intrinsic theory uncertainty}~(Th$_\mt{Intr}$).}
\end{itemize}
Of course, in the interpretation of any measurement in a particular extension of the SM, there are also errors associated with the missing corrections in the 
expansion(s) including the new physics parameters. In the particular case of the EFT framework, these would come from NLO
corrections in the perturbative expansion including dimension-6 interactions or, from the point of view of the EFT expansion, from $q^4/\Lambda^4$ effects coming from either the square of the dimension-6 contributions to the amplitudes, or the SM interference with amplitudes involving dimension-8 operators or double insertions of the dimension-6 ones. 
Note that all these corrections affect the interpretation of a measurement in terms of pinpointing what is the source of the deformation from the SM, i.e. which particular operator and how large its coefficient can be, but not on the size of the overall deformation per se. 
The latter is only controlled by the SM theoretical uncertainty.
Because of that, and in the absence of a fully developed program including such contributions in the SMEFT framework, we restrict the discussion in this section to SM uncertainties only. 

In the previous sections the results for future colliders after the HL/HE-LHC era were presented taking into account parametric uncertainties only.
This was done to illustrate the final sensitivity to BSM deformations in Higgs couplings, as given directly by the experimental measurements of the different inputs (i.e. Higgs rates, diBoson measurements, EWPO or the processes used to determine the values of the SM input parameters).
On the other hand, for this scenario to be meaningful, 
it is crucial to also study the effect in such results of the projections for the future intrinsic errors. This is needed to be able to quantify how far we will be from the assumption that such intrinsic errors become subdominant and, therefore,
which aspects of theory calculations should the theory community focus on to make sure we reach the maximum experimental sensitivity at future colliders.

In this section we discuss more in detail the impact of the two types of SM theory errors described above, from the point of view of the calculations of the predictions for Higgs observables. 
This will be done both within the $\kappa$ framework and also in the context of the EFT results.
For the results from the $\kappa$-framework we will use the most general scenario considered in Section~\ref{kappa}, i.e.~kappa-3, which allows non-SM decays. 
On the EFT side, we will use the scenario SMEFT$_{\rm PEW}$, where the uncertainty associated with the precision of EWPO has already been ``factorized''. In this scenario each fermion coupling is also treated separately, thus being sensitive to the uncertainties in the different $H\to f\bar{f}$ decay widths. Finally, we will also restrict the study in this subsection to the case of future lepton colliders only (we always consider them in combination with the HL-LHC projections. For the latter we keep the theory uncertainties as reported by the WG2 studies~\cite{Cepeda:2019klc}). 

\begin{figure}[t]
\centering
\includegraphics[width=\linewidth]{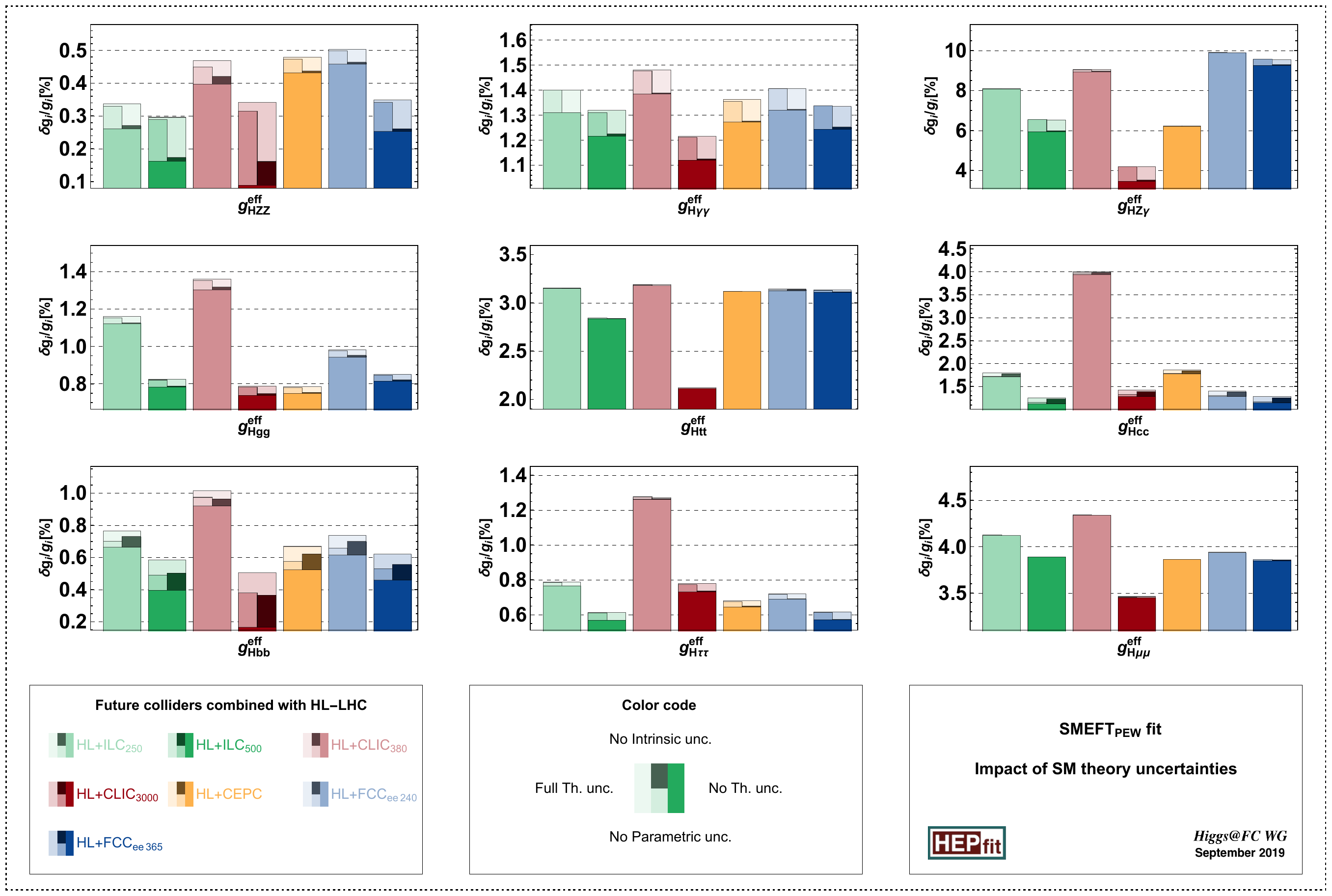}
\caption{\label{fig:SMEFTThunc}
Impact of the different sources of SM theory uncertainties in the  coupling reach at the different lepton-collider projects based on the SMEFT$_{\rm PEW}$ fit. Using dark to light shades we show the results without SM theory uncertainties (darkest shade), only with the intrinsic uncertainty (medium), and the full SM error (lightest shade). The solid line indicates the result with SM parametric uncertainties only. The most significant differences are found for the effective coupling to vector bosons (dominated by intrinsic uncertainties) and to bottom quarks (controlled by the parametric error associated with $m_b$). See Table~\ref{tab:eft-SMTh} and text for details.}
\end{figure}

In Table~\ref{tab:kappa-SMTh} we show the results of the $\kappa$ fit for the benchmark scenario kappa-3, indicating the results obtained including/excluding the different sources of SM theory uncertainties. 
Similarly, Table~\ref{tab:eft-SMTh} shows the results of the EFT fit for the benchmark scenario SMEFT$_{\rm PEW}$. For the EFT results the impact of the different theory uncertainties is also illustrated in Figure~\ref{fig:SMEFTThunc}.
As can be seen, if the SM errors were reduced to a level where they become sub-dominant, the experimental precision would allow to test deviations in some of the couplings at the one per-mille level, e.g. the coupling to vector bosons at CLIC in the SMEFT framework (the presence of extra decays would however reduce the precision to the $0.4\%$ level, as shown in the kappa-3 results).
The assumed precision of the SM theory calculations and inputs, however, prevents reaching this level of sensitivity. The most notable obstacle to achieve this close to per-mille level of precision are the intrinsic uncertainties for the $e^+ e^- \to ZH$ and, especially, in $e^+ e^- \to H\bar{\nu}\nu$, estimated to be $\sim$0.5\%. In reaching this level of theoretical precision it was assumed that 
predictions at NNLO in the EW coupling for both processes will be available. 
This is within reach for $ZH$ production, but it may be more challenging for $e^+ e^- \to H\bar{\nu}\nu$ (and $H\to V V^* \to 4f$).
However, with enough effort on the theory side~\cite{Lepage:2014fla,Blondel:2019qlh,Blondel:2018mad}, this type of uncertainties can be reduced. If the necessary resources are dedicated to develop these types of calculations, it should be possible to achieve, or even surpass, the required level of precision.
This is not the case for the SM parametric errors, which depend on the experimental measurements of the corresponding input parameters. From the results of the fits, the largest effect of this type of uncertainty on the determination of the fermion couplings affects the effective coupling of the bottom to the Higgs. The corresponding SM error in $H\to b\bar{b}$ depends on the precision of the bottom quark mass, whose projected future determination was assumed to be $\sim 13$\,MeV. Taking into account the projected improvements from Lattice QCD calculations, this should be a conservative estimate~\cite{Lepage:2014fla}.
Other parametric uncertainties, e.g. in $H\to c\bar{c}, gg$ and associated with $m_c$ and $\alpha_S$, are larger than the one for $H\to b\bar{b}$ but have a smaller effect in the results due to the also larger experimental errors expected in the corresponding channels. 
From the point of view of the Higgs decays into vector bosons, the predictions of $H\to ZZ^*, WW^*$ have a strong dependence on the value of the Higgs mass. It it therefore important to accompany the precise measurements of the Higgs couplings with equally precise measurements of the Higgs mass, to the level of 10 MeV. This would be possible at 240/250 GeV lepton colliders but more challenging at CLIC, where the final precision on $M_H$ is expected at the level of 20-30 MeV (see Section~\ref{mass}). In the kappa-framework, the fact that the dependence of the production $e^+ e^-$ Higgs cross sections on $M_H$ is less severe helps to reduce the impact of the $M_H$ uncertainty in the CLIC results. This is no longer the case once we move to the more general description of the SMEFT. In that case, non-SM like interactions contribute to the effective $HZZ$ and $HWW$ couplings, and the information on $H\to WW^*$ becomes relevant to determine $g_{HZZ}^{\rm eff}$. The measurement of $M_H$ at the HL-LHC at the 10-20 MeV level prevents this from becoming an issue at the lower energy stages at CLIC. But there is still a factor $\sim 2$ deterioration in the precision of the $g_{HZZ}^{\rm eff}$ coupling in the final CLIC results, emphasising again the necessity of a precise determination of $M_H$.

\begin{table}[ht]
\caption{
Comparison of the sensitivity at 68\% probability to deviations in the different Higgs couplings modifiers in the kappa-3 fit, under different assumptions for the SM theory uncertainties. 
We compare the results obtained neglecting both intrinsic and parametric uncertainties, including each of them separately, and adding the full SM uncertainty. 
\label{tab:kappa-SMTh}}
\centering
{\scriptsize
\begin{tabular}{| c | c | c c c | c c c | c | c c |}
\toprule
        &   Benchmark    &\multicolumn{9}{c}{HL-LHC +}  \\
   &kappa-3   &\multicolumn{3}{c |}{ILC}   &\multicolumn{3}{c |}{CLIC}   &CEPC   &\multicolumn{2}{c}{FCC-ee}   \\
   &     &250   &500   &1000   &380   &1500   &3000   &   &240   &365   \\
\midrule
$\kappa_{W}[\%]$& Exp$_\mt{Stat}$   & $1.$   & $0.28$   & $0.24$   & $0.73$   & $0.4$   & $0.38$   & $0.87$   & $0.87$   & $0.4$  \\
 & Exp$_\mt{Stat}$ + Th$_\mt{Par}$   & $1.$   & $0.29$   & $0.24$   & $0.73$   & $0.4$   & $0.38$   & $0.88$   & $0.88$   & $0.41$  \\
 & Exp$_\mt{Stat}$ + Th$_\mt{Intr}$   & $1.$   & $0.51$   & $0.47$   & $0.82$   & $0.53$   & $0.49$   & $0.89$   & $0.89$   & $0.56$  \\
 & Exp$_\mt{Stat}$ + Th   & $1.$   & $0.51$   & $0.47$   & $0.81$   & $0.53$   & $0.63$   & $0.89$   & $0.89$   & $0.56$  \\
\crowcolorA$\kappa_{Z}[\%]$& Exp$_\mt{Stat}$   & $0.28$   & $0.22$   & $0.22$   & $0.44$   & $0.4$   & $0.39$   & $0.17$   & $0.19$   & $0.16$  \\
\crowcolorA & Exp$_\mt{Stat}$ + Th$_\mt{Par}$   & $0.29$   & $0.23$   & $0.22$   & $0.44$   & $0.4$   & $0.39$   & $0.18$   & $0.2$   & $0.17$  \\
\crowcolorA & Exp$_\mt{Stat}$ + Th$_\mt{Intr}$   & $0.32$   & $0.27$   & $0.26$   & $0.46$   & $0.42$   & $0.41$   & $0.23$   & $0.24$   & $0.22$  \\
\crowcolorA & Exp$_\mt{Stat}$ + Th   & $0.32$   & $0.27$   & $0.27$   & $0.46$   & $0.42$   & $1.2$   & $0.23$   & $0.24$   & $0.23$  \\
$\kappa_{g}[\%]$& Exp$_\mt{Stat}$   & $1.3$   & $0.83$   & $0.58$   & $1.5$   & $1.1$   & $0.83$   & $1.$   & $1.1$   & $0.87$  \\
 & Exp$_\mt{Stat}$ + Th$_\mt{Par}$   & $1.4$   & $0.85$   & $0.63$   & $1.5$   & $1.1$   & $0.86$   & $1.$   & $1.2$   & $0.9$  \\
 & Exp$_\mt{Stat}$ + Th$_\mt{Intr}$   & $1.4$   & $0.97$   & $0.8$   & $1.6$   & $1.1$   & $0.95$   & $1.1$   & $1.2$   & $1.$  \\
 & Exp$_\mt{Stat}$ + Th   & $1.4$   & $0.99$   & $0.82$   & $1.6$   & $1.1$   & $2.1$   & $1.1$   & $1.2$   & $1.$  \\
\crowcolorA$\kappa_{ \gamma}[\%]$& Exp$_\mt{Stat}$   & $1.4$   & $1.2$   & $1.1$   & $1.4$   & $1.3$   & $1.2$   & $1.3$   & $1.3$   & $1.3$  \\
\crowcolorA & Exp$_\mt{Stat}$ + Th$_\mt{Par}$   & $1.4$   & $1.2$   & $1.1$   & $1.4$   & $1.3$   & $1.2$   & $1.3$   & $1.3$   & $1.3$  \\
\crowcolorA & Exp$_\mt{Stat}$ + Th$_\mt{Intr}$   & $1.4$   & $1.3$   & $1.1$   & $1.5$   & $1.3$   & $1.2$   & $1.3$   & $1.3$   & $1.3$  \\
\crowcolorA & Exp$_\mt{Stat}$ + Th   & $1.4$   & $1.3$   & $1.1$   & $1.4$   & $1.3$   & $5.9$   & $1.3$   & $1.4$   & $1.3$  \\
$\kappa_{Z \gamma}[\%]$& Exp$_\mt{Stat}$   & $10.$   & $10.$   & $10.$   & $10.$   & $8.2$   & $5.7$   & $6.3$   & $10.$   & $10.$  \\
 & Exp$_\mt{Stat}$ + Th$_\mt{Par}$   & $10.$   & $10.$   & $10.$   & $10.$   & $8.2$   & $5.7$   & $6.3$   & $10.$   & $10.$  \\
 & Exp$_\mt{Stat}$ + Th$_\mt{Intr}$   & $10.$   & $10.$   & $10.$   & $10.$   & $8.2$   & $5.7$   & $6.3$   & $10.$   & $10.$  \\
 & Exp$_\mt{Stat}$ + Th   & $10.$   & $10.$   & $10.$   & $10.$   & $8.2$   & $17.$   & $6.3$   & $10.$   & $10.$  \\
\crowcolorA$\kappa_{c}[\%]$& Exp$_\mt{Stat}$   & $1.9$   & $1.1$   & $0.74$   & $4.$   & $1.8$   & $1.3$   & $1.9$   & $1.4$   & $1.2$  \\
\crowcolorA & Exp$_\mt{Stat}$ + Th$_\mt{Par}$   & $2.$   & $1.2$   & $0.9$   & $4.1$   & $1.9$   & $1.4$   & $2.$   & $1.5$   & $1.3$  \\
\crowcolorA & Exp$_\mt{Stat}$ + Th$_\mt{Intr}$   & $1.9$   & $1.2$   & $0.84$   & $4.$   & $1.8$   & $1.4$   & $1.9$   & $1.5$   & $1.2$  \\
\crowcolorA & Exp$_\mt{Stat}$ + Th   & $2.$   & $1.3$   & $0.99$   & $4.1$   & $1.9$   & $3.6$   & $2.$   & $1.5$   & $1.3$  \\
$\kappa_{t}[\%]$& Exp$_\mt{Stat}$   & $3.1$   & $2.8$   & $1.4$   & $3.2$   & $2.1$   & $2.1$   & $3.1$   & $3.1$   & $3.1$  \\
 & Exp$_\mt{Stat}$ + Th$_\mt{Par}$   & $3.1$   & $2.8$   & $1.4$   & $3.2$   & $2.1$   & $2.1$   & $3.1$   & $3.1$   & $3.1$  \\
 & Exp$_\mt{Stat}$ + Th$_\mt{Intr}$   & $3.2$   & $2.9$   & $1.4$   & $3.2$   & $2.1$   & $2.1$   & $3.1$   & $3.1$   & $3.1$  \\
 & Exp$_\mt{Stat}$ + Th   & $3.1$   & $2.8$   & $1.4$   & $3.2$   & $2.1$   & $7.$   & $3.1$   & $3.1$   & $3.1$  \\
\crowcolorA$\kappa_{b}[\%]$& Exp$_\mt{Stat}$   & $1.1$   & $0.47$   & $0.36$   & $1.2$   & $0.5$   & $0.41$   & $0.82$   & $0.91$   & $0.56$  \\
\crowcolorA & Exp$_\mt{Stat}$ + Th$_\mt{Par}$   & $1.1$   & $0.56$   & $0.47$   & $1.2$   & $0.59$   & $0.52$   & $0.9$   & $0.98$   & $0.64$  \\
\crowcolorA & Exp$_\mt{Stat}$ + Th$_\mt{Intr}$   & $1.1$   & $0.64$   & $0.54$   & $1.2$   & $0.64$   & $0.54$   & $0.86$   & $0.94$   & $0.68$  \\
\crowcolorA & Exp$_\mt{Stat}$ + Th   & $1.2$   & $0.71$   & $0.62$   & $1.3$   & $0.71$   & $0.87$   & $0.93$   & $1.$   & $0.76$  \\
$\kappa_{\mu}[\%]$& Exp$_\mt{Stat}$   & $4.2$   & $3.9$   & $3.6$   & $4.4$   & $4.1$   & $3.5$   & $3.9$   & $4.$   & $3.9$  \\
 & Exp$_\mt{Stat}$ + Th$_\mt{Par}$   & $4.2$   & $3.9$   & $3.6$   & $4.4$   & $4.1$   & $3.5$   & $3.9$   & $4.$   & $3.9$  \\
 & Exp$_\mt{Stat}$ + Th$_\mt{Intr}$   & $4.2$   & $3.9$   & $3.6$   & $4.4$   & $4.1$   & $3.5$   & $3.9$   & $4.$   & $3.9$  \\
 & Exp$_\mt{Stat}$ + Th   & $4.2$   & $4.$   & $3.6$   & $4.4$   & $4.1$   & $4.4$   & $3.9$   & $4.$   & $3.9$  \\
\crowcolorA$\kappa_{\tau}[\%]$& Exp$_\mt{Stat}$   & $1.1$   & $0.64$   & $0.53$   & $1.4$   & $0.99$   & $0.81$   & $0.91$   & $0.94$   & $0.65$  \\
\crowcolorA & Exp$_\mt{Stat}$ + Th$_\mt{Par}$   & $1.1$   & $0.64$   & $0.54$   & $1.4$   & $1.$   & $0.82$   & $0.91$   & $0.94$   & $0.66$  \\
\crowcolorA & Exp$_\mt{Stat}$ + Th$_\mt{Intr}$   & $1.1$   & $0.74$   & $0.64$   & $1.4$   & $1.$   & $0.85$   & $0.93$   & $0.96$   & $0.74$  \\
\crowcolorA & Exp$_\mt{Stat}$ + Th   & $1.1$   & $0.75$   & $0.65$   & $1.4$   & $1.$   & $3.3$   & $0.94$   & $0.96$   & $0.75$  \\
$\mathrm{BR}_{\mathrm{inv}}^{95\%}<$& Exp$_\mt{Stat}$   & $0.26$   & $0.22$   & $0.23$   & $0.63$   & $0.62$   & $0.62$   & $0.27$   & $0.22$   & $0.19$  \\
 & Exp$_\mt{Stat}$ + Th$_\mt{Par}$   & $0.26$   & $0.23$   & $0.22$   & $0.63$   & $0.62$   & $0.62$   & $0.27$   & $0.22$   & $0.19$  \\
 & Exp$_\mt{Stat}$ + Th$_\mt{Intr}$   & $0.26$   & $0.23$   & $0.23$   & $0.63$   & $0.62$   & $0.62$   & $0.27$   & $0.22$   & $0.19$  \\
 & Exp$_\mt{Stat}$ + Th   & $0.26$   & $0.23$   & $0.23$   & $0.63$   & $0.62$   & $0.69$   & $0.27$   & $0.22$   & $0.19$  \\
\crowcolorA$\mathrm{BR}_{\mathrm{unt}}^{95\%}<$& Exp$_\mt{Stat}$   & $1.8$   & $1.4$   & $1.4$   & $2.8$   & $2.4$   & $2.4$   & $1.1$   & $1.2$   & $1.$  \\
\crowcolorA & Exp$_\mt{Stat}$ + Th$_\mt{Par}$   & $1.8$   & $1.4$   & $1.4$   & $2.7$   & $2.4$   & $2.4$   & $1.1$   & $1.2$   & $1.$  \\
\crowcolorA & Exp$_\mt{Stat}$ + Th$_\mt{Intr}$   & $1.8$   & $1.4$   & $1.4$   & $2.8$   & $2.4$   & $2.4$   & $1.1$   & $1.2$   & $1.$  \\
\crowcolorA & Exp$_\mt{Stat}$ + Th   & $1.8$   & $1.4$   & $1.4$   & $2.7$   & $2.4$   & $2.4$   & $1.1$   & $1.2$   & $1.1$  \\
\bottomrule
\end{tabular}
}
\end{table}

\begin{table}[p]
\caption{
Comparison, within the SMEFT$_{\rm PEW}$ fit, 
of the sensitivity at 68\% probability to deviations in the different effective Higgs couplings and aTGC under different assumptions for the SM theory uncertainties. 
We compare the results obtained neglecting both intrinsic and parametric uncertainties, including each of them separately,  and finally adding the full SM uncertainty. 
\label{tab:eft-SMTh}}
\centering
{\scriptsize
\begin{tabular}{| c | c | c c c | c c c | c | c c|}
\toprule
        &   Benchmark    &\multicolumn{9}{c}{HL-LHC +}  \\
   & SMEFT$_{\rm PEW}$ &\multicolumn{3}{c |}{ILC}   &\multicolumn{3}{c |}{CLIC}   &CEPC   &\multicolumn{2}{c}{FCC-ee}   \\
   &     &250   &500   &1000   &380   &1500   &3000   &   &240   &365   \\
\midrule
$g_{HZZ}^{\mathrm{eff}}[\%]$& Exp$_\mt{Stat}$   & $0.26$   & $0.16$   & $0.12$   & $0.4$   & $0.14$   & $0.089$   & $0.43$   & $0.46$   & $0.25$  \\
 & Exp$_\mt{Stat}$ + Th$_\mt{Par}$   & $0.27$   & $0.17$   & $0.13$   & $0.42$   & $0.19$   & $0.16$   & $0.44$   & $0.46$   & $0.26$  \\
 & Exp$_\mt{Stat}$ + Th$_\mt{Intr}$   & $0.33$   & $0.29$   & $0.28$   & $0.45$   & $0.33$   & $0.32$   & $0.47$   & $0.5$   & $0.34$  \\
 & Exp$_\mt{Stat}$ + Th   & $0.34$   & $0.3$   & $0.29$   & $0.47$   & $0.36$   & $0.34$   & $0.48$   & $0.5$   & $0.35$  \\
\crowcolorA$g_{HWW}^{\mathrm{eff}}[\%]$& Exp$_\mt{Stat}$   & $0.28$   & $0.17$   & $0.13$   & $0.41$   & $0.14$   & $0.091$   & $0.41$   & $0.45$   & $0.26$  \\
\crowcolorA & Exp$_\mt{Stat}$ + Th$_\mt{Par}$   & $0.29$   & $0.18$   & $0.14$   & $0.43$   & $0.18$   & $0.15$   & $0.41$   & $0.45$   & $0.27$  \\
\crowcolorA & Exp$_\mt{Stat}$ + Th$_\mt{Intr}$   & $0.37$   & $0.32$   & $0.31$   & $0.48$   & $0.36$   & $0.34$   & $0.47$   & $0.5$   & $0.37$  \\
\crowcolorA & Exp$_\mt{Stat}$ + Th   & $0.37$   & $0.33$   & $0.32$   & $0.5$   & $0.38$   & $0.36$   & $0.47$   & $0.51$   & $0.38$  \\
$g_{H \gamma \gamma}^{\mathrm{eff}}[\%]$& Exp$_\mt{Stat}$   & $1.3$   & $1.2$   & $1.1$   & $1.4$   & $1.3$   & $1.1$   & $1.3$   & $1.3$   & $1.2$  \\
 & Exp$_\mt{Stat}$ + Th$_\mt{Par}$   & $1.3$   & $1.2$   & $1.1$   & $1.4$   & $1.3$   & $1.1$   & $1.3$   & $1.3$   & $1.3$  \\
 & Exp$_\mt{Stat}$ + Th$_\mt{Intr}$   & $1.4$   & $1.3$   & $1.1$   & $1.5$   & $1.4$   & $1.2$   & $1.4$   & $1.4$   & $1.3$  \\
 & Exp$_\mt{Stat}$ + Th   & $1.4$   & $1.3$   & $1.2$   & $1.5$   & $1.4$   & $1.2$   & $1.4$   & $1.4$   & $1.3$  \\
\crowcolorA$g_{HZ \gamma}^{\mathrm{eff}}[\%]$& Exp$_\mt{Stat}$   & $8.1$   & $5.9$   & $5.4$   & $8.9$   & $4.3$   & $3.5$   & $6.2$   & $9.9$   & $9.3$  \\
\crowcolorA & Exp$_\mt{Stat}$ + Th$_\mt{Par}$   & $8.1$   & $6.$   & $5.5$   & $9.$   & $4.4$   & $3.5$   & $6.2$   & $9.9$   & $9.3$  \\
\crowcolorA & Exp$_\mt{Stat}$ + Th$_\mt{Intr}$   & $8.1$   & $6.6$   & $6.1$   & $9.$   & $5.3$   & $4.2$   & $6.2$   & $9.9$   & $9.6$  \\
\crowcolorA & Exp$_\mt{Stat}$ + Th   & $8.1$   & $6.5$   & $6.1$   & $9.1$   & $5.3$   & $4.2$   & $6.2$   & $9.9$   & $9.5$  \\
$g_{Hgg}^{\mathrm{eff}}[\%]$& Exp$_\mt{Stat}$   & $1.1$   & $0.78$   & $0.54$   & $1.3$   & $0.95$   & $0.74$   & $0.75$   & $0.94$   & $0.81$  \\
 & Exp$_\mt{Stat}$ + Th$_\mt{Par}$   & $1.1$   & $0.79$   & $0.54$   & $1.3$   & $0.96$   & $0.75$   & $0.75$   & $0.95$   & $0.82$  \\
 & Exp$_\mt{Stat}$ + Th$_\mt{Intr}$   & $1.2$   & $0.82$   & $0.6$   & $1.4$   & $0.99$   & $0.78$   & $0.78$   & $0.98$   & $0.85$  \\
 & Exp$_\mt{Stat}$ + Th   & $1.2$   & $0.82$   & $0.6$   & $1.4$   & $1.$   & $0.79$   & $0.78$   & $0.98$   & $0.85$  \\
\crowcolorA$g_{Htt}^{\mathrm{eff}}[\%]$& Exp$_\mt{Stat}$   & $3.1$   & $2.8$   & $1.5$   & $3.2$   & $2.1$   & $2.1$   & $3.1$   & $3.1$   & $3.1$  \\
\crowcolorA & Exp$_\mt{Stat}$ + Th$_\mt{Par}$   & $3.2$   & $2.8$   & $1.5$   & $3.2$   & $2.1$   & $2.1$   & $3.1$   & $3.1$   & $3.1$  \\
\crowcolorA & Exp$_\mt{Stat}$ + Th$_\mt{Intr}$   & $3.1$   & $2.8$   & $1.5$   & $3.2$   & $2.1$   & $2.1$   & $3.1$   & $3.1$   & $3.1$  \\
\crowcolorA & Exp$_\mt{Stat}$ + Th   & $3.2$   & $2.8$   & $1.5$   & $3.2$   & $2.1$   & $2.1$   & $3.1$   & $3.1$   & $3.1$  \\
$g_{Hcc}^{\mathrm{eff}}[\%]$& Exp$_\mt{Stat}$   & $1.7$   & $1.1$   & $0.72$   & $3.9$   & $1.8$   & $1.3$   & $1.8$   & $1.3$   & $1.1$  \\
 & Exp$_\mt{Stat}$ + Th$_\mt{Par}$   & $1.8$   & $1.2$   & $0.88$   & $4.$   & $1.9$   & $1.4$   & $1.8$   & $1.4$   & $1.3$  \\
 & Exp$_\mt{Stat}$ + Th$_\mt{Intr}$   & $1.7$   & $1.2$   & $0.77$   & $4.$   & $1.8$   & $1.3$   & $1.8$   & $1.3$   & $1.2$  \\
 & Exp$_\mt{Stat}$ + Th   & $1.8$   & $1.3$   & $0.92$   & $4.$   & $1.9$   & $1.4$   & $1.9$   & $1.4$   & $1.3$  \\
\crowcolorA$g_{Hbb}^{\mathrm{eff}}[\%]$& Exp$_\mt{Stat}$   & $0.66$   & $0.4$   & $0.29$   & $0.92$   & $0.3$   & $0.17$   & $0.52$   & $0.61$   & $0.46$  \\
\crowcolorA & Exp$_\mt{Stat}$ + Th$_\mt{Par}$   & $0.73$   & $0.5$   & $0.42$   & $0.96$   & $0.44$   & $0.37$   & $0.62$   & $0.7$   & $0.56$  \\
\crowcolorA & Exp$_\mt{Stat}$ + Th$_\mt{Intr}$   & $0.7$   & $0.49$   & $0.41$   & $0.97$   & $0.45$   & $0.38$   & $0.57$   & $0.66$   & $0.53$  \\
\crowcolorA & Exp$_\mt{Stat}$ + Th   & $0.76$   & $0.58$   & $0.52$   & $1.$   & $0.56$   & $0.5$   & $0.67$   & $0.74$   & $0.62$  \\
$g_{H\tau\tau}^{\mathrm{eff}}[\%]$& Exp$_\mt{Stat}$   & $0.77$   & $0.57$   & $0.48$   & $1.3$   & $0.92$   & $0.73$   & $0.64$   & $0.69$   & $0.57$  \\
 & Exp$_\mt{Stat}$ + Th$_\mt{Par}$   & $0.77$   & $0.57$   & $0.48$   & $1.3$   & $0.93$   & $0.74$   & $0.65$   & $0.69$   & $0.57$  \\
 & Exp$_\mt{Stat}$ + Th$_\mt{Intr}$   & $0.79$   & $0.61$   & $0.53$   & $1.3$   & $0.95$   & $0.77$   & $0.67$   & $0.72$   & $0.61$  \\
 & Exp$_\mt{Stat}$ + Th   & $0.79$   & $0.61$   & $0.53$   & $1.3$   & $0.95$   & $0.78$   & $0.68$   & $0.72$   & $0.62$  \\
\crowcolorA$g_{H\mu\mu}^{\mathrm{eff}}[\%]$& Exp$_\mt{Stat}$   & $4.1$   & $3.9$   & $3.5$   & $4.3$   & $4.1$   & $3.5$   & $3.9$   & $3.9$   & $3.8$  \\
\crowcolorA & Exp$_\mt{Stat}$ + Th$_\mt{Par}$   & $4.1$   & $3.9$   & $3.5$   & $4.3$   & $4.1$   & $3.4$   & $3.9$   & $3.9$   & $3.9$  \\
\crowcolorA & Exp$_\mt{Stat}$ + Th$_\mt{Intr}$   & $4.1$   & $3.9$   & $3.5$   & $4.3$   & $4.1$   & $3.5$   & $3.9$   & $3.9$   & $3.9$  \\
\crowcolorA & Exp$_\mt{Stat}$ + Th   & $4.1$   & $3.9$   & $3.5$   & $4.3$   & $4.1$   & $3.5$   & $3.9$   & $3.9$   & $3.9$  \\
$\delta g_{1Z}[\times 10^{2}]$& Exp$_\mt{Stat}$   & $0.039$   & $0.015$   & $0.013$   & $0.03$   & $0.0034$   & $0.0012$   & $0.087$   & $0.085$   & $0.036$  \\
 & Exp$_\mt{Stat}$ + Th$_\mt{Par}$   & $0.039$   & $0.015$   & $0.013$   & $0.03$   & $0.0034$   & $0.0012$   & $0.087$   & $0.085$   & $0.036$  \\
 & Exp$_\mt{Stat}$ + Th$_\mt{Intr}$   & $0.039$   & $0.015$   & $0.013$   & $0.031$   & $0.0034$   & $0.0012$   & $0.087$   & $0.086$   & $0.037$  \\
 & Exp$_\mt{Stat}$ + Th   & $0.039$   & $0.015$   & $0.013$   & $0.031$   & $0.0034$   & $0.0012$   & $0.088$   & $0.086$   & $0.037$  \\
\crowcolorA$\delta \kappa_{ \gamma}[\times 10^{2}]$& Exp$_\mt{Stat}$   & $0.056$   & $0.019$   & $0.015$   & $0.043$   & $0.0073$   & $0.0026$   & $0.089$   & $0.086$   & $0.049$  \\
\crowcolorA & Exp$_\mt{Stat}$ + Th$_\mt{Par}$   & $0.056$   & $0.019$   & $0.015$   & $0.043$   & $0.0074$   & $0.0026$   & $0.089$   & $0.086$   & $0.049$  \\
\crowcolorA & Exp$_\mt{Stat}$ + Th$_\mt{Intr}$   & $0.056$   & $0.02$   & $0.016$   & $0.044$   & $0.0074$   & $0.0026$   & $0.09$   & $0.086$   & $0.05$  \\
\crowcolorA & Exp$_\mt{Stat}$ + Th   & $0.056$   & $0.02$   & $0.016$   & $0.044$   & $0.0074$   & $0.0026$   & $0.09$   & $0.086$   & $0.05$  \\
$\lambda_{Z}[\times 10^{2}]$& Exp$_\mt{Stat}$   & $0.041$   & $0.019$   & $0.014$   & $0.042$   & $0.0053$   & $0.0018$   & $0.11$   & $0.1$   & $0.05$  \\
 & Exp$_\mt{Stat}$ + Th$_\mt{Par}$   & $0.041$   & $0.019$   & $0.014$   & $0.042$   & $0.0053$   & $0.0018$   & $0.11$   & $0.1$   & $0.05$  \\
 & Exp$_\mt{Stat}$ + Th$_\mt{Intr}$   & $0.041$   & $0.019$   & $0.014$   & $0.043$   & $0.0053$   & $0.0018$   & $0.11$   & $0.1$   & $0.05$  \\
 & Exp$_\mt{Stat}$ + Th   & $0.041$   & $0.019$   & $0.014$   & $0.042$   & $0.0053$   & $0.0018$   & $0.11$   & $0.1$   & $0.05$  \\
\bottomrule
\end{tabular}
}
\end{table}

\clearpage

\section{The Higgs boson self-coupling} \label{selfcoupling}

The Higgs field is responsible for the spontaneous breaking of the electroweak symmetry, and for the generation of all the SM particle masses,  because its potential features 
a global minimum away from the origin. Within the SM, this potential is fully characterised by two parameters, the Higgs mass $m_h$, and $v$, which can be experimentally inferred from the measurements of the Fermi constant ($v=1/\sqrt{\sqrt{2}G_F}\approx 246$\,GeV).
\begin{equation}\label{eq:SMpotential}
V(h)= \frac{1}{2} m_H^2 h^2 + \lambda_3 v h^3 + \frac{1}{4}\lambda_4 h^4,\qquad \textrm{with}\quad \lambda_3^{\rm SM}= \lambda_4^{\rm SM}= \frac{m_H^2}{2 v^2}.
\end{equation}

 However, the Higgs potential could show sizeable departures from the SM form, described in eq.~(\ref{eq:SMpotential}).
 The understanding of EW symmetry breaking will remain hypothetical until experimental measurements reconstruct the shape of the Higgs potential. The \textit{measurement} of the Higgs potential is therefore a high priority goal on the physics programme of all future colliders.

Unfortunately, the Higgs self-interactions, apart from the simple kinematical 2-point interaction that corresponds to the Higgs boson mass, are not physical observables.  Therefore, a theoretical framework is needed to infer their values from experimental measurements.  One needs a general parametrisation of the departures from the SM that allows the various Higgs couplings to vary continuously. Within this framework, one makes accurate predictions of various observables as a function of the modified Higgs couplings and a global fit then leads to a determination of all these couplings. Effective Field Theory offers us such a theoretically sound framework in which higher order calculations can be performed to provide solid and improvable predictions able to cope with systematic and statistic experimental uncertainties. As in Section~\ref{eft}, we will focus our attention on EFT where the EW symmetry is linearly realised, i.e. under the assumption that no new heavy degree of freedom acquires its mass from the Higgs expectation value. In that case, there are only two dimension-6 operators that induce a deviation of the Higgs self-couplings
\begin{eqnarray}
\mathcal{L}&=&\mathcal{L}^{\rm SM} + \frac{c_\phi}{2 \Lambda^2} \partial_\mu |\phi|^2 \partial^\mu |\phi|^2 - \frac{c_6\, \lambda_3^{\rm SM}}{\Lambda^2} |\phi|^6 
\nonumber\\ 
&\Rightarrow & \kappa_3 \equiv \frac{\lambda_3}{\lambda_3^{\rm SM}} = 1+ \left(c_6-\frac{3}{2}c_\phi\right) \frac{v^2}{\Lambda^2}, 
\quad \kappa_4 \equiv \frac{\lambda_4}{\lambda_4^{\rm SM}} = 1+ \left(6 c_6-\frac{25}{3}c_\phi\right) \frac{v^2}{\Lambda^2}.
\label{eq:kappa3}
\end{eqnarray}

In particular, the operator proportional to $c_\phi$ requires a non-linear field definition to keep the Higgs boson kinetic term canonically normalised. The modifications of the cubic and quartic self-interactions are related 
in this model. Independent modifications are only obtained when operators of dimension 8 are considered.

\begin{figure}[ht]
\centering
\includegraphics[width=.95\linewidth]{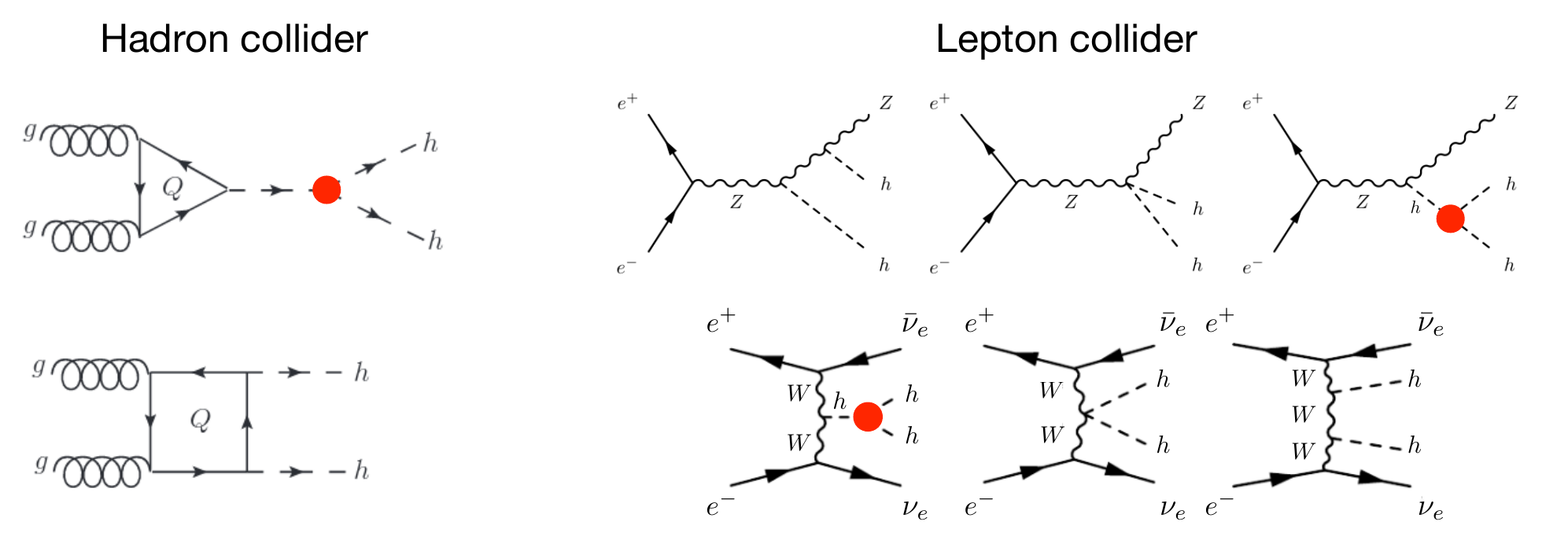}
\caption{\label{fig:HHproduction}
Representative Feynman diagrams for the leading contribution to double Higgs production at hadron (left) and lepton (right) colliders. Extracting the value of the Higgs self-coupling, in red, requires a knowledge of the other Higgs couplings that also contribute to the same process. See Table~\ref{tab:xsecs} for the SM rates. At lepton colliders, double Higgs production can also occur via vector boson fusion with neutral currents but the rate is about ten times smaller.  
The contribution proportional to the cubic Higgs self-coupling involves an extra Higgs propagator that dies off at high energy. Therefore, the kinematic region close to threshold is more sensitive to the Higgs self-coupling. 
}
\end{figure}

\begin{figure}[ht]
\centering
\includegraphics[width=.95\linewidth]{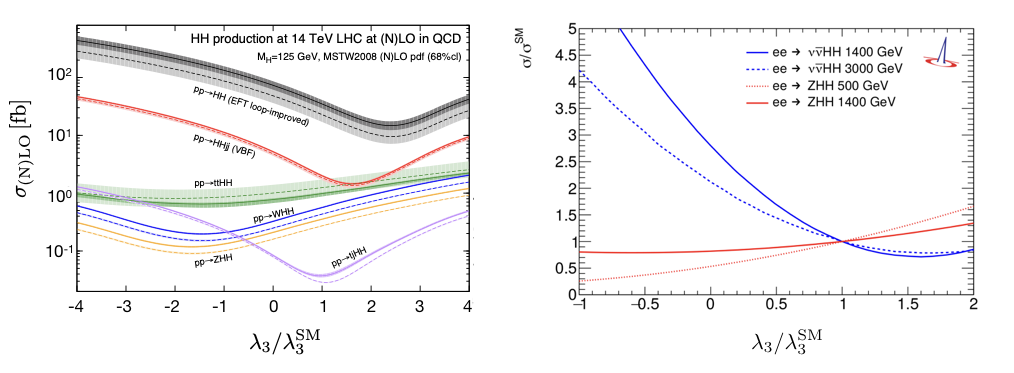}
\caption{\label{fig:HHxs}
Double Higgs production at hadron (left)~\cite{Frederix:2014hta} and lepton (right)~\cite{WHIZARD} colliders as a function of the modified Higgs cubic self-coupling. See Table~\ref{tab:xsecs} for the SM rates. 
At lepton colliders, the production cross sections do depend on the polarisation but this dependence drops out in the ratios to the SM rates (beam spectrum and QED ISR effects have been included).
}
\end{figure}

The most direct way to assess the Higgs cubic self-interaction is through the measurement of double Higgs production  either at hadron colliders, where the production is dominated by gluon fusion, $gg\to HH$, or at lepton colliders via double Higgs-strahlung,  $e^+e^-\to ZHH$, particularly relevant at low energies, or via vector boson fusion (VBF), $e^+e^-\to HH\nu_e\bar{\nu}_e$, more important at centre-of-mass energies of 1\,TeV and above. At leading order, double Higgs production receives a contribution proportional to the cubic coupling, for both $pp$ and $e^+e^-$ collisions, as shown in Fig.~\ref{fig:HHproduction}. Figure~\ref{fig:HHxs} shows the dependence of the inclusive double Higgs production cross section when the value of the Higgs cubic coupling is varied, assuming no other deviation from the SM. 
Gluon fusion production at a hadron collider has been computed within the SM at NNLO accuracy in the infinite top mass limit\cite{deFlorian:2013uza,deFlorian:2013jea,Grigo:2014jma,deFlorian:2016uhr} and at NLO with the full top mass dependence\cite{Borowka:2016ehy,Borowka:2016ypz,Baglio:2018lrj}, leading to a prediction whose theoretical and parametric uncertainties are of the order of a few percent. 

For the LHC at 14\,TeV, the cross section is predicted to be $36.69^{+2.1\%}_{-4.9\%}$\,fb, about three orders of magnitude smaller than the single Higgs production, which makes the double Higgs channel a challenging process to observe. The most up-to-date analysis relies on the combination of the $b\bar{b}\gamma\gamma$ and $b\bar{b}\tau\tau$ decay channels to reach almost 5 standard deviation evidence for double Higgs production at HL-LHC (see Table~55 and Fig.~65 of Ref~\cite{Cepeda:2019klc}), which can be translated into a 68\% CL bound of order 50\% on the deviation of the Higgs cubic coupling relative to the SM prediction. Note that the mapping of the inclusive $gg\to HH$ cross section onto a value of the Higgs cubic self-coupling is not unique: for instance, at 14\,TeV LHC, a value of the cross section equal to the SM prediction corresponds either to $\kappa_3=1$ or to $\kappa_3\approx6.2$. This ambiguity can however be resolved by analysing the shape of 
the invariant mass distribution of the reconstructed two Higgs boson system: the larger the value of $\kappa_3$, the closer to threshold the $m_{HH}$ distribution is peaked.
This kinematic information is a crucial element of Boosted Decision Trees (BDT) based analysis performed at HL-LHC. However the BDT and the final selection cuts are often devised to optimise the significance of the SM cross section for double Higgs production 
and therefore it is not necessarily optimised for the determination of the Higgs self-coupling directly, leaving room for possible improvement towards an even higher sensitivity. At lepton colliders, double Higgs-strahlung, $e^+e^-\to ZHH$, gives stronger constraints on positive deviations ($\kappa_3>1$), while VBF is better in constraining negative deviations, ($\kappa_3<1$). 
While at HL-LHC, values of $\kappa_3>1$, as expected in models of strong first order phase transition, result in a smaller double-Higgs production cross section due to the destructive interference, at lepton colliders for the $ZHH$ process they actually result in a larger cross section, and hence into an increased precision. For instance at ILC$_{500}$, the sensitivity around the SM value is 27\% but it would reach 18\% around $\kappa_3=1.5$.

Modified Higgs self-interactions can also affect, at higher orders, the single Higgs processes~\cite{McCullough:2013rea,Degrassi:2016wml,Bizon:2016wgr}  and even the electroweak precision observables~\cite{vanderBij:1985ww, Degrassi:2017ucl, Kribs:2017znd}. Since the experimental sensitivities for these observables are better than for double Higgs production, one can devise alternative ways to assess the value of the Higgs self-interactions. For a 240\,GeV lepton collider, the change of the $ZH$ production cross section at NLO induced by a deviation of the Higgs cubic coupling amounts to
\begin{equation}
\sigma^{\rm NLO}_{ZH} \approx \sigma^{\rm NLO, SM}_{ZH}  ( 1+ 0.014 \, \delta \kappa_3).
\end{equation}

Thus, to be competitive with the HL-LHC constraint, the $ZH$ cross section needs to be measured with an accuracy below 1\%, but this is expected to be achieved by $e^+e^-$ Higgs factories at 240/250\,GeV. 
However, one needs to be able to disentangle a variation due to a modified Higgs self-interaction from variations due to another deformation of the SM. This cannot always be done relying only on inclusive measurements~\cite{DiVita:2017eyz,DiVita:2017vrr} and it calls for detailed studies of kinematical distributions with an accurate estimate of the relevant uncertainties~\cite{Maltoni:2017ims}. Inclusive rate measurements performed at two different energies also help lifting the degeneracy among the different Higgs coupling deviations (see for instance the $\kappa_3$ sensitivities reported in Table~\ref{tab:h3} for FCC-ee$_{240}$ vs FCC-ee$_{365}$; it is the combination of the two runs at different energies that improve the global fit, a single run at 365\,GeV alone would not improve much compared to a single run at 240\,GeV). 

In principle, large deformations of $\kappa_3$ could also alter the fit of single Higgs processes often performed at leading order, i.e. neglecting the contribution of $\kappa_3$ at next-to-leading order. The results presented in Section~\ref{eft-results} were obtained along that line.
It was shown in~\cite{DiVita:2017eyz} that a 200\% uncertainty on $\kappa_3$ could for instance increase the uncertainty in $g_{Htt}$ or $g_{Hgg}^{\rm eff}$ by around 30--40\%. The fact that HL-LHC from the double Higgs channel analysis will limit the deviations of $\kappa_3$ to 50\% prevents such a large deterioration of the global fits to single Higgs couplings when also allowing $\kappa_3$ to float. In the effective coupling basis we are considering in this report, the effect of $\kappa_3$ would be mostly in the correlations among the single Higgs couplings. In other bases, like the Warsaw basis, there would be a deterioration up to 15-20\% in the sensitivity of the operator $\mathcal{O}_{\phi\Box}$. Anyway, one should keep in mind that such a deterioration only concerns specific models where the deviations of the Higgs self-coupling is parametrically larger than the deviations of the single Higgs couplings and in generic situations, the results of Section~\ref{eft-results} hold. 

In order to set quantitative goals in the determination of the Higgs self-interactions, it is useful 
to understand how large the deviations from the SM could be while remaining compatible with the 
existing constraints on the different single Higgs couplings. From an agnostic point of view, the Higgs cubic coupling can always be linked to the independent higher dimensional operator $|H|^6$ that does not alter any 
other Higgs couplings.  Still, theoretical considerations set an upper bound on the deviation of the trilinear Higgs couplings.
Within the plausible linear EFT assumption discussed above, perturbativity imposes a maximum deviation of the Higgs cubic self-interaction, relative to the SM value, of the order of~\cite{DiVita:2017eyz,Falkowski:2019tft}
\begin{equation}
  |\kappa_3 |  \lsim {\rm Min}(600\, \xi, 4\pi) \, ,
\end{equation}
where $\xi$ is the typical size of the deviation of the single Higgs couplings to other SM particles~\cite{Giudice:2007fh}.
However, the stability condition of the EW vacuum, i.e. the requirement that no other deeper minimum results from the inclusion of higher dimensional operators in the Higgs potential, 
gives the bound~\cite{Falkowski:2019tft,DiLuzio:2017tfn}
\begin{equation}
| \kappa_3 | \lsim 70\, \xi\, .
\end{equation}

At HL-LHC, $\xi$ can be determined with a precision of 1.5\% at best, corresponding to a sensitivity on the Higgs self-coupling of about 100\%, and thus somewhat inferior but roughly comparable to the direct sensitivity of 50\%~\cite{Cepeda:2019klc}.
Parametric enhancements of the deviations of Higgs cubic self-coupling relative to the single Higgs couplings require a particular dynamics for the new physics. An example is encountered in Higgs portal models where the Higgs boson mixes with a SM neutral scalar field, possibly contributing to the dark matter relic abundance~\cite{Azatov:2015oxa, DiVita:2017eyz}. In more traditional scenarios addressing the hierarchy problem, such as supersymmetric or composite models, the deviation of $\kappa_3$ is expected to be of the order $\xi$ and is likely to remain below the experimental sensitivity. 

The sensitivity of the various future colliders to the Higgs cubic coupling can be obtained using five different methods (1, 2(a), 2(b), 3, and 4):
\begin{enumerate}
\item an exclusive analysis of HH production, i.e., a fit of the double Higgs cross section considering only deformation of the Higgs cubic coupling;
\item a global analysis of HH production, i.e., a fit of of the double Higgs cross section considering also all possible deformations of the single Higgs couplings that are constrained by single Higgs processes;
	\begin{enumerate}
	\item  the global fit does not consider the effects at higher order of the modified Higgs cubic coupling to single Higgs production and to Higgs decays;
	\item  these higher order effects are included;
	\end{enumerate}
\item an exclusive analysis of single Higgs processes at higher order, i.e., considering only deformation of the Higgs cubic coupling; technically, this will be a one-dimensional EFT fit where only the linear combination of the two operators of Eq.~(\ref{eq:kappa3}) corresponding to the $\kappa_3$ deformation is turned on;
\item a global analysis of single Higgs processes at higher order, i.e., considering also all possible deformations of the single Higgs couplings.
Technically, this will be a 30-parameter EFT fit done within the scenario SMEFT$_{\rm ND}$ scenario of Eq.~(\ref{eq:SMEFT_ND}).
The contribution of $\kappa_3$ to EWPO at 2-loop could also be included but for the range of $\kappa_3$ values discussed here, the size of effects would be totally negligible. 
\end{enumerate}

Most of the studies of the Higgs self-couplings at Future Colliders were done following Method~(1). In order to maximize the sensitivity to $\lambda_3$, the analyses rely on sophisticated BDTs, and a simple recasting within an EFT framework is not an easy task. A pragmatic approach was followed along the line of what was proposed in~\cite{DiVita:2017eyz}: different bins in $m_{HH}$ are considered and the experimental uncertainty on the total rate is distributed in the different bins according to their number of expected events. This certainly ignores the bin-to-bin correlations and it does not take into account either that the background itself has a non-trivial shape as a function of $m_{HH}$. Nevertheless, the results  obtained that way are in good agreement with those quoted by the different collaborations. This approach has the advantage that it can be easily generalised to a global EFT analysis that considers all the operators modifying also the single Higgs couplings, Methods~(2). One should keep in mind that the bounds derived that way represent a crude estimate that waits for a proper experimental study\footnote{A detailed $m_{HH}$ binned analysis was not available for HE-LHC, hence we could not estimate the $\kappa_3$ sensitivity along Method~(2) for that collider. Similarly, for CLIC$_{3000}$, the granularity of the available information was not sufficient to match the announced sensitivity, and therefore we did not venture into a complete study along Method~(2) either. In both cases, our checks led to the conclusion that there will not be any noticeable difference between the sensitivity obtained in Methods~(1) and~(2).}.

For most colliders, the single Higgs constraints are strong enough that they give a contribution to the double Higgs production below its experimental sensitivity. And Method~(1) and Method~(2) lead to rather similar bound on $\kappa_3$. A notable exception is at FCC-hh where the 1\% uncertainty on the top Yukawa coupling results in a deviation of the double Higgs production rate at a level comparable to the one induced by a shift of $\kappa_3$ by 5\%. 
While a parametric enhancement of the deviation in $\kappa_3$ compared to the other Higgs couplings deviations could make its higher order contributions to single Higgs processes as important as the leading order ones and thus could in principle modify the global fit, in practice, the constraints set by the double Higgs production are strong enough that there is hardly any difference in the results obtained using Methods~(2a) and~(2b). 
Methods (3) and (4) are particularly relevant for low-energy colliders below the double Higgs production threshold. Above this threshold, these methods can still be relevant to complement results from the double Higgs analysis, for instance by helping to resolve the degeneracy between the SM and a second minimum of the likelihood. While this does not modify the 1$\sigma$ bound on $\kappa_3$, it can impact the bound starting at the 2$\sigma$ level due to the non-Gaussian profile of the likelihood. It should be remembered that the single Higgs data used in Methods~(3) and (4) have not been optimised for the extraction of the Higgs self-coupling that would benefit from further differential information. Therefore, the bounds on $\kappa_3$ should be considered as conservative and are certainly improvable.

Table~\ref{tab:h3} reports the sensitivity at the various colliders for the Higgs cubic coupling determination. For the global EFT fits, we limit ourselves to the SMEFT$_{\rm ND}$ scenario, see Eq.~(\ref{eq:SMEFT_ND}), extended with Eq.~(\ref{eq:Lh3}). For all results a simple combination with the HL-LHC results is done, i.e.~by using a 50\% uncertainty on $\kappa_{3}$. It is seen that the results for Methods~(1) and~(2) are very similar, showing that the determination of $\kappa_3$ is dominated by the di-Higgs measurements when these are included. When comparing Methods~(3) and~(4) one observes that the exclusive results appear to be more constraining that the global results. But they overestimate the sensitivity as \textit{a priori} it is not known which operator coefficients to fix and the same single-Higgs data should be used to constrain all operators. Method~(4) is significantly more robust than Method~(3). In the following we focus on Methods~(1) and~(4). 

\begin{table}[!ht]
\caption{\label{tab:h3} Sensitivity at 68\% probability on the Higgs cubic self-coupling at the various future colliders. All the numbers reported correspond to a simplified combination of the considered collider with HL-LHC, which is approximated by a 50\% constraint on $\kappa_3$. The numbers in the first column (i.e. "di-H excl." or Method~(1)) correspond to the results given by the future collider collaborations and in parenthesis, we report our derived estimate obtained in the binned analysis described in the text. In the three last columns, i.e. Methods~(2a),~(3) and~(4), we report the results computed by the Higgs@FC working group. For the leptonic colliders, the runs are considered in sequence. For the colliders with $\sqrt{s}\lesssim 400$\,GeV, Methods~(1) and~(2.a) cannot be used, hence the dash signs in the corresponding cells. No sensitivity was computed along Method~(2.a) for HE-LHC and CLIC$_{3000}$ but our initial checks do not show any difference with the sensitivity obtained for Method~(1). In the global analyses, Methods~(2.a) and~(4), we consider the flavour scenario of \textit{Neutral Diagonality} (the results show little difference compared to the ones reported in the first version of this report within the \textit{Neutral Diagonality} scenario). 
Due to the lack of results available for the $ep$ cross section in SMEFT, we do not present any result for LHeC nor HE-LHeC, and only results with Method~(1) for FCC-eh. 
For Method~(3) results are shown with and without combination with HL-LHC for many of the colliders (in several cases, the fit for Method~(4) does not converge for the standalone collider without HL-LHC input).
}
\centering
{\small
\begin{tabular}{ |c| c |c |c c | c |}
\toprule
collider       &    \multicolumn{2}{c|}{di-Higgs}  & \multicolumn{3}{c|}{single-Higgs}              \\
&  (1) excl.  & (2.a) glob.    & \multicolumn{2}{c}{(3) excl.}    & (4) glob.              \\
 & & & with HL-LHC & w/o HL-LHC & \\
\midrule
HL-LHC         &  $_{-50}^{+60}$\% (50\%) & 52\%    & 47\%  & --  & 50\%                 \\
\crowcolorA 
HE-LHC         &  10-20\% (n.a.)          &  n.a.   & 40\%  & 80\%  & 50\%  \\
ILC$_{250}$    &  $-$                     & $-$     & 29\%  & 37\%  & 49\%   \\
\crowcolorA
ILC$_{350}$    &  $-$                     & $-$     &28\%  & 37\%   & 46\%   \\
ILC$_{500}$    &  27\% (27\%)             & 27\%    &27\%  & 32\%    & 38\%   \\
\crowcolorA
ILC$_{1000}$   &  10\%  (n.a.)                  & 10\%    &25\%  & 30\%   & 36\%   \\
CLIC$_{380}$   &  $-$                     & $-$     &46\%  & 120\%   & 50\%   \\
\crowcolorA
CLIC$_{1500}$  &  36\% (36\%)             & 36\%    &41\%  & 78\%    & 49\%   \\
CLIC$_{3000}$  & $_{-7}^{+11}$\% (n.a.)   & n.a.    &35\%  & 63\%    & 49\%   \\
\crowcolorA
FCC-ee$_{240}$ &  $-$                     & $-$     &19\%  & 21\%   & 49\%   \\
FCC-ee$_{365}$ &  $-$                     & $-$     &19\% & 21\%     & 33\%   \\
\crowcolorA
FCC-ee$_{365}^{\mathrm{4IP}}$ &$-$                  & $-$  & 14\% & 14\%    & 24\%   \\
FCC-eh   &  17-24\% (n.a.)               & n.a.     &n.a.  & n.a.    & n.a.   \\
\crowcolorA
FCC-ee/eh/hh   &  5\% (5\%)               & 6\%     &18\%  &19\%      & 25\%   \\
LE-FCC & 15\% (n.a) & n.a & n.a. & n.a. & n.a.\\
\crowcolorA
CEPC           & $-$                      & $-$     &17\%  & 18\%   & 49\%   \\
\bottomrule
\end{tabular}
}
\end{table}

\begin{figure}[!ht]
\centering
\includegraphics[width=.8\linewidth]{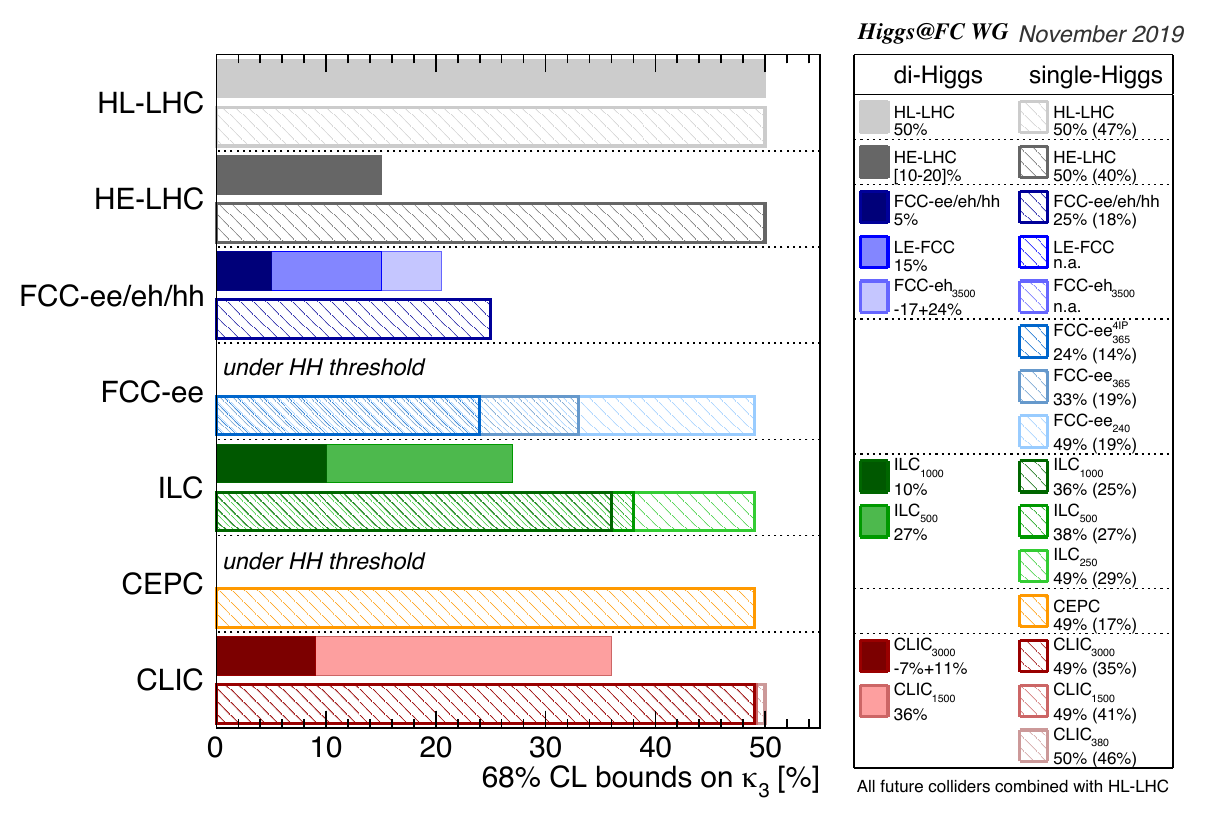}
\caption{\label{fig:h3-summary}
Sensitivity at 68\% probability on the Higgs cubic self-coupling at the various FCs. All values reported correspond to a simplified combination of the considered collider with HL-LHC. Only numbers for  Method~(1), i.e. "di-H excl.", corresponding to the results given by the future collider collaborations, and for Method~(4), i.e. "single-H glob." are shown (the results for Method~(3) are reported in parenthesis). For Method~(4) we report the results computed by the Higgs@FC working group. For the leptonic colliders, the runs are considered in sequence. 
For the colliders with $\sqrt{s}\lesssim 400$\,GeV, Method (1) cannot be used, hence the dash signs. Due to the lack of results available for the $ep$ cross section in SMEFT, we do not present any result for LHeC nor HE-LHeC, and only results with Method (1) for FCC-eh. When uncertainties are asymmetric (CLIC and FCC-eh) or a range is given (HE-LHC) the mid value is displayed.
}
\end{figure}

The results are also summarised in Fig.~\ref{fig:h3-summary} for these two methods.
Even though the likelihood is not a symmetric function of $\kappa_3$, the current level  of precision in this EFT analysis is not good enough to robustly assess an asymmetric error and we report only symmetrised bounds in the Figure.

Based on di-Higgs measurements, with a 50\% sensitivity on $\kappa_3$, HL-LHC will exclude the absence of the Higgs self-interaction ($\kappa_3$ = 0) at 95\%CL. 
Several of the proposed FCs (HE-LHC, LE-FCC and LHeC) will reach a sensitivity of order 20\% based on di-Higgs production, thus establishing the existence of the self-interaction at 5$\sigma$.
Even more remarkable, CLIC$_{3000}$ and ILC$_{1000}$ are expected to reach a sensitivity of order 10\% and FCC$_{hh}$ of the order of 5\%, where one could start probing the size of the quantum corrections to the Higgs potential directly. 

With single-Higgs production at FCC-ee and ILC$_{500}$ and ILC$_{1000}$, in combination with di-Higgs results from HL-LHC, a sensitivity of $\sim 30\%$ can be reached. For FCC-ee with 4 interaction points (IPs) this is reduced to 24\%. For the other collider options with $\sqrt{s}<400$~GeV no improvement w.r.t. the HL-LHC result is seen.

Even though we do not report any sensitivity on $\kappa_3$ at muon-collider, we note that preliminary studies~\cite{muoncprepmeeting} indicate that a 10\,TeV (resp. 30\,TeV) machine could reach a 3\% (resp. 1\%) sensitivity.

\section{Rare Higgs boson decays} \label{raredecays}

There are many reasons for the interest in rare Higgs boson decays. First, they provide access to Higgs couplings which are expected to be small in the SM and have not yet been directly probed. A leading example is the coupling to second and first generation fermions, whose determination would test the hypothesis that the same Higgs doublet is responsible for the mass generation of the lighter states of the SM. More specifically, the measurement of several Yukawa couplings will allow the comparison of ratios of couplings with ratios of masses on the one hand, and test constants of proportionality on the other. The second motivation is that processes which are predicted to be rare in the SM, offer enhanced sensitivity to new physics residing at high scales. 
A leading example is the search for flavour-changing neutral interactions, which are extremely suppressed in the SM and if detected would reliably point to the existence of new physics. Third, peculiar and rare final state signatures can have a special connection with BSM scenarios. One example is $H$ decaying to invisible particles, which is used to constrain scenarios featuring DM candidates. In the SM, the Higgs boson can decay invisibly via $H\to 4 \nu$ with a branching ratio of $0.11\%$. 
Finally, Yukawa interactions with first generation fermions are the cornerstone of the low-energy constraints on $CP$ violation of the couplings on the third generation. The typical example here are limits obtained by the EDM's on the  CP-odd interaction of the third generation fermions (Section~\ref{higgscp}).  

The reach of various colliders for rare decays, depends in the first place on the available statistics of the Higgs bosons being produced. The expected rates are presented in the Appendix~\ref{app:xs}, Table ~\ref{tab:xsecs}. 

In the following, we restrict ourselves to a summary of the prospects to bound or determine the size of the interactions of the Higgs to the other SM particles through decays. These can occur either directly, through a process which is proportional to a tree-level coupling squared, i.e. all decays $H \to \bar f f$, where $f$ is any SM fermion of the first or second generation,  or indirectly, i.e. through interfering amplitudes or loops, such as $H \to \gamma \gamma$ and $H\to \gamma Z$.  We will also briefly present results on very rare exclusive decays, which  could provide indirect information on the light-quark Yukawa couplings.  We follow the notation introduced in the $\kappa$-framework and consider the rescaling factors $\kappa_i= y_i/y^{SM}_i$ introduced previously for the couplings to quarks $\kappa_u, \kappa_d, \kappa_c, \kappa_s$  and for $\kappa_\mu$, and for the loop induced processes, $\kappa_\gamma$ and $\kappa_{Z\gamma}$. The values of $\kappa_\mu,\kappa_\gamma,\kappa_{Z\gamma}, \kappa_c$ have been obtained from the kappa-3,-4 fits presented in Section~\ref{kappa-results} and we do not reproduce them here, while the upper bounds on $\kappa_u, \kappa_d, \kappa_s$ ($\kappa_c$ for hadron colliders) are obtained from the upper limits on ${\rm BR}_{\rm unt}$. Constraints on flavour-changing Higgs boson interactions are not reported here. 

\begin{figure}[ht]
\centering
\includegraphics[width=.62\linewidth]{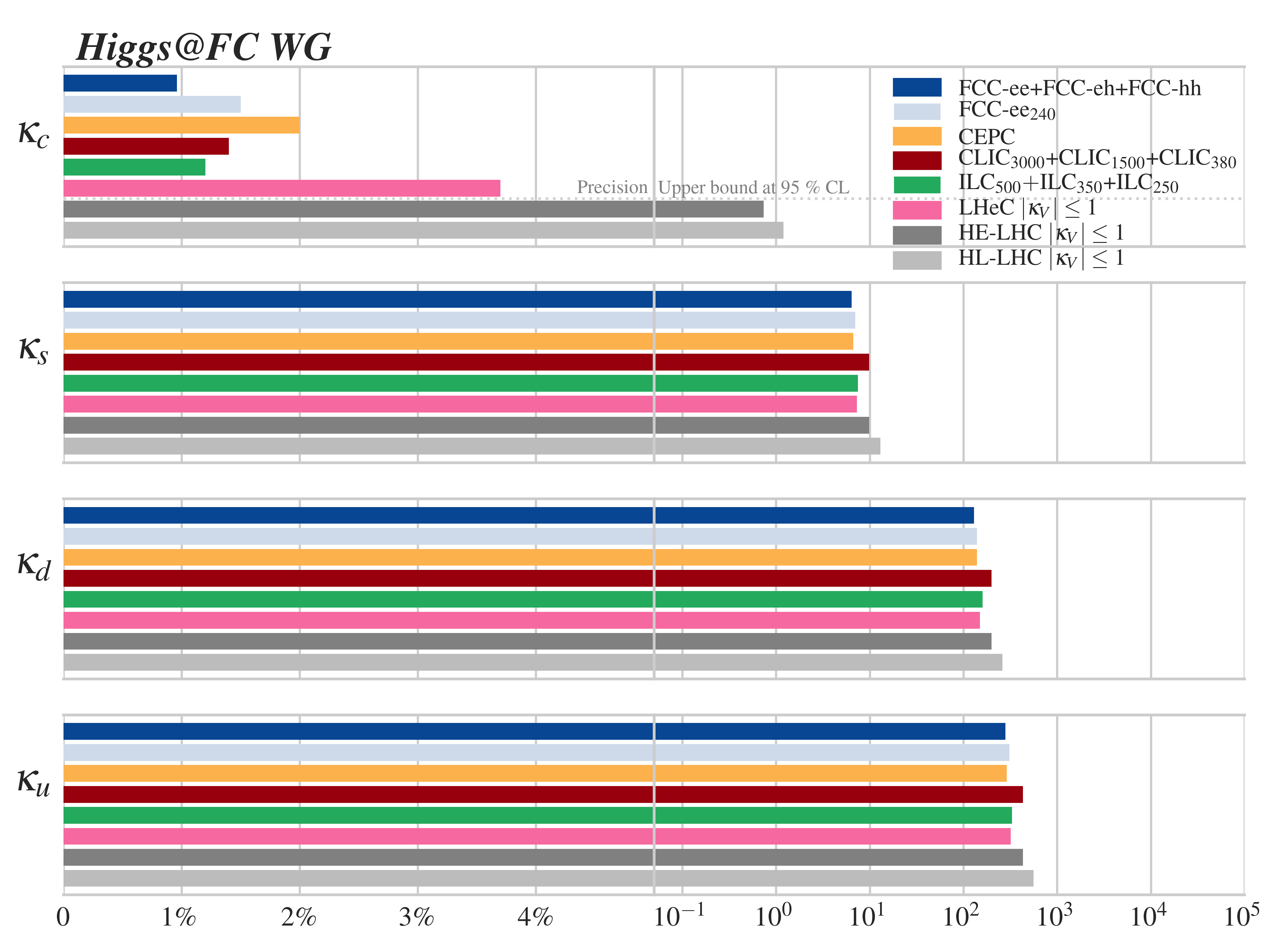}
\caption{\label{fig:raredecays} Summary plot illustrating the limits that can be obtained from rare Higgs decays on the couplings. 
}
\end{figure}

The constraints of the couplings to first and second generation quarks are given in Table~\ref{tab:rare} and displayed in Fig.~\ref{fig:raredecays}, based on the results on BR$_\textrm{unt}$. For $\kappa_c$ the hadron colliders reach values of ${\cal{O}}(1)$, and lepton colliders and LHeC are expected to  improve the precision by about two orders of magnitude, to a 1-2\%. For the strange quarks the constraints are about 5-10$\times$ the SM value while for the first generation it ranges between 100-600$\times$ the SM value. For the latter, future colliders could improve the limits obtained at  the HL-LHC by about a factor of two. For HL-LHC, HE-LHC and LHeC, the determination of BR$_\textrm{unt}$ relies on assuming $\kappa_V\leq 1$. For $\kappa_\gamma$, $\kappa_{Z\gamma}$ and $\kappa_\mu$ the lepton colliders do not significantly improve the precision compared to HL-LHC but the higher energy hadron colliders, HE-LHC and FCC$_{hh}$, achieve improvements of factor of 2-3 and 5-10, respectively, in these couplings.

For the electron Yukawa coupling, the current limit $\kappa_e<611$~\cite{Khachatryan:2014aep} is based on the direct search for $H \to e^{+} e^{-}$. A preliminary study at the FCC-{ee}~\cite{Abada:2019zxq} has assessed the reach of a dedicated run at $\sqrt{s}=m_H$. At this energy the cross section for $e^+ e^- \to H$ is 1.64\,fb, which reduces to 0.3 with an energy spread equal to the SM Higgs width. According to the study, with 2\,ab$^{-1}$ per year achievable with an energy spread of 6\,MeV, a significance of 0.4 standard deviations could be achieved, equivalent to an upper limit of $2.5$ times the SM value, while the SM sensitivity would be reached in a five year run.

\begin{table}
\caption{\label{tab:rare} Upper bounds on the $\kappa_i$ for $u$, $d$, $s$ and $c$ (at hadron colliders) at 95\% CL, obtained from the upper bounds on BR$_\textrm{unt}$ in the kappa-3 scenario.}
\centering
\small
\begin{tabular}{| c | c| c c | c c c c c|}
\toprule
 &  & \multicolumn{7}{c|}{HL-LHC+}\\
   &HL-LHC  &  LHeC  &  HE-LHC &  ILC$_{500}$ 
&  CLIC$_{3000}$  &  CEPC & FCC-ee$_{240}$&  FCC-ee/eh/hh \\
\hline
$\kappa_u$  & 560.   & 320. & 430.  &   330. &   430.    &    290.   &   310.   &    280.\\
$\kappa_d$   & 260.  &  150.& 200.  &   160. &   200.    &    140.   &   140.   &    130.\\
$\kappa_s$   & 13.   & 7.3  & 9.9   &   7.5  &   9.9     &     6.7   &     7.   &    6.4\\
$\kappa_c$ & 1.2     & 
&  0.87     &     \multicolumn{5}{c|}{measured directly}\\
\bottomrule
\end{tabular}
\end{table}

\begin{table}[ht]
\caption{\label{tab:invisible} 
Expected upper limits on the invisible and untagged BRs of the Higgs boson. The SM decay, $H\to 4 \nu$, has been subtracted as a background. 
Given are the values of the direct searches using missing (transverse) momentum searches, the constraint derived from the coupling fit (see Table~\ref{tab:resultsKappa3}) in the kappa-3 scenario, and the result from a fit in the $\kappa$ framework where only modifications of ${\rm BR}_{\rm inv}$ are allowed. The last two columns show the corresponding information for untagged BR of the Higgs, ${\rm BR}_{\rm unt}$. For all fits the direct search for invisible decays is included. 
}
\begin{center}
\begin{tabular}{|l|c|c|c || c  | c |}
\toprule
  Collider & \multicolumn{5}{c|}{95\% CL upper bound on } \\
  & \multicolumn{3}{c||}{${\rm BR}_{\rm inv}$ [\%]}& \multicolumn{2}{c|}{ ${\rm BR}_{\rm unt}$ [\%]}  \\
           & Direct & kappa-3 &  ${\rm BR}_{\rm inv}$ only  & kappa-3 & ${\rm BR}_{\rm unt}$ only\\
\midrule
  HL-LHC 			 			&2.6 		& 1.9 & 1.9 		& 4.0 & 3.6\\
  HL-LHC + HE-LHC($S_2^\prime$) &  		& 1.5 & 1.5 		& 2.4 & 1.9 \\
  FCC-hh 			 			&0.025 	& 0.024 & 0.024 	& 1.0 & 0.36\\
\midrule
 HL-LHC + LHeC 	 			&2.3 		& 1.1 & 1.1 		& 1.3 & 1.3 \\
\midrule
 HL-LHC + CEPC 						 &0.3 	& 0.27 & 0.26 		& 1.1 & 0.49\\
 HL-LHC + FCC-ee$_{240}$    			&0.3 		& 0.22 & 0.22 		& 1.2 & 0.62 \\
 HL-LHC + FCC-ee$_{365}$    			& 		& 0.19 & 0.19 		& 1.0  & 0.54 \\
 HL-LHC + ILC$_{250}$           			&0.3 		& 0.26 & 0.25 		& 1.8 & 0.85\\
 HL-LHC + ILC$_{500}$ 		 			&  		& 0.23 & 0.22 		& 1.4 & 0.55\\
 HL-LHC + ILC$_{1000}$ 	 			&  		& 0.22 & 0.20 		& 1.4  & 0.43\\
 HL-LHC + CLIC$_{380}$ 	 			&0.69 	& 0.63 & 0.56 		& 2.7 & 1.0\\
 HL-LHC + CLIC$_{1500}$ 	 			& 		& 0.62 & 0.40 		& 2.4  & 0.51\\
 HL-LHC + CLIC$_{3000}$ 	 			&  		& 0.62 & 0.30 		& 2.4 & 0.33\\
\bottomrule
\end{tabular}
\end{center}
\end{table}

While the limits quoted on $\kappa_c$ from hadron colliders (see Table~\ref{tab:rare}) have been obtained indirectly, we mention that  progress in inclusive direct searches for $H\rightarrow c\bar{c}$ at the LHC has been reported from ATLAS together with a projection for the  HL-LHC. Currently the upper bound on the charm coupling is $\kappa_c \lsim 10$ \cite{Aaboud:2018fhh}. 
With HL-LHC, it is expected to improve the sensitivity to values of $\kappa_c<2.1$ (based on Ref.~\cite{Perez:2015lra}), while LHCb, with the foreseen detector improvement, could reach a sensitivity on $\kappa_c$ of 2-3~\cite{Cepeda:2019klc}. 

Exclusive Higgs decays to a vector meson ($V$) and a photon, $H\to V\gamma$, $V=\rho,\omega,\phi,J/\psi,\Upsilon$ directly probe the Higgs bottom, charm
strange, down and up quark Yukawas~\cite{Bodwin:2013gca,Bodwin:2014bpa,Kagan:2014ila}. 
Within the LHC, the Higgs exclusive decays are the only direct probe of the $u$ and $d$ Yukawa couplings, while if $s$-tagging  could be implemented at the LHC~\cite{Kagan:2014ila}, then the strange Yukawa could be probed both inclusively and exclusively.
On the experimental side, both ATLAS and CMS have reported  upper bounds on 
$H\to J/\psi\gamma$~\cite{Aaboud:2018txb,Sirunyan:2018fmm},  
$H\to\phi\gamma$ and $h\to\rho\gamma$~\cite{Aaboud:2017xnb,Khachatryan:2015lga}. 
These processes receive contributions from two amplitudes, only one of which is proportional to the Yukawa coupling. Since the contribution proportional to the Yukawa is smaller, the largest sensitivity to the Higgs $q$-quark coupling is via the interference between the two diagrams.
The prospects for probing light quark Yukawas within future LHC runs employing the direct probe from exclusive decays are not competitive with indirect limits that can be set from production or global fit or inclusive search for $c$-Yukawa~\cite{Perez:2015lra,Cepeda:2019klc}. However, the information coming from exclusive decays will be relevant regardless of the global fit sensitivity. For example, a limit of $|y_s/y_b| \lesssim 50$ could be set HL-LHC~\cite{Cepeda:2019klc} and $y_s/y_b \lesssim 25$ at FCC$_{hh}$~\cite{Abada:2019lih}. 

The constraints on invisible and untagged BRs to new particles are reported in Table~\ref{tab:invisible}. For the invisible decays the SM  $H\to 4\nu$ process  (${\rm BR}^{\rm SM}_{\rm inv}= {\rm BR}(H\to 4 \nu)=0.11 \%$) is treated as background.
Shown are the estimated projections for direct searches for invisible decays using signatures of missing transverse or total energy, and the results from the kappa-3 fit presented earlier in Table~\ref{tab:resultsKappa3}. Also shown is a kappa-fit where all SM BR values are fixed and only BR$_\textrm{inv}$ is free in the fit. It is seen that the $e^+e^-$ colliders generally improve the sensitivity by about a factor 10 compared to HL-LHC. FCC-hh improves it by another order of magnitude and will probe values below that of the SM.
Comparing the three determinations of the BR$_\textrm{inv}$ for the various colliders, it is seen that in most cases the difference is small, indicating that the BR$_\textrm{inv}$ is mostly constrained by the direct search. An exception is LHeC where the kappa-fits improve the direct search result by a factor two. 

Finally, comparing the bounds on the invisible and untagged BR one notices the latter are always weaker as the untagged BR is not constrained by any direct search here. For the untagged BR, the kappa-3 fit sensitivity is significantly worse than that obtained by fitting only BR$_\textrm{unt}$ as the kappa-3 fit implicitly takes into account the experimental uncertainties on all other BR values.


\section{Sensitivity to Higgs CP} \label{higgscp}

Barring the strong-CP problem, in the SM the only source of CP violation stems from fermion mixing in the charged currents, while the Higgs boson is predicted to have CP-even, flavour-diagonal interactions. Detecting non-zero CP-odd components in the Higgs interactions with the SM particles, would therefore clearly point to physics beyond the Standard Model. Departures from the SM can be efficiently parametrised in terms of a limited set of (flavour conserving) dimension-6 operators. Employing the Higgs basis, the (P-violating/C-conserving) CP-violating (CPV) 
$HVV$ couplings are given by 
\begin{eqnarray}
\delta{\cal L}^{hVV}_{\rm CPV} & = &  \frac{h}{v} \Big[
{\tilde c_{gg}} \frac{g_s^2}{4} G^a_{\mu\nu}\tilde{G}^a_{\mu\nu} 
+{\tilde c_{aa}} \frac{e^2}{4} A_{\mu\nu}\tilde{A}_{\mu\nu}  \nonumber\\
&&+{\tilde c_{za}} \frac{e\sqrt{g^2 + g'^2}}{2} Z_{\mu\nu}\tilde{A}_{\mu\nu} 
+ {\tilde c_{zz}}  \frac{g^2 + g'^2}{4} Z_{\mu\nu}\tilde{Z}_{\mu\nu} 
+ {\tilde c_{ww}} \frac{g^2}{2}  W^+_{\mu\nu} \tilde W^{-}_{\mu\nu}\Big]\,,
\label{eqn:hVV-CPV}
\end{eqnarray}  
where, $g_s$, $g$ and $g'$ are the $SU(3)$, $SU(2)_L$ and $U(1)_Y$
gauge coupling constants and $\tilde V_{\mu\nu}= \frac12 \epsilon^{\mu \nu \rho\sigma} V_{\rho\sigma}$.  Out of the four electroweak parameters, only three are independent at this order in the EFT expansion. In particular,
\begin{equation}
{\tilde c_{ww}} = {\tilde c_{zz}} + 2 \sin^2{\theta_w}~\! {\tilde c_{za}} + \sin^4{\theta_w}~\! {\tilde c_{aa}}.
\end{equation}
The (P-violating/C-violating) CP-violating (yet flavour-diagonal) interactions of the Higgs boson with fermions can be parametrised as   
\begin{equation}
{\cal L}^{hff}_{\rm CPV}   = -\bar \kappa_f m_f \frac{h}{v} 
\bar{\psi}_f (\cos \alpha + i  \gamma_5 \sin \alpha) \psi_f\,,
\label{eqn:hff-CPV}
\end{equation}
where the angle $\alpha$ parametrizes the departure from the CP-even case. Another, equivalent parametrization employs $\kappa_f = \bar \kappa_f \cos \alpha$ and $\tilde\kappa_f = \bar \kappa_f \sin \alpha$, where $\kappa_f=1+\delta y_f$ in the notation used for the CP conserving cases in the $\kappa$-framework (with $\kappa>0$). The pure scalar coupling corresponds to $\alpha=0$ ($\tilde\kappa_f=0$),  a pure pseudoscalar coupling to $\alpha=90^\circ$ ($\kappa_f=0$), while CP violation occurs in all other intermediate cases. 

Sensitivity to the CP-odd operators can arise from two distinct classes of observables. The first class includes CP-even observables, such as  total cross sections or single particle inclusive distributions. In this case, CP-odd operators contribute in a way that is analogous to CP-even operators, i.e. affecting rates and shapes.  The second class includes observables that are built to be directly sensitive to CP violation, i.e. they are zero (at the lowest order) if CP is conserved. Limits obtained from this second class are therefore automatically insensitive to the presence of higher-dimensional $CP$-conserving operators and deviations from zero would uniquely point to CP violation. 

Sensitivity to the CP-odd $hgg$ interaction comes from gluon fusion processes at the inclusive level, while direct sensitivity to CP violation can arise only starting from final states featuring at least two jets in the final state. Studies performed at the LHC exist, yet no dedicated investigation for future colliders has been documented. Sensitivity to the CP-odd $hVV$ weak operators comes from Higgs-strahlung processes ($WH$ and $ZH$), the vector boson fusion and the Higgs decay into four charged leptons ($H\to4\ell$). Studies have been performed both at the level of rates/distributions and via CP-sensitive observables~\cite{Cepeda:2019klc}.

CP-violation effects in the couplings to fermions have been considered for the top quark and the tau lepton. Proposals to access information on CP violation in top quark interactions exist for both classes of observables, yet studies at future colliders have been mostly based on rates and distributions. These focus on $ttH$ at hadron colliders and on $ttH$ and $tH$
final states at $e^+e^-$ colliders and $ep$ colliders, respectively, which are also sensitive to the absolute signs of CP-even and CP-odd interactions through interference effects. For example, by studying distributions in $ttH$, the HL-LHC will be able to exclude a CP-odd Higgs at 95\%CL with about 200 fb$^{-1}$ of integrated luminosity.
CLIC 1.5\,TeV foresees to measure the mixing angle for the top quark, $\alpha_t$ in $t\bar t H$ to better than 15$^\circ$. At LHeC, a Higgs interacting with the top quarks with a CP-odd coupling can be excluded at $3\sigma$ with 3 ab$^{-1}$.  At FCC$_{eh}$ a precision of 1.9\% could be achieved on $\alpha_t$. 

The most promising direct probe of CP violation in fermionic Higgs decays is the $\tau^+ \tau^-$ decay channel, which benefits from a relatively large branching fraction ($6.3\%$). Accessing the CP violating phase requires a measurement of the linear polarisations of both $\tau$ leptons and the azimuthal angle between them. This can be done by analysing the angular distribution of the various components of the tau decay products and by building suitable CP sensitive quantities (such as triple products of three-vectors or acoplanarities).
The estimated sensitivities for the CP-violating phases,  $\alpha_\tau$  of the $\tau$ Yukawa coupling and $\tilde{c}_{zz}$ extracted from CP-sensitive variables are collected in Table~\ref{tab:alpha}.

\begin{table}[ht]
\caption{\label{tab:alpha} 
Upper bounds on the CP phase $\alpha$ of the Yukawa coupling for $\tau$ leptons and the CP-violating coefficient $\tilde{c}_{zz}$ entering the HZZ coupling. 
The result in parenthesis for the HL-LHC is obtained with the same method used for the CEPC study.}
\begin{center}
\begin{tabular}{|c|c|c|c|}
\toprule
\rule{0pt}{1.0em}%
  Name &  $\alpha_\tau$ & $\tilde{c}_{zz}$ & Ref. \\
\midrule
 HL-LHC &   8$^\circ$       & 0.45 (0.13) & \cite{Cepeda:2019klc}\\
 HE-LHC & -- & 0.18 & \cite{Cepeda:2019klc}\\
\midrule
   CEPC            &    --          & 0.11 & \cite{CEPCStudyGroup:2018ghi} \\
   FCC-ee$_{240}$  &   10$^\circ$   & --   & \cite{Abada:2019lih}\\
   ILC$_{250}$     &    4$^\circ$   &0.014 & \cite{Bambade:2019fyw} \\
\bottomrule
\end{tabular}
\end{center}
\end{table}

Before concluding this section, we recall that CP-violating Yukawa couplings are well constrained from bounds on the electric dipole moments~(EDMs)\cite{Brod:2013cka,Chien:2015xha,Altmannshofer:2015qra,Egana-Ugrinovic:2018fpy,Brod:2018pli,Brod:2018lbf} under the assumptions of i) no cancellation with other contributions to EDMs, ii) SM values for the CP-even part of the Yukawa couplings.

CP violation in the top quark sector can be constrained by the EDM of the electron, giving $\tilde\kappa_t<0.001$ once the latest limits of the ACME collaboration are considered~\cite{Andreev:2018ayy}. For the bottom and charm Yukawas the strongest limits come from the neutron EDM, $\tilde\kappa_b<5$ and $\tilde\kappa_c <21$ when theory errors are taken into account.  For the light quark CPV Yukawas, measurements of the neutron EDM give a rather weak constraint on the strange quark Yukawa of $\tilde\kappa_s <7.2$, while the bound on the mercury EDM translates into strong bounds 
on the up and down Yukawas of $\tilde \kappa_u <0.11$ and 
$\tilde\kappa_d <0.05$  (no theory errors, $90\%$ CL).
For the $\tau$ Yukawa coupling, using the latest ACME measurement gives $\tilde\kappa_\tau<0.3$, while for the electron Yukawa, provides an upper bound of $\tilde\kappa_e <1.9\times 10^{-3}$. 

Assuming a SM Yukawa coupling of the Higgs to the electron, one can easily compare the indirect limits from EDMs with the prospects for direct ones.
Using the relations between $(\bar\kappa,\alpha)$ and $(\kappa,\tilde\kappa)$
one can convert the results for both the top quark (given above) and for the $\tau$ lepton (collected in Table~\ref{tab:alpha}).  
One finds that the direct top quark limits are not competitive with the indirect ones, while those on the $\tau$ lepton are comparable with the current indirect ones. 


\section{The Higgs boson mass and full width} \label{mass}

The current best measurement of the Higgs boson mass, based on the ATLAS and CMS analyses of $H \rightarrow ZZ^*$ and $H \rightarrow \gamma \gamma$ events in the LHC Run-2 data is $125.18 \pm 0.16$\,GeV\cite{PhysRevD.98.030001}. Future accelerators are expected to substantially improve the precision of this mass measurement. 

The mass measurements at lepton colliders in the centre-of-mass energy range 240-350\,GeV analyse the recoil mass of the Higgs boson in $e^+e^- \rightarrow ZH$ events, with
$Z \rightarrow e^+e^-, \mu^+\mu^-$. 
Only the statistical uncertainties on the mass measurements are shown, as systematic uncertainties in this recoil mass analysis are expected to be negligible. The CLIC
mass measurements at higher centre-of-mass energies analyse the $H \rightarrow b\bar{b}$ invariant mass distribution in $e^+e^- \rightarrow H(\rightarrow b\bar{b}) \nu \nu$ events. The quoted mass resolutions based on $m_{bb}$ measurements account only for statistical uncertainties, but are sensitive to $b$-jet energy scale uncertainties. This systematic uncertainty can be constrained with a $e^+e^- \rightarrow Z(\rightarrow b\bar{b}) \nu \nu$ calibration sample which is expected to yield comparable statistics to the Higgs sample. The mass measurement at HL-LHC is based on the analysis of $H \rightarrow Z(l^+l^-)Z(l^+l^-)$ events. While the calibration of lepton momentum scales has not been studied in detail, a resolution of 10-20\,MeV is projected to be plausibly in reach with the assumption that the higher statistics can help to significantly improve muon $p_T$ systematic uncertainties. 

Table \ref{tab:higgs_mass} summarizes the expected precision of Higgs boson mass measurements of future accelerators. Also shown is the impact of the $m_H$ uncertainty on the $H\to ZZ^*$ partial decay width. Already with HL-LHC, it will be possible to reduce this impact to the level of about 0.2\%.  
At this value, the parametric uncertainty on Higgs partial widths, (primarily on $ZZ^*,WW^*$) is much smaller than the expected precision at any hadron collider.
For the $e^+e^-$ colliders the precision on the $W$ and $Z$ couplings is of that order, so that the $m_H$ precision needs to be further improved to about 10\,MeV to avoid any limitations on the Higgs coupling extraction precision (assuming the uncertainty due to higher order processes gets improved in the future, see Table~\ref{tab:widths}). 
\begin{table}[hbt]
\caption{\label{tab:higgs_mass} Overview of expected precision of Higgs boson mass measurements for future accelerators scenarios. For the lepton colliders (ILC, CLIC, CEPC) the projected uncertainties listed are statistical only. The impact of $\delta m_H$ on $\delta \Gamma_{ZZ^*}$ reported in this table is calculated as $1.2\% \cdot ( \delta m_H / 100~ \rm{MeV} )$, following Ref.~\cite{LHCHXSWG4}. 
}
\begin{center}
\begin{tabular}{|l|c|cll|}
\toprule
Collider  & Strategy & $\delta m_H$ (MeV) & Ref. & $\delta (\Gamma_{ZZ^*})$ [\%] \\
\midrule
LHC Run-2 &  $m(ZZ), m(\gamma\gamma)$ & 160 & \cite{PhysRevD.98.030001} & 1.9 \\
HL-LHC &  $m(ZZ)$ & 10-20 & \cite{Cepeda:2019klc} & 0.12-0.24 \\
\midrule
ILC$_{250}$ & $ZH$ recoil  & 14 & \cite{Bambade:2019fyw} & 0.17 \\
CLIC$_{380}$ &  $ZH$ recoil & 78 & \cite{Robson:2018zje} & 0.94 \\
CLIC$_{1500}$ &  $m(bb)$  in $H\nu\nu$ & 30\tablefootnote{In Ref.~\cite{Robson:2018zje} the values are 36 MeV (for $\sqrt{s}=1.5$ TeV) and 28 MeV (for $\sqrt{s}=3$ TeV) are based on unpolarized beams. The values quoted here are for the default scenario of 80\% electron polarisation assumed throughout.} & \cite{Robson:2018zje}  & 0.36 \\
CLIC$_{3000}$ &  $m(bb)$  in $H\nu\nu$ & 23 &  \cite{Robson:2018zje} & 0.28 \\
\midrule
FCC-ee &$ZH$ recoil  & 11 & \cite{Azzi:2012yn} & 0.13\\
CEPC &  $ZH$ recoil  & 5.9 & \cite{CEPCStudyGroup:2018ghi} & 0.07\\
\bottomrule
\end{tabular}
\end{center}
\end{table}

In the SM, the width of a 125 GeV $H$ boson  is predicted to be around  $4$ MeV, i.e. three orders of magnitude smaller than that of the weak bosons and of the top quark. It is therefore very challenging to measure it directly. All methods considered so far at colliders are in fact indirect and model dependent to various degrees.  Three methods have been proposed at the LHC, and are considered for future hadron colliders. 

The most direct method involves the diphoton decay mode and it is based on the measurement of the shape of the invariant mass of the diphoton close to the Higgs boson mass. This observable has a dependence on the width from signal-background interference effects. The foreseen sensitivity, however, will not allow to probe values close to the SM predictions, and can provide constraints of about $8-22\times\Gamma_\textrm{SM}$~\cite{Cepeda:2019klc}.  

A second method extracts the width indirectly from a global fit of the Higgs boson couplings by employing specific assumptions. For example, in the $\kappa$-framework, assuming $\kappa_Z\leq 1$ and ${\rm BR}_{\rm unt}=0$ one can determine the width from the fit. \footnote{In fact, the width and the branching ratio to undetected final states are not independent observables. In the analysis presented in Section~\ref{kappa-results} we opted to fit ${\rm BR}_{\rm unt}$ and calculate $\Gamma_H$ from Eq.~(\ref{eq:width}).}

A third method is based on the combination of two independent measurements in gluon fusion production of a $H$ boson with subsequent decay into a $ZZ$ final state: $gg \to H \to ZZ^*$, where the $H$ boson is on shell (and at least one of the final state $Z$ off shell)  and $gg \to ZZ$ with two on-shell $Z$ bosons, where the $H$ boson contribution is off shell~\cite{Kauer:2012hd}. The ratio of the off-shell over the on-shell rate is directly proportional to the total width~\cite{Caola:2013yja}. Even though in generic BSM scenarios including the EFT,  the interpretation of the off-shell/on-shell ratio  as an extraction of the width is model dependent,  this ratio can provide  useful information on other key aspects of the Higgs couplings, e.g. their energy dependence~\cite{Azatov:2014jga}.  
It is foreseen that, with the HL-LHC and improvements in the theoretical calculations, $\Gamma_H$ can be measured with a precision of up to 20\% using this method~\cite{Cepeda:2019klc}.

At lepton colliders, the mass recoil method allows to measure the inclusive cross section of the $ZH$ process directly, without making any assumption about the Higgs BR's.  This possibility is unique to lepton colliders as it relies on the precise knowledge of the total initial energy of the event.
In combination with measurements of exclusive Higgs decay cross sections, it allows to extract the total width $\Gamma_H$ with a mild model dependence. The simplest way is to consider the ratio of the $ZH$ cross section (from the recoil method) with the $H\to ZZ$ branching ratio (extracted from the $ZH, H \to ZZ^*$ rate) 
\begin{equation}
\frac{\sigma(e^+e^- \to ZH)}{{\rm BR}(H\to ZZ^*)}= \frac{\sigma(e^+e^- \to ZH)}{\Gamma(H\to ZZ^*)/\Gamma_H}\simeq \left[\frac{\sigma(e^+e^- \to ZH)}{\Gamma(H\to ZZ^*)}\right]_{\rm SM}\times\Gamma_H\,,
\end{equation}
where the last approximate equality assumes a cancellation of new physics effects, which holds, for instance, in the $\kappa$-framework. This method is limited by the relatively poor statistical precision of the $H\to ZZ$ BR measurement. More in general, even in scenarios where such a cancellation does not hold, {\it e.g.} in an EFT, a global fit can be performed to extract information on the width,  using other decays (particularly the $bb$ and $WW$ decays) and channels ($e^+e^- \to H\nu\bar{\nu}$). This method is used for CEPC. For FCC-ee and CLIC the $\kappa$-formalism is used to extract the width, similar to what is done in this report for Table~\ref{tab:resultsKappa3}. 
For ILC, the width reported here was extracted using an EFT formalism that does not assume that there is only one operator that governs the interactions between the Higgs boson and the Z boson (as is done implicitly in the $\kappa$-framework).
In this determination of $\Gamma_H$, angular distributions and polarisation asymmetries are used to constrain the free parameters that result from relaxing this assumption~\cite{Barklow:2017suo}, in addition to the parameters used by the $\kappa$-formalism for the other lepton colliders. This fit is different from the EFT fits performed in Section~\ref{eft-results}.

Table \ref{tab:higgs_width} summarizes the expected relative precision that can be reached on the Higgs width at future lepton colliders, comparing the estimates of the standalone estimates of the future lepton colliders to the results of the kappa-3 scenario fits performed in this article (with HL-LHC data included).
It is seen that the result obtained in the kappa-3 fit is generally more constraining than the results quoted in the references, primarily as this result also includes the constraint from the HL-LHC data, and, in some cases, uses a different approach to modelling changes to the total width. In both cases, the best precision is obtained for the ILC$_{500}$ and FCC-ee$_{365}$ scenarios.

\begin{table}[hbt]
\begin{center}
\caption{\label{tab:higgs_width} Overview of expected precision of Higgs boson width measurements for future accelerator scenarios. The result given in the second column refers to the width extraction as performed by the future lepton colliders using the stated technique, and as provided in the references given. The last column of the table lists the width extracted from the kappa-3 scenario fit. It also includes the HL-LHC measurements (but excludes the constraint $\kappa_V<1$ that is used in HL-LHC-only fits).}
\begin{tabular}{|l|cl|c|}
\toprule
Collider &   $\delta \Gamma_H$ [\%] & Extraction technique &  $\delta\Gamma_H$ [\%] \\
 & from Ref. & for standalone result  &  kappa-3 fit  \\
\midrule
ILC$_{250}$ & 2.3 &  EFT fit \cite{Bambade:2019fyw,Fujii:2019zll} & 2.2\\
ILC$_{500}$ & 1.6  & EFT fit \cite{Fujii:2017vwa,Bambade:2019fyw,Fujii:2019zll} & 1.1 \\
ILC$_{1000}$ & 1.4  & EFT fit \cite{Fujii:2019zll} & 1.0 \\
CLIC$_{380}$ &  4.7 & $\kappa$-framework \cite{Robson:2018zje} & 2.5 \\
CLIC$_{1500}$  & 2.6 &  $\kappa$-framework \cite{Robson:2018zje}& 1.7\\
CLIC$_{3000}$ & 2.5 & $\kappa$-framework \cite{Robson:2018zje}  & 1.6\\
\midrule
CEPC &  2.8 
& $\kappa$-framework \cite{CEPCPhysics-DetectorStudyGroup:2019wir,An:2018dwb} & 1.7 \\
FCC-ee$_{240}$ & 2.7 &  $\kappa$-framework \cite{Abada:2019lih} & 1.8 \\
FCC-ee$_{365}$ & 1.3 &  $\kappa$-framework \cite{Abada:2019lih} & 1.1 \\
\bottomrule
\end{tabular}
\end{center}
\end{table}


\section{Future studies of the Higgs sector, post-European Strategy}
\label{futureprospects}


\subsection{Higgs prospects at the muon collider} \label{muoncollider}

Electron-positron colliders offer  a well-defined value of the collision energy of the hard-scattering process and a relatively clean event, as opposed to hadron collisions where the underlying event and the high-level of event pileup challenge the reconstruction of the hard scattering event and its measurement.

The main limitation to the collision energy of circular electron-positron colliders is due to the low mass of the electrons/positrons which leads to large fraction of their energy emitted as synchrotron radiation.
The solutions pursued so far to reach high lepton collision energies are based on limiting the energy loss by synchrotron radiation by reducing the curvature either by increasing the radius of the circular colliders or by employing linear colliders. However, the beam acceleration does require a number of RF cavities imposing a machine of large dimensions.

With a mass of about two hundred times that of electrons, muons do not suffer significant energy losses due to synchrotron radiation (the loss goes as the inverse of the fourth power of the mass) and therefore could be accelerated up to multi-TeV collision energies. For example, if the LHC ring were used, with the proposed HE-LHC dipoles (Nb$_3$Sn, 16\,T), muons would collide at an energy close to $\sqrts$= 14\,TeV, compared to the 0.2 to 0.4\,TeV of an electron-positron collider.

Alternatively, a collider with $\sqrt{s}=125$\,GeV could be a very compact (diameter $\sim 60$~m) Higgs factory using $s$-channel production of Higgs bosons~\cite{Barger:2001mi}. 
However, it should be noted that the expected rate of produced Higgs bosons by $s$-channel is small, given the instantaneous luminosity possible at this machine \cite{muoncollhadr}, and the limited production cross section (taking into account both the beam energy spread and the initial state radiation effects)~\cite{Greco:2016izi, Han:2012rb, Alexahin:2013ojp}. Estimates of the achievable precision on Higgs couplings for such a machine  
are given in~\cite{Janot}.

Muon production, cooling, lifetime and physics background \cite{Bartosik:2019dzq} pose severe challenges to the accelerator and detector technologies. Although the study of a Muon Collider (machine and physics prospects) is not as mature as those of other future proposed colliders, its physics potential certainly merits consideration.

Currently, two different configurations have been proposed for the muon collider. In the first configuration, muons are produced by the decay of hadronically produced charged pions or kaons, and cooled before they undergo the acceleration~\cite{muoncollhadr}. In the second configuration, muons are produced at threshold (in the centre of mass frame) by high energy positron collisions with atomic electrons~\cite{muoncollemma}. The first configuration has been originally proposed for $\mu^+ \mu^-$ collision at the Higgs boson pole ($\sqrts\sim$125\,GeV), while the second is mainly considered for very high energy collisions, in the range of $O(10)$\,TeV. 

At muon collision c.m. energies $\sqrt s\gtrsim 10$\,TeV, assuming the {\it point cross section} $\sigma\simeq 4\pi\alpha^2/(3s)\simeq 1~$fb$\cdot (10\,$TeV$/\sqrt{s})^2$, the requirement of a percent statistical precision in the measurement of heavy particle pair production would imply  an integrated luminosity of the order  $L\sim 10$ ab$^{-1}(\sqrt {s}/10\,$TeV$)^2)$.
This could correspond to a 10-year physics run with an instantaneous luminosity of the order 
10$^{35}(\sqrt {s}/10 $TeV$)^2$cm$^{-2}$s$^{-1}$  \cite{Delahaye:2019omf}.
At such large values of $\sqrt{s}$,  both the  single-Higgs and the multi-Higgs production mechanisms are dominated by vector-boson fusion (VBF) processers, which provides very large statistical Higgs samples \cite{muoncprepmeeting}.
As an example, at $\sqrt s\sim 14$\,TeV, with 20\,ab$^{-1}$, one would produce about 20 million single Higgs, 90,000 Higgs pairs, and 140 triple Higgs final states, with presumably quite moderate background. 
Although there is currently only preliminary analysis of the Higgs production in such an environment this would be a robust basis to considerably advance on the Higgs couplings determination.
The Higgs self-coupling sector might be explored with unprecedented precision. In particular, with the above Higgs production statistics, and no unexpectedly difficult background, an accuracy of few percent for the trilinear Higgs coupling, and a few tens of percent for the quadrilinear Higgs coupling might be reached at $\sqrt{s}\sim 14$\,TeV, with 20\,ab$^{-1}$, assuming all the remaining Higgs and EW parameters at their SM value.
Many other investigations of the Higgs properties might significantly benefit from such collider configuration \cite{Delahaye:2019omf,muoncprepmeeting}.

\subsection{Higgs physics at multi-TeV electron-positron colliders}
\label{sec:MultiTeVee}

The length of linear accelerators proposed today, is largely determined by the electric field gradients that can be achieved with RF cavities. For the superconducting RF technology used by ILC the limit is about 35\,MV/m while for the drive-beam technology, envisaged for CLIC, it is about 100\,MV/m. 

Much higher gradients (up to 1000 times more acceleration compared to RF) can be achieved using plasma-wakefield acceleration, where laser pulses~\cite{Mangles:2004ta, Lee:2000yy, Krall:1995kon, Whittum:1998at}, electron~\cite{Blumenfeld:2007ph, Litos:2014yqa} or proton~\cite{Adli:2018end} bunches (called drivers) can excite ultra-high fields in plasma devices. 
Thus this is a very promising technique for future high energy $e^+e^-$ and $\gamma\gamma$ colliders~\footnote{For $\gamma\gamma$ colliders it is sufficient to accelerate two $e^-$ beams which is technically less demanding than accelerating positrons}. The ALEGRO Collaboration~\cite{ALEGRO:2019alc} has been formed with the goal of designing an Advanced Linear Collider (ALIC) based on this technology. A summary of the facilities operating today and planned for the future, as well as the R\&D needed, are given in~\cite{ALEGRO:2019alc}. The physics opportunities of an $e^+e^-$ collider with $\sqrt{s}$ up to $100$\,TeV are also discussed there. 

The minimum instantaneous luminosity that needs to be achieved for probing cross sections of new particles interacting weakly at energies in the $10-100$\,TeV is found to be $10^{36}$\,cm$^{-2}$s$^{-1}$. With such a collider, an integrated luminosity of 30\,ab$^{-1}$ could be collected within a few years. With this dataset, the Higgs physics programme is similar to that of a Multi-TeV muon collider outlined above. It is also being considered to have such a collider at lower collision energies, in the range between $m_Z$ and 3\,TeV. Here, it would have the same physics programme as the other proposed colliders, assuming that comparable luminosity values can be achieved and background conditions are similar. 

The proposed ALIC collider~\cite{ALEGRO:2019alc} would achieve $\sqrt{s}=30$\,TeV with a peak luminosity of $10^{36}$~cm$^{-2}$s$^{-1}$ in a tunnel of 9\,km length. While the principle of acceleration has been proven, there are many issues that need to be resolved before a collider based on plasma-technology can be achieved, but none are considered to be show-stoppers at present. The primary focus of the R\&D programme is the beam quality which is addressed at lower-energy applications (e.g. free-electron lasers, fixed target experiments) and will benefit the development of a collider based on this technology. 


\subsection{What and Why: Higgs prospect studies beyond this report} \label{futurestudies}

The purpose of this subsection is to place the Higgs coupling measurements in perspective with other new physics studies performed at future colliders with the aim of providing answers to the following two questions:  {\it What are we going to learn?},  {\it What can we possibly discover?}. 
The unknown territory of energy and precision to be explored may  have different discoveries in store, including unexpected  ones. Given the scope of this document, a discussion of the various options would hardly be self-contained, and would  miss, by definition, the most exciting case of unexpected discoveries. On the other hand, by focusing on some open problems in particle physics, it is possible to structure a self-contained discussion at least around the first question. The hierarchy problem (HP),  dark matter (DM) and the electroweak phase transition (EW$\Phi$T) are  issues on which we shall definitely make progress. Flavour could also be added to this list, but mostly in the measure in which it is connected to the HP.

In view of its centrality, and of the controversial regard in which it is sometimes held, a succinct but modern appraisal of the HP is needed. The HP is a paradox  challenging  the modern effective field theory (EFT) view of particle physics. 
The challenge is presented by the clash between infrared (IR) Simplicity and Naturalness.
IR Simplicity is an unavoidable feature of  any EFT when making observations at energies much below its fundamental  scale $\luv$. In practice that is due to the decoupling of the infinite (complicated) set of  non-renormalizable couplings in favor of the finite (simple) set of renormalizable ones. Naturalness instead arises by viewing EFT parameters  as  functions of more fundamental ones: in this point of view it is expected that any specific structure, like  the presence of a very small parameter, should  be accounted for by symmetries and selection rules rather than by accidents.   Now, the structure of the SM is such that several crucial experimental facts like approximate baryon and lepton numbers, lightness of neutrinos, GIM suppression of FCNC, custodial symmetry {\it all } remarkably and  beautifully follow from IR Simplification. That is by assuming  $\luv \gg m_{weak}$. However when considering the Higgs mass parameter, one famously finds  that $\luv \gg m_{H}$ is inconsistent with the predicate of Naturalness. In the SM, IR Simplicity can thus only be obtained at the price of un-Naturalness. But this is only half of the problem. The other half is that models  realizing Naturalness, like supersymmetry (SUSY) or Composite Higgs (CH), invariably sacrifice Simplicity. Indeed all these natural extensions
have concrete structural difficulties in reproducing the observed simplicity in flavour, CP violating and electroweak observables.
  In order to meet the corresponding experimental constraints, these scenarios must rely on {\it clever} constructions mostly associated with {\it ad hoc}  symmetries, like flavour symmetries or  custodial symmetry, which in the SM are either not needed or automatic.  The paradoxical tension between Simplicity and Naturalness is what defines the hierarchy problem: no win-win scenario seems to be available. 
  
The  paradox could already be formulated before LEP, and gained in importance with more and more precise flavour and electroweak data that demands a more elaborate structure in natural models. Futhermore, the ever stronger bounds from direct searches for `Natural' agents at Tevatron and LHC imply the need for some amount of un-Naturalness, or fine tuning, even in models like SUSY or CH that aimed at full Naturalness.
Depending on the scenario, the finesse of the cancellation in the Higgs mass parameter  needed 
to lift new physics out of LHC reach can be  quantified to roughly range from $1/10$  
to $ 1/10^3$.

The test of Naturalness vs. Simplicity offers one concrete criterion to  compare future machines across their reach in three different sets of measurements: direct searches,  Higgs couplings, EW precision tests (EWPT). 
\begin{itemize}
\item{\bf Direct searches:} Natural models all possess computational control over the Higgs squared mass. The result varies in a finite range, and a small or vanishing result can only be achieved by tuning different contributions against one another.  Indicating by $\Delta m_H^2$  the most sizeable  contribution,  the  tuning is simply measured by 
\begin{equation}
\label{eq:DeltamH}
\epsilon \equiv m_H^2/\Delta m_H^2\, ,
\end{equation}
 with $m_H$ the observed Higgs mass. Because of its large Yukawa coupling, the most sizeable effects come from  coloured states associated with the top, the so-called top-partners.  Models can be  broadly classified  into three classes according to the dependence  of $\Delta m_H^2$ on the top partner mass $m_T$:
\begin{itemize} 
\item { \textbf {\textit {Soft:}}} $\Delta m_H^2 \sim m_T^2$. This situation is realized in SUSY   with soft terms generated at a high scale. In the absence of any tuning   $m_T\sim m_H \sim 100$\,GeV, within the energy range of LEP and Tevatron.
\item { \textbf {\textit {SuperSoft:}}} $\Delta m_H^2 \sim (3y_t^2)/(4\pi^2)\, m_T^2$.
This situation is realized in SUSY with low scale mediation  and in CH. Without  any tuning one expects $m_T \sim m_H/\sqrt {3y_t^2/4\pi^2}\sim 450$\,GeV, within the reach of the LHC.
\item { \textbf {\textit {HyperSoft:}}} $\Delta m_H^2 \sim (3 \lambda_h)/(16\pi^2)\, m_T^2$.  The mechanism of Neutral Naturalness is a prime example. The top partner mass is naturally pushed around 1.5\,TeV.
\end{itemize}
A $\sim 10$ TeV reach on $m_T$  like offered by FCC-hh or  muon-collider (the top partners have often EW quantum numbers) will thus  probe Naturalness down to $\epsilon= 10^{-4}, 10^{-3}, 10^{-2}$ in respectively Soft,  SuperSoft and HyperSoft scenarios.
\item{\bf Higgs couplings:} The  deviations $\delta g_h$ from the SM in single and multi-Higgs couplings satisfy
\begin{equation}
{\delta g_H}/{g_H}^{\rm SM} \sim c\,\epsilon\, ,\label{gh}
\end{equation}
with $c$ a coupling-dependent coefficient, and $\epsilon$ the Higgs mass correction defined in~(\ref{eq:DeltamH}). In basically all models, there always exists a set of couplings where $c\sim O(1)$. The only exception is strictly supersoft SUSY, where one can cleverly go down to $c\sim 0.1$.
Not surprisingly full Naturalness basically mandates $O(1)$ deviations in Higgs couplings.
 
 The best measurements that will be carried out at future machines aim at $10^{-3}$ precision on some of the Higgs couplings, in particular $g_{HWW}$ and $g_{HZZ}$. This should be compared to the reach in $\epsilon$ in direct searches. In particular, Higgs couplings probe less than direct searches in the simplest high scale SUSY models. But one must not forget that these models admit countless variants, with additional states, in particular SM singlets, and with a spread spectra. So one cannot completely discount the relevance of Higgs couplings to probe these models. 
 In any case, one should not underestimate the value of Higgs precision programme that can measure the Higgs couplings with a $10^{-3}$ precision.
 The equal relevance of Higgs studies and direct searches for CH models seems robust.
  
 In view of parametric uncertainties, $10^{-3}$ seems like a limiting (or at least a  critical) sensitivity to
 BSM deviations  in  single Higgs production near threshold.
However  these deviations   are all associated with operators of dimension $\geq 6$, whose effects grow with energy when considering processes with a sufficient number of legs, like $gg\to HH, hV_L$ or $VV\to HH, V_LV_L$. And one must then consider the possibility of obtaining a better sensitivity by measuring such processes.
For instance, FCC-hh can probe $gg\to HV_L$  and $VV\to VV$ up to $\epsilon\sim 1\%$~\cite{Abada:2019zxq}. Lepton machines might compete better:  while CLIC can reach a sensitivity to $\epsilon \sim 10^{-2}$ in $VV\to HH$~\cite{Contino:2013gna}, still one order of magnitude poorer than single Higgs measurements,  a recent analysis~\cite{muoncprepmeeting} of a muon collider shows the $\epsilon=10^{-3}$ wall is beaten for a CM energy of 10\,TeV.  A $\mu$-dream machine running at 30\,TeV could go down to $\epsilon = 2\times 10^{-4}$~\cite{muoncprepmeeting}, which competes well even with the reach on top partners ($\sim 14$\,TeV) for such a machine.
 \item{\bf EWPT:}  While Higgs couplings are prime sensors of Naturalness, EWPT sense the dynamics of EWSB indirectly, via loops. To make this concrete, 
consider the $\hat S$ parameter, defined with the normalization of~\cite{Barbieri:2004qk}.  For all models, encompassing supersymmetry, CH or technicolor one can write a parametric formula
\begin{equation}
\widehat S\sim \frac{\alpha_W}{4\pi} \frac{g_*^2 v^2}{m_*^2} N\lsim \frac{m_W^2}{m_*^2},
\end{equation}
where $m_*, g_*$ indicate  overall mass and coupling of the new dynamics (the most obvious expectation being $m_*\sim m_T$), while $N$ measures the number of new degrees of freedom. 
Theoretical considerations   set the upper bound $g_* \sqrt N\lsim 4\pi$, which is saturated in CH and technicolor where $\widehat S \sim m_W^2/m_*^2$. In these models a measurement of $\widehat S$  translates into an indirect measurement of the scale $m_*$.  In the case of CH, one obtains $\widehat S \equiv 3\times 10^{-2} \epsilon$,
indicating that a sensitivity to $\widehat S\sim\, {\rm few}\, \times 10^{-5}$ corresponds to $10^{-3}$ sensitivity in Higgs couplings/fine tuning.
Supersymmetric models are instead well below the saturation of the upper bound, as in that case the $g_*$ is of the order of SM couplings, principally $g_W$ and $y_t$, while the multiplicity $N$ is $O(1)$~\cite{Marandella:2005wc}. One can then very roughly write $\widehat S\sim (\alpha_W/4\pi) (m_{weak}^2/m_*^2)$ implying $m_*\gsim 1$\,TeV is enough to make $\widehat S\lsim 10^{-5}$, below the wildest dreams of an FCC-ee.

Very much like for Higgs couplings, we can consider the sensitivity to the same class of dim-6 operators contributing to $\widehat S$ in processes
with more legs, where the growth with energy can be exploited
 A crucial comparison here is that between the reach of a $Z$-pole machine like the FCC-ee and CLIC which can study processes such as $e^+e^-\to hZ, h\gamma, WW$ at higher energies\footnote{The latter processes are sensitive to slightly different combinations of operator coefficients than $\widehat S$ at low energy, but in well motivated models like CH, this difference is often subdominant, and at worse they represent equally interesting but different combinations.}. The available CLIC studies estimate its reach 
as $\widehat S\simeq 0.5 \times 10^{-5}$. This should be compared to the estimated reach of $5 \times 10^{-5}$ at FCC-ee. Again the systematics of the two measures would be drastically different, with CLIC dominated by statistics and with FCC-ee dominated by parametric and intrinsic systematics.
\end{itemize}

The above analysis also offers the starting point for the consideration of other motivations and other viewpoints. As we mentioned at the beginning the  EW phase transition and DM offer alternative motivations. We will discuss them briefly in what follows. 

The interest in the order of the EW phase transition is largely related to baryogenesis. A strongly first order transition with  sizeable sources of CP violation from BSM dynamics could generate the observed cosmological baryon asymmetry. The corresponding new physics would impact 
 both future colliders and  precision low energy studies. In particular a first order phase transition implies a $O(1)$ change in the Higgs potential at finite temperature, indicating the possibility for important effects also at zero temperature, the regime we can test at colliders.
The connection between $T\not =0$ and $T=0$ is however model dependent and one can broadly distinguish two scenarios.  In the absence of new symmetries $T\not =0$ and $T=0$ are directly connected and the Higgs trilinear is expected to be $O(1)$ off its SM prediction (see e.g. \cite{Grojean:2004xa}).  
On the other hand in the presence of extra symmetries \cite{Espinosa:2011ax} there could be a further phase separation between 
$T\not =0$ and $T=0$ in which case deviations can be smaller but hardly smaller than a few $\%$ \cite{Curtin:2014jma}. The  low energy implications of EW baryogenesis concern electric dipole moments (edms) from new sources of CP violation. Here it must be noticed that the present bound $d_e< 10^{-29}$ e$\cdot$cm on the electron edm is already very strongly constraining many scenarios. Moreover there are serious plans to improve the sensitivity on $d_e$ by a few orders of magnitude in the future. A thorough analysis is not available to our knowledge, but it would be interesting in order to better appreciate the impact of such improved measurements on the space of possibilities, as that  conditions the importance of Higgs trilinear studies. It should however be kept in mind that the EW$\Phi$T could be viewed as interesting {\it per se}, regardless of baryogenesis, as it is an essential part of the history of our Universe. It also offers a new bridge between Higgs physics and gravitational physics: in case of a strong first order phase transition, the EW$\Phi$T would be the source of a stochastic gravitational wave (GW) background and future GW experiments like LISA could proffer complementary probes of the Higgs potential~\cite{Grojean:2006bp}.

The search for Dark Matter (DM) at future colliders is a broad field whose implications cannot be even partially described here. One crucial strength of a machine with a reach in the multi-TeV range is that it should be able to discover the minimal incarnation of DM, the one which arises from the purely radiatively split EW multiplets. Amazingly, this simple and well motivated class of models is hardly directly detectable in view of its loop suppressed spin independent cross section (see Table 1 in \cite{Farina:2013mla}). Focussing on Higgs studies
the basic question is: to what extent can an invisible Higgs width be associated with DM?
The answer is given in Fig. 4 of ref. \cite{deSimone:2014pda}: considering all present constraints there remains a very small region around $m_{DM}= m_H/2$, and part of this region will be explored, by direct DM detection experiments a long time before the future precision Higgs studies. The chance that DM sits in this region seems slim but a discovery in direct detection would certainly add to the Higgs programme. One should also keep in mind that the Higgs boson can decay invisibly to new particles that are not by themselves stable with the right relic abundance, but that would be part of a more complex DM sector whose abundance would not set by its interactions with the Higgs but rather by its own internal dynamics. Twin sectors of Mirror Twin Higgs models~\cite{Chacko:2018vss} and strongly coupled dark sectors~\cite{Contino:2018crt} are examples of this type and they offer a motivation to search for invisible decays of the Higgs.


\section{Summary} \label{summary}
The precision study of the Higgs boson will be the primary guaranteed
deliverable of any future collider facility. The apparent uniqueness
of the Higgs boson, as the only fundamental scalar boson thus far
discovered, justifies the importance which it is accorded in planning
for future facilities.

Several future colliders have been proposed in the context of the
update of the European Strategy for Particle Physics.
The potential of these machines for Higgs boson physics has been
intensely studied by the proponents of these machines.  The purpose of
the Higgs@FutureCollider Working Group (Higgs@FC WG) and this report
is to provide a coherent comparison of the reach of these machines in
the exploration of the Higgs sector.  We have assumed the baseline
provided by the approved programme of the HL-LHC and quantified the
additional information that would be provided by each of the future
facilities.

Taking into account the inputs submitted to the strategy process and
our dedicated discussions with proponents of future colliders, we
provide this report on the comparisons achieved, using both
the simplified kappa framework and an alternative EFT framework. The
comparisons are made in reasonable frameworks developed based on
current knowledge, with the prime objective to allow a clear and
coherent comparison.  Where relevant we note the potential caveats in
the approaches taken.  We have also reported on the rare decays of the
Higgs boson, on measurements of its mass and width, and on the expectations for CP violation studies.



\acknowledgments
We would like to express our appreciation to the numerous colleagues who have presented 
and discussed material to our working group, and in addition for the valuable comments made on a draft of
this report by representatives of the future collider projects.

We wish to express special thanks to J.~Gu for discussions, for providing some of the inputs for the EFT studies in this report, and for his help in producing some of the results presented in Section~\ref{selfcoupling}. We are also very grateful to J.~Tian for his cross-checks of the EFT fit results and to J.~Reuter for his results on the double Higgs cross-sections at lepton colliders.

We thank S.~Heinemeyer and collaborators for kindly sharing an updated version of their study of the theory uncertainties in electroweak and Higgs precision observables, whose results were also include in this report.
We also benefited from helpful discussions with R.~Contino, G.~Durieux, G.F.~Giudice, M.~Kumar, M.~Mangano, M.~McCullough,  G.~Panico, A.~Paul, M.E.~Peskin, A.~Pomarol, J.~Reuter, F.~Riva, A.~Shivaji, Y.~Soreq, E.~Vryonidou, A.~Wulzer and X.~Zhao.

The work of CG and BH was in part funded by the Deutsche Forschungsgemeinschaft under Germany's Excellence Strategy--EXC 2121 ``Quantum Universe" --390833306. 
The work of JDH and FM was partly supported by F.R.S.-FNRS under the ``Excellence of Science -- EOS" -- be.h project n. 30820817.


\appendix

\section{Mandate agreed by RECFA in consultation with the PPG “Higgs physics with future colliders in parallel and beyond the HL-LHC”}\label{app:mandate}

In the context of exploring the Higgs sector, provide a coherent comparison of the reach with all future collider programmes proposed for the European Strategy update, and to project the information on a timeline.
\begin{itemize}
    \item For the benefit of the comparison, motivate the choice for an adequate interpretation framework (e.g. EFT, $\kappa$, ...) and apply it, and map the potential prerequisites related to the validity and use of such framework(s).
    \item For at least the following aspects, where achievable, comparisons should be aim for:
    \begin{itemize}
        \item Precision on couplings and self-couplings (through direct and indirect methods);
        \item Sensitivities to anomalous and rare Higgs decays (SM and BSM), and precision on the total width; 
        \item Sensitivity to new high-scale physics through loop corrections;
        \item Sensitivities to flavour violation and CP violating effects.
    \end{itemize}
\item In all cases the future collider information is to be combined with the expected HL-LHC reach, and the combined extended reach is to be compared with the baseline reach of the HL-LHC.
\item In April 2019, provide a comprehensive and public report to inform the community.
\item ECFA helps in the creation of a working group relevant for the Strategy process, especially for the Physics Preparatory Group (PPG).
\item Towards the Open Symposium the working group will work together with the PPG to provide a comprehensive and public report to inform the community, i.e. this is not an ECFA report.
\item The working group has a scientific nature, i.e. not a strategic nature; it uses the input submitted to the Strategy process to map the landscape of Higgs physics at future colliders.
\item The convenors in the PPG who are connected to this specific topic (Beate Heinemann and Keith Ellis) and the ECFA chair will be included as {\it ex-officio} observers.
\end{itemize}
 
\clearpage

 
\section{Theoretical Cross Sections and Partial Width Uncertainties}\label{app:xs}
\begin{table}[ht]
\small
\caption{\label{tab:xsecs} 
Cross sections for the main production channels 
expected for Higgs boson production at the different types of colliders (as defined in Table~\ref{tab:colliders}).}
\begin{center}
\begin{tabular}{ c
|c|c c c c c c  c}
\toprule
\rule{0pt}{1.0em}%
&\multicolumn{7}{c}{~~~~~~~~~Cross Section $\sigma$ [pb]}\\
$pp$ collider 
& Total   & $ggH$ & VBF   & $WH$ & $ZH$ & $t \bar t H$ & $tH$ & $ggHH$ \\
\midrule
         LHC   (13 TeV)          
         & 56   & 48.6 & 3.77       & 1.36 & 0.88 & 0.510  & 0.074 & 0.031\\
         HL-LHC          
         & 62   & 54.7 & 4.26       & 1.50 & 0.99 & 0.613  & 0.090 & 0.037 \\
         HE-LHC          
         & 168 & 147  & 11.8       & 3.40 & 2.47 & 2.86   & 0.418 & 0.140 \\
         FCC$_{hh}$      
         &  936 & 802  & 69         & 15.7 & 11.4 & 32.1   & 4.70  & 1.22  \\
\bottomrule
\end{tabular}
\end{center}

\begin{center}
\begin{tabular}{ c
|c|c c c  c}
\multicolumn{5}{c}{ }  \\
\toprule
\rule{0pt}{1.0em}%
&\multicolumn{5}{c}{~~~~~~~~~Cross Section $\sigma$ [fb]}\\
$e^+ e^-$ collider ($\mathcal{P}_{e^-}/\mathcal{P}_{e^+}$) 
& Total    & VBF   & $ZH$ & $t \bar t H$  & $ZHH$ \\
                    & 
                    &CC/NC&      &      & (CC VBF) \\
\midrule
         CEPC           
         & 199    & 6.19/0.28  &  192.6  & & \\
         FCC-{ee}     
         & 199    & 6.19/0.28  &  192.6  & & \\
\midrule
         ILC$_{250}$ (-80/30)   
         & 313    &  15.4/0.70 &  297  & & \\
         ILC$_{500}$ (-80/30)   
         & 262    &  158/7.8    &  96  & 0.41  & 0.2  \\
\midrule
         CLIC$_{380}$ (0/0) 
         & 160        & 40/7.4 &  113  & --    & 0.029 (0.0020)\\
         CLIC$_{1500}$(0/0)  
         & 329        & 290/30 &  7.5  &  1.3  & 0.082 (0.207)\\ 
         CLIC$_{3000}$(0/0)  
         & 532        & 480/49 &  2    & 0.48  &   0.037 (0.77)\\
         CLIC$_{380}$ (-80/0)
         & 209        & 68/8.7  &  133   & --    & 0.034 (0.0024)  \\
         CLIC$_{1500}$(-80/0)
         & 574        & 528/35  &  8.8   &  1.70   & 0.97 (0.37) \\ 
         CLIC$_{3000}$(-80/0)
         & 921         & 860/57  &  2.4    & 0.61    &   0.043 (1.38) \\ 
         CLIC$_{380}$ (+80/0)
         & 112        & 13/6.0  &  93   & --    & 0.024 (0.0016)  \\
         CLIC$_{1500}$(+80/0)
         & 91        & 59/24  &  6.2   &  0.89   & 0.068 (0.045) \\ 
         CLIC$_{3000}$(+80/0)
         & 138         & 96/40  &  1.7    & 0.34    &   0.30 (1.56) \\

\bottomrule
\end{tabular}
\end{center}
%
\begin{center}
\begin{tabular}{ c
|c|c c  c}
\multicolumn{5}{c}{ }  \\
\toprule
\rule{0pt}{1.0em}%
&\multicolumn{3}{c}{~~~~~~~~~Cross Section $\sigma$ [fb]}\\
$e^-p$ collider ($\mathcal{P}_{e^-}$) 
& Total    & VBF   &  $tH$ & $HH$ \\
                    & 
                    &   CC/NC                  &      & (CC VBF) \\
\midrule
    LHeC (0)         
    &130  &110/20   & 0.07  & 0.01 \\ 
    HE-LHeC (0)      
    &247  &206/41   &  0.37 & 0.04  \\
    FCC$_{eh}$ (0)   
    &674  &547/127  &  4.2  & 0.26  \\
    LHeC (-80)       
    &221  &197/24   & 0.12   &0.02  \\
    HE-LHeC (-80)    
    &420  &372/48   &0.67   &0.07 \\
    FCC$_{eh}$ (-80) 
    &1189 &1040/149 & 7.6   &0.47 \\ 
\bottomrule
\end{tabular}
\end{center}
\end{table}

\renewcommand{\arraystretch}{2}

\begin{table}[ht]
\small
\caption{\label{tab:widths} 
Partial decay widths for the Higgs boson to specific final states and the uncertainties in their calculation~\cite{LHCHXSWG4}.
The uncertainties arise either from intrinsic limitations in the theoretical calculation (Th$_\mt{Intr}$) and parametric uncertainties (Th$_\mt{Par}$). The parametric uncertainties are due to the finite precision on the quark masses, Th$_\mt{Par}$($m_q$), on the strong coupling constant, Th$_\mt{Par}$($\alpha_s$), and on the Higgs boson mass, Th$_\mt{Par}$($M_H$). The columns labelled "partial width" and "current uncertainty" and refer to the current precision~\cite{LHCHXSWG4}, while the predictions for the future are taken from ref.~\cite{Heinemeyer}. For the future uncertainties, the parametric uncertainties assume a precision of $\delta m_b=13$~MeV, $\delta m_c=7$~MeV, $\delta m_t=50$~MeV, $\delta \alpha_s=0.0002$ and $\delta M_H=10$~MeV. 
}
\begin{center}
\rotatebox{90}{
\begin{tabular}{ |c|c|cccc|cccc|}
\toprule
Decay  & $\Gamma$ & \multicolumn{4}{c|}{current unc. $\Delta \Gamma/\Gamma$ [\%]} & \multicolumn{4}{c|}{future unc. $\Delta \Gamma/\Gamma$ [\%]}\\
    & [keV] & Th$_\mt{Intr}$ & Th$_\mt{Par}$($m_q$) & Th$_\mt{Par}$($\alpha_s$) & Th$_\mt{Par}$($\mhiggs$)& Th$_\mt{Intr}$ & Th$_\mt{Par}$($m_q$) & Th$_\mt{Par}$($\alpha_s$) & Th$_\mt{Par}$($\mhiggs$) \\     
\midrule
$H\to b\bar{b}$ & 2379 & $<0.4$ & $1.4$ & $0.4$ & $-$ & $0.2$ & $0.6$ & $<0.1$ & $-$\\
$H\to \tau^+\tau^-$ & 256& $<0.3$ & $-$ & $-$ & $-$& $<0.1$ & $-$ & $-$ & $-$\\\
$H\to  c\bar{c} $ & 118 & $<0.4$ & $4.0$ & $0.4$ & $-$& $0.2$ & $1.0$ & $<0.1$ & $-$\\
$H\to \mu^+\mu^-$ & 0.89 & $<0.3$ & $-$ & $-$ & $-$ & $<0.1$ & $-$ & $-$ & $-$ \\
$H\to W^+W^-$ & 883 & $0.5$ & $-$ & $-$ & $2.6$ & $0.4$ & $-$ & $-$ & $0.1$\\
$H\to gg$ & 335 & $3.2$ & $<0.2$ & $3.7$ & $-$ & $1.0$ & $-$ & $0.5$ & $-$\\
$H\to ZZ$ & 108 & $0.5$ & $-$ & $-$ & $3.0$ & $0.3$ & $-$ & $-$ & $0.1$\\
$H\to \gamma\gamma$ & 9.3 & $<1.0$ & $<0.2$ & $-$ & $-$ & $<1.0$ & $-$ & $-$ & $-$\\
$H\to Z\gamma$ & 6.3 & $5.0$ & $-$ & $-$ & 2.1 & $1.0$ & $-$ & $-$ & $0.1$\\
\bottomrule
\end{tabular}
}
\end{center}
\end{table}
\renewcommand{\arraystretch}{1}

\newpage
\clearpage 


\section{Inputs}
\label{app:inputs}
\begin{figure}[hbt]
\centering
\includegraphics[width=\linewidth]{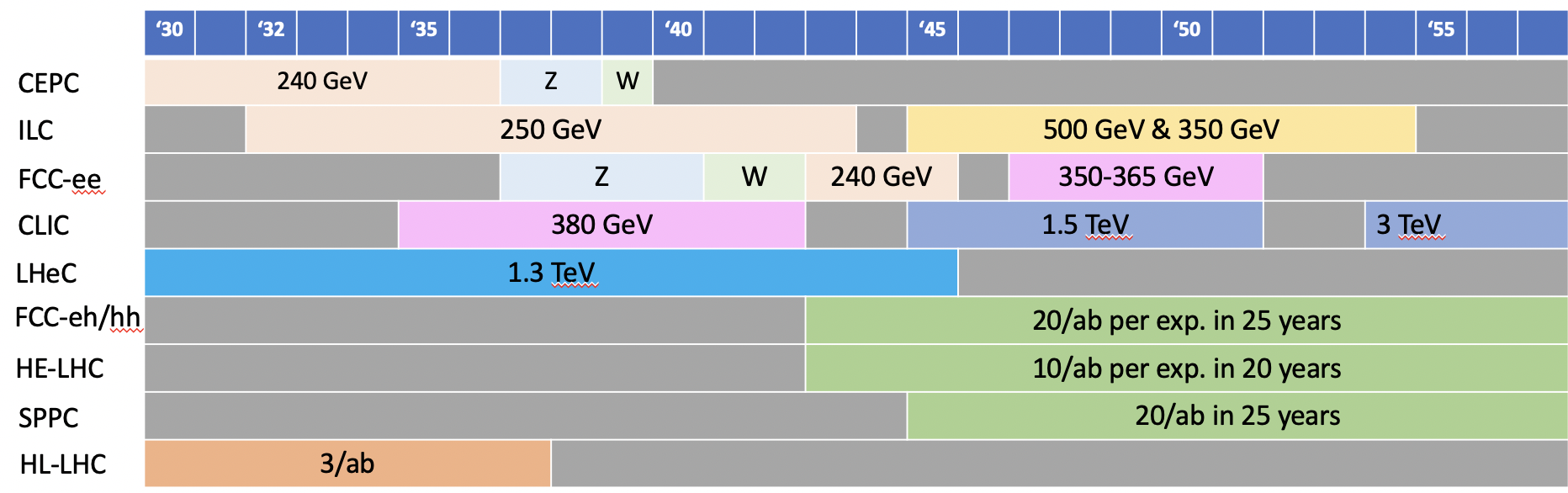}
\caption{\label{fig:timelineabs}
Sketch of timeline of various collider projects starting at the "earliest start time" stated in the respective documents. For FCC-eh/hh this figure assumes that it is not preceeded by FCC-ee. If it comes after FCC-ee it would start in the early 2060s. Only scenarios as submitted to the European Strategy by Dec. 2018 are displayed. Additional scenarios considered in this report (e.g. Giga-Z at ILC/CLIC, ILC at 1~TeV, LE-FCC) are not shown. 
}
\end{figure}

In this section we report some information relative to the inputs to the
strategy process. Figure~\ref{fig:timelineabs} shows the start date and extent
of the runs of proposed future projects, using the earliest start time
provided in the submitted documentation.


\subsection{Inputs for Higgs studies}
\label{app:inputsHiggs}
The uncertainties on inputs for all the colliders used in our analysis are listed
in Tables~\ref{Table-first}-\ref{Table-last}.
In all cases the relative uncertainty on the measurement is given corresponding to a Gaussian $1$-$\sigma$ uncertainty.

\begin{table}[hbt]
\small
  \caption{Inputs used for CEPC and FCC-ee projections. All uncertainties are given as fractional 68\% CL intervals and are taken to be symmetric. The upper limits are given at 68\% CL. A dash indicates the absence of a projection for the corresponding channel.}
  \label{Table-first}
  \begin{center}
    \begin{tabular}{llll}
      \toprule
      &   FCC-ee$_{240}$    & FCC-ee$_{365}$ & CEPC \\
      \hline
 $\delta\sigma_{ZH}$                	&	0.005	&	0.009	&	0.005		\\
\hline								
 $\delta\mu_{ZH,bb}$                	&	0.003	&	0.005	&	0.0031	\\	
 $\delta\mu_{ZH,cc}$                	&	0.022	&	0.065	&	0.033	\\	
 $\delta\mu_{ZH,gg}$                	&	0.019	&	0.035	&	0.013	\\	
 $\delta\mu_{ZH,WW}$                	&	0.012	&	0.026	&	0.0098	\\	
 $\delta\mu_{ZH,ZZ}$                	&	0.044	&	0.12	&	0.051	\\	
 $\delta\mu_{ZH,\tau\tau}$          	&	0.009	&	0.018	&	0.0082	\\	
 $\delta\mu_{ZH,\gamma\gamma}$      	&	0.09	&	0.18	&	0.068	\\	
 $\delta\mu_{ZH,\mu\mu}$	&	0.19	&	0.40	&	0.17	\\	
 $\delta\mu_{ZH,Z\gamma}  $	& $-$		& $-$		&	0.16	\\	
 \hline
 $\delta\mu_{\nu\nu H,bb}$           	&	0.031	&	0.009	&	0.030	\\
 $\delta\mu_{\nu\nu H,cc}$            	&	$-$	&	0.10	& $-$		\\	
 $\delta\mu_{\nu\nu H,gg}$            	&	$-$	&	0.045	& $-$		\\	
 $\delta\mu_{\nu\nu H,ZZ}$            	&	$-$	&	0.10	& $-$		\\	
 $\delta\mu_{\nu\nu H,\tau\tau}$      	&	$-$	&	0.08	& $-$		\\	
 $\delta\mu_{\nu\nu H,\gamma\gamma}$  	&	$-$	&	0.22	& $-$		\\	
 \hline
$\rm{BR}_{\rm inv}$	&	$<$0.0015	&	$<$0.003	&	$<$0.0015	\\	

      \bottomrule
    \end{tabular}
  \end{center}
\end{table}

\begin{table}[hbt]
\small
  \caption{Inputs used for ILC projections at the 250 and 350 GeV energy stages and two polarisations. All uncertainties are given as fractional 68\% CL intervals and are taken to be symmetric. The upper limits are given at 68\% CL.}
  \begin{center}
    \begin{tabular}{lll}
    \toprule
     \multicolumn{3}{c}{ILC$_{250}$} \\
      \hline
    Polarization:  & $e^-$: -80\%  $e^+$: +30\% & $e^-$: +80\%  $e^+$: -30\%\\
    \hline
      $\delta\sigma_{ZH}/\sigma_{ZH}$&0.011 & 0.011 \\
      \hline
      $\delta\mu_{ZH,bb}$& 0.0072 & 0.0072\\
      $\delta\mu_{ZH,cc}$&0.044 & 0.044\\
      $\delta\mu_{ZH,gg}$& 0.037 & 0.037 \\
      $\delta\mu_{ZH,ZZ}$& 0.095 & 0.095\\
      $\delta\mu_{ZH,WW}$& 0.024 & 0.024\\
      $\delta\mu_{ZH,\tau\tau}$ & 0.017 & 0.017\\
      $\delta\mu_{ZH,\gamma\gamma}$   & 0.18 & 0.18\\
      $\delta\mu_{ZH,\mu\mu}$      & 0.38 & 0.38\\
      \hline
      $\delta\mu_{\nu\nu H,bb}$& 0.043 & 0.17\\
      \hline
      $\rm{BR}_{\rm inv}    $&        $<$0.0027&    $<$0.0021   \\
%
\bottomrule
 & &\\
\toprule
      \multicolumn{3}{c}{ILC$_{350}$} \\
      \hline
    Polarization:  & $e^-$: -80\%  $e^+$: +30\% & $e^-$: +80\%  $e^+$: -30\%\\
    \hline
      $\delta\sigma_{ZH}/\sigma_{ZH}$&0.025 & 0.042 \\
      \hline
      $\delta\mu_{ZH,bb}$             & 0.021 & 0.036 \\
      $\delta\mu_{ZH,cc}$             & 0.15 & 0.26 \\
      $\delta\mu_{ZH,gg}$             & 0.11 & 0.20 \\
      $\delta\mu_{ZH,ZZ}$  & 0.34 & 0.59 \\
      $\delta\mu_{ZH,WW}$ & 0.076 & 0.13\\
      $\delta\mu_{ZH,\tau\tau}$ & 0.054 & 0.094\\
      $\delta\mu_{ZH,\gamma\gamma}$   & 0.53 & 0.92 \\
      $\delta\mu_{ZH,\mu\mu}$      & 1.2 & 2.1 \\
      \hline
      $\delta\mu_{\nu\nu H,bb}$     & 0.025 & 0.18 \\
      $\delta\mu_{\nu\nu H,cc}$     & 0.26 & 1.9 \\
      $\delta\mu_{\nu\nu H,gg}$     & 0.10 & 0.75\\
      $\delta\mu_{\nu\nu H,ZZ}$    & 0.27 & 1.9 \\
      $\delta\mu_{\nu\nu H,WW}$  &  0.078 & 0.57 \\
      $\delta\mu_{\nu\nu H,\tau\tau}$  &  0.22 & 1.6 \\
      $\delta\mu_{\nu\nu H,\gamma\gamma}$  &  0.61 & 4.2 \\
      $\delta\mu_{\nu\nu H,\mu\mu}$  & 2.2 & 16 \\
      \hline
      $\rm{BR}_{\rm inv}    $&   $<$0.0096     &  $<$0.015    \\
      \bottomrule
    \end{tabular}
  \end{center}
\end{table}

\begin{table}[hbt]
\small
  \caption{Inputs used for ILC projections at the 500 and 1000 GeV energy stages and two polarisations. All uncertainties are given as fractional 68\% CL intervals and are taken to be symmetric. The upper limits are given at 68\% CL.}
  \begin{center}
    \begin{tabular}{lll}
    \toprule
      \multicolumn{3}{c}{ILC$_{500}$} \\
      \hline
      Polarization:  & $e^-$: -80\%  $e^+$: +30\% & $e^-$: +80\%  $e^+$: -30\%\\
      \hline
      
      $\delta\sigma_{ZH}/\sigma_{ZH}$&0.017 & 0.017\\
      \hline
      $\delta\mu_{ZH,bb}$             & 0.010& 0.010\\
      $\delta\mu_{ZH,cc}$             & 0.071 &  0.071\\
      $\delta\mu_{ZH,gg}$             & 0.059 & 0.059 \\
      $\delta\mu_{ZH,ZZ}           $  & 0.14 & 0.14 \\
      $\delta\mu_{ZH,WW}         $ & 0.030 & 0.030 \\
      $\delta\mu_{ZH,\tau\tau}     $ & 0.024 & 0.024\\
      $\delta\mu_{ZH,\gamma\gamma}$   & 0.19 & 0.19\\
      $\delta\mu_{ZH,\mu\mu}$      & 0.47 & 0.47 \\
       \hline
      $\delta\mu_{\nu\nu H,bb}$     & 0.0041 & 0.015\\
      $\delta\mu_{\nu\nu H,cc}$     & 0.035 & 0.14\\
      $\delta\mu_{\nu\nu H,gg}$     & 0.023 & 0.095 \\
      $\delta\mu_{\nu\nu H,ZZ}$    & 0.047 & 0.19\\
      $\delta\mu_{\nu\nu H,WW}$  & 0.014 & 0.055 \\
      $\delta\mu_{\nu\nu H,\tau\tau}$  & 0.039 & 0.16 \\
      $\delta\mu_{\nu\nu H,\gamma\gamma}$  & 0.11 & 0.43\\
      $\delta\mu_{\nu\nu H,\mu\mu}$  & 0.4 & 1.7 \\
       \hline
      $\delta\mu_{ttH,bb}$&0.20 & 0.20 \\
      \hline
      $\rm{BR}_{\rm inv}$&        $<$0.0069&    $<$0.0050   \\
      \hline
      Direct constraint on Higgs self-interaction& & \\
      $\delta\kappa_3$ & \multicolumn{2}{c}{0.27} \\
%
\bottomrule
 & &\\
\toprule
      \multicolumn{3}{c}{ILC$_{1000}$} \\
      \hline
      Polarization:  & $e^-$: -80\%  $e^+$: +20\% & $e^-$: +80\%  $e^+$: -20\%\\
      \hline
      $\delta\mu_{\nu\nu H,bb}$     & 0.0032 & 0.010\\
      $\delta\mu_{\nu\nu H,cc}$     & 0.017 & 0.064\\
      $\delta\mu_{\nu\nu H,gg}$     & 0.013 & 0.047\\
      $\delta\mu_{\nu\nu H,ZZ}$     & 0.023 & 0.084\\
      $\delta\mu_{\nu\nu H,WW}$     & 0.0091 & 0.033 \\
      $\delta\mu_{\nu\nu H,\tau\tau}$  & 0.017 & 0.064 \\
      $\delta\mu_{\nu\nu H,\gamma\gamma}$  & 0.048 & 0.17 \\
      $\delta\mu_{\nu\nu H,\mu\mu}$  & 0.17 & 0.64 \\    
       \hline  
      $\delta\mu_{tt H,bb}$     & 0.045 & 0.045\\
      \hline
      Direct constraint on Higgs self-interaction& & \\
      $\delta\kappa_3$ & \multicolumn{2}{c}{0.10} \\
      \bottomrule
    \end{tabular}
  \end{center}
\end{table}

\begin{table}[hbt]
\small
  \caption{Inputs used for CLIC projections at the three energy stages and two polarisations. All uncertainties are given as fractional 68\% CL intervals and are taken to be symmetric. The upper limits are given at 68\% CL.}
  \begin{center}
    \begin{tabular}{lll}
      \toprule
     \multicolumn{3}{c}{CLIC$_{380}$} \\
      \hline
      Polarization:  & $e^-$: -80\%  $e^+$: 0\% & $e^-$: +80\%  $e^+$: 0\%\\
           \hline
      $\delta\sigma_{ZH,Z\to ll}/\sigma_{ZH,Z\to ll}$&0.036&0.041\\
      $\delta\sigma_{ZH,Z\to qq}/\sigma_{ZH,Z\to qq}$&0.017&0.020\\
      \hline
      $\delta\mu_{ZH,bb}          $&0.0081&0.0092\\
      $\delta\mu_{ZH,cc}          $&0.13&0.15\\
      $\delta\mu_{ZH,gg}          $&0.057&0.065\\
      $\delta\mu_{ZH,WW}         $&0.051&0.057\\
      $\delta\mu_{ZH,\tau\tau}     $ &0.059&0.066\\
      \hline
      $\delta\mu_{\nu\nu H,bb}    $&0.014&0.041\\
      $\delta\mu_{\nu\nu H,cc}      $&0.19&0.57\\
      $\delta\mu_{\nu\nu H,gg}     $&0.076&0.23\\
       \hline
      $\rm{BR}_{\rm inv}    $&        $<$0.0027&       $<$0.003\\
%
\bottomrule
 & &\\
\toprule
      \multicolumn{3}{c}{CLIC$_{1500}$} \\
      \hline
      Polarization:  & $e^-$: -80\%  $e^+$: 0\% & $e^-$: +80\%  $e^+$: 0\%\\
      \hline
      $\delta\mu_{ZH,bb}$        &0.028&0.062\\
      \hline
      $\delta\mu_{\nu\nu H,bb}$ &0.0025&0.015\\
      $\delta\mu_{\nu\nu H,cc}$  &0.039&0.24\\
      $\delta\mu_{\nu\nu H,gg}$   &0.033&0.20\\
      $\delta\mu_{\nu\nu H,WW}$    &0.0067&0.04\\
      $\delta\mu_{\nu\nu H,ZZ}$     &0.036&0.22\\
      $\delta\mu_{\nu\nu H,\gamma\gamma}$ &0.1&0.6\\
      $\delta\mu_{\nu\nu H,Z\gamma}$     &0.28&1.7\\
      $\delta\mu_{\nu\nu H,\tau\tau}$     &0.028&0.17\\
      $\delta\mu_{\nu\nu H,\mu\mu}$      &0.24&1.5\\
      \hline
      $\delta\mu_{eeH,bb}$       &0.015&0.033\\
       \hline
      $\delta\mu_{ttH,bb}$       &0.056&0.15\\
%
\bottomrule
 & &\\
\toprule
      \multicolumn{3}{c}{CLIC$_{3000}$} \\
      \hline
      Polarization:  & $e^-$: -80\%  $e^+$: 0\% & $e^-$: +80\%  $e^+$: 0\%\\
      \hline
      $\delta\mu_{ZH,bb}$                 &0.045&0.10\\
      \hline
      $\delta\mu_{\nu\nu H,bb}$            &0.0017&0.01\\
      $\delta\mu_{\nu\nu H,cc}$            &0.037&0.22\\
      $\delta\mu_{\nu\nu H,gg}$            &0.023&0.14\\
      $\delta\mu_{\nu\nu H,WW}$            &0.0033&0.02\\
      $\delta\mu_{\nu\nu H,ZZ}$            &0.021&0.13\\
      $\delta\mu_{\nu\nu H,\gamma\gamma}$  &0.05&0.3\\
      $\delta\mu_{\nu\nu H,Z\gamma}$       &0.16&0.95\\
      $\delta\mu_{\nu\nu H,\tau\tau}$      &0.023&0.14\\
      $\delta\mu_{\nu\nu H,\mu\mu}$        &0.13&0.8\\
      \hline
      $\delta\mu_{eeH,bb}$                &0.016&0.036\\
      \hline
      Direct constraint on Higgs self-interaction& & \\
      $\delta\kappa_3$ &\multicolumn{2}{c}{0.11}\\
      \bottomrule
    \end{tabular}
  \end{center}
\end{table}

\begin{table}[hbt]
\small
  \caption{Inputs used for LHeC and FCC-eh projections. All uncertainties are given as fractional 68\% CL intervals and are taken to be symmetric.}
  \begin{center}
    \begin{tabular}{lll}
    \toprule
     Observable & LHeC & FCC-eh \\
     \hline
      $\delta\mu_{WBF, bb}$ & 0.008 & 0.0025\\
      $\delta\mu_{WBF, cc}$ & 0.071 & 0.022\\
      $\delta\mu_{WBF, gg}$ & 0.058 & 0.018 \\      
      $\delta\mu_{ZBF, bb}$ & 0.023 & 0.0065\\
      $\delta\mu_{WBF, WW}$ & 0.062 & 0.019 \\
      $\delta\mu_{WBF, ZZ}$ & 0.120 & 0.038 \\
      $\delta\mu_{WBF,\tau\tau}$ & 0.052 & 0.016 \\
      $\delta\mu_{WBF,\gamma\gamma}$ & 0.15 & 0.046\\
      \hline
      $\delta\mu_{ZBF, cc}$ & 0.200 & 0.058 \\
      $\delta\mu_{ZBF, gg}$ & 0.160 & 0.047 \\
      $\delta\mu_{ZBF, WW}$ & 0.170  & 0.050 \\      
      $\delta\mu_{ZBF, ZZ}$ & 0.350  &0.100\\   
      $\delta\mu_{ZBF,\tau\tau}$ & 0.15  & 0.042 \\   
      $\delta\mu_{ZBF,\gamma\gamma}$ & 0.42 & 0.120 \\   
      \bottomrule
    \end{tabular}
  \end{center}
\end{table}

\begin{table}[hbt]
\small
  \caption{Left) Inputs used for FCC-hh. All uncertainties are given as fractional 68\% CL intervals and are taken to be symmetric. Right) Extra inputs used in the $\kappa$ fit studies. }
  \begin{center}
  \begin{tabular}{c c}
    \begin{tabular}{ccc}
      \toprule
            \multicolumn{3}{c}{FCC-hh} \\
      \hline
      $\delta\mu_{ggF,4\mu}$        &0.019&\\
      $\delta\mu_{ggF,\gamma\gamma}$  &0.015&\\
      $\delta\mu_{ggF,Z\gamma}$       &0.016&\\
      $\delta\mu_{ggF,\mu\mu}$        &0.012&\\
      \hline
      $\delta (\rm{BR}_{\mu\mu}/\rm{BR}_{4\mu})$        &0.013&\\
      $\delta (\rm{BR}_{\gamma\gamma}/\rm{BR}_{2e2\mu})$&0.008&\\
      $\delta (\rm{BR}_{\gamma\gamma}/\rm{BR}_{\mu\mu}) $&0.014&\\
      $\delta (\rm{BR}_{\mu\mu\gamma}/\rm{BR}_{\gamma\gamma})$&0.018&\\
      \hline
      $\delta(\sigma_{ttH}^{bb}/\sigma_{ttZ}^{bb})$&0.019&\\
      \hline
      Invisible decays & & \\
      $ \rm{BR}_{\rm inv}$        &$<$0.00013&\\
      \hline
      Direct constraint on Higgs self-interaction & & \\
      $\delta\kappa_3 $&0.05&\\
      \bottomrule
    \end{tabular}
    &
    \begin{tabular}{cc}
      \toprule
            \multicolumn{2}{c}{FCC-hh} \\
            \multicolumn{2}{c}{(Extra inputs used in $\kappa$ fits)} \\ 
      \hline
      $\delta(\sigma_{WH}^{H \to \gamma\gamma}/\sigma_{WZ}^{Z\to e^+ e^-})$&0.014\\
      $\delta(\sigma_{WH}^{H \to \tau\tau}/\sigma_{WZ}^{Z\to \tau\tau})$&0.016\\
      $\delta(\sigma_{WH}^{H \to bb}/\sigma_{WZ}^{Z\to bb})$&0.011\\
      $\delta(\sigma_{WH}^{H \to WW}/\sigma_{WH}^{H\to \gamma\gamma})$&0.015\\
      \bottomrule
    \end{tabular}
    \end{tabular}
  \end{center}
\end{table}

\begin{table}[hbt]
\small
  \caption{Inputs used for a low-energy FCC-hh running at 37.5 TeV (LE-FCC). All uncertainties are given as fractional 68\% CL intervals and are taken to be symmetric. }
  \label{Table-last}
  \begin{center}
    \begin{tabular}{ccc}
      \toprule
            \multicolumn{3}{c}{LE-FCC} \\
      \hline
      $\delta (\rm{BR}_{\mu\mu}/\rm{BR}_{4\mu})$        &0.029&\\
      $\delta (\rm{BR}_{\gamma\gamma}/\rm{BR}_{2e2\mu})$&0.015&\\
      $\delta (\rm{BR}_{\gamma\gamma}/\rm{BR}_{\mu\mu}) $&0.028&\\
      $\delta (\rm{BR}_{\mu\mu\gamma}/\rm{BR}_{\gamma\gamma})$&0.06&\\
      \hline
      $\delta(\sigma_{ttH}^{bb}/\sigma_{ttZ}^{bb})$&0.04-0.06&\\
      \hline
      Direct constraint on Higgs self-interaction & & \\
      $\delta\kappa_3 $&0.15&\\
      \bottomrule
    \end{tabular}
  \end{center}
\end{table}


\clearpage
\newpage

\subsection{Inputs for electroweak precision observables}
\label{app:inputsEWPO}

The uncertainties on several electroweak precision observables related to the properties (masses and couplings) of the electroweak vector bosons are presented in Table~\ref{tab:ewkpar}. We also report the expected uncertainties on the top-quark mass, which enters in the analysis as an input of the global electroweak fit. 

For the extraction of $m_\textrm{top}$ from a $t\bar{t}$ scan threshold at $e^+ e^-$ colliders the current theoretical uncertainty is $\sim 40$~MeV~\cite{Simon:2016pwp}. As it was done for the other intrinsic theory uncertainties, this is expected to improve in the future and is neglected in the baseline fits. We therefore use a common statistical uncertainty of $\sim 20$ MeV for all lepton colliders running at the $t\bar{t}$ threshold. 

For the circular colliders, the asymmetries $A_f$ and partial-width ratios $R_f$ are not given for $\sqrt{s}\gg 90$~GeV as the statistical precision is much higher at the $Z$ pole, and the $Z$-pole run is part of the default programme. For the linear colliders a Giga-$Z$ run is not part of the running plan submitted to the EPPSU, but it is used in some of the results presented in this report to illustrate the impact of such a run in the EW and Higgs programmes. For the ILC and CLIC Giga-$Z$ option, an integrated luminosity of 100~fb$^{-1}$ is assumed, and polarisations as stated in Table~\ref{tab:colliders}. Note that the asymmetry parameters $A_f$ can be extracted in different ways depending on the access of 
polarization of the initial and/or final states. For instance, at linear colliders with polarized beams $A_b$ can be directly extracted from a left-right forward-backward asymmetry. Without polarized beams, circular colliders can access that quantify via a forward-backward asymmetry $A_{FB}^{b}=\frac 34 A_e A_b$, but it requires to know $A_e$. On the other hand, both circular and linear colliders could access $A_e$ and $A_\tau$ separately, measuring the polarization of the final states in $e^+ e^- \to \tau^+ \tau^-$.
We refer to the discussion in Section~\ref{sec:SMEFTfit} for the assumptions adopted in the treatment of systematic uncertainties for the heavy flavor observables included the fits.

For ILC all values are taken from Ref.~\cite{Fujii:2019zll}. For CLIC all values are taken from Refs.~\cite{gigazclic,Charles:2018vfv}. For CEPC they are taken from either Ref. ~\cite{CEPCStudyGroup:2018ghi} or from Ref.~\cite{cepcewk}. For FCC-ee they are taken from Refs.~\cite{Mangano:2018mur,Abada:2019zxq,Blondel:2019yqr,FCCeeEWTalk}.

\begin{table}[htbp]
\scriptsize
    \centering
    \caption{Uncertainty on several observables related to the properties of the electroweak vector bosons. We also list the uncertainty on the top mass. For dimensionful quantities the absolute uncertainty is given, while relative errors are listed for dimensionless quantities.
    A few comments on some particular numbers are in order:
    a) For hadron colliders the top mass is not the pole mass. 
    b) For the top mass all lepton colliders require a dedicated top threshold scan to achieve the uncertainty given here. (For ILC the quoted value comes from a dedicated run at 350 GeV.)
    c) From direct reconstruction in the $ZH$ run 2-3 MeV can be achieved~\cite{CEPCStudyGroup:2018ghi}. 
    d) In a 4-year dedicated run 2~MeV can be achieved by ILC~\cite{Wilson:2016hne}. 
    e) From $\tau$ polarization measurements.
    f) At circular colliders, for $A_b$ and $A_c$ previous measurement uncertainties were dominated by the physics modelling~\cite{ALEPH:2005ab} and the systematic uncertainty arising from this was only estimated by FCC-ee~\cite{Blondel:2019yqr}. When these systematics are set to zero in the measurements of $A_{FB}^{b}$ and $A_{FB}^{c}$ the uncertainty in $A_b$ and $A_c$ is controlled by the statistical errors plus the uncertainty on $A_e$. This is the setup used for the baseline fits. See discussion in Section~\ref{sec:SMEFTfit} for details.
    g) $R_\nu\equiv \Gamma_{Z\to {\rm inv}}/\Gamma_{Z\to {\rm had}}$ and $R_{\rm inv}\equiv \Gamma_{Z\to {\rm inv}}/\Gamma_{Z\to \ell\ell}$.
    \label{tab:ewkpar}}
    \begin{tabular}{l|r|r|r|r|rr|rr}
    \toprule
Quantity & Current & HL-LHC & FCC-ee & CEPC & \multicolumn{2}{c|}{ILC} & \multicolumn{2}{c}{CLIC} \\
 & & & & & Giga-Z & 250 GeV & Giga-Z & 380 GeV \\
 \midrule
     $\delta m_\textrm{top}$ [MeV] & $\sim$500 $^{a)}$ & $\sim$400 $^{a)}$ & 20 $^{b)}$ & $-$ & $-$ & 17 $^{b)}$ & $-$ & 20-22 $^{b)}$  \\ 
     \midrule
    
    $\delta M_Z$ [MeV] & 2.1 & $-$ & $0.1$ & $0.5$ & $-$ & $-$ & $-$ & $-$ \\
    $\delta \Gamma_Z$ [MeV] & 2.3 & $-$ & $0.1$ & $0.5$ & $1$ & $-$ & $1$ & $-$ \\
    $\delta \Gamma_{Z\to {\rm had}}$ [MeV] & 2.0 & $-$ & $-$ & $-$ & $0.7$ & $-$ & $0.7$ & $-$ \\
    
    $\delta \sigma_{{\rm had}}^{0}$ [pb] & 37 & $-$ & $4$ & $5$ & $-$ & $-$ & $-$ & $-$ \\
    
     \midrule
     
    $\delta M_W$ [MeV] & 12 & 7 & 0.7 & 1.0 (2-3) $^{c)}$ & $-$ & 2.4 $^{d)}$ & $-$ & 2.5\\
    $\delta \Gamma_W$ [MeV] & 42 & $-$ & 1.5 & 3 & $-$ & $-$ & $-$ & $-$\\
     \midrule
    $\delta {\rm BR}_{W\to e\nu}$[$10^{-4}$] & 150 & $-$ & 3 & 3 & $-$  & 4.2 & $-$  & 11\\
    $\delta {\rm BR}_{W\to \mu\nu}$[$10^{-4}$] & 140 & $-$  & 3 & 3 & $-$ & 4.1 & $-$  & 11\\
    $\delta {\rm BR}_{W\to \tau\nu}$[$10^{-4}$] & 190 & $-$  & 4 & 4 & $-$ & 5.2 & $-$ & 11 \\
    $\delta {\rm BR}_{W\to{\rm had}}$[$10^{-4}$] & 40 & $-$ & 1 & 1 & $-$ & $-$ & $-$ & $-$\\
     \midrule    
    $\delta A_e$ [$10^{-4}$] & 140 & $-$ & 1.1 $^{e)}$ & 3.2 $^{e)}$ & 5.1 & 10 & 10 & 42\\
    $\delta A_\mu$ [$10^{-4}$] & 1060 & $-$ & $-$ & $-$ & 5.4 & 54 & 13 & 270 \\
    $\delta A_\tau$ [$10^{-4}$] & 300 & $-$ & 3.1 $^{e)}$ & 5.2 $^{e)}$ & 5.4 & 57 & 17 & 370\\
    $\delta A_b$ [$10^{-4}$] & 220 & $-$ & $-$ & $-$ & 5.1 & 6.4 & 9.9 & 40\\
    $\delta A_c$ [$10^{-4}$] & 400 & $-$ & $-$ & $-$ & 5.8 & 21 & 10 & 30 \\
    
    $\delta A^{\mu}_{\rm FB}$ [$10^{-4}$] & 770 & $-$ & 0.54 & 4.6  & $-$ & $-$ & $-$ & $-$\\
    $\delta A^b_{\rm FB}$ [$10^{-4}$] & 160 & $-$ & 30 $^{f)}$ & 10 $^{f)}$ & $-$ & $-$ & $-$ & $-$\\
    $\delta A^c_{\rm FB}$ [$10^{-4}$] & 500 & $-$ & 80 $^{f)}$ & 30 $^{f)}$ & $-$ & $-$ & $-$ & $-$\\
    
    \midrule
    $\delta R_e$ [$10^{-4}$] & 24 & $-$ & 3 & 2.4 & 5.4 & 11 & 4.2 & 27\\
    $\delta R_\mu$ [$10^{-4}$] & 16 & $-$ & 0.5 & 1 & 2.8& 11 & 2.2 & 27\\
    $\delta R_\tau$ [$10^{-4}$] & 22 & $-$ & 1 & 1.5 & 4.5 & 12 & 4.3 & 60\\
    $\delta R_b$ [$10^{-4}$] & 31 & $-$ & 2 & 2 & 7 & 11 & 7 & 18\\
    $\delta R_c$ [$10^{-4}$] & 170 & $-$ & 10 & 10 & 30 & 50 & 23 & 56 \\
    \midrule
    $\delta R_\nu$ [$10^{-3}$] $^{g)}$ & $-$ & $-$ & $-$ & $-$ & $-$ & $-$ & $-$ & 9.4 \\
    $\delta R_{\rm inv}$ [$10^{-3}$] $^{g)}$  & $-$ & $-$ & 0.27 & 0.5 & $-$ & $-$ & $-$ & $-$ \\
    \bottomrule
    \end{tabular}
\end{table}

\clearpage
\newpage


\section{Correlation matrices~\label{kappacorrelations}}

\begin{figure}[t]
\centering
\includegraphics[width=0.48\linewidth]{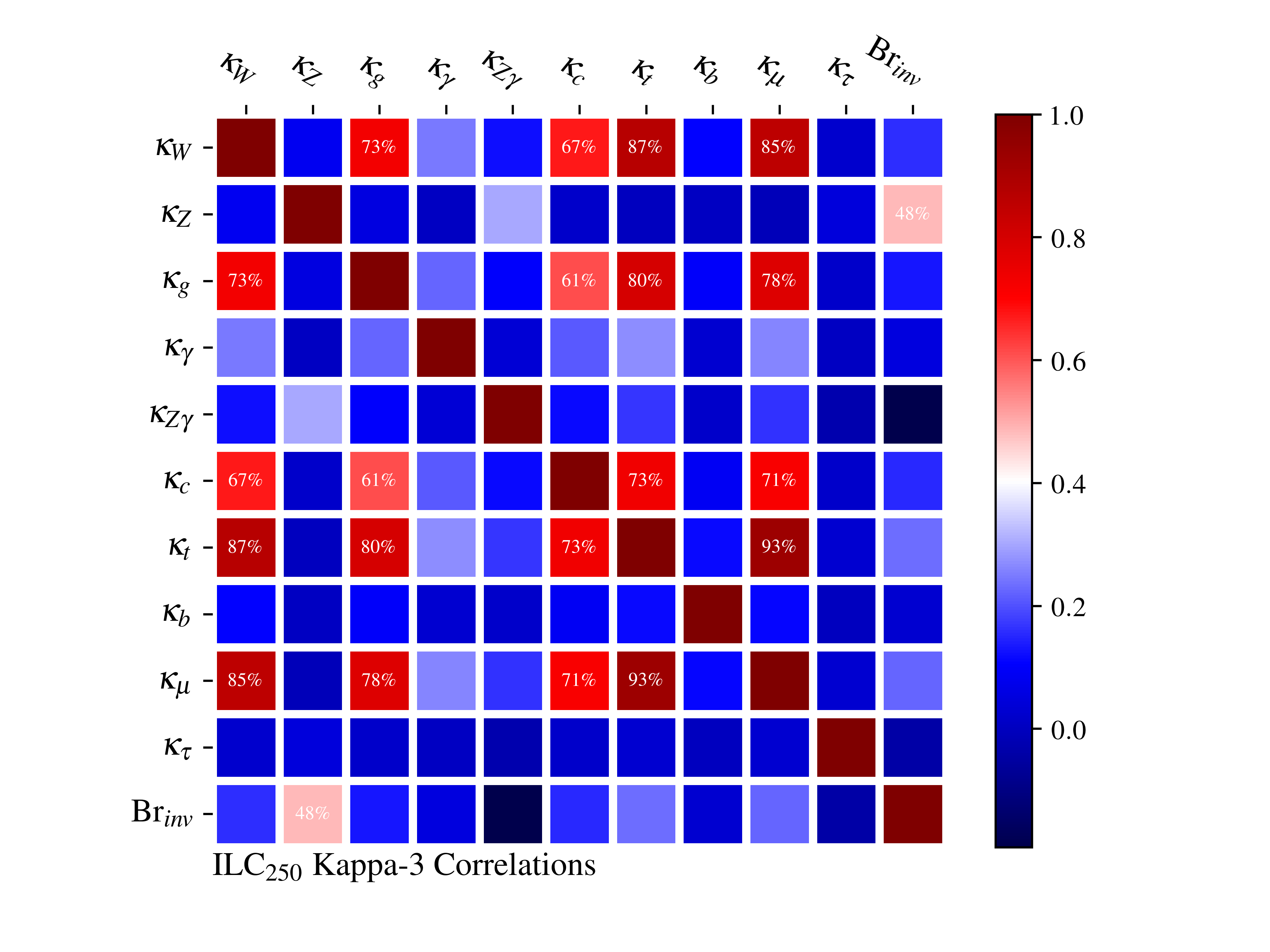}
\includegraphics[width=0.48\linewidth]{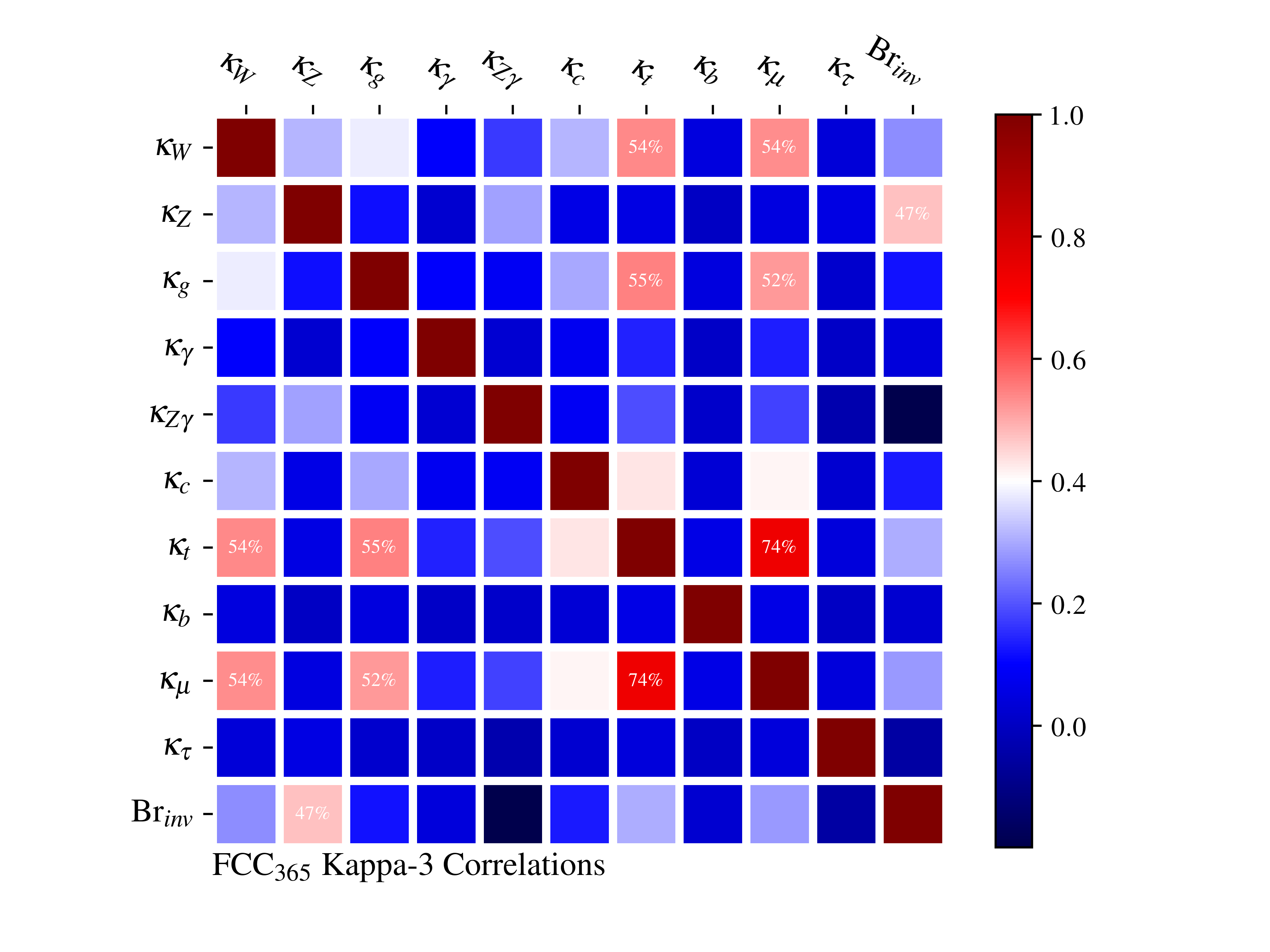}
\includegraphics[width=0.48\linewidth]{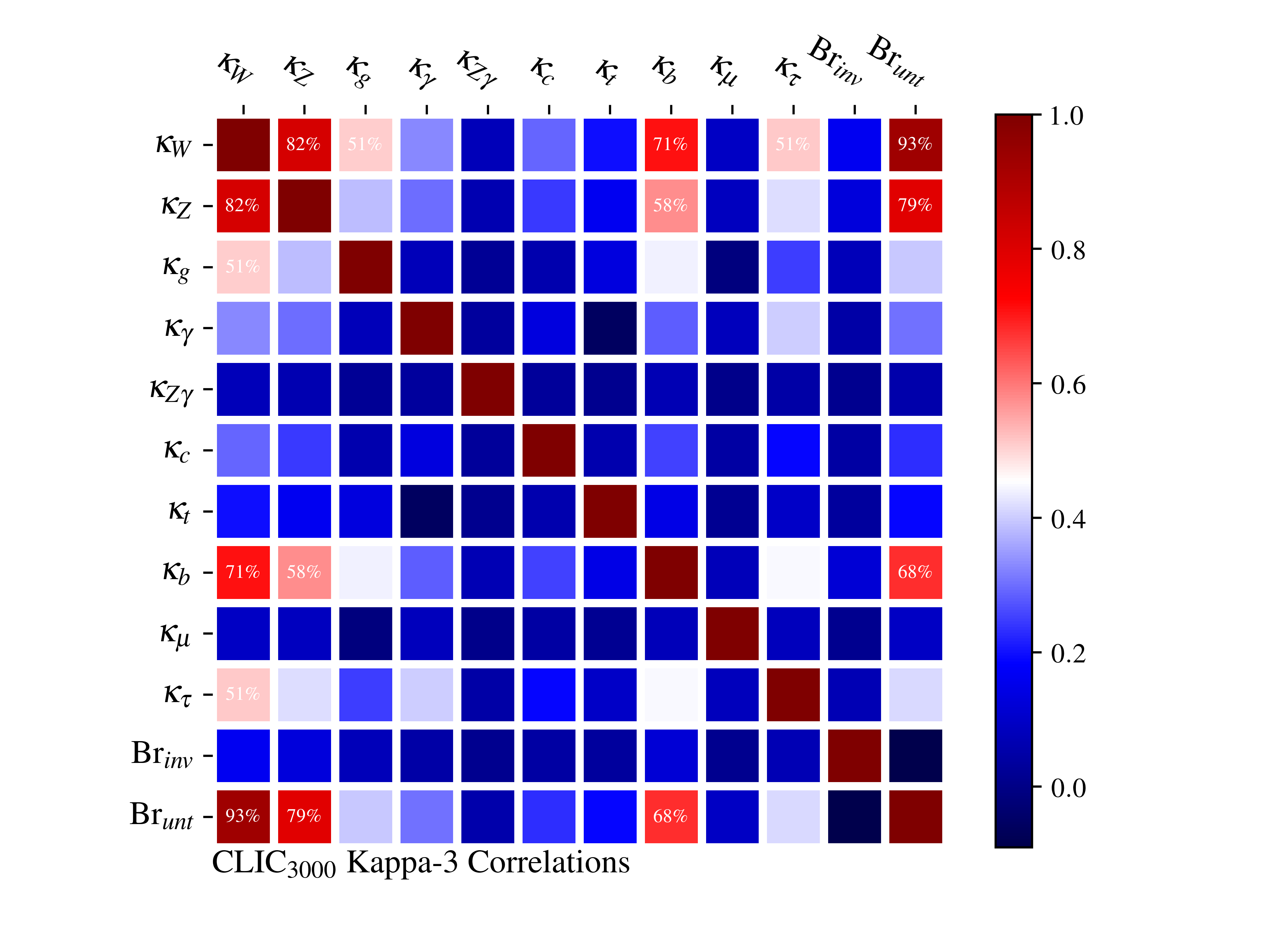}
\caption{\label{fig:Kappa3Correlation}
Correlation seen in the kappa-3 scenario fit for three future colliders as discussed in section~\ref{kappa-results}. Top left: ILC$_{250}$. Top right: FCC$_{365}$. Bottom: CLIC$_{3000}$.}
\end{figure}

The correlations of three of the lepton collider kappa-3 fits, discussed in Section~\ref{kappa-results}, are shown in Figure~\ref{fig:Kappa3Correlation}.

\section{Additional Kappa scenario fits~\label{kappabenchmarks}}

This appendix contains additional kappa scenarios to complement the results shown in section~\ref{kappa-results}. Tables~\ref{tab:resultsLHCKappa2} and~\ref{tab:resultsKappa2} present the results of the different colliders in the kappa-2 scenario, in which BSM decays are allowed and future colliders are considered independently and not fitted together with the HL-LHC prospects. 

\begin{table}[hbt]
\centering
\scriptsize
\caption{ \label{tab:resultsLHCKappa2}
Expected relative precision (\%) of the $\kappa$ parameters in the kappa-2
(standalone) scenario  described in
section~\ref{tab:kappa_scenarios} for the HL-LHC,LHeC, and HE-LHC. A bound on $|\kappa_V|\leq 1$ is applied since no direct access to the Higgs width is possible, thus the uncertainty on $\kappa_W$ and $\kappa_{Z}$ is one-sided. For the remaining kappa parameters one standard deviation is provided in $\pm$. 
The corresponding 95\%CL upper limit on $\mathrm{BR}_{\mathrm{inv}}$
is also given. In this scenario $\mathrm{BR}_{\mathrm{unt}}$ is a floating parameter in the fit, to propagate the effect of an assumed uncertain total width on the measurement of the other $\kappa_{i}$. Based on this constraint the reported values on $\mathrm{BR}_{\mathrm{unt}}$ are inferred. The 95\% CL upper limits are given for these.  Cases in which a particular parameter has been fixed to the SM value due to lack of sensitivity are shown with a dash ($-$). In the case of $\kappa_t$ sensitivity at the LHeC, note that the framework relies as input on $\mu_{ttH}$, and does not take into consideration $\mu_{tH}$.  A star ($\star$) indicates the cases in which a parameter has been left free
in the fit due to lack of input in the reference documentation. The integrated luminosity and running conditions considered for each collider in this comparison are described in Table~\ref{tab:colliders}.
}
\begin{tabular}{ c | c  | c | c c }
\toprule
kappa-2    &HL-LHC   &LHeC   & HE-LHC (S2) & HE-LHC (S2')  \\
\bottomrule
$1\geq\kappa_{W}>$ (68\%)  & 0.985 & 0.994 & 0.988 & 0.992 \\
\crowcolorA$1\geq\kappa_{Z}>$ (68\%)  & 0.987 & 0.988 & 0.990 & 0.990\\  
$\kappa_{g}$ [$\%$]   & $\pm 2.$   & $\pm3.9$    & $\pm1.6$  &  $\pm1.1$    \\
\crowcolorA$\kappa_{\gamma}$ [$\%$]  & $\pm1.6$   & $\pm7.8$   & $\pm1.3$ & $\pm0.96$     \\
$\kappa_{Z\gamma}$ [$\%$]  & $\pm10.$   & $-$      & $\pm5.6$  & $\pm3.8$    \\
\crowcolorA$\kappa_{c}$ [$\%$]  & $-$   & $\pm 4.3$    & $-$    &$-$  \\
$\kappa_{t}$ [$\%$]  & $\pm3.2$   & $-$   & $\pm2.6$   &  $\pm1.6$     \\
\crowcolorA$\kappa_{b}$ [$\%$]  & $\pm2.5$   & $\pm2.3$    & $\pm2.0$   & $\pm1.5$    \\
$\kappa_{\mu}$ [$\%$]  & $\pm4.4$   & $-$   & $\pm2.3$    & $\pm1.5$   \\
\crowcolorA$\kappa_{\tau}$ [$\%$]  & $\pm1.6$   & $\pm3.6$   & $\pm1.2$      & $\pm0.85$    \\
\arrayrulecolor{black}\bottomrule
$\mathrm{BR}_{\mathrm{inv}}$ ($<$\%)   & $1.9$   & $2.2$   & $3.2\star$    & $2.4\star$      \\
\crowcolorA $\mathrm{BR}_{\mathrm{unt}}$ ($<$\%)     & \multicolumn{4}{c}{inferred using constraint $|\kappa_V|\leq 1$} \\
\crowcolorA &  $4.$   & $2.2$     & $3.2$  & 2.1\\
\arrayrulecolor{black}\bottomrule
\end{tabular}
\end{table}

\begin{table}[ht]
\centering
{\footnotesize
\caption{ \label{tab:resultsKappa2}
Expected relative precision (\%) of the $\kappa$ parameters in the kappa-2
(standalone collider) scenario  described in
section~\ref{tab:kappa_scenarios} for future accelerators beyond the LHC era.
The corresponding 95\%CL upper limits on $\mathrm{BR}_{\mathrm{unt}}$ and $\mathrm{BR}_{\mathrm{inv}}$
and the derived constraint on the Higgs width (in \%) are also given.
Cases in which a particular parameter has been fixed to the SM value due to lack of
sensitivity are shown with a dash (-). An asterisk ($\ast$) indicates the cases in which a parameter
has been left free in the fit due to lack of input in the reference documentation.
The integrated luminosity and running conditions considered for each collider in this comparison are described in Table~\ref{tab:colliders}.
FCC-ee/eh/hh corresponds to the combined performance of FCC-ee$_{240}$+FCC-ee$_{365}$, FCC-eh and FCC-hh.}
\begin{tabular}{ |c |  c c c | c c c | c | c c c| }
\toprule
kappa-2 & ILC   & ILC & ILC  & CLIC & CLIC & CLIC & CEPC & FCC-ee &FCC-ee & FCC     \\
 & 250 & 500 & 1000 & 380 & 1500 & 3000 & & 240 & 365& ee/eh/hh \\
  \bottomrule
$\kappa_{W}$ [$\%$]                               & 1.8   & 0.31 &0.26 & 0.86  & 0.39  & 0.38  & 1.3   & 1.3   & 0.44  & 0.2   \\
\crowcolorA$\kappa_{Z}$ [$\%$]                               & 0.3   & 0.24 &0.24 & 0.5   & 0.39  & 0.39  & 0.19  & 0.21  & 0.18  & 0.17  \\
$\kappa_{g}$  [$\%$]                              & 2.3   & 0.98 &0.67 & 2.5   & 1.3   & 0.96  & 1.5   & 1.7   & 1.0    & 0.52  \\
\crowcolorA$\kappa_{\gamma}$  [$\%$]                         & 6.8   & 3.5  &1.9  & 88.$\ast$   & 5.    & 2.3   & 3.7   & 4.8   & 3.9   & 0.32  \\
$\kappa_{Z\gamma}$ [$\%$]                         & 87.$\ast$   & 75.$\ast$   & 74.$\ast$  & 110.$\ast$  & 15.   & 7.    & 8.2   & 71.$\ast$   & 66.$\ast$   & 0.71  \\
\crowcolorA$\kappa_{c}$ [$\%$]                               & 2.5   & 1.3  & 0.91 & 4.4   & 1.9   & 1.4   & 2.2   & 1.8   & 1.3   & 0.96  \\
$\kappa_{t}$ [$\%$]                               & -     & 6.9  & 1.6 & -     & -     & 2.7   & -     & -     & -     & 1.0    \\
\crowcolorA$\kappa_{b}$ [$\%$]                               & 1.8   & 0.6  & 0.5  & 1.9   & 0.6  & 0.52  & 1.3   & 1.3   & 0.69  & 0.48   \\
$\kappa_{\mu}$ [$\%$]                             & 15.   & 9.4  & 6.3  & 290.$\ast$  & 13.   & 5.9   & 9.    & 10.   & 8.9   & 0.43  \\
\crowcolorA$\kappa_{\tau}$ [\%]                            & 1.9   & 0.72 & 0.58  & 3.1   & 1.3   & 0.95  & 1.4   & 1.4   & 0.74  & 0.49  \\
\arrayrulecolor{black}\bottomrule
$\mathrm{BR}_{\mathrm{inv}}$ ($<$\%)   & 0.26  & 0.23 & 0.23 & 0.65  & 0.65  & 0.65  & 0.28  & 0.22  & 0.19  & 0.024 \\
\crowcolorA$\mathrm{BR}_{\mathrm{unt}}$ ($<$\%)  & 1.8   & 1.4 & 1.3  & 2.7   & 2.4   & 2.4   & 1.1   & 1.2   & 1.1   & 1.0    \\
\arrayrulecolor{black}\toprule
\end{tabular}
}
\end{table}



\section{Additional Comparisons}
\label{app:addons}
In this section additional potential scenarios for accelerators are compared. The inputs for these were mostly provided after the European Strategy meeting in Granada. 

\begin{figure}[ht]
\centering
\begin{tabular}{c c}
\includegraphics[width=1\linewidth]{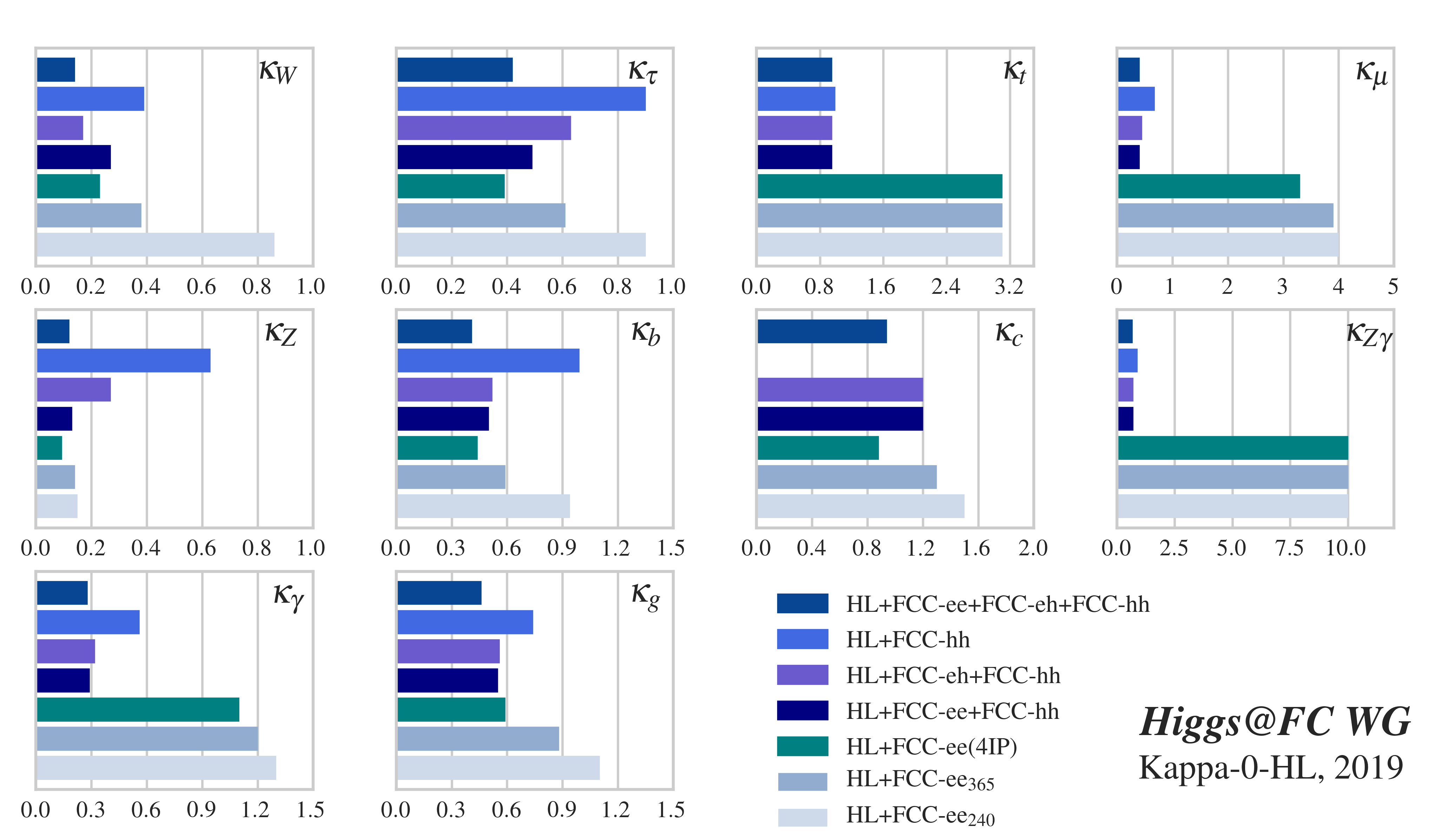}
\end{tabular}
\caption{\label{fig:Kappa0_FCCOptions}
Comparison of the different FCC scenarios in the kappa-0-HL scenario (similar to kappa-0 in that it does not allow any BSM decay, but including HL-LHC data).
}
\end{figure}

Table~\ref{tab:fcconly} and Fig.~\ref{fig:Kappa0_FCCOptions} show the results of the kappa-0-HL fit for various FCC scenarios. This fit is a replica of the already described kappa-0 one, which does not allow any BSM decay of the Higgs, but incorporating the HL-LHC information in a combined fit for completeness.
With 4 instead of 2 IPs the uncertainties reduce by a factor of up to 1.4 due to the increased statistics. With the FCC-hh only, the uncertainties all increase by factors of 2-5. When omitting FCC-eh, the uncertainty on $\kappa_W$ increases by a factor of two and that on $\kappa_b$ increases by 20\%, the others are mostly unaffected. When omitting FCC-ee, most uncertainties increase by about 20\% and that on $\kappa_Z$ increases by more than a factor of two. 

\begin{table}[ht]
\centering
{\scriptsize
\caption{\label{tab:fcconly} Results of kappa-0-HL fit for various scenarios of the FCC. In all cases the FCC data are combined with HL-LHC. The "4 IP" option considers 4 experiments instead of the 2 experiments considered in the CDR. For the FCC-hh scenario constraints on the $b$, $\tau$ and $W$ couplings come from measurements of ratios of $WH$ to $WZ$ production with the $H$ and $Z$ decaying to $b$-quarks or $\tau$ leptons, see Ref.~\cite{mlmreffcchhonly}.}
\begin{tabular}{| c |c c c c c c c|}
\toprule
kappa-0-HL & \multicolumn{7}{c|}{HL-LHC +}\\ 
 &FCC-ee$_{240}$   &FCC-ee   &FCC-ee (4 IP)   &FCC-ee/hh   &FCC-eh/hh   &FCC-hh   &FCC-ee/eh/hh  \\
\midrule
$\kappa_{W}[\%]$   & $0.86$   & $0.38$   & $0.23$   & $0.27$   & $0.17$   & $0.39$   & $0.14$  \\
\crowcolorA$\kappa_{Z}[\%]$   & $0.15$   & $0.14$   & $0.094$   & $0.13$   & $0.27$   & $0.63$   & $0.12$  \\
$\kappa_{g}[\%]$   & $1.1$   & $0.88$   & $0.59$   & $0.55$   & $0.56$   & $0.74$   & $0.46$  \\
\crowcolorA$\kappa_{ \gamma}[\%]$   & $1.3$   & $1.2$   & $1.1$   & $0.29$   & $0.32$   & $0.56$   & $0.28$  \\
$\kappa_{Z \gamma}[\%]$   & $10.$   & $10.$   & $10.$   & $0.7$   & $0.71$   & $0.89$   & $0.68$  \\
\crowcolorA$\kappa_{c}[\%]$   & $1.5$   & $1.3$   & $0.88$   & $1.2$   & $1.2$   & $-$   & $0.94$  \\
$\kappa_{t}[\%]$   & $3.1$   & $3.1$   & $3.1$   & $0.95$   & $0.95$   & $0.99$   & $0.95$  \\
\crowcolorA$\kappa_{b}[\%]$   & $0.94$   & $0.59$   & $0.44$   & $0.5$   & $0.52$   & $0.99$   & $0.41$  \\
$\kappa_{\mu}[\%]$   & $4.$   & $3.9$   & $3.3$   & $0.41$   & $0.45$   & $0.68$   & $0.41$  \\
\crowcolorA$\kappa_{\tau}[\%]$   & $0.9$   & $0.61$   & $0.39$   & $0.49$   & $0.63$   & $0.9$   & $0.42$  \\
\midrule
$\Gamma_{H}[\%]$   & $1.6$   & $0.87$   & $0.55$   & $0.67$   & $0.61$   & $1.3$   & $0.44$  \\
\bottomrule
\end{tabular}
}
\end{table}

\begin{figure}[ht]
\centering
\includegraphics[width=1\linewidth]{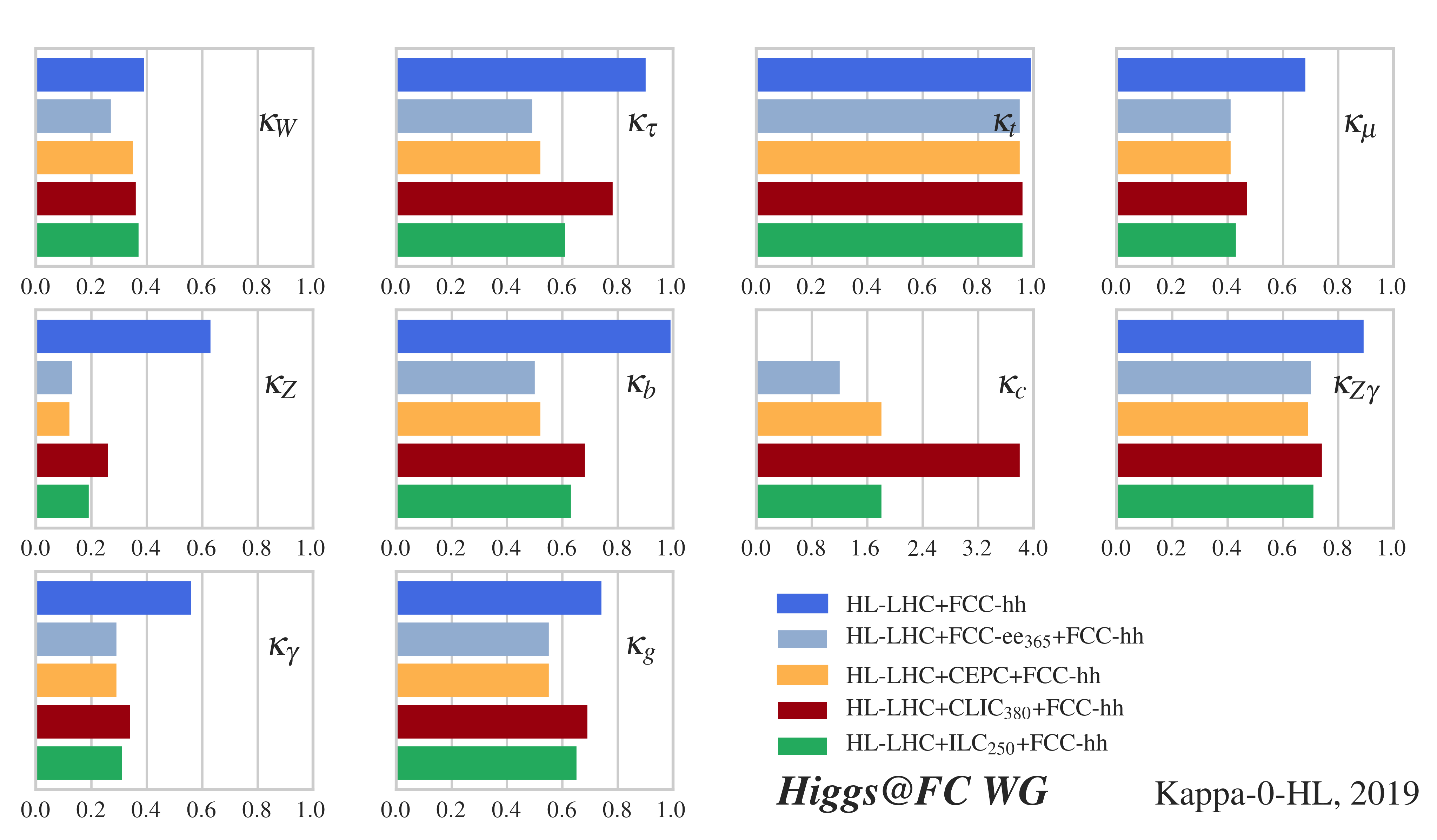}
\caption{\label{fig:Kappa0_FCCPlusEEColliders}
Combination of the different future ee colliders with FCC-hh and HL-LHC, in an extension of the kappa-0-HL scenario. Note that ILC$_{250}$ and CLIC$_{380}$ (first stages) are shown in comparison with CEPC (240) and FCC-ee$_{365}$. 
}
\end{figure}

Another interesting questions is what uncertainties are obtained when combining any of the lower energy stages of $e^+e^-$ colliders with the FCC-hh. This is shown in Table~\ref{tab:kappa0fcchhplusee} for the kappa-0-HL fit and in Table~\ref{tab:eft-global-fcchh} for the EFT fit. 
The results for the various 1st-stage $e^+e^-$ colliders  are comparable within a factor of about two for the Higgs couplings in the kappa-fits.
 Figure~\ref{fig:Kappa0_FCCPlusEEColliders} shows this comparison graphically. For the EFT framework, the differences are a bit larger, in particular for the aTGC values, see Table~\ref{tab:eft-global-fcchh}.

\begin{table}[ht]
\centering
{\small
\caption{\label{tab:kappa0fcchhplusee} Results for the kappa-0-HL fit for FCC-hh combined with any of the four $e^+e^-$ colliders proposed.} 
\begin{tabular}{| c |c c c c|}
\toprule
kappa-0-HL & \multicolumn{4}{c|}{HL-LHC + FCC-hh +}\\ 
   &ILC$_{250}$  &CLIC$_{380}$   &CEPC   &FCC-ee$_{365}$  \\
\midrule
$\kappa_{W}[\%]$   & $0.37$   & $0.36$   & $0.35$   & $0.27$  \\
\crowcolorA$\kappa_{Z}[\%]$   & $0.19$   & $0.26$   & $0.12$   & $0.13$  \\
$\kappa_{g}[\%]$   & $0.65$   & $0.69$   & $0.55$   & $0.55$  \\
\crowcolorA$\kappa_{\gamma}[\%]$   & $0.31$   & $0.34$   & $0.29$   & $0.29$  \\
$\kappa_{Z\gamma}[\%]$   & $0.71$   & $0.74$   & $0.69$   & $0.7$  \\
\crowcolorA$\kappa_{c}[\%]$   & $1.8$   & $3.8$   & $1.8$   & $1.2$  \\
$\kappa_{t}[\%]$   & $0.96$   & $0.96$   & $0.95$   & $0.95$  \\
\crowcolorA$\kappa_{b}[\%]$   & $0.63$   & $0.68$   & $0.52$   & $0.5$  \\
$\kappa_{\mu}[\%]$   & $0.43$   & $0.47$   & $0.41$   & $0.41$  \\
\crowcolorA$\kappa_{\tau}[\%]$   & $0.61$   & $0.78$   & $0.52$   & $0.49$  \\
\midrule
$\Gamma_{H}[\%]$   & $0.90$   & $0.98$   & $0.74$   & $0.67$  \\
\bottomrule
\end{tabular}
}
\end{table}

\begin{table}[ht]
\caption{\label{tab:eft-global-fcchh} 
Results for the global EFT fit for FCC-hh combined with any of the four $e^+e^-$ colliders proposed, also shown in Table~\ref{tab:kappa0fcchhplusee}. 
}
\centering
{\small
\begin{tabular}{ |c |c c c c |}
\toprule
SMEFT$_{\rm ND}$  & \multicolumn{4}{c|}{HL-LHC + FCC-hh +}\\
  &ILC$_{250}$   &CLIC$_{380}$   &CEPC  &FCC-ee$_{365}$  \\
\midrule
$g_{HZZ}^{\mathrm{eff}}[\%]$   & $0.35$   & $0.46$   & $0.38$   & $0.21$  \\
\crowcolorA$g_{HWW}^{\mathrm{eff}}[\%]$   & $0.36$   & $0.46$   & $0.36$   & $0.21$  \\
$g_{H \gamma \gamma}^{\mathrm{eff}}[\%]$   & $0.47$   & $0.55$   & $0.48$   & $0.38$  \\
\crowcolorA$g_{HZ \gamma}^{\mathrm{eff}}[\%]$   & $0.78$   & $0.83$   & $0.76$   & $0.72$  \\
$g_{Hgg}^{\mathrm{eff}}[\%]$   & $0.73$   & $0.88$   & $0.54$   & $0.56$  \\
\crowcolorA$g_{Htt}[\%]$   & $3.1$   & $2.2$   & $3.1$   & $1.7$  \\
$g_{Hcc}[\%]$   & $1.8$   & $3.9$   & $1.8$   & $1.2$  \\
\crowcolorA$g_{Hbb}[\%]$   & $0.75$   & $0.95$   & $0.58$   & $0.51$  \\
$g_{H\tau\tau}[\%]$   & $0.78$   & $1.2$   & $0.61$   & $0.54$  \\
\crowcolorA$g_{H\mu\mu}[\%]$   & $0.54$   & $0.61$   & $0.53$   & $0.46$  \\
\midrule
$\delta g_{1Z}[\times 10^{2}]$   & $0.078$   & $0.04$   & $0.08$   & $0.028$  \\
\crowcolorA$\delta \kappa_{ \gamma}[\times 10^{2}]$   & $0.12$   & $0.079$   & $0.089$   & $0.048$  \\
$\lambda_{Z}[\times 10^{2}]$   & $0.042$   & $0.043$   & $0.1$   & $0.047$  \\
\bottomrule
\end{tabular}
}
\end{table}

The ILC and CLIC documents submitted to the European Strategy did not contain any explicit analysis related to the $Z$ boson properties. In the meantime, it was explored what can be done either using radiative $Z$ boson events during the standard running or by a dedicated running at $\sqrt{s}\approx M_Z$ for a period of 1-3 years, collecting 100~fb$^{-1}$.
The results shown in the main body of this paper now include the radiative return analysis. Here, we present the additional improvement which can be made when a dedicated $Z$ running period is considered. Table ~\ref{tab:eft-global-gigazh} shows the effective Higgs boson couplings with and without Giga-Z running for the ILC and CLIC. It is seen that for ILC$_{250}$ a Giga-Z running improves the $H$ couplings to vector bosons by about 30\%, and for other couplings the improvement is much smaller. For ILC$_{500}$ and CLIC$_{380}$ the impact of dedicated Giga-Z running is low, except for the precision on the TGC parameter $\delta\kappa_\gamma$. 

\begin{table}[ht]
\caption{\label{tab:eft-global-gigazh} Comparison of the effective Higgs coupling sensitivities for the ILC and CLIC with and without a dedicated running at $\sqrt{s}\approx M_Z$.}
\centering
{\small
\begin{tabular}{ |c |c c |c c |c c|}
\toprule
SMEFT$_{\rm ND}$  & \multicolumn{6}{c|}{HL-LHC + }\\
   &ILC$_{250}$   &ILC$_{250}$  &ILC$_{500}$   &ILC$_{500}$  &CLIC$_{380}$   &CLIC$_{380}$  \\
 & & +GigaZ & & +GigaZ & & +GigaZ \\
\midrule
$g_{HZZ}^{\mathrm{eff}}[\%]$   & $0.39$   & $0.31$   & $0.22$   & $0.19$   & $0.5$   & $0.47$  \\
\crowcolorA$g_{HWW}^{\mathrm{eff}}[\%]$   & $0.41$   & $0.32$   & $0.22$   & $0.2$   & $0.5$   & $0.47$  \\
$g_{H \gamma \gamma}^{\mathrm{eff}}[\%]$   & $1.3$   & $1.3$   & $1.2$   & $1.2$   & $1.4$   & $1.4$  \\
\crowcolorA$g_{HZ \gamma}^{\mathrm{eff}}[\%]$   & $9.6$   & $9.4$   & $6.8$   & $6.7$   & $9.7$   & $9.5$  \\
$g_{Hgg}^{\mathrm{eff}}[\%]$   & $1.1$   & $1.1$   & $0.79$   & $0.79$   & $1.3$   & $1.3$  \\
\crowcolorA$g_{Htt}[\%]$   & $3.2$   & $3.1$   & $2.9$   & $2.8$   & $3.2$   & $3.2$  \\
$g_{Hcc}[\%]$   & $1.8$   & $1.8$   & $1.2$   & $1.2$   & $4.$   & $4.$  \\
\crowcolorA$g_{Hbb}[\%]$   & $0.78$   & $0.74$   & $0.52$   & $0.51$   & $0.99$   & $0.98$  \\
$g_{H\tau\tau}[\%]$   & $0.81$   & $0.78$   & $0.59$   & $0.58$   & $1.3$   & $1.3$  \\
\crowcolorA$g_{H\mu\mu}[\%]$   & $4.1$   & $4.1$   & $3.9$   & $3.9$   & $4.4$   & $4.3$  \\
\midrule
$\delta g_{1Z}[\times 10^{2}]$   & $0.091$   & $0.071$   & $0.047$   & $0.031$   & $0.045$   & $0.044$  \\
\crowcolorA$\delta \kappa_{ \gamma}[\times 10^{2}]$   & $0.12$   & $0.087$   & $0.076$   & $0.047$   & $0.079$   & $0.066$  \\
$\lambda_{Z}[\times 10^{2}]$   & $0.042$   & $0.042$   & $0.021$   & $0.021$   & $0.043$   & $0.043$  \\
\bottomrule
\end{tabular}
}
\end{table}

Table ~\ref{tab:eft-global-gigazew} shows the impact of the Giga-Z running on the precision on the effective couplings of the $Z$ boson to fermions. In many cases, the impact is significant, improving the precision by up to a factor of $\sim 4$. Also shown are the results expected for CEPC and FCC-{ee}. In most cases, 
CEPC and FCC-{ee} achieve the highest precision. A notable exception is the top quark coupling which is best constrained by the ILC$_{500}$. 

\begin{table}[ht]
\caption{\label{tab:eft-global-gigazew} 
Comparison of the effective $Z$-boson coupling sensitivities for the ILC and CLIC with and w/o a dedicated running at $\sqrt{s}\approx M_Z$. Also shown are the values for CEPC and FCC-{ee} In all cases, the combination with HL-LHC is shown but the sensitivity is dominated by the $e^+e^-$ collider.
}
\centering
{\scriptsize
\begin{tabular}{ |c |c c |c c |c c| c| c|}
\toprule
SMEFT$_{\rm ND}$  & \multicolumn{8}{c|}{HL-LHC + }\\
   &ILC$_{250}$   &ILC$_{250}$  &ILC$_{500}$   &ILC$_{500}$  &CLIC$_{380}$   &CLIC$_{380}$  & CEPC & FCC-{ee} \\
 & & +GigaZ & & +GigaZ & & +GigaZ & &  \\
\midrule
$g_{L}^{\nu_{e}}[\%]$   & $0.082$   & $0.058$   & $0.048$   & $0.032$   & $0.027$   & $0.024$   & $0.032$   & $0.028$  \\
\crowcolorA$g_{L}^{\nu_{\mu}}[\%]$   & $0.11$   & $0.075$   & $0.088$   & $0.064$   & $0.16$   & $0.11$   & $0.036$   & $0.03$  \\
$g_{L}^{\nu_{\tau}}[\%]$   & $0.12$   & $0.079$   & $0.095$   & $0.071$   & $0.16$   & $0.11$   & $0.038$   & $0.034$  \\
\crowcolorA$g_{L}^{e}[\%]$   & $0.048$   & $0.025$   & $0.037$   & $0.023$   & $0.035$   & $0.021$   & $0.014$   & $0.0073$  \\
$g_{R}^{e}[\%]$   & $0.055$   & $0.028$   & $0.041$   & $0.025$   & $0.047$   & $0.025$   & $0.016$   & $0.0089$  \\
\crowcolorA$g_{L}^{\mu}[\%]$   & $0.072$   & $0.023$   & $0.07$   & $0.022$   & $0.17$   & $0.022$   & $0.031$   & $0.007$  \\
$g_{R}^{\mu}[\%]$   & $0.09$   & $0.025$   & $0.087$   & $0.024$   & $0.24$   & $0.026$   & $0.047$   & $0.0092$  \\
\crowcolorA$g_{L}^{\tau}[\%]$   & $0.076$   & $0.027$   & $0.073$   & $0.026$   & $0.17$   & $0.029$   & $0.016$   & $0.0076$  \\
$g_{R}^{\tau}[\%]$   & $0.094$   & $0.031$   & $0.091$   & $0.03$   & $0.24$   & $0.034$   & $0.018$   & $0.0094$  \\
\crowcolorA$g_{L}^{u=c}[\%]$   & $0.24$   & $0.13$   & $0.23$   & $0.12$   & $0.26$   & $0.11$   & $0.051$   & $0.05$  \\
$g_{R}^{u=c}[\%]$   & $0.35$   & $0.15$   & $0.35$   & $0.15$   & $0.43$   & $0.16$   & $0.08$   & $0.066$  \\
\crowcolorA$g_{L}^{t}[\%]$   & $11.$   & $11.$   & $0.84$   & $0.84$   & $2.4$   & $2.3$   & $11.$   & $1.5$  \\
$g_{R}^{t}[\%]$   & $-$   & $-$   & $2.$   & $2.$   & $6.$   & $6.$   & $-.$   & $3.5$  \\
\crowcolorA$g_{L}^{d=s}[\%]$   & $0.23$   & $0.14$   & $0.21$   & $0.13$   & $0.25$   & $0.14$   & $0.056$   & $0.051$  \\
$g_{R}^{d=s}[\%]$   & $4.8$   & $3.2$   & $3.9$   & $2.5$   & $5.9$   & $3.9$   & $1.4$   & $1.1$  \\
\crowcolorA$g_{L}^{b}[\%]$   & $0.071$   & $0.034$   & $0.068$   & $0.033$   & $0.13$   & $0.041$   & $0.017$   & $0.011$  \\
$g_{R}^{b}[\%]$   & $0.51$   & $0.31$   & $0.51$   & $0.32$   & $3.$   & $0.75$   & $0.29$   & $0.088$  \\
\bottomrule
\end{tabular}
}
\end{table}

It is also interesting to compare the highest energy options closely. This is done in Table~\ref{tab:eft-global-highenergy}. In all cases, it is assumed that the colliders also include a Giga-Z run of 1-3 years~\cite{Fujii:2019zll,gigazclic}. 

\begin{table}[ht]
\caption{\label{tab:eft-global-highenergy} 
Effective Higgs couplings precision for the EFT fit for a selection of colliders at high energy. For the linear colliders it is assumed that 100~fb$^{-1}$ of dedicated running on the Z-pole, corresponding to 1-3 years of data taking, are part of the programme.}
\centering
{\small
\begin{tabular}{ |c| c c c c c c|}
\toprule
SMEFT$_{\rm ND}$  & \multicolumn{6}{c|}{HL-LHC + }\\
 &ILC$_{500}$    &ILC$_{1000}$    &CLIC$_{1500}$   &CLIC$_{3000}$   &FCC-ee/hh   &FCC-ee/eh/hh  \\
\midrule
$g_{HZZ}^{\mathrm{eff}}[\%]$   & $0.19$   & $0.15$   & $0.2$   & $0.16$   & $0.21$   & $0.13$  \\
\crowcolorA$g_{HWW}^{\mathrm{eff}}[\%]$   & $0.2$   & $0.16$   & $0.18$   & $0.15$   & $0.21$   & $0.13$  \\
$g_{H \gamma \gamma}^{\mathrm{eff}}[\%]$   & $1.2$   & $1.1$   & $1.3$   & $1.1$   & $0.38$   & $0.34$  \\
\crowcolorA$g_{HZ \gamma}^{\mathrm{eff}}[\%]$   & $6.7$   & $6.6$   & $4.5$   & $3.6$   & $0.72$   & $0.7$  \\
$g_{Hgg}^{\mathrm{eff}}[\%]$   & $0.79$   & $0.55$   & $0.97$   & $0.75$   & $0.56$   & $0.49$  \\
\crowcolorA$g_{Htt}[\%]$   & $2.8$   & $1.5$   & $2.2$   & $2.1$   & $1.7$   & $1.7$  \\
$g_{Hcc}[\%]$   & $1.2$   & $0.88$   & $1.8$   & $1.4$   & $1.2$   & $0.95$  \\
\crowcolorA$g_{Hbb}[\%]$   & $0.51$   & $0.43$   & $0.44$   & $0.37$   & $0.51$   & $0.44$  \\
$g_{H\tau\tau}[\%]$   & $0.58$   & $0.49$   & $0.92$   & $0.74$   & $0.54$   & $0.46$  \\
\crowcolorA$g_{H\mu\mu}[\%]$   & $3.9$   & $3.5$   & $4.1$   & $3.4$   & $0.46$   & $0.42$  \\
\midrule
$\delta g_{1Z}[\times 10^{2}]$   & $0.031$   & $0.03$   & $0.012$   & $0.0099$   & $0.028$   & $0.018$  \\
\crowcolorA$\delta \kappa_{ \gamma}[\times 10^{2}]$   & $0.047$   & $0.044$   & $0.022$   & $0.018$   & $0.048$   & $0.047$  \\
$\lambda_{Z}[\times 10^{2}]$   & $0.021$   & $0.014$   & $0.0053$   & $0.0018$   & $0.047$   & $0.045$  \\
\bottomrule
\end{tabular}
}
\end{table}

After the Granada meeting, it was also studied what could be achieved with a hadron-hadron collider with $\sqrt{s}=37.5$~TeV and ${\cal L}=15$~ab$^{-1}$, in conjunction with one of the $e^+e^-$ colliders~\cite{lowefcc}. This is shown in Table.~\ref{tab:resultskappa0FCC37p5hhLeptonFit} compared to the nominal FCC-hh in combination with the various $e^+e^-$ colliders. For most coupling parameters the sensitivity of the $37.5$~TeV collider is degraded by about a factor $1.5-2$ w.r.t. the 100~TeV collider, except for $Z\gamma$ where it is a factor of 5. For $\kappa_Z$ and $\kappa_c$ there is no difference as both are very much dominated by the lepton collider sensitivity.

\begin{table}[ht]
\centering
\caption{ \label{tab:resultskappa0FCC37p5hhLeptonFit} Expected relative precision (\%) of the $\kappa$ parameters in the kappa-0-HL scenario  described in
section~\ref{tab:kappa_scenarios} for future lepton colliders combined with the HL-LHC and the FCC-hh$_{37.5}$ (top part) and with HL-LHC and FCC-hh (bottom part). No BSM width is allowed in the fit: both $\mathrm{BR}_{\mathrm{unt}}$ and $\mathrm{BR}_{\mathrm{inv}}$ are set to 0.}
\begin{tabular}{ c | c  | c | c |  c}
\toprule
\multirow{2}{*}{kappa-0-HL} & \multicolumn{4}{c}{HL-LHC+FCC-hh$_{37.5}$+} \\
   &ILC$_{250}$   &CLIC$_{380}$   &CEPC  &FCC-ee$_{365}$  \\
\midrule
$\kappa_{W}[$\%$]$   & $0.94$   & $0.62$   & $0.81$   & $0.38$  \\
\crowcolorA$\kappa_{Z}[\%]$   & $0.21$   & $0.33$   & $0.13$   & $0.14$  \\
$\kappa_{g}[\%]$   & $1.3$   & $1.3$   & $0.97$   & $0.87$  \\
\crowcolorA$\kappa_{ \gamma}[\%]$   & $0.64$   & $0.68$   & $0.62$   & $0.62$  \\
$\kappa_{Z \gamma}[\%]$   & $3.$   & $3.1$   & $2.8$   & $3.$  \\
\crowcolorA$\kappa_{c}[\%]$   & $1.9$   & $3.9$   & $1.9$   & $1.3$  \\
$\kappa_{t}[\%]$   & $1.9$   & $1.9$   & $1.9$   & $1.9$  \\
\crowcolorA$\kappa_{b}[\%]$   & $0.99$   & $0.94$   & $0.81$   & $0.58$  \\
$\kappa_{\mu}[\%]$   & $1.$   & $1.1$   & $1.$   & $1.$  \\
\crowcolorA$\kappa_{\tau}[\%]$   & $0.96$   & $1.2$   & $0.83$   & $0.6$  \\
\bottomrule
\end{tabular}
\begin{tabular}{ c | c  | c | c |  c}
\toprule
\multirow{2}{*}{kappa-0-HL} & \multicolumn{4}{c}{HL-LHC+FCC-hh+} \\
   &ILC$_{250}$   &CLIC$_{380}$   &CEPC  &FCC-ee$_{365}$  \\
\midrule
$\kappa_{W}[\%]$   & $0.37$   & $0.36$   & $0.35$   & $0.27$  \\
\crowcolorA$\kappa_{Z}[\%]$   & $0.19$   & $0.26$   & $0.12$   & $0.13$  \\
$\kappa_{g}[\%]$   & $0.65$   & $0.69$   & $0.55$   & $0.55$  \\
\crowcolorA$\kappa_{\gamma}[\%]$   & $0.31$   & $0.34$   & $0.29$   & $0.29$  \\
$\kappa_{Z\gamma}[\%]$   & $0.71$   & $0.74$   & $0.69$   & $0.7$  \\
\crowcolorA$\kappa_{c}[\%]$   & $1.8$   & $3.8$   & $1.8$   & $1.2$  \\
$\kappa_{t}[\%]$   & $0.96$   & $0.96$   & $0.95$   & $0.95$  \\
\crowcolorA$\kappa_{b}[\%]$   & $0.63$   & $0.68$   & $0.52$   & $0.5$  \\
$\kappa_{\mu}[\%]$   & $0.43$   & $0.47$   & $0.41$   & $0.41$  \\
\crowcolorA$\kappa_{\tau}[\%]$   & $0.61$   & $0.78$   & $0.52$   & $0.49$  \\
\midrule
$\Gamma_{H}[\%]$   & $0.90$   & $0.98$   & $0.74$   & $0.67$  \\
\bottomrule
\end{tabular}
\end{table}

\clearpage
\newpage


\section{Electroweak precision constraints on oblique parameters~\label{Obliquebenchmarks}}

In this section we will focus on the constraints on heavy new physics that can be obtained by precise measurement of the on-shell  $W$ and $Z$ properties. We will focus on universal effects that can be fully encapsulated in the vector boson propagators, with no direct correction to the interaction vertices with fermions. This assumption allows on the one hand  to limit the analysis to a few parameters. One the other hand, it should be noted,   motivated models, like the minimal composite Higgs, often satisfy this assumption to a very good approximation. When considering universal deviations from the SM one must distinguish between the number of on-shell $Z$ and $W$ observables and the  total number of parameters,
which corresponds to the total number of on- and off-shell observables.

Under the assumption of universality, the relevant on-shell observables of $W$ and $Z$ physics reduce to three quantities: the relative normalization of charged and neutral currents (or the $Z$'s axial coupling), and the two relative differences among the three possible definitions of the Weinberg angle (from $\alpha(M_Z), G_F, M_Z$, from the  $Z$'s vector coupling, and from $M_W/M_Z$). These can be nicely encapsulated in the $\epsilon$'s of Altarelli and Barbieri~\cite{Altarelli:1990zd}:
\begin{eqnarray}
\varepsilon_1&=&\Delta \rho ,\nonumber\\
\varepsilon_2&=&\cos^2{\theta_w} \Delta \rho + \frac{\sin^2{\theta_w}}{\cos^2{\theta_w}-\sin^2{\theta_w}} \Delta r_W - 2 s_0^2 \Delta \kappa,\nonumber\\
\varepsilon_3&=& \cos^2{\theta_w} \Delta \rho + (\cos^2{\theta_w}-\sin^2{\theta_w}) \Delta \kappa.
\end{eqnarray}
where we define the weak angle from $\sin^2{\theta_w} \cos^2{\theta_w}\equiv\frac{\pi \alpha(M_Z)}{\sqrt{2}G_F M_Z^2}$ and the $\Delta r_W$, $\Delta \rho$ and $\Delta \kappa$ parameters are defined from the masses and effective vector and axial couplings of the electroweak bosons :
\begin{eqnarray}
%
\frac{M_W^2}{M_Z^2}&=&\frac 12 \sqrt{1+\sqrt{1-\frac{4\pi \alpha(M_Z)}{\sqrt{2}G_F M_Z^2 (1-\Delta r_W)}}},\nonumber\\
%
g_V^f&=&\sqrt{1+\Delta \rho}~\!(T_3^f-2 Q_f (1+\Delta \kappa) \sin^2{\theta_w}),\nonumber\\
g_A^f&=&\sqrt{1+\Delta \rho}~\! T_3^f.
\end{eqnarray}
($T_3^f$ and $Q_f$ are weak isospin and charge of the corresponding fermion.) Notice that $\epsilon_2$ relies on the measurement of the $W$ mass while  $\epsilon_{1,3}$ do not. 

The number of parameters describing universal new physics in the $W, Z$ channel on- and off-shell is instead four\footnote{For a good part of the history of EW precision tests, the community 
has mostly relied on a set of three quantities, $S, T, U$. These are however inadequate in any realistic new physics scenario: they are always  either redundant or incomplete. 
Indeed in technicolor models, it was understood that $U$ is negligible and the set is redundant in that case~\cite{Peskin:1990zt,Holdom:1990tc, Golden:1990ig,Peskin:1991sw}. On the other hand $S, T, U$ are insufficient to describe even the simplest sequential $Z'$ models that fall into the class of universal theories. That the relevant set should be consist of 4 quantities was first realized in~\cite{Grinstein:1991cd} in the context of linearly realized EW symmetry, i.e. SMEFT. The generality of this counting, and therefore its validity also in technicolor/HEFT scenarios, has been clarified in~\cite{Barbieri:2004qk}.}. 
They correspond to the leading effects in a derivative expansion of the vector boson self-energies, $\Pi_{VV^\prime}(q^2)$. More precisely they correspond to the leading effects in each independent channel (with the channels characterized by the relevant quantum numbers: electric charge, electroweak and custodial symmetry). Considering the vector boson self-energies, using the constraints from $U(1)_{EM}$ unbroken gauge invariance (massless photon), and subtracting the quantities that play the role of SM inputs ($v$, $g$, $g'$), one is left with these four leading quantities 
%
\begin{eqnarray}
\hat{S}&=&g^2 \Pi_{W_3 B}^\prime (0),\nonumber\\
\hat{T}&=&\frac{g^2}{M_W^2}\left( \Pi_{W_3 W_3}(0)-\Pi_{W^+ W^-}(0)\right),\nonumber\\
W&=&\frac 12 g^2 M_W^2 \Pi_{W_3 W_3}^{\prime\prime} (0),\nonumber\\
Y&=&\frac 12 g^{\prime~2} M_W^2 \Pi_{BB}^{\prime\prime} (0),
\end{eqnarray}
which can be mapped to four linear combinations of operators in the SMEFT Lagrangian. Finally, considering their effect on $W, Z$ on-shell propagation, one finds, writing $\epsilon_i=\epsilon_{i,{\rm SM}}+\delta \epsilon_i$,
%
\begin{eqnarray}
\delta \varepsilon_1&=&\hat{T}-W -Y 
\tan^2 \theta_w,\nonumber\\
%
\delta \varepsilon_2&=&-W,\nonumber\\
%
\delta \varepsilon_3&=&\hat{S} -W -Y.
\end{eqnarray}

Strongly coupled models come with a parametric enhancement of $\hat{S}, \hat{T}$ over $W, Y$, such that in the class of models, one can simplify further the analysis of EW data and perform a two-dimensional fit.

\begin{table}[t]
\caption{\label{tab:ST-global} 
Comparison of the sensitivity at 68\% probability to new physics contributions to EWPO in the form of the oblique $S$ and $T$ parameters, under different assumptions for the SM theory uncertainties. We express the results in terms of the usually normalised parameters: $S=4 \sin^2 \theta_w \hat{S}/\alpha$ and $T=\hat{T}/\alpha$. 
}
\centering
{\scriptsize
\begin{tabular}{ |c | c | c | c c | c c | c c |}
\toprule
  &  & HL-LHC & \multicolumn{6}{c|}{HL-LHC+}\\
   &   &  &CLIC$_{380}$   &CLIC$_{380}$  &ILC$_{250}$   &ILC$_{250}$  &CEPC   &FCC-ee  \\
   &   &   &   & (+GigaZ)   &   & (+GigaZ)   &   &  \\
\midrule
$S$& Full Th$_\mt{Intr}$ Unc.   & $0.053$   & $0.032$   & $0.013$   & $0.015$   & $0.012$   & $0.01$   & $0.0079$  \\
 & No Th$_\mt{Intr}$ Unc.   & $0.053$   & $0.032$   & $0.011$   & $0.012$   & $0.009$   & $0.0068$   & $0.0038$  \\
 & No Th$_\mt{Par+Intr}$ Unc.   & $0.052$   & $0.031$   & $0.0091$   & $0.011$   & $0.0067$   & $0.0031$   & $0.0013$  \\
\crowcolorA$T$& Full Th$_\mt{Intr}$ Unc.   & $0.041$   & $0.023$   & $0.013$   & $0.015$   & $0.014$   & $0.0094$   & $0.0058$  \\
\crowcolorA & No Th$_\mt{Intr}$ Unc.   & $0.041$   & $0.023$   & $0.012$   & $0.014$   & $0.013$   & $0.0072$   & $0.0022$  \\
\crowcolorA & No Th$_\mt{Par+Intr}$ Unc.   & $0.039$   & $0.022$   & $0.01$   & $0.011$   & $0.0091$   & $0.0041$   & $0.0019$  \\
\bottomrule
\end{tabular}
}
\end{table}

\begin{figure}[t]
\centering
\begin{tabular}{c c}
\includegraphics[width=0.47\linewidth]{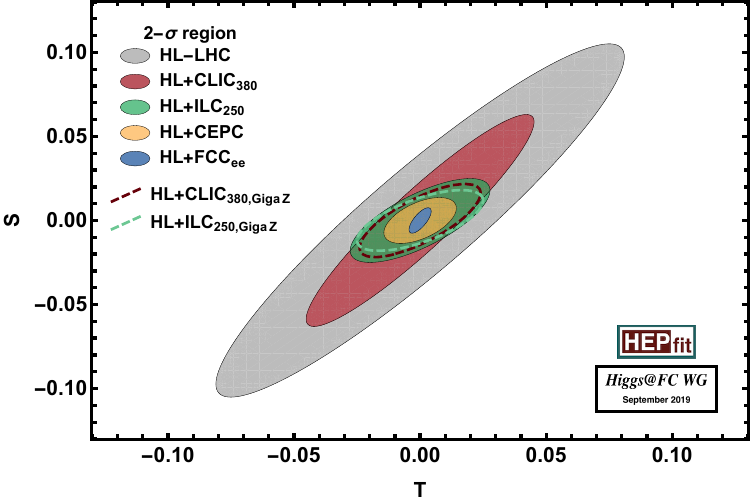}
&
\includegraphics[width=0.47\linewidth]{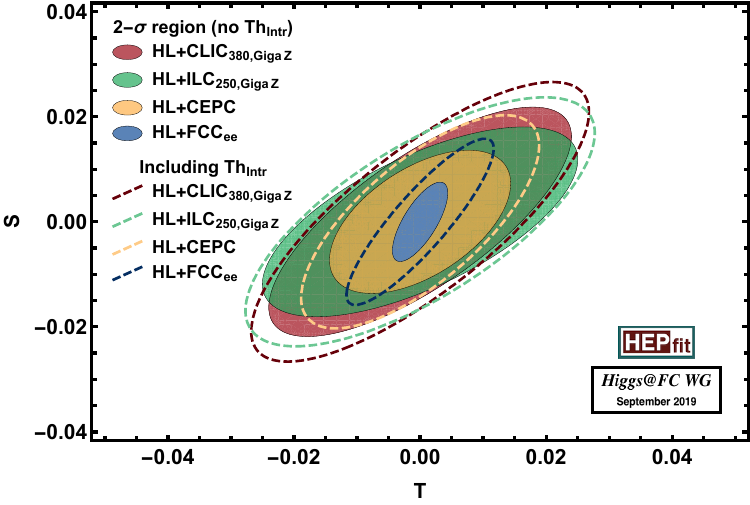}
\end{tabular}
\caption{\label{fig:ST_2sigma}
(Left) 2-$\sigma$ regions in the $S-T$ plane at the different future colliders, combined with the HL-LHC (including also the LEP/SLD EWPO programme). We express the results in terms of the usually normalised parameters: $S=4 \sin^2 \theta_w \hat{S}/\alpha$ and $T=\hat{T}/\alpha$. 
The results include the future projected parametric uncertainties in the SM predictions of the different EWPO, but not the intrinsic ones. (Right) The same illustrating the impact of neglecting such intrinsic theory errors. For each project (including the Giga-Z option for linear colliders) the solid regions show the results in the left panel, to be compared with the regions bounded by the dashed lines, which include the full projected theory uncertainty.
}
\end{figure}

The results of this fit setup are presented in Table~\ref{tab:ST-global} and Figure~\ref{fig:ST_2sigma}, for the different future lepton-collider options, where the largest improvement in terms of measurements of the EW precision observables (EWPO) is expected.
In the table and figures we also show the impact of the SM theory uncertainties in the results. The results are presented assuming the projected future improvements in SM theory calculations ({\it Full Th$_\mt{Intr}$ Unc.}), neglecting the intrinsic theory uncertainties associated to such calculations ({\it No Th$_\mt{Intr}$ Unc.}) and, finally, 
also assuming that parametric uncertainties become subdominant ({\it No Th$_\mt{Par+Intr}$ Unc.}). Since several of the SM EW inputs are to be measured at the future collider under consideration, the latter scenario goes beyond the physics potential of these machines. This scenario is presented only to illustrate whether the precision of the measurements of such inputs can become a limiting factor in terms of the reach of $\hat{S}$ and $\hat{T}$. This seems to be the case for the circular colliders and, to a less extent, the linear collider Giga-Z options.

\clearpage
\newpage

\section{Consistency of electroweak precision data~\label{smcheck}}
\begin{figure}[ht]
\centering
\begin{tabular}{c c c}
\hspace{-0.5cm}\includegraphics[width=0.32\linewidth]{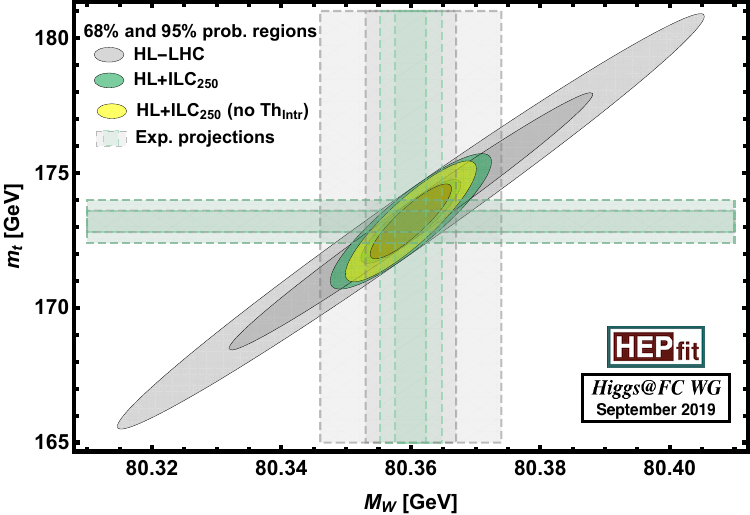}
&
\includegraphics[width=0.32\linewidth]{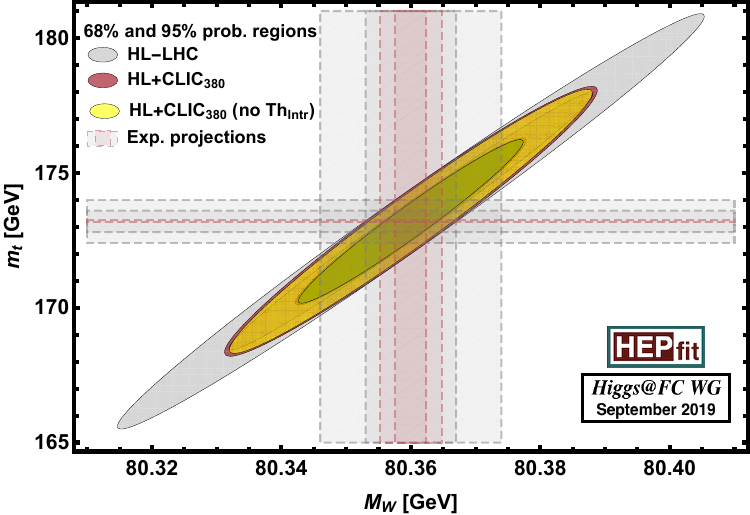}
&
\includegraphics[width=0.32\linewidth]{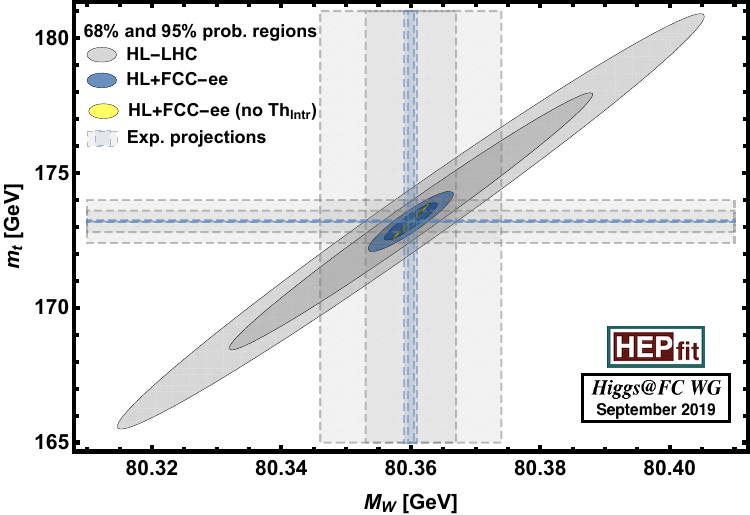}
\\[0.25cm]
\hspace{-0.5cm}\includegraphics[width=0.32\linewidth]{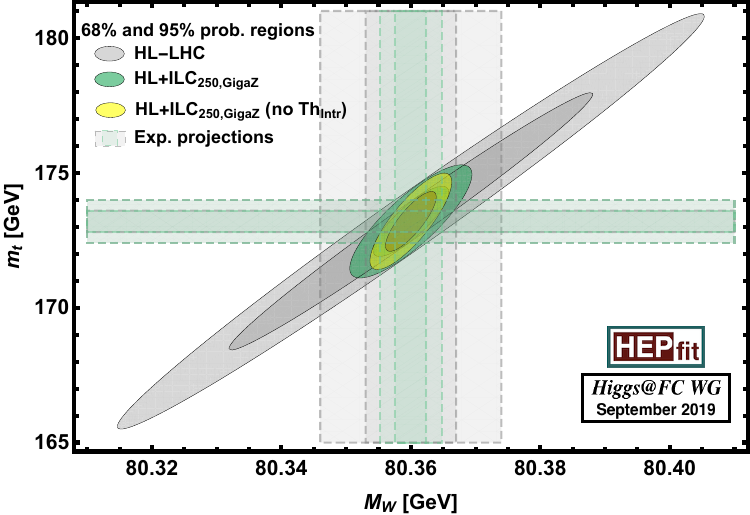}
&
\includegraphics[width=0.32\linewidth]{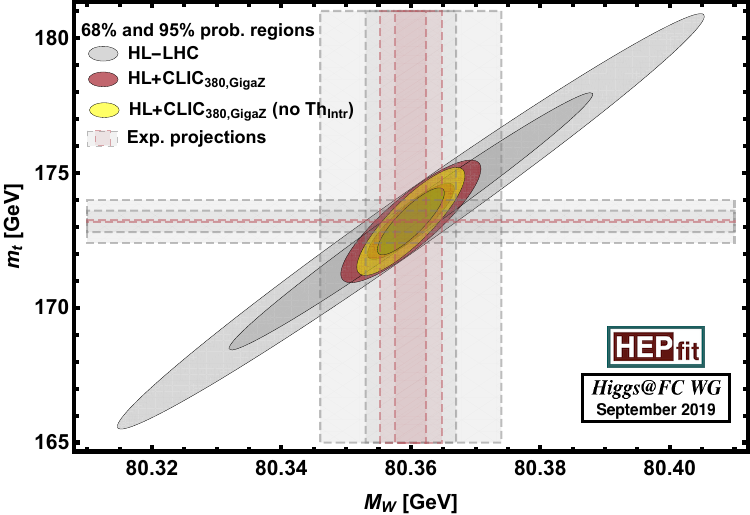}
&
\includegraphics[width=0.32\linewidth]{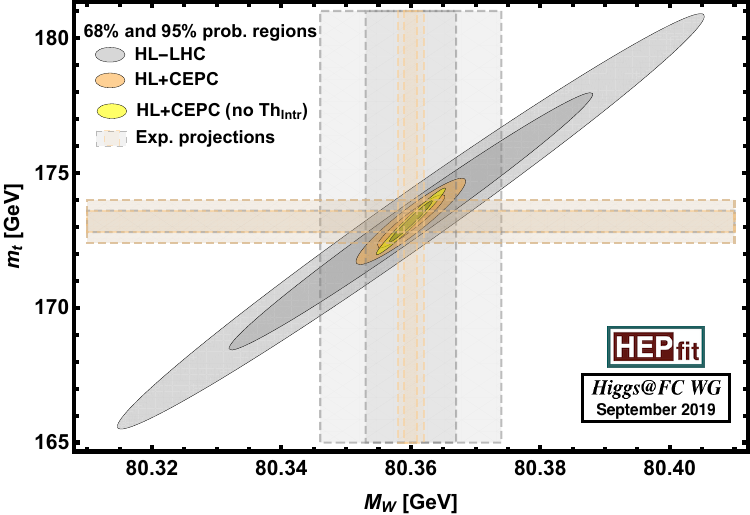}
\end{tabular}
\caption{\label{fig:ewkprec}
Constraints on $m_W$ and $m_\textrm{top}$ from direct measurements (horizontal and vertical lines) and indirect constraints (ellipses). In all cases the constraints from current data plus HL-LHC are compared to the ones expected for the $e^+e^-$ collider. For ILC and CLIC the result is shown without (top row) and with a Giga-Z (bottom row) run.
}
\end{figure}

Before the discovery of a Higgs boson, the consistency of the SM has often been illustrated by comparing the direct measurement of $m_W$ and $m_\textrm{top}$ with the indirect constraints derived from precision measurement at the $Z$-pole and at low-energy experiments. Fig.~\ref{fig:ewkprec} for the future $e^+e^-$ colliders. 

\vspace{1.5cm}

\clearpage
\newpage
\section{Improvement with respect to HL-LHC }

\begin{figure}[ht]
\centering
\includegraphics[width=1.0\linewidth]{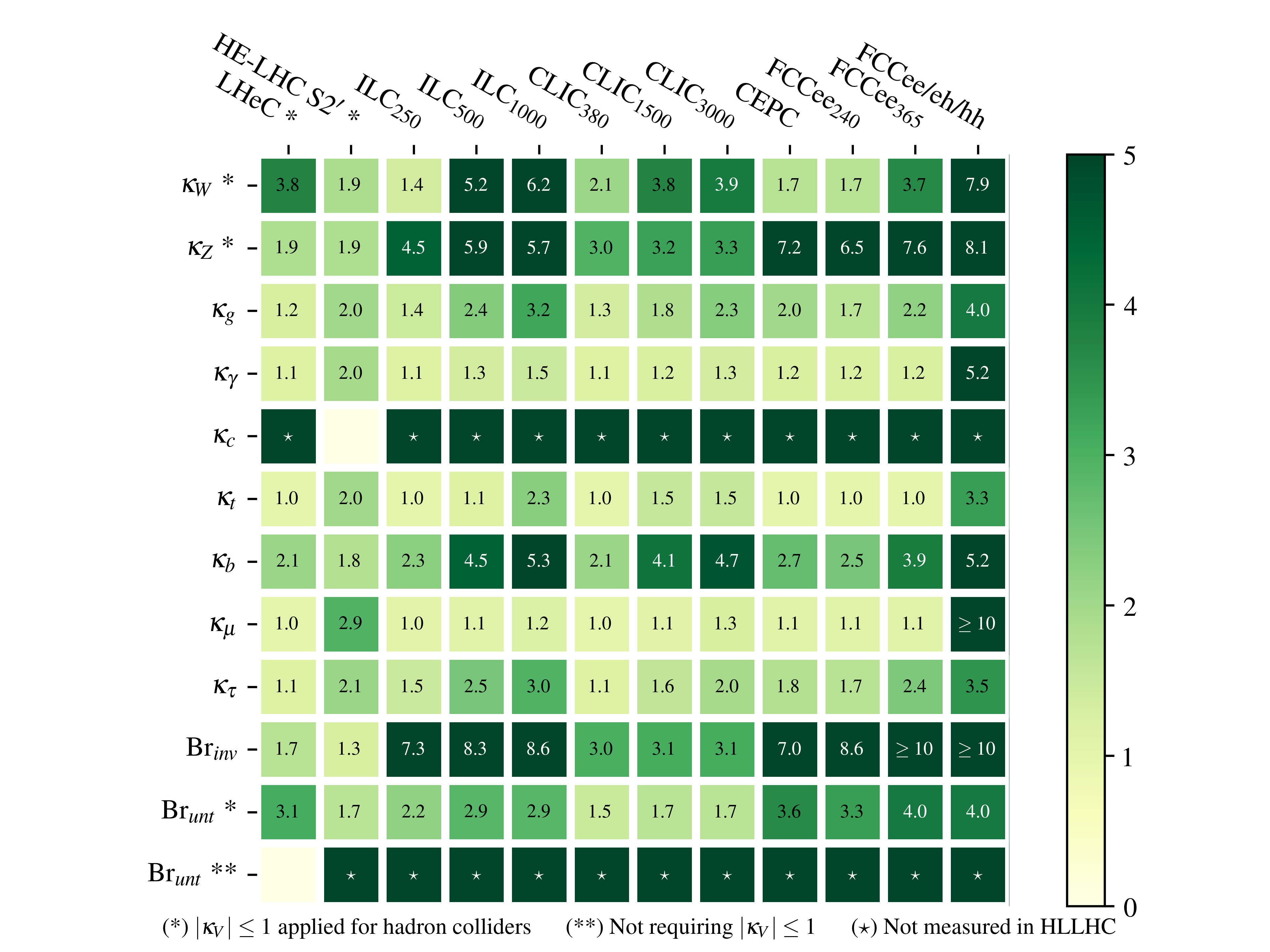}
\caption{\label{fig:improvement_heatmap_Kappa}Graphic comparison of the improvement with respect to HL-LHC in the Kappa-3 framework. }
\end{figure}

\begin{figure}[ht]
\centering
\includegraphics[width=1.0\linewidth]{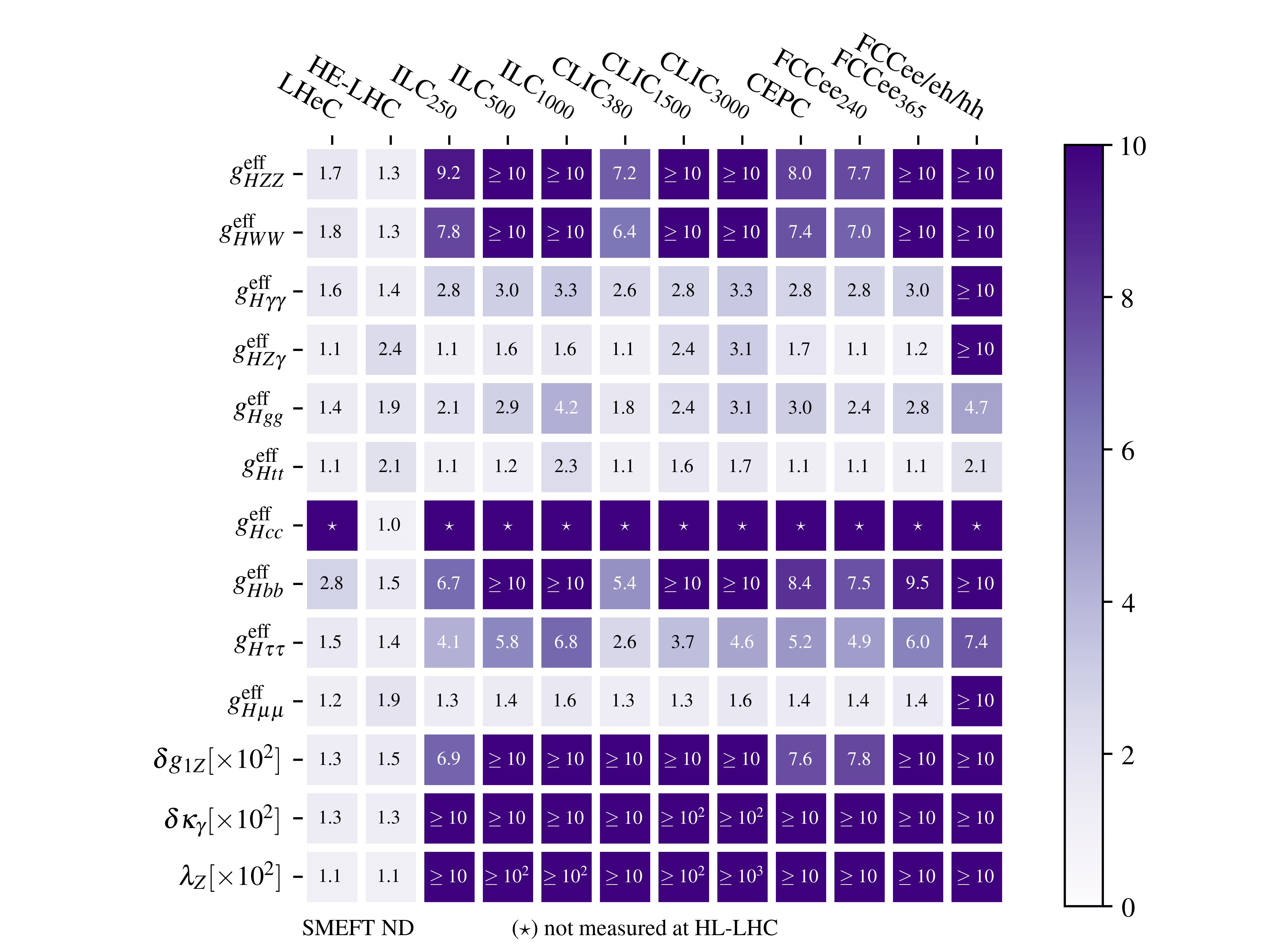}
\caption{\label{fig:improvement_heatmap_SMEFT}Graphic comparison of the improvement with respect to HL-LHC in the  SMEFT-ND framework. }
\end{figure}

Figures~\ref{fig:improvement_heatmap_Kappa} and ~\ref{fig:improvement_heatmap_SMEFT} give a graphic comparison of the improvement with respect to HL-LHC in the Kappa-3 and SMEFT-ND frameworks. This improvement is shown as the ratio of the precision at the HL-LHC over the precision at the future collider, with more darker colors corresponding to larger improvement factors. The kappa-3 result shows large improvements, up to an order of magnitude, for all future ee colliders for the measurement of the couplings to Z, W and b and the limits on the invisible branching ratio, and an 'infinite' improvement in the case of the coupling to charm, immeasurable at the HL-LHC. Rare, statistically dominated, couplings, as well as the coupling to the top quark are shown to be markedly improved with respect to HL-LHC only with FCC-hh. The more complete SMEFT-ND fit highlights more clearly the improvement in precision, with improvements of the order of an order of magnitude in the measurement of Z, W and b couplings for all future ee colliders. The aTGC results show an even more dramatic improvement, with factors over 100 and 1000 for the last stages of the linear colliders.

\providecommand{\href}[2]{#2}\begingroup\raggedright\endgroup

\end{document}